\documentclass{jkas0}
\usepackage{amsmath}

\def\beginpage{21} 
\def\received{August 25, 2014} 
\def\accepted{December 15, 2014} 
\date{Received \received ; accepted \accepted}

\usepackage{flushend}

\newcommand{\farcs}{\mbox{\ensuremath{.\!\!^{\prime\prime}}}}
\newcommand{\arcs}{\mbox{\ensuremath{^{\prime\prime}}}}

\newcommand{\mnras}{MNRAS}
\newcommand{\aap}{A\&A}
\newcommand{\apj}{ApJ}
\newcommand{\aaps}{A\&A Suppl.}
\newcommand{\aj}{AJ}
\newcommand{\nat}{Nature}
\newcommand{\apjs}{ApJS}
\newcommand{\apjl}{ApJL}
\newcommand{\pasj}{PASJ}
\newcommand{\prd}{PRD}
\newcommand{\physrep}{PRep}

\title{Massive Structures of Galaxies at High Redshifts in the Great Observatories Origins Deep Survey Fields}

\author[1,2]{Eugene Kang}
\author[1]{Myungshin Im}

\affil[1]{CEOU, Astronomy Program, Department of Physics and Astronomy, Seoul National University, 1 Gwanak-ro, Gwanak-gu, Seoul 151-742, Korea; \email{mim@astro.snu.ac.kr} }
\affil[2]{Korea Educational Broadcasting System, 2748, Nambusunhwanro, Gangnam-Gu, Seoul 135-854, Korea}

\def\corrauthor{M. Im}
\def\runningauthor {Kang \& Im}
\def\runningtitle{Massive Structures of Galaxies at High Redshifts in GOODS Fields}
\def\keywords{cosmology: observations --- galaxies: evolution --- galaxies: clusters --- galaxies: high-redshift}

\def\abstracttext{
If the Universe is dominated by cold dark matter and dark energy as in the 
currently popular $\Lambda$CDM cosmology, it is expected that large scale structures form gradually, with galaxy clusters of mass $M \gtrsim 10^{14}\, M_{\odot}$ appearing at around 6 Gyrs after the Big Bang ($z \sim 1$).
 Here, we report the discovery of 59 massive structures of galaxies with masses
 greater than a few times $10^{13} M_{\odot}$
 at redshifts between $z=0.6$ and $4.5$ in the Great Observatories Origins Deep Survey fields.
  The massive structures are identified by running top-hat filters on the two dimensional spatial distribution of magnitude-limited samples of galaxies
 using a combination of spectroscopic and photometric redshifts.
 We analyze the Millennium simulation data in a similar way to the analysis of the observational data in order to test the $\Lambda$CDM cosmology.
  We find that there are too many massive structures ($M > 7 \times 10^{13} M_{\odot}$) observed at $z > 2$ in comparison with the simulation predictions 
  by a factor of a few, giving a probability of $< 1/2500$ of the observed data being consistent with the simulation.
  Our result suggests that massive structures have emerged early, but the reason for the discrepancy with the simulation
  is unclear. It could be due to the limitation of the simulation such as
  the lack of key,
  unrecognized ingredients (strong non-Gaussianity or other baryonic physics),
  or simply a difficulty in the halo mass estimation from observation,
  or a fundamental problem of the $\Lambda$CDM cosmology.
  On the other hand, the over-abundance of massive structures at high redshifts does not favor heavy neutrino mass of $\sim 0.3$ eV or larger, 
  as heavy neutrinos make the discrepancy between 
  the observation and the simulation more pronounced by a factor of 
  3 or more.}

\begin{document}
\jkashead


\section{Introduction\label{sec:intro}}

Currently, the energy density of the Universe is believed to be dominated by dark energy (73\%), and cold dark matter (23\%), based on various pieces of observational evidence  (\citealt{a1,chiba97, a2, a3, chae02, a5,a4,35, a7}).
 In such a Universe, called the $\Lambda$CDM cosmology, the formation of galaxies and large scale structures of galaxies
proceeds from less massive objects to more massive objects through hierarchical mergers.
  Thus,
  based on the $\Lambda$CDM model, it is expected that massive galaxies and most groups and clusters of galaxies appear relatively late, at $z<1$
  (e.g., \citealt{allen11, kravstov12, park12, 33, im02}).
 Analytical calculations of halo mass functions and numerical simulations of galaxy formation support this expectation.
  Therefore, the evolution of the number density of massive structures can serve as a useful test of the $\Lambda$CDM cosmology (\citealt{a9, a10}).
 Furthermore, the epoch of the emergence of massive structures depends on other physical conditions that are poorly constrained so far such as the initial density fluctuations (\citealt{r1}) and the neutrino mass (Bond et al. 1980; reviewed in Lesgourgues \& Pastor 2006), and hence its study can provide new insights on these physical conditions as well.

  Current searches for high redshift clusters are unveiling massive clusters and proto-clusters at high redshifts, making it possible to investigate the evolution of massive structures of galaxies throughout the age of the Universe and to test the predictions of the $\Lambda$CDM models.
  Various cluster-search techniques exist, such as searches using extended X-ray sources, red-sequence galaxies, the Sunyaev-Zeldovich (SZ) effect, and narrow-band imaging around high redshift radio galaxies, AGNs and Lyman break galaxies (LBGs).
 The cluster searches using X-ray emissions are based on the fact that
 a massive, gravitationally bound object can be identified as an extended X-ray source since
 deep gravitational potential wells of galaxy clusters provide the necessary conditions
 for the X-ray emitting hot gas to exist.
 Spectroscopic observation of galaxies associated with extended
 X-ray sources has revealed clusters of galaxies out to $z \sim 2$  (e.g., Stanford et al. 2006; Hilton et al. 2007; Fassbender et al. 2011).
  The red-sequence technique, on the other hand, utilizes the fact that galaxies (mostly early-type) in a cluster form a tight sequence in the color-magnitude parameter space. If early-type galaxies in clusters formed the bulk of their stars at a similar epoch at high redshift ($z > 2$), one would expect to find a red sequence in high redshift clusters too. Therefore, one can identify cluster/proto-clusters at high redshift by searching for areas that are populated by red galaxies with a certain color that is expected for passively evolving galaxies at high redshift. Groups/clusters/proto-clusters of galaxies are identified in this way out to $z \sim 2$  (\citealt{gy1,mc1,go1,mu1,bi1,a12}), although in most cases, the identified structures are at $z < 1$ (\citealt{mu08,gi1,li1}).
  Going out to higher redshifts, radio galaxies or quasars are being used as signposts for massive structures. Radio galaxies at low redshift are often massive early-type galaxies that are in cluster regions, and if a similar trend holds at high redshift, one can expect to see massive structures around radio galaxies. A similar argument applies to quasars too. Overdense areas are searched within a certain projected radius around high redshift radio galaxies or quasars by looking for excessive number of extremely red objects (EROs), BzK galaxies, LBGs, or emission line galaxies through optical/NIR broad band imaging or narrow-band imaging centered at the wavelength where emission lines (e.g., Ly$\alpha$, H$\alpha$) show up at the radio galaxy's redshift.  Such studies reveal possible large scale structures as far as $z \sim 6$  (\citealt{lef96,kur00,pen00,wol03,kaj06,kod07,ven07,doh10,cap11,kui11,tos12}).
 Recently, the Sunyaev-Zeldovich effect has been utilized to uncover clusters.
 These searches are now revealing clusters
 out to $z \sim 1.9$, but mostly at $z < 1$, with reliable estimates on cluster mass
 (Foley et al. 2011; Sehgal et al. 2011; Benson et al. 2013; Williamson et al. 2011; Menanteau et al. 2012; Stalder et al. 2013; Bayliss et al. 2014).

 Using the above techniques, a handful of massive clusters of galaxies have been identified at $z\sim1.5$, and as several previous works emphasize, the implication of the existence of massive clusters at high redshift is quite tantalizing. While the $\Lambda$CDM models can successfully explain the abundance of galaxy clusters at $z < 1$ (e.g., Benson et al. 2013; Williamson et al. 2011)
 the situation is more controversial at higher redshift.
   Several studies suggest that there are too many massive clusters at $z \sim 1.5$ compared to theoretical expectations based on the $\Lambda$CDM cosmology (\citealt{a11, a12, a13, a14, holz12, gonzalez12, menanteau12}). This does not necessarily mean the failure of the $\Lambda$CDM model, but could reflect
   either our poor understanding of the physics involved in the growth of massive structures and the galaxies within them, or limitations of current observational studies.  For example, the studies disfavoring the $\Lambda$CDM cosmology using clusters mainly focus on a small number of the most massive structures. Thus, they could be subject to bias from small number statistics (\citealt{a15}). Other explanations exist which try to match
   the over-abundance of massive clusters at high redshift using exotic forms of dark energy
  (\citealt{baldi12, carlesi12}).
 On the other hand, different studies suggest that there is no significant lack of high redshift massive clusters (\citealt{cap11,co1,williamson11,stalder13, bayliss14}), and that the discovery of massive high redshift clusters so far is consistent with
 the $\Lambda$CDM cosmology (Mortonson et al. 2011; Hotchkiss 2011; Waizmann et al. 2012, 2013; Harrison et al. 2012, 2013).
Extending these kinds of studies to higher redshifts can place stronger constraints on the $\Lambda$CDM cosmology, but there is a general lack of unbiased searches for high redshift clusters/proto-clusters at $z > 2$. Searches based on radio galaxies or quasars could miss massive structures where radio galaxies or quasars are not found.

\begin{table*}[t!]
\caption{Summary of the multiwavelength imaging data}
\centering
\begin{tabular}{lcccccc}
\toprule
 Telescope/  & Filter & $\lambda_c^{\rm a}$ & Pixel scale     & Limiting mag.   & Field &  Reference \\
 Instrument  &        &  (micron)             & ($\arcs$/pix) & (5$\sigma$, AB) &       &   \\
\midrule
    VLT/VIMOS & $U$ & 0.39 & 0.20 & 28.0 & South & b \\
    KPNO/MOSAIC & $U$ & 0.37 & 0.30 & 27.1 & North & c \\
    HST/ACS & $B_{435}$ & 0.43 & 0.03 & 26.5 & Both & d \\
    {} & $V_{606}$ & 0.60 & 0.03 & 26.7 & {} & {} \\
    {} & $i_{775}$ & 0.77 & 0.03 & 26.2 & {} & {} \\
    {} & $z_{850}$ & 0.91 & 0.03 & 26.9 & {} & {} \\
    VLT/ISAAC & $J$ & 1.24 & 0.15 & 25.2 & South & e \\
    {} & $H$ & 1.65 & 0.15 & 24.4 & {} & {}\\
    {} & $K_s$ & 2.17 & 0.15 & 24.4 & {} & {}\\
    CFHT/WIRCam & $J$ & 1.25 & 0.30 & 24.6 (v2.0) & North & f \\
    {} & $K_s$ & 2.15 & 0.30 & 24.0 (v2.0) & {} \\
    SPITZER/IRAC  & ch1 & 3.56 & 0.60 & 25.0 & Both & g \\
    {}  & ch2 & 4.51 & 0.60 & 24.6 & {} & {} \\
    {}  & ch3 & 5.76 & 0.60 & 22.7 & {} & {} \\
    {}  & ch4 & 7.96 & 0.60 & 22.6 & {} & {} \\
\bottomrule
\end{tabular}
\tabnote{
  a: Filter central wavelength; b: Capak et al. (2004); c: Nonino et al. (2009);
  d: Giavalisco et al. (2004a) and references therein.; \\
 e: Retzlaff et al. (2010); f: Wang et al. (2010);
  g: M. Dickinson et al. in preparation, R. Chary (private communication)}
\label{t1}
\end{table*}

  One promising way to uncover high redshift massive structures in an unbiased way is to utilize spectroscopic/photometric redshift information from multi-wavelength data (e.g., Kang \& Im 2009; Cucciati et al. 2014). Mixing spectroscopic redshift information with photometric redshifts has disadvantages, such as the dilution of signals from over-dense areas and the confusion of foreground and background structures into a single structure. However, such disadvantages can be overcome if one also applies a similar observational conditions to simulation data in order to interpret the observational results.

 In order to place constraints on the formation of large scale structures at high redshift and see how well $\Lambda$CDM models can reproduce the observed trend, we embarked on a blind search of massive structures of galaxies (hereafter, MSGs) out to $z\sim 4.5$, extending the redshift range of MSG searches well beyond the redshift limit of existing works ($z = 2$). Here, we use the term ``massive structures of galaxies" instead of terms such as clusters, massive halos, or proto-clusters that are commonly used in the literature, since it is difficult to separate each halo or cluster when they are undergoing hierarchical build-up. The search is performed on the Great Observatories Origins Deep Survey (GOODS) fields, which include two fields, each covering 160 arcmin$^2$. These fields offer a wealth of ultra-deep, multi-wavelength imaging and spectroscopic datasets that are necessary to identify and physically characterize distant objects. The area coverage is large enough to mitigate the cosmic variance problem, if MSGs are searched over a sufficiently large depths along the redshift direction ($\sim$ a few hundred Mpc). The GOODS-South is the field where we previously identified an overdensity of galaxies at $z \sim 3.7$ (Kang \& Im 2009), by using photometric and spectroscopic redshifts derived from the multi-wavelength data ranging from $U-$band through 8 $\mu$m band of the Spitzer. Our work basically builds on the previous work of Kang \& Im (2009), extending both the redshift range and the mass range of the search for MSGs. We will derive the number density of MSGs out to $z \sim 4.5$, and compare them to the Millennium Simulation (Springel et al. 2005) where we impose the same selection criteria as those used in the analysis of the observed data. The use of the simulation data is essential, to minimize selection effects associated with the MSG identification process. In the analysis of the observed data, we will adopt the cosmological parameter values identical to those adopted by the Millennium Simulation, i.e., $\Omega_{m} = \Omega_{dm} + \Omega_{b} = 0.25$,
 $\Omega_{b} = 0.045$, $h = 0.73$, and $\Omega_{\Lambda} = 0.75$
(\citealt{col01,spe03,sel05}).
Note that magnitudes are in AB units throughout the paper.

\section{Data}

\subsection{Imaging Data and Catalog}

 The GOODS fields cover two separate regions in the northern and the southern hemispheres centered on
 $12^{\mathrm{h}}36^{\mathrm{m}}55^{\mathrm{s}}$,
 +$62^{\mathrm{\circ}}14^{\mathrm{m}}15^{\mathrm{s}}$ (J2000) of the Hubble Deep Field-North (HDF-N)
  and $3^{\mathrm{h}}32^{\mathrm{m}}30^{\mathrm{s}}$,
  -$27^{\mathrm{\circ}}48^{\mathrm{m}}20^{\mathrm{s}}$ (J2000)
   of the Chandra Deep Field-South (CDF-S), respectively. Each field provides an area coverage of 160 arcmin$^2$.
 A wealth of multi-wavelength imaging data is available in the GOODS fields, and we utilize these for our study.

 For the optical imaging data, we take the version 1.1 of the $HST$ Advanced Camera for Surveys (ACS) multi-band source catalog
 (F435W, F606W, F775W, F850LP) released by the GOODS team (\citealt{2}).
The HST data are augmented by ground-based, deep, $U$-band data taken with VIMOS on the VLT with a depth of 28.0 mag at $5 \sigma$ within a 1$^{\prime\prime}$ radius aperture (\citealt{1}), or  with the MOSAIC prime focus camera of the Kitt Peak National Observatory (KPNO) 4m telescope to the depth of $U=27.1$ mag (\citealt{cap04}).

 For the GOODS-South area, we use the publicly available deep $J$-,$H$- and $K_s$-band images (\citealt{3}). These images were taken by the ESO-GOODS team using Infrared Spectrometer And Array Camera (ISAAC) on the Very Large Telescope (VLT).  The depth of the VLT/ISAAC data reaches to $J=25.2$ mag, $H$ and $K_s=24.4$ mags at $5 \sigma$.
 We use the version 2.0 released images which are made of 24 fields in the $J$-, $H$-bands and 26 fields in $K_s$-band.

 For the GOODS-North area, we use the NIR source catalog ($J$, $K_s$; e.g., Wang et al. 2010)  that is based on images taken by the
 Wide Field Infrared Camera (WIRCam) on the Canada-France-Hawaii Telescope (CFHT).
The depth of the CFHT/WIRCam data reaches to $J=24.6$ mag
and $K_s=24.0$ mag at $5 \sigma$, which are comparable to the depths of the GOODS-South NIR images.

 The imaging data at 3.6, 4,5, 5.8, and 8.0 $\mu$m are taken by the $Spitzer$ Infrared Array Camera (IRAC), and we use the photometry in the catalog created by R. Chary (private communication) in both the northern and the southern fields.
 Table \ref{t1} summarizes the available multi-wavelength imaging data, along with their depths and angular resolutions.

 To derive photometry of objects in $U$, $J$, $H$, and $K_{s}$,
we ran SExtractor (\citealt{bar96}).
 The source detection threshold was set to a minimum contiguous area of 16 pixels, with each pixel value above $1.0\sigma$. For deblending parameter, we use 0.0075 of minimum contrast, and the background mesh size is set to 128 pixels.
  We used $K_{s}$-band detections as a reference, and derived photometry in the other bands using SExtractor in dual-mode at the $K_{s}$-band detection position.
 Auto-magnitudes, generally representing total magnitudes of sources, are used for the photometry, since the multi-wavelength images have a wide range of spatial resolutions.

  The cross-matching of the objects in the multi-wavelength data is done by matching $U$, $K_{s}$ and IRAC-band catalogs against the objects in the HST ACS catalog using 0$\farcs$7, 0$\farcs$7 and 1$\farcs$5 matching radii, respectively.
 The matching radius is defined as the radius where the ACS sources start to be multiply matched beyond $\sim 50$\% of cases.
 We find that our adopted cross-match radius provides counterparts in 98\% and 87\% of the $K_{s}$-band limited objects against the ACS-band and IRAC-band sources, respectively.
 Multiple matches are examined visually on the image to determine which of the multiple matches should be chosen as a good match. The objects that have the smallest positional difference between the different wavelength images are generally chosen to be good matches. In some cases, multiple objects in an ACS image are blended together in $K_{s}$ or IRAC images, making it impossible to construct a reliable spectral energy distribution (SED). Such cases are excluded from the analysis, which
 represent 4\% and 5\% of the $K_s$-band and IRAC-band detected objects, respectively, for a $K$-band limited sample at $K_s < 24$ mag.

\subsection{Spectroscopic Redshifts}

\begin{figure}[t!]
\centering
\includegraphics[width=82mm]{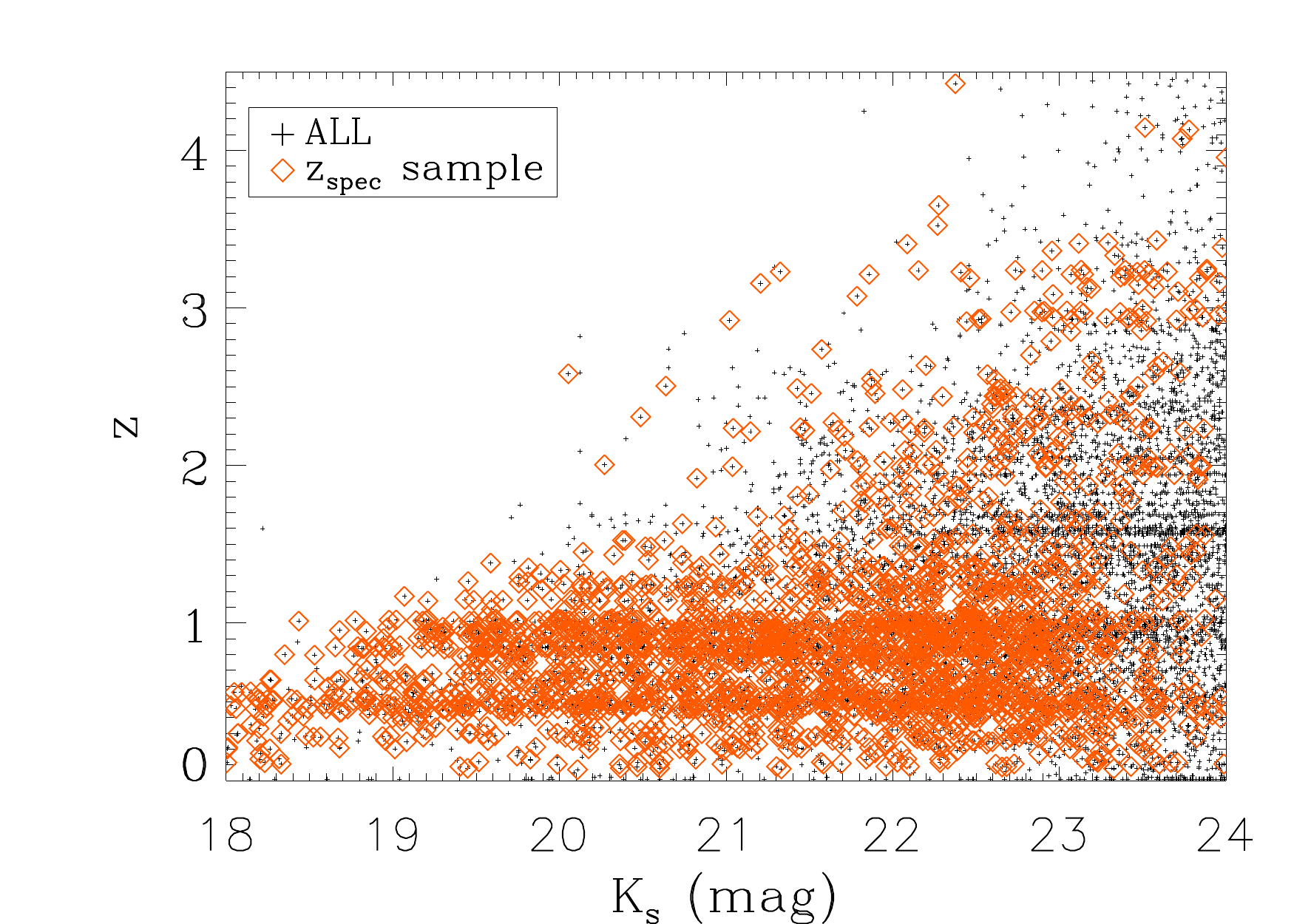} \\
\includegraphics[width=82mm]{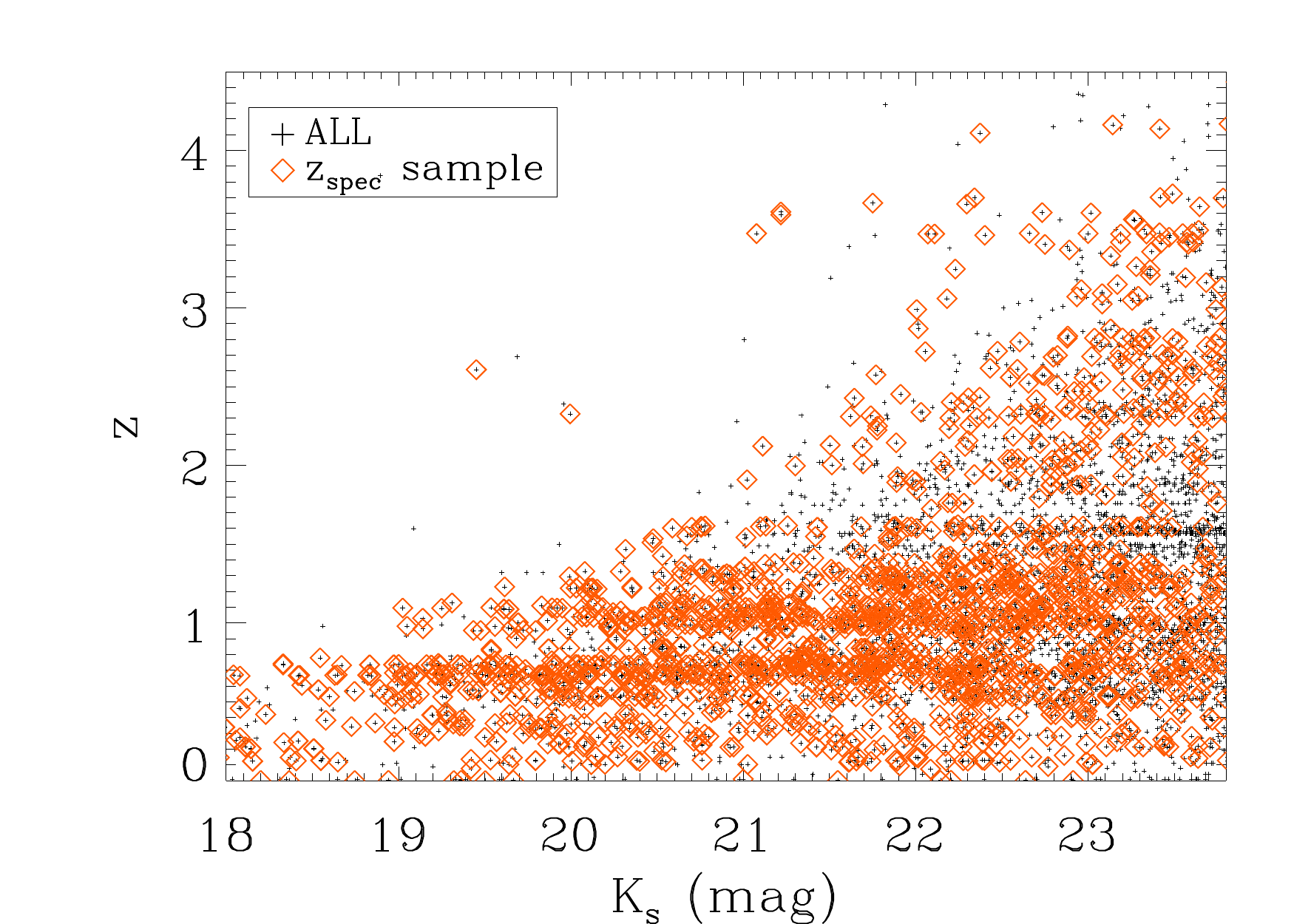}
\caption{The $K_s$-band magnitude versus spectroscopic redshift (crosses
enclosed by red diamonds) and  photometric redshifts (crosses)
for the GOODS-Norh (top) and -South (bottom).
}
\label{f01}
\end{figure}

\begin{figure}[t!]
\centering
\includegraphics[width=82mm]{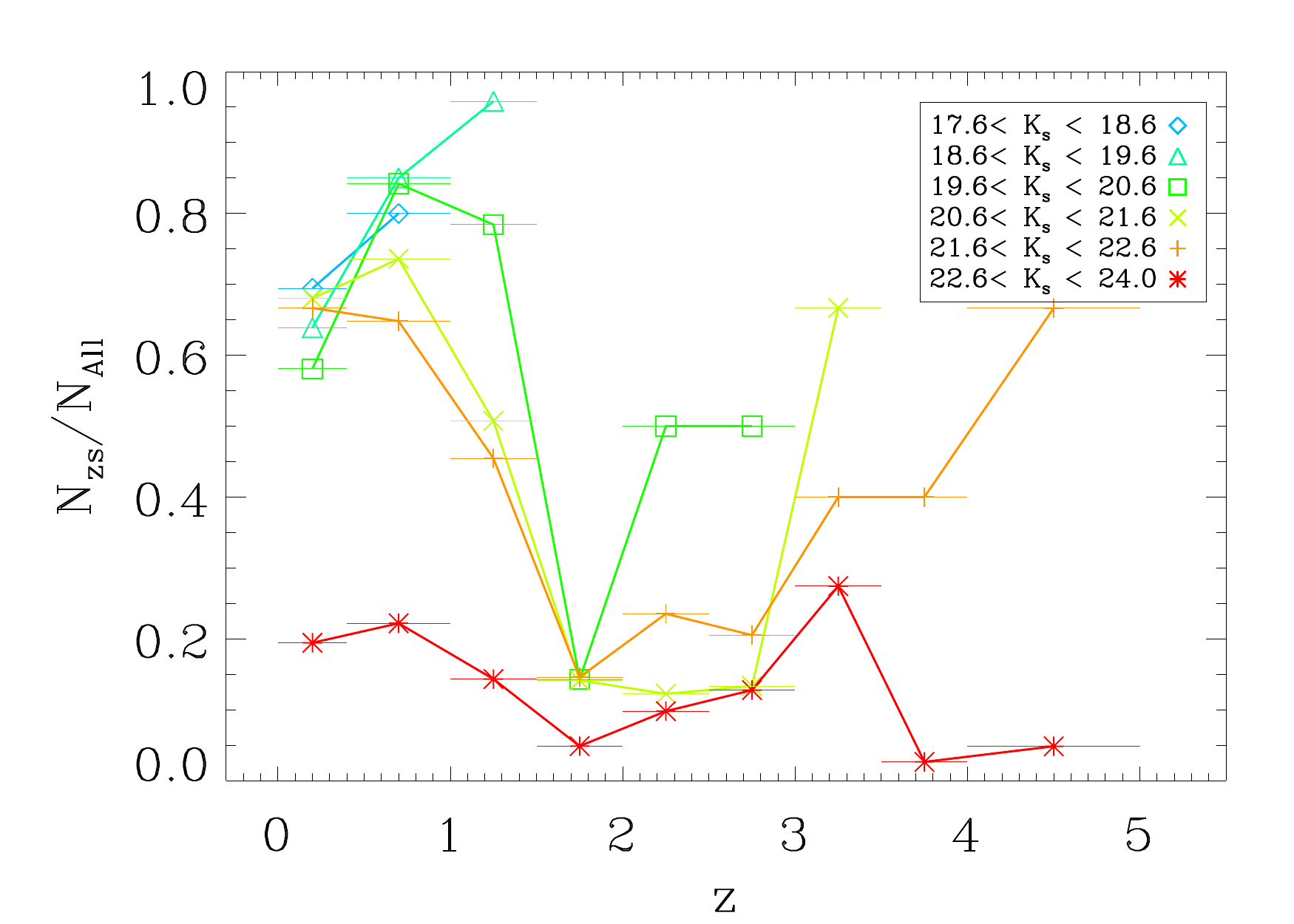} \\
\includegraphics[width=82mm]{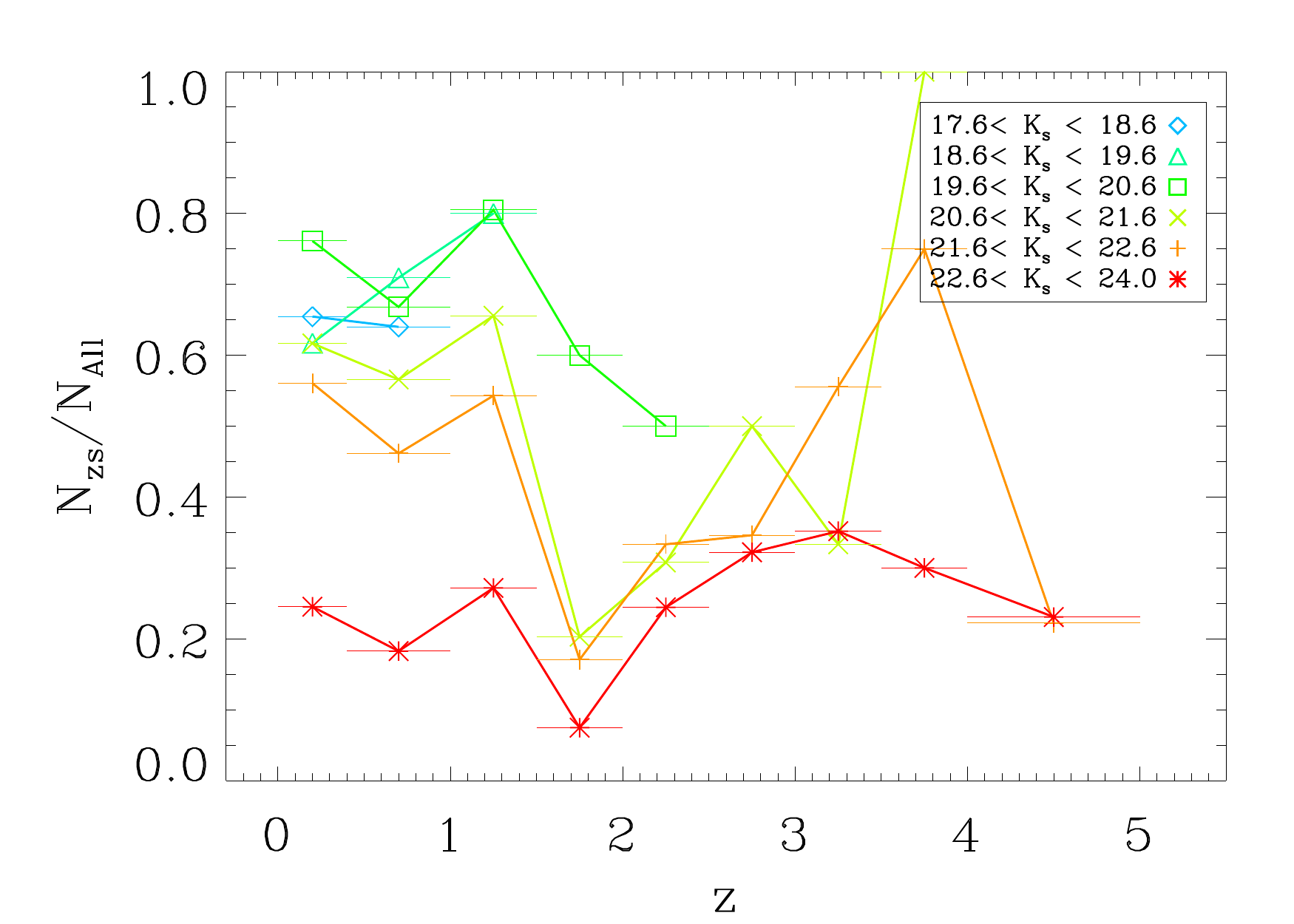}
\caption{The completeness of the spectroscopic identification (the ratio of the number of spectroscopically confirmed galaxies to all galaxies) at different $K_s$ magnitude bins as a function of redshift for the GOODS-Norh (top) and the GOODS-South (bottom).
}
\label{f02}
\end{figure}

Galaxies in both the GOODS-South and the GOODS-North fields have been subject to
extensive spectroscopic follow-up, and a fair number of spectroscopic redshifts are available. The observations were carried out on a heterogeneous conglomeration of samples, including
 magnitude-limited samples in specific bands, or color-selected objects such as LBGs.
 Nevertheless, we gather all available redshifts for our analysis.

  In the GOODS-South, a total of 3166 spectroscopic redshifts are available. These are mostly taken with
 the FOcal Reducer and low dispersion Spectrograph (FORS2) and with the
 VIsible Multi-Object Spectrograph (VIMOS) on the VLT (\citealt{6, 7, 8, 9, 10, 11}), by the ESO-GOODS team.
The sample selection ranges from $z$-band selected, $BzK$ selected, $U$-, $B$-, $V$-, $i$-band dropouts and X-ray sources.
 Many spectroscopic observations were carried out in the GOODS-North field too, with the number of spectroscopic redshifts being 2843. Barger et al. (2008) assembled a catalog containing various spectroscopic studies including redshift surveys of $z$-band or $r$-band limited objects (TKRS, e.g., Wirth et al. 2004; Cooper et al. 2011), and color-selected galaxies at high redshift (Reddy et al. 2006). To the redshifts available in the literature, they added redshifts from their own observations in an effort to increase the completeness of the spectroscopic coverage of the GOODS-North field. The redshift completeness is over 90\% at $z_{850}$ = 23.3 mag, or
 at $K_{s}= 21.6$ mag.

\begin{table}[t!]
\centering
\caption{Summary of the spectroscopic data}
\begin{tabular}{llcc}
\toprule
Telescope/ & Sample Selection & \# of  & Ref. \\
Instrument &                  & $z_{\rm spec}$ &      \\
\midrule
  &  GOODS-South  &     &  \\
\hline
      VLT/FORS2   & $U$, $B$, $V$, $i$-dropouts  & 3166  & a \\
              &  ($z_{850} < 26$)            &       &   \\
     VLT/VIMOS & $BzK$-color selection & {} & {} \\
     {} &  X-ray sources & {} & {} \\
     {} &  $B < 24.5$, $i_{775} < 25$, & {} & {} \\
     {} &  \& $R < 24.5$  &  {}  & {}  \\
     {} &  {} & {} & {} \\
\midrule
   &   GOODS-North    &   &   \\
\hline
        Keck/DEIMOS &  24$\mu$m sources & 2843  & b \\
     Keck/LRIS &  $K_{AB} < 24.5$,   & {} & {} \\
     {}  &  \& $NUV_{AB} < 25$ &  &  \\
     Subaru/MOIRCS &  $FUV_{AB} < 25.5$,  & {} & {} \\
     {}  &  \& $R_{AB} < 24.4$  &   &   \\
     {} &  X-ray and 20 cm  & {} & {} \\
\bottomrule
\end{tabular}
\tabnote{a: Vanzella et al. (2005, 2006, 2008, 2009), Popesso et al. (2009), Balestra et al. (2010) \\
b: Barger et al. (2008), Cooper et al. (2011), and references therein.}
\label{t2}
\end{table}

 Table \ref{t2} summarizes the spectroscopic redshift data in the GOODS fields.

 Figure \ref{f01} shows the $K_s$-band magnitudes versus spectroscopic redshifts of
 galaxies in the GOODS fields. Also plotted in the same figure are photometric redshifts, which are derived from the multi-wavelength imaging data as described in the next section. Using the photometric redshifts, we estimate the completeness of the spectroscopic redshift measurements in the redshift versus apparent magnitude parameter space, which is presented in Figure \ref{f02} for each GOODS field separately (see also Table \ref{t2a}). Later, this information will be used to add photometric redshift noise in the simulation data.

\begin{table*}[!t]
\centering
\caption{Completeness of the spectroscopic identification}
\begin{tabular}{cccccccc}
\toprule
$K_s$ range &  17.6 -- 18.6 & 18.6 -- 19.6 & 19.6 -- 20.6 & 20.6 -- 21.6 & 21.6 -- 22.6 & 22.6 -- 24.0 \\
\midrule
            & $f_{N}^{\rm a}$/$f_{S}^{\rm b}$ & $f_{N}$/$f_{S}$ & $f_{N}$/$f_{S}$ & $f_{N}$/$f_{S}$ &  $f_{N}$/$f_{S}$  & $f_{N}$/$f_{S}$ \\
\midrule
    $0.0 < z < 0.4$  & $ 0.69\,\,/\,\,0.65$ & $0.64\,\,/\,\,0.62$ & $0.58\,\,/\,\,0.76$ & $0.68\,\,/\,\,0.62$ & $0.67\,\,/\,\,0.56$ & $0.19\,\,/\,\,0.25$  \\
    $0.4 < z < 1.0$  & $0.80\,\,/\,\,0.64$ &  $0.85\,\,/\,\,0.71$ & $0.84\,\,/\,\,0.67$ & $0.74\,\,/\,\,0.57$ & $0.65\,\,/\,\,0.46$ & $0.22\,\,/\,\,0.18$  \\
    $1.0 < z < 1.5$  & $-\,\,/\,\,-$ &  $0.96\,\,/\,\,0.80$ & $0.78\,\,/\,\,0.81$ & $0.51\,\,/\,\,0.66$ & $0.46\,\,/\,\,0.54$ & $0.14\,\,/\,\,0.27$ \\
    $1.5 < z < 2.0$  & $-\,\,/\,\,-$ &  $-\,\,/\,\,-$ & $0.14\,\,/\,\,0.60$ & $0.14\,\,/\,\,0.20$ & $0.15\,\,/\,\,0.17$ & $0.05\,\,/\,\,0.08$ \\
    $2.0 < z < 2.5$  & $-\,\,/\,\,-$ & $-\,\,/\,\,-$ & $0.50\,\,/\,\,0.50$ & $0.12\,\,/\,\,0.31$ & $0.24\,\,/\,\,0.33$ & $0.10\,\,/\,\,0.24$  \\
    $2.5 < z < 3.0$  & $-\,\,/\,\,-$ & $-\,\,/\,\,-$ & $0.50\,\,/\,\,\,-\,\,\,\,$ & $0.13\,\,/\,\,0.50$ & $0.21\,\,/\,\,0.35$ & $0.13\,\,/\,\,0.32$  \\
    $3.0 < z < 3.5$  & $-\,\,/\,\,-$ & $-\,\,/\,\,-$ & $-\,\,/\,\,-$ & $0.67\,\,/\,\,0.33$ & $0.40\,\,/\,\,0.56$ & $0.27\,\,/\,\,0.35$  \\
    $3.5 < z < 4.0$  & $-\,\,/\,\,-$ & $-\,\,/\,\,-$ & $-\,\,/\,\,-$ & $\,\,\,-\,\,\,\,/\,\,1.00$ & $0.40\,\,/\,\,0.75$ & $0.03\,\,/\,\,0.30$  \\
    $4.0 < z < 5.0$  & $-\,\,/\,\,-$ & $-\,\,/\,\,-$ & $-\,\,/\,\,-$ & $-\,\,/\,\,-$ & $0.67\,\,/\,\,0.22$ & $0.05\,\,/\,\,0.23$  \\
\bottomrule
\end{tabular}
\tabnote{a: The ratio of the number of spectroscopically confirmed galaxies to all galaxies at different $K_s$ magnitude/redshift bins for the GOODS-North field. \\ b: The ratio of the number of spectroscopically confirmed galaxies to all galaxies at different $K_s$ magnitude/redshift bins for the GOODS-South field.}
\label{t2a}
\end{table*}

\section{Photometric Redshifts and SED Fitting}

  We derive photometric redshifts of galaxies to aid the selection of MSGs.
  Galaxies are primarily drawn from a $K_{s}$-band limited sample, but are
supplemented by $z$-band selected galaxies, the latter being basically $U$- and $B$-band dropout galaxies (\citealt{2,1}).
 In the $K_{s}$-band limited sample, galaxies with ${K_s} \leq 23.8$ mag (GOODS-South) and ${K_s} \leq 24$ mag (GOODS-North) are used. Spectroscopic redshifts are available for 30\% of the magnitude-limited sample, and for galaxies without spectroscopic redshifts, we derive photometric redshifts using $U$, $B_{435}$, $V_{606}$, $i_{775}$, $z_{850}$, $J$, $H$, $K_s$ bands, and the Spitzer photometry at $3.6$ through $8.0$ $\mu$m if available.

\begin{figure}[t!]
\centering
\includegraphics[trim=12mm 2mm 4mm 2mm, clip, width=82mm]{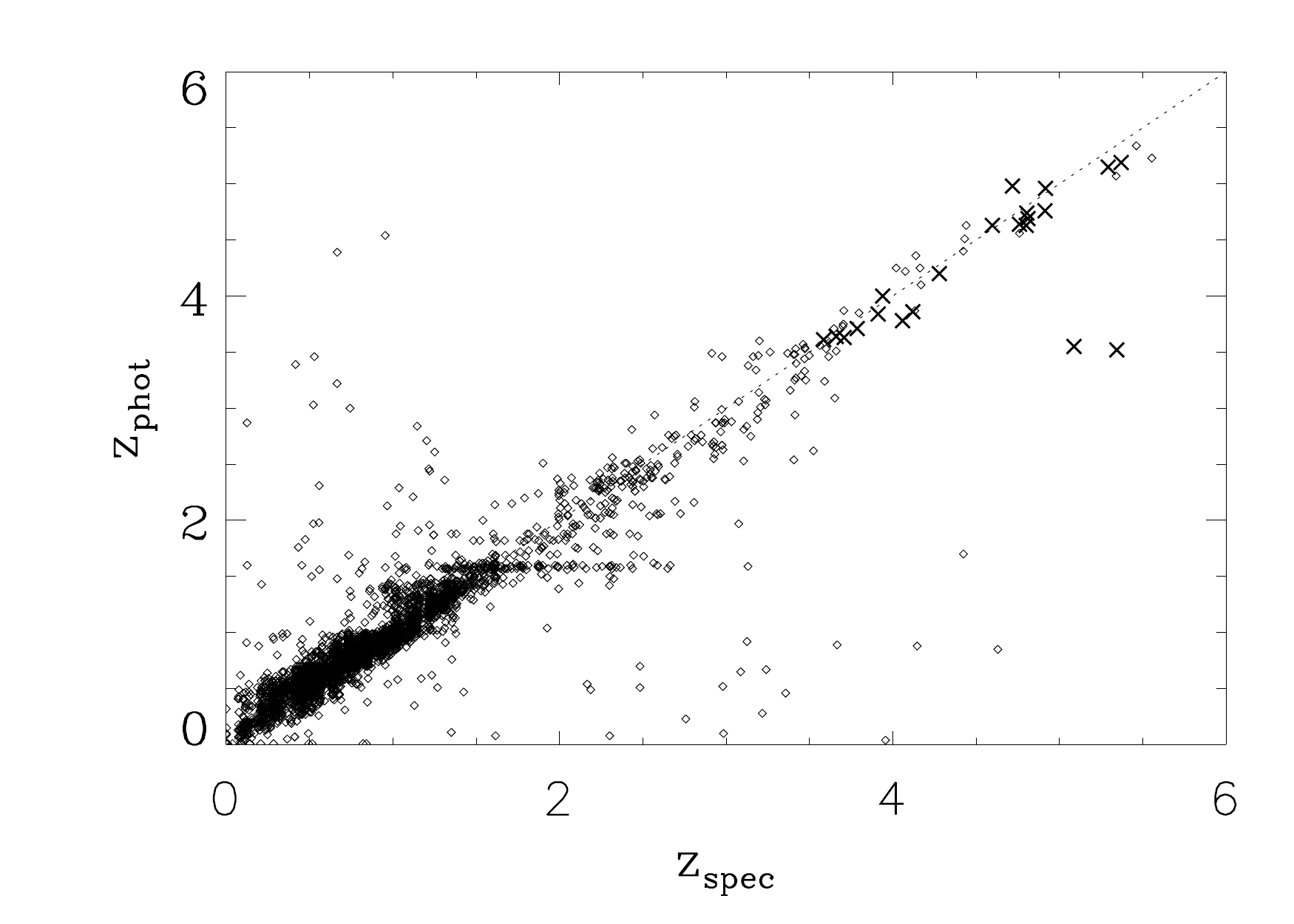}
\caption{
Photometric redshift ($z_{\rm phot}$) versus spectroscopic redshift ($z_{\rm spec}$) over
the combined areas of the two GOODS fields. The diamonds and crosses indicate the
 $K_s$-band and $z$-band limited samples, respectively.
}
\label{f1}
\end{figure}
\begin{figure}[t!]
\centering
\includegraphics[trim=12mm 2mm 4mm 2mm, clip, width=82mm]{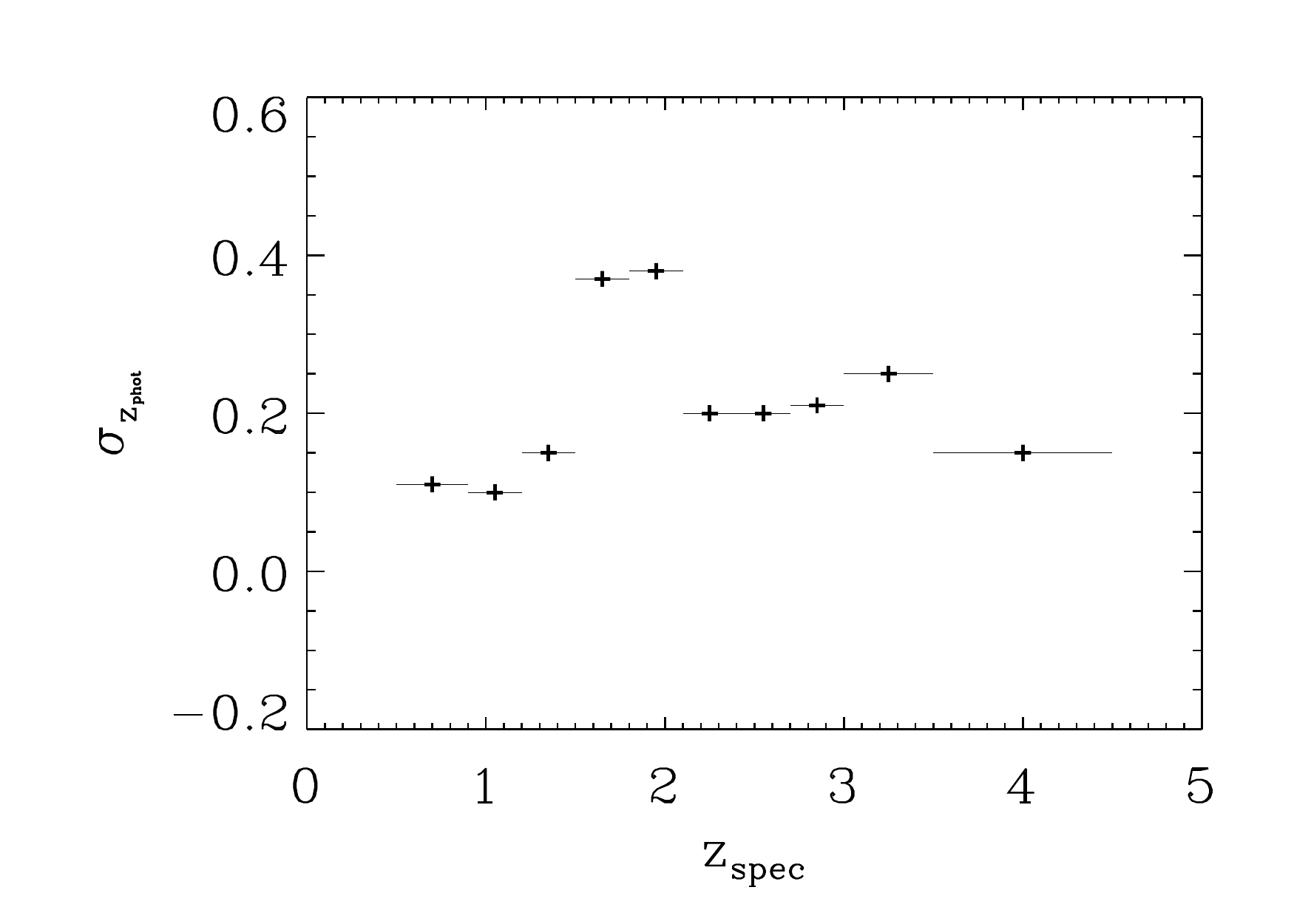}
\caption{
The rms scatter, $\sigma_{z}$, of the photometric redshift ($\sigma_{z}$) as a function of spectroscopic redshift.
The $\sigma_{z}$ values stay at around 0.1 to 0.2 for most redshift intervals, except for
$z_{\rm spec} = 1.5 - 2.0$ where the $\sigma_{z}$ value is quite large at around 0.4.
}
\label{f2}
\end{figure}

 To derive photometric redshifts, we generated 742 SED templates using the stellar population synthesis code GALAXEV (\citealt{14}). The assumed parameters for these SED templates are the following. For the star formation rates, we use a single burst, or a constant star formation rate (SFR) or an exponentially declining SFR in the form of $SFR \propto$ exp(-t/$\tau$) for $\tau$ =  1, 2, 3, 5, 15, or 30 Gyr. For the metallicity, we assumed values of $0.4 Z_{\odot}$, $1 Z_{\odot}$ or $2 Z_{\odot}$, and we adopted the Salpeter initial mass function with a mass range of 0.1 to 100 $M_{\odot}$. The SED ages range from 100 Myr to 10 Gyr with an upper limit (the age of the Universe at the corresponding redshift). The reddening parameter $A_{V}$ is allowed to change from 0 to 1.8
with the optical depth, $\tau_{\lambda}$, following the two
component power law model of
Charlot \& Fall (2000). This model fits the extinction properties of star-forming galaxies well, and adopts two different optical depths depending on the age of stars (younger or older
than $10^{7}$ years).
The Bayesian photometric redshift estimation (BPZ; Ben$\acute{i}$tez 2000) code is used for deriving photometric redshifts.
 During the photometric redshift derivation, the best-fit templates are identified, which are used for estimating stellar masses of galaxies.
 We also obtain stellar masses of galaxies with spectroscopic redshifts by fitting their SEDs  using the same code. The age of the template SEDs is restricted to the age of the Universe at the corresponding redshift when performing the SED fit.

  To test the accuracy of the photometric redshifts, we compare our photometric redshifts ($z_{\rm phot}$) with the available spectroscopic redshifts ($z_{\rm spec}$). Here, we use only objects having spectroscopic redshifts with high quality flags (B or better: Vanzella et al. 2006, 2008, 2009; Poppesso et al. 2009; Balestra et al. 2010, 3 or better: Barger et al. 2008; Cooper et al. 2011).
 The cross-match of the photometric redshift sample with the $z_{\rm spec}$ data
 gives 6009 galaxies for this comparison. Figure \ref{f1} shows the comparison of the spectroscopic redshifts versus the photometric redshifts. In most cases, photometric
 redshifts are reasonably accurate with a small number of outliers, where the root-mean square (rms) scatter of $z_{\rm phot}$ versus $z_{\rm spec}$, $\sigma_{z}$, is $\sigma_{z} \simeq 0.12$ at low redshift ($z < 1$), and $\sigma_{z} \simeq 0.2$ at high redshift ($z > 2$). At the intermediate redshift range, the photometric redshift accuracy is poor with a value of
  $\sigma_{z} \simeq 0.4$.
 Figure \ref{f2} shows the rms scatter of $z_{\rm phot}$ against $z_{\rm spec}$ as a function of $z_{\rm spec}$.
 We find that the average absolute scatter is $\langle|\Delta z|\rangle=\langle(|z_{\rm spec}-z_{\rm phot})/(1+z_{\rm spec})|\rangle$ of 0.06 after excluding several clear outliers, which is slightly worse than, but comparable to the values ($\langle|\Delta z|\rangle$) from other photometric redshift studies (\citealt{a16, a17}).

\begin{figure*}[t!]
\centering
\includegraphics[width=160mm]{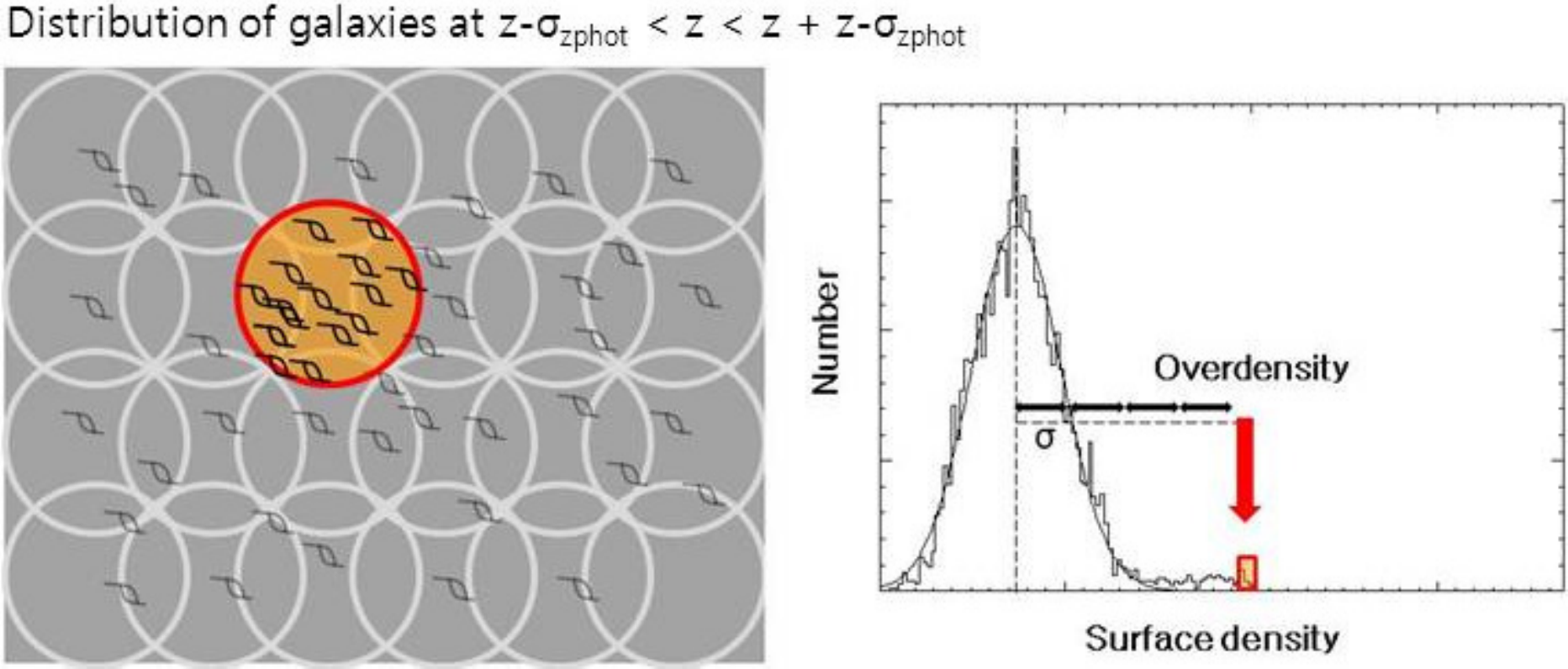}
\caption{
 This figure illustrates how MSGs are identified. The left panel shows
 top-hat filters (circles) that are moved around over the two-dimensional distribution of galaxies in a specific redshift bin. We obtain the surface number
 density of galaxies in each circle by counting the number of objects within each redshift slice and dividing by the area covered by the circle.
 The surface number densities are computed over the entire image,
 and the distribution of the surface number density is plotted (right panel). The rms scatter, $\sigma_{rms}$
 of the distribution is determined from the distribution, and the significance
 of the overdensity, the overdensity factor, is determined as the number of $\sigma$'s
 from the mean surface number density. MSGs are defined as the location where
 the surface density exceeds 3.5 $\sigma_{rms}$ (right panel).
}
\label{f3}
\end{figure*}

\section{MSGs in GOODS Fields}

\subsection{Search Method}

 We searched for MSGs by running circular top-hat filters with a 1 Mpc diameter in physical size on a two-dimensional spatial distribution of galaxies within a fixed redshift range. The procedure is described below, and graphically illustrated in Figure \ref{f3}.
Note that the length scale of the analysis is set to physical scale
 rather than the comoving scale throughout the paper, unless mentioned otherwise.

 The redshift bins are chosen to have a width of $\pm2000 \, $km s$^{-1}$ for galaxies with spectroscopic redshifts, or a width corresponding to the median photometric redshift accuracy at a certain redshift range ($\sigma_{z}$) as presented in Figure \ref{f2}.


 The center of each successive redshift bin is shifted by $\Delta v = 3000$ km s$^{-1}$ from 0.65 to 4.5. The photometric redshift uncertainty of 0.1 at $z=1$ corresponds to a velocity
 uncertainty of 15000 km s$^{-1}$. So, a shift of $\Delta v = 3000$ km s$^{-1}$ is
 fine enough not to miss any overdensities between two successive redshift bins.
 For the spectroscopic redshift sample, this shift allows
 an overlap between two successive redshift bins and the identification of MSGs that are situated between two successive redshift bins, since the velocity dispersion of a typical galaxy cluster is $\sigma_{v} \sim 1000$ km s$^{-1}$.

 The density contour maps are generated by counting the number of objects around a point within a 1 Mpc diameter circle in the image.
 The top-hat filter is moved around over a 100 by 100 grid for each GOODS field. The grid dimension gives a separation of each contour map center of 0.2 arcmin or 0.1 Mpc at $z=2$. The separation between the grid nodes is sufficiently smaller than the typical scales of groups/clusters.
 The distributions of the projected number densities are created, and the rms of the distribution ($\sigma_{rms}$) is determined after a $\sigma$ clipping,
with the clipping at 3-$\sigma$ over 3 iterations.
Note that we also obtain the standard deviation $\sigma_{\rm G}$ of the same projected number density distribution by performing a Gaussian fit, and the $\sigma_{\rm G}$  values derived in this way agree with the values of $\sigma_{rms}$.
 Finally we identify MSGs as the regions where the projected number density exceeds the mean projected number density by more than $3.5 \, \sigma_{rms}$. Here, we define the overdensity factor as the number of $\sigma_{rms}$ units
 the number density of an overdense region deviates from the mean number density.

 The top-hat filter diameter may affect the search result by changing the overdensity factors. If the search radius is too large or too small, the overdensity factors may not reach the MSG identification threshold. Therefore,
 we performed the analysis by varying the top-hat filter diameter from 0.5 Mpc to 2 Mpc to check for any dependence of our result on the adopted diameter of the top-hat filter, but our test shows no significant difference in the overdensity factor of MSGs
  over this range in scale (Figure \ref{f4}).
  We also checked how the number of MSGs changes depending on the filter diameter.
 With a 0.5 Mpc diameter, the number of MSGs increases by $\sim$10\% at $z>2.9$, while there is no change in the number of MSGs at lower redshifts.
The increase in the MSG number is mostly at lower mass regions, i.e., we tend to find less massive MSGs when using a top-hat filter with a smaller diameter.
On the other hand, if a large diameter of 2.0 Mpc is used, we miss $\sim 30$\% of MSGs at $z < 2$, meaning that larger search diameters tend to miss concentrated structures. Hence, the 1 Mpc diameter appears to be optimal for finding MSGs.

 Foreground or background halos or filamentary structures along the line of sight could produce strong clustering signal that can be identified as MSGs, especially when photometric redshifts are the main source of the redshift identification. We examined such a possibility using a simulation dataset (see Section 5 and Figure 12) that is similarly constructed to mimic the observation. We find that
the identification of MSGs is mostly due to one strong massive halo (see the top and middle panels of Figure 12), rather than many halos with similar masses scattered
along the line of sight. About 30\% of the identified MSGs consist of several distinctive peaks along the line of sight, but even in such cases, the distinctive halos are closely located in
 velocity
and real space (the bottom panel of Figure 12). However, the numerous minor halos along the line of sight can contribute significantly to the mass estimate of MSGs, but such effects
are accounted for by estimating the interloper contribution (see Sections 4.2 and 5).

\begin{figure}[t!]
\centering
\includegraphics[width=82mm]{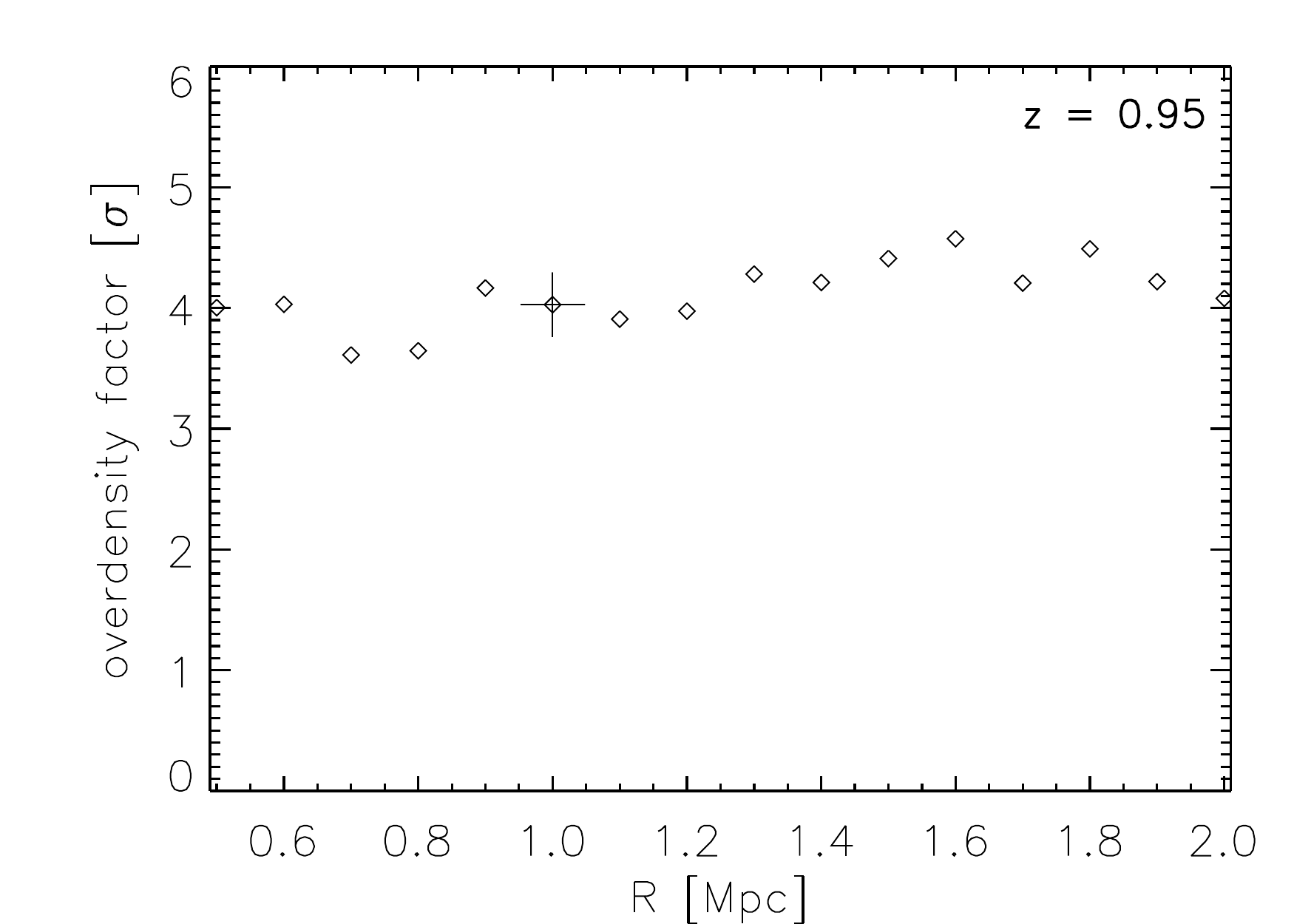}
\includegraphics[width=82mm]{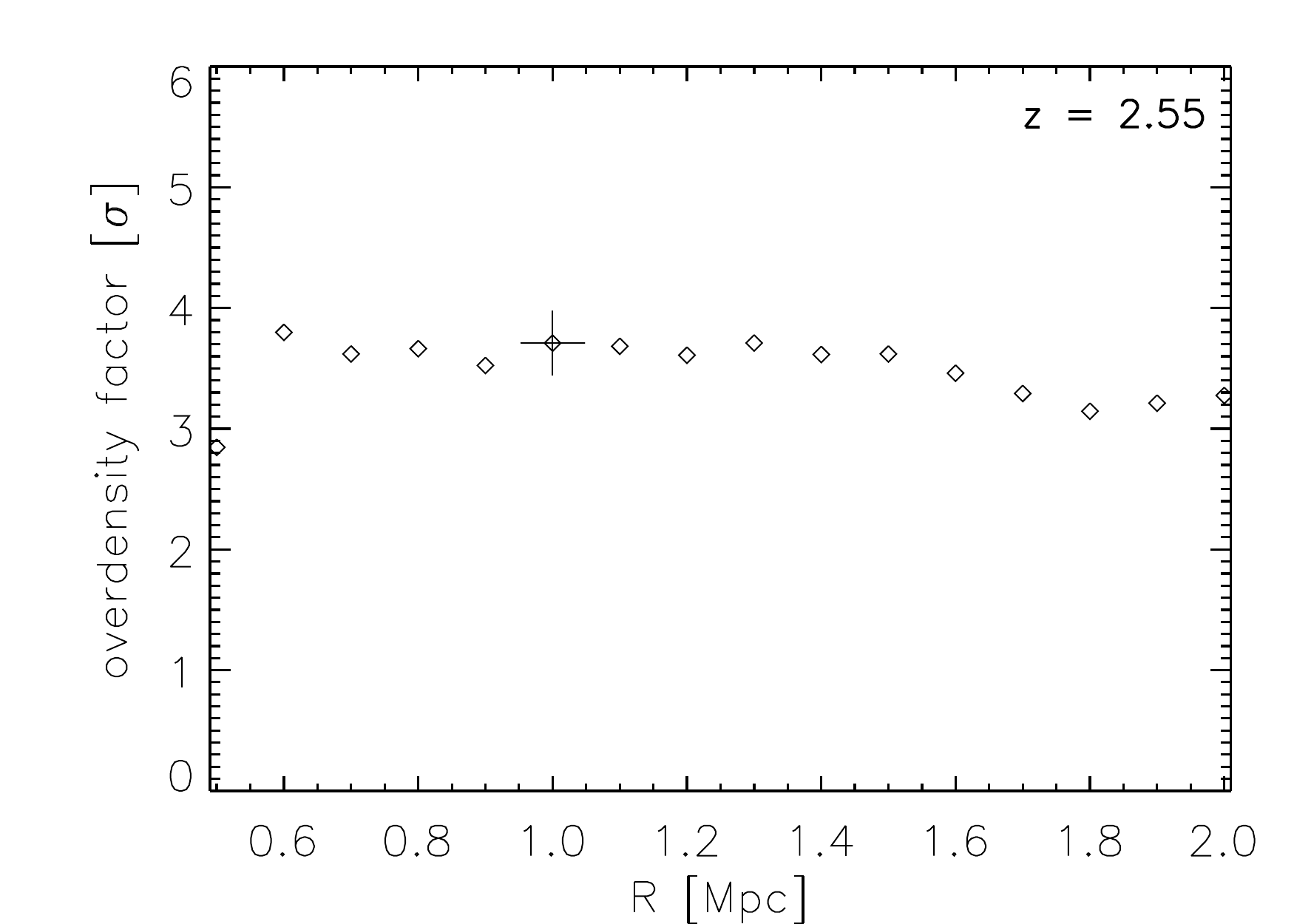}
\includegraphics[width=82mm]{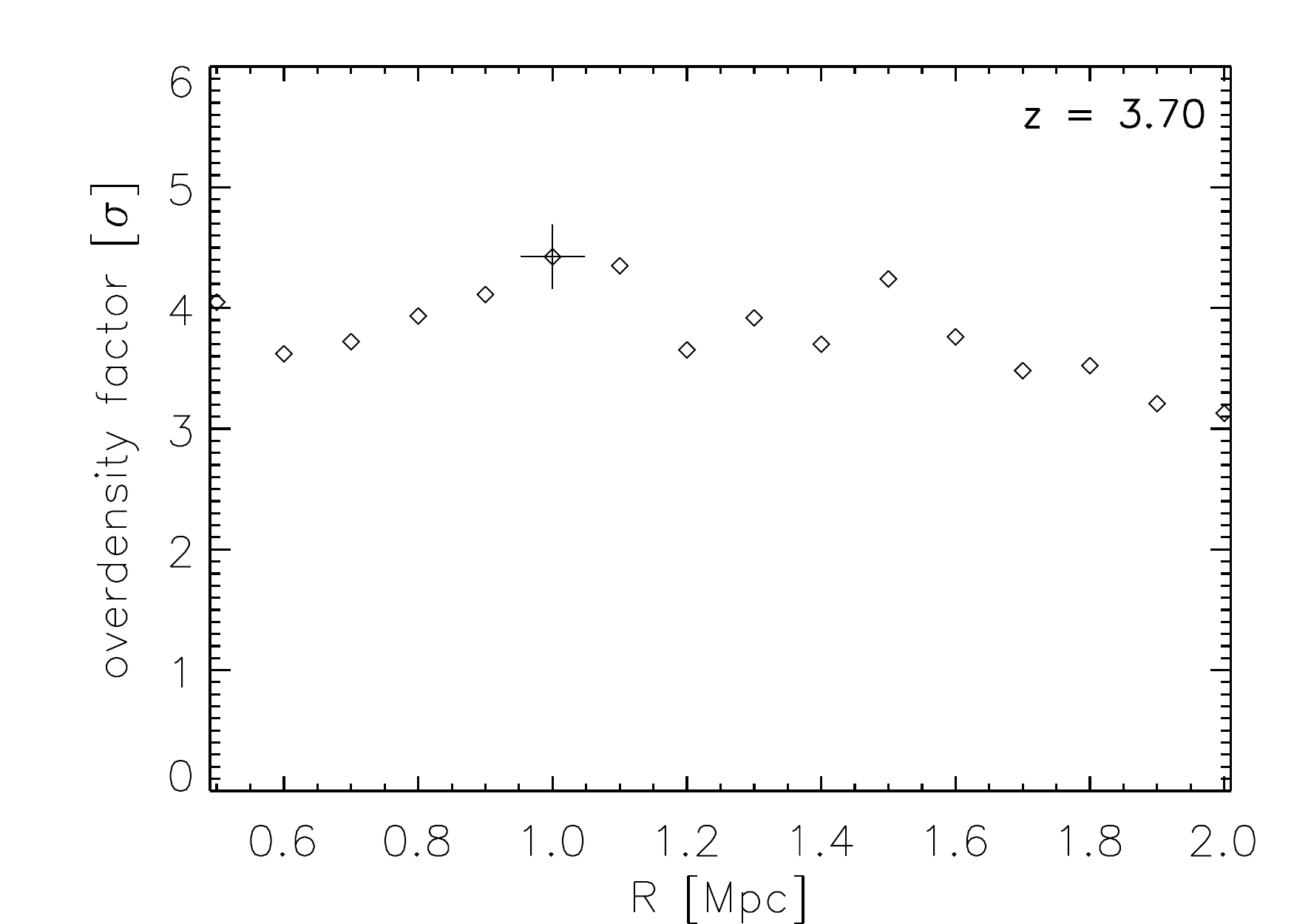}
\caption{Examples showing how the overdensity factor
 changes as a function of a diameter of the top-hat circular filter.
 The overdensity factor on the y-axis
 is in units of $\sigma_{rms}$.
 From 0.5 Mpc to 2.0 Mpc, we find that the overdensity factor is not very sensitive to the assumed diameter, although the adopted 1 Mpc seems to give an optimal result. The examples are given for typical MSGs at $z = 0.95, 2.55$ and $3.7$ in GOODS-South.}
\label{f4}
\end{figure}

 The redshift of each MSG is chosen to be the mean spectroscopic redshift of the galaxies in the MSG,
within 1 Mpc of the projected distance from the peak of the distribution.
 If no $z_{\rm spec}$ is available, we pick the central redshift of the redshift bin where the MSG is identified.

 For this procedure, note that we used a 1 Mpc radius, rather than a 1 Mpc diameter, as used for the MSG search. While a top-hat filter with a 1 Mpc diameter is efficient for identifying MSGs, MSGs are in general more extended than 1 Mpc diameter. The larger circle includes more member galaxies while minimizing the number of interlopers.

\subsection{MSGs in GOODS Fields}

  In Figures \ref{f5a} and \ref{f6a}, we present the projected surface number density contour maps of the MSGs (left panel) and the redshift distributions of galaxies that falls within 1 Mpc radius from each MSG center (right panel). The crosses indicate the spectroscopically confirmed galaxies within a redshift slice of $|v| < 2000$ km s$^{-1}$ at each redshift. The shaded histograms represent the spectroscopically identified galaxies within the same redshift window. When multiple MSGs are identified at the same redshift, MSGs are categorized as ``A", ``B", and ``C", which denote the significance level in alphabetical order (A being the most significant). The numerical digits distinguish MSGs of the same significance. Overall, we identify 34 and 25 MSGs in the GOODS-North and the GOODS-South, respectively. The identified MSGs are listed in Table \ref{t3}.

 These 59 MSGs have overdensity factors of $3.5-8\sigma$, and some of them have already been reported in the literature as clusters/proto-clusters (\citealt{28, 18, 7, 29, 20,27, 30, 26, 11}).
 In Figures 7 an 8, we also mark radio galaxies (yellow open triangles), submm galaxies (blue
 open diamonds), and AGNs (red filled stars)
  if there are such objects in the redshift range.
  The MSGs at $z \sim 0.85$, 1.02 and 1.14 in the GOODS-North and $z \sim 0.67$, 0.74, 1.09 and 1.61 in the GOODS-South include radio galaxies at the same redshift,
 and the MSGs at $z=0.95$, 1.23, 1.61 and $z = 3.7$ are associated with AGNs in the GOODS-South.
  However,
  only 20\%
   of the identified MSGs are found to be spatially associated with known AGN/radio sources.
  Therefore, MSG searches radio galaxies and AGNs as signposts can miss a large number of overdense areas.


\begin{figure*}[tp!]
\setcounter{figure}{6}
\centering
\includegraphics[width=75mm]{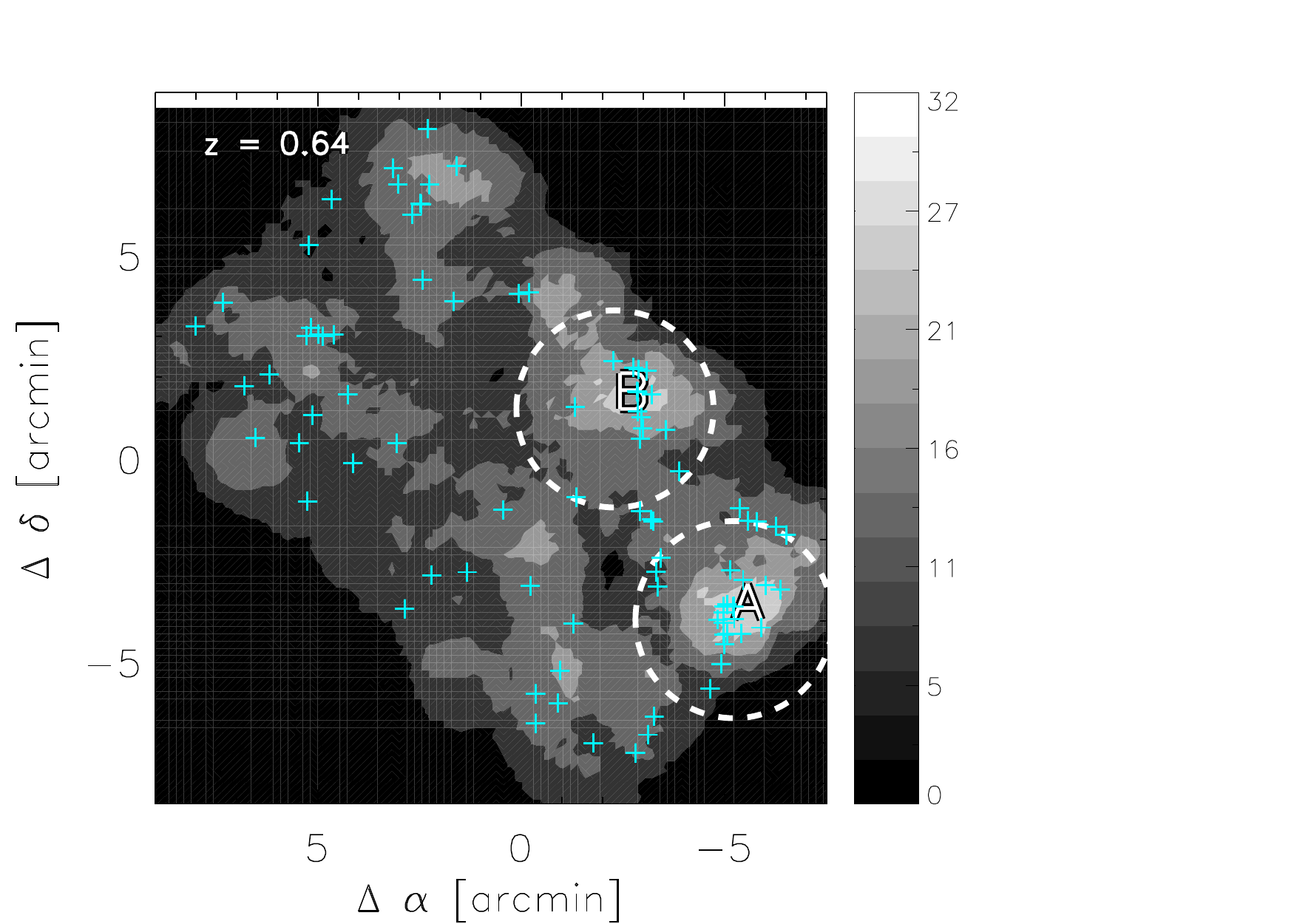}
 \hspace{-0.85in}
       \includegraphics[width=75mm]{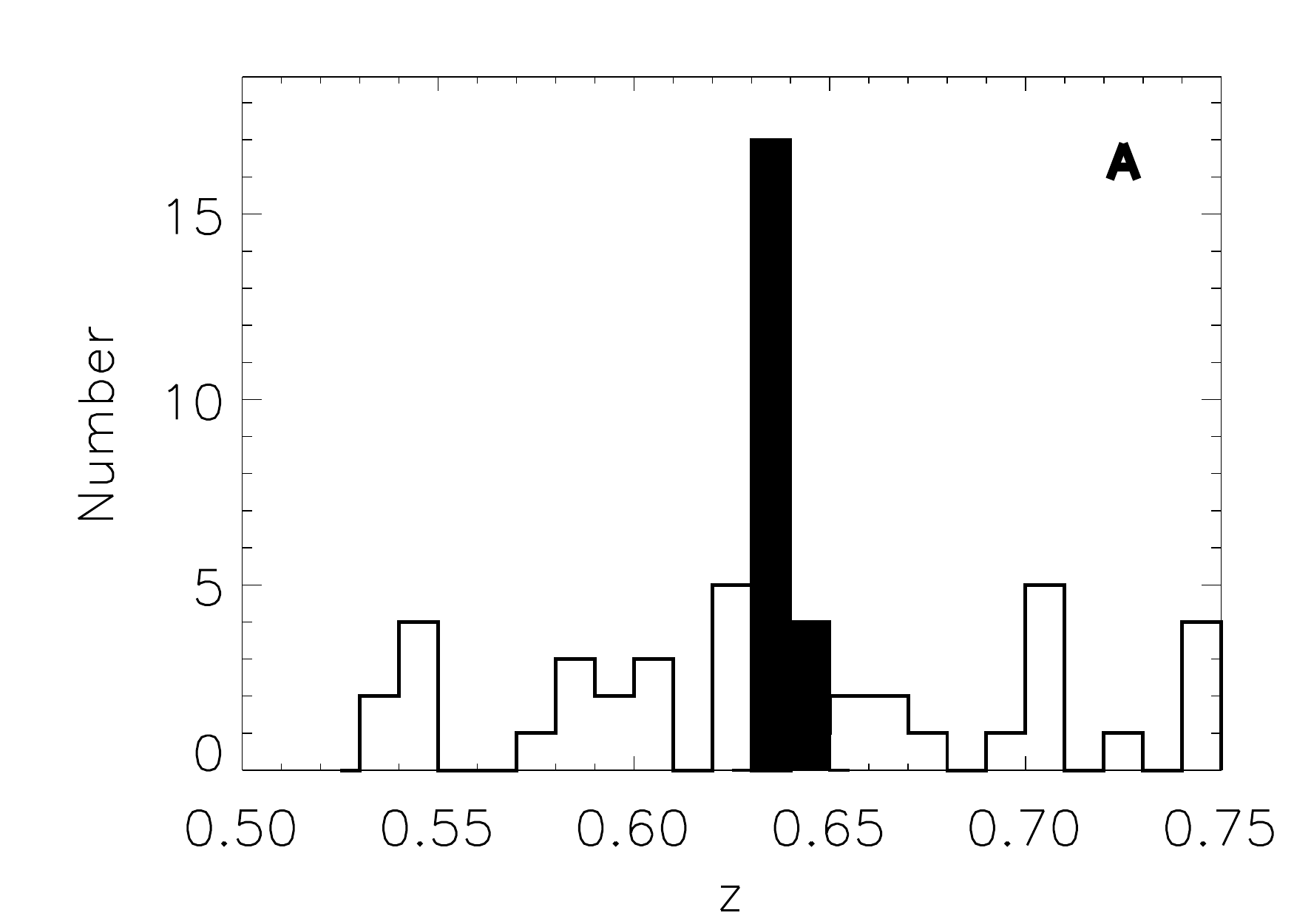} \\
\includegraphics[width=75mm]{f7a1-eps-converted-to.pdf}
 \hspace{-0.85in}
       \includegraphics[width=75mm]{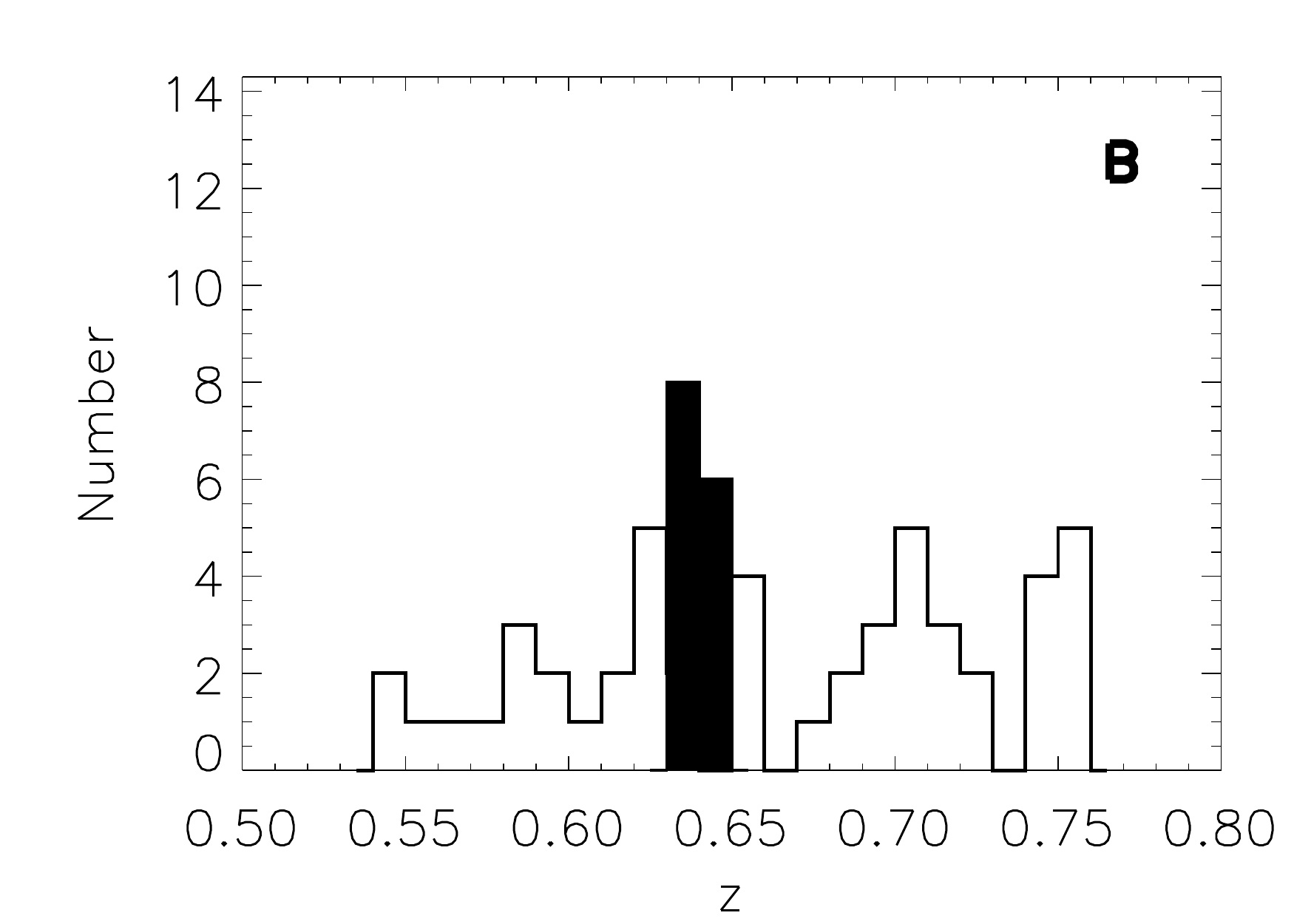}
\\
\includegraphics[width=75mm]{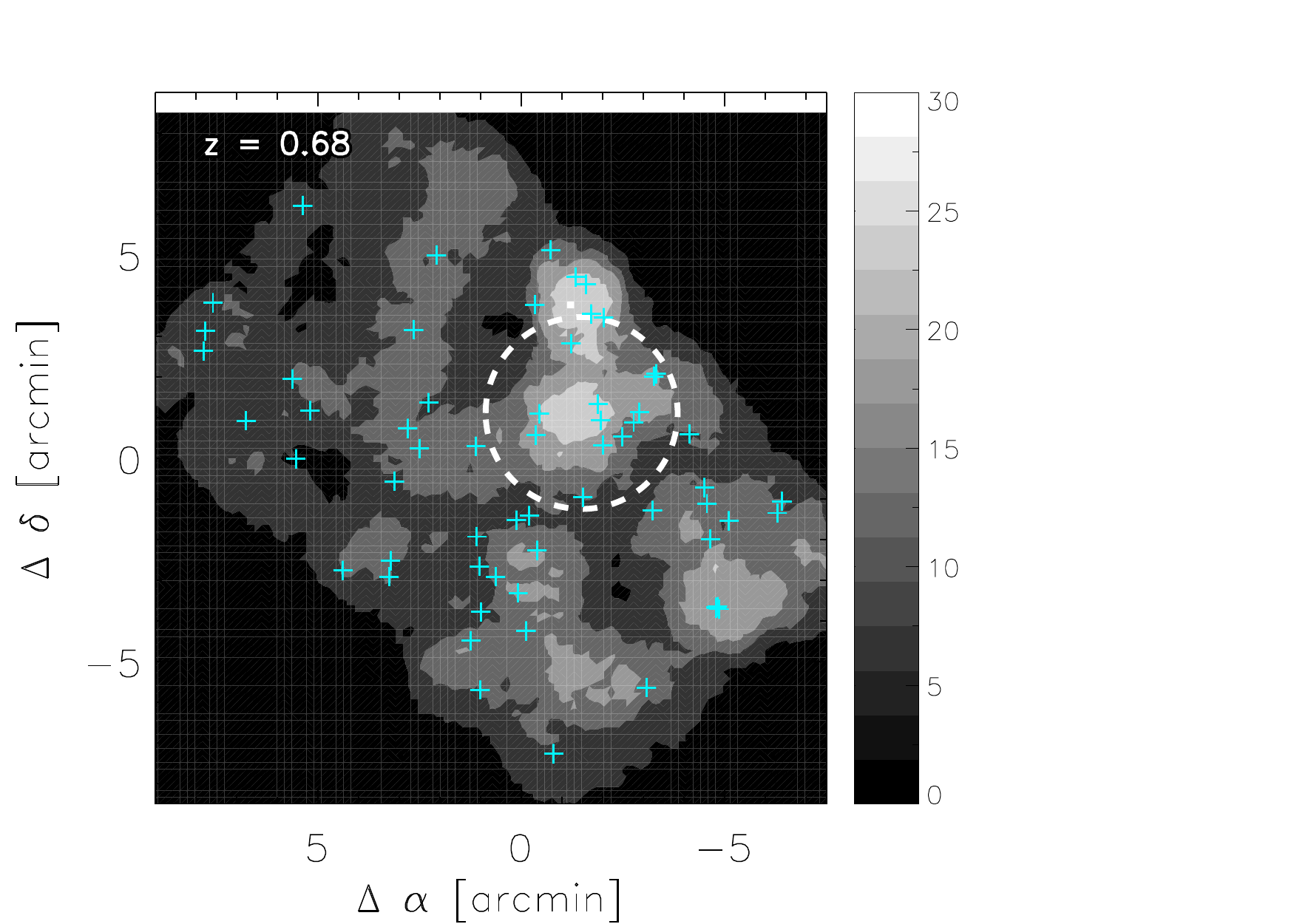} \hspace{-0.85in}
   \includegraphics[width=75mm]{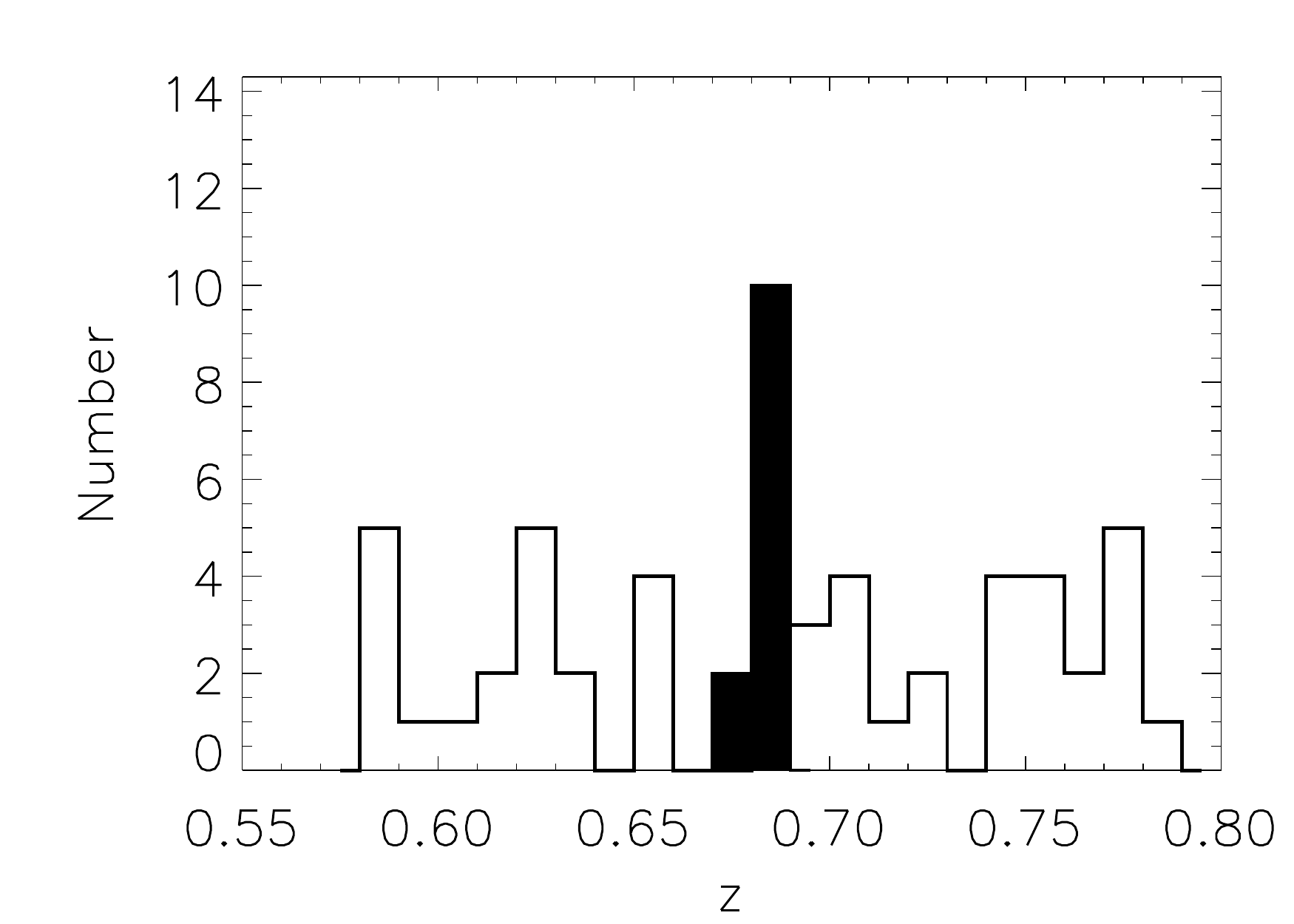}
\\
\includegraphics[width=75mm]{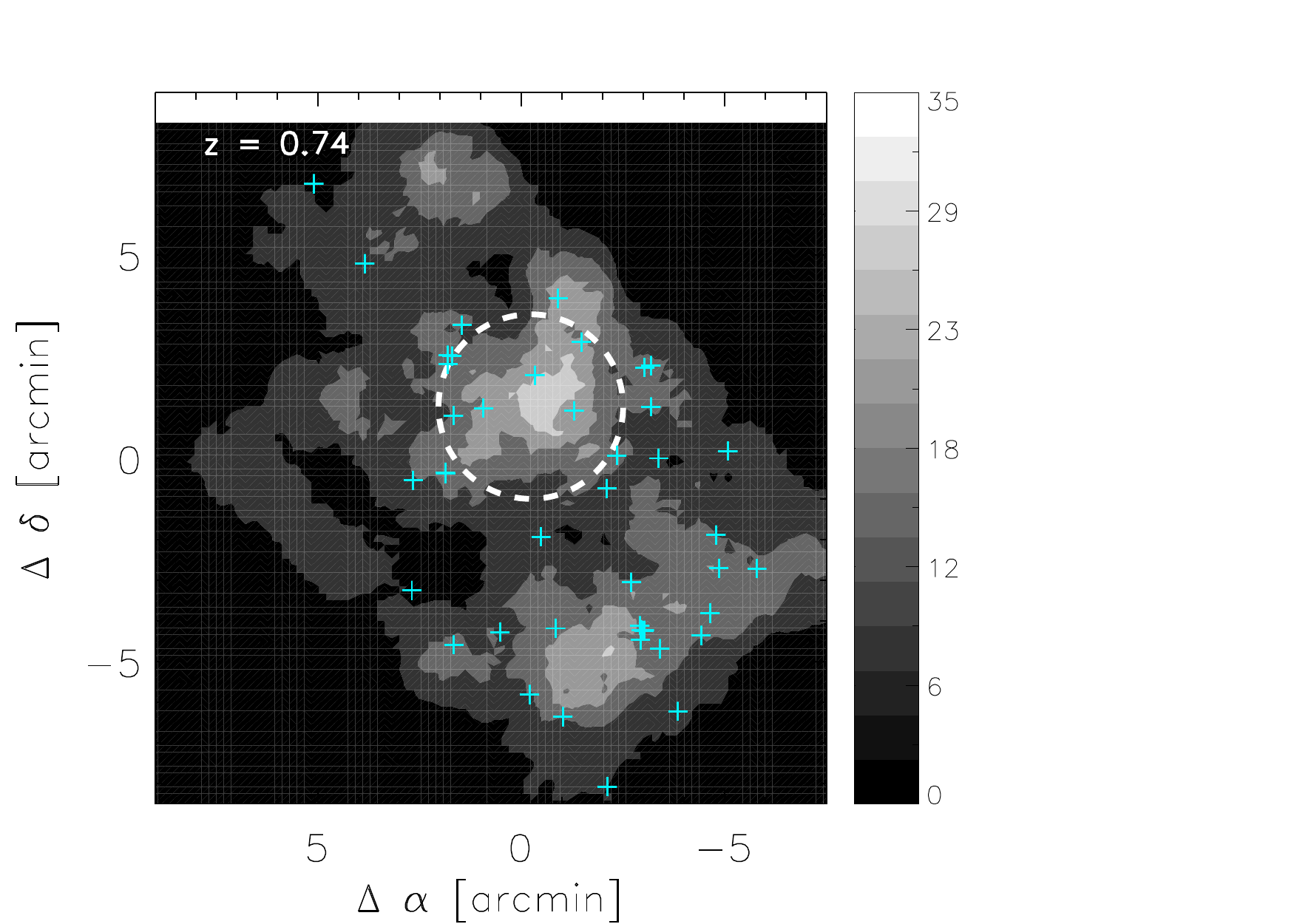} \hspace{-0.85in}
\includegraphics[width=75mm]{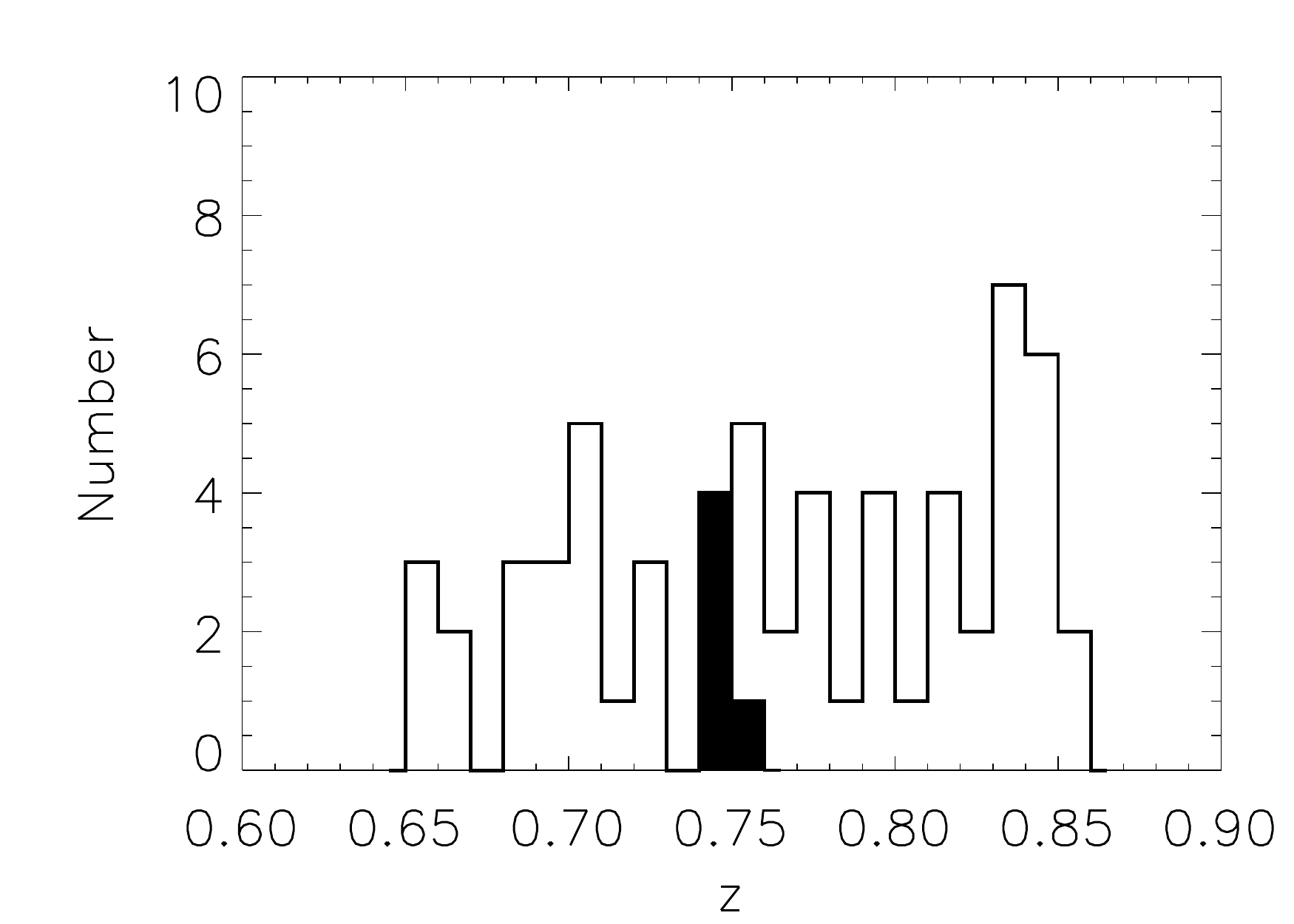}
\caption[The spatial distribution and redshift distributions of MSGs in the GOODS-North.]
{$Left$: MSGs in the GOODS-North field. The contours indicate the surface number density of galaxies
 in each redshift slice in units of number arcmin$^{-2}$.
 Both galaxies with $z_{\rm spec}$ and $z_{\rm phot}$ are included to make the surface density map.
 The crosses represent the spectroscopically confirmed galaxies in the rest-frame redshift interval of $\pm$2000 km s$^{-1}$ at each redshift.
 Member galaxies of MSGs are considered as those lying within 2 Mpc diameter circle from the number density peak (the dashed circles) at the MSG redshift.
$Right$: The redshift distribution of MSG member galaxies.
 Galaxies with $z_{\rm spec}$ are indicated using the shaded histograms,
 while galaxies with $z_{\rm phot}$ are indicated with the open histograms.
}
\label{f5a}
\end{figure*}


\begin{figure*}[tp!]
 \setcounter{figure}{6}
\centering
\includegraphics[width=81mm]{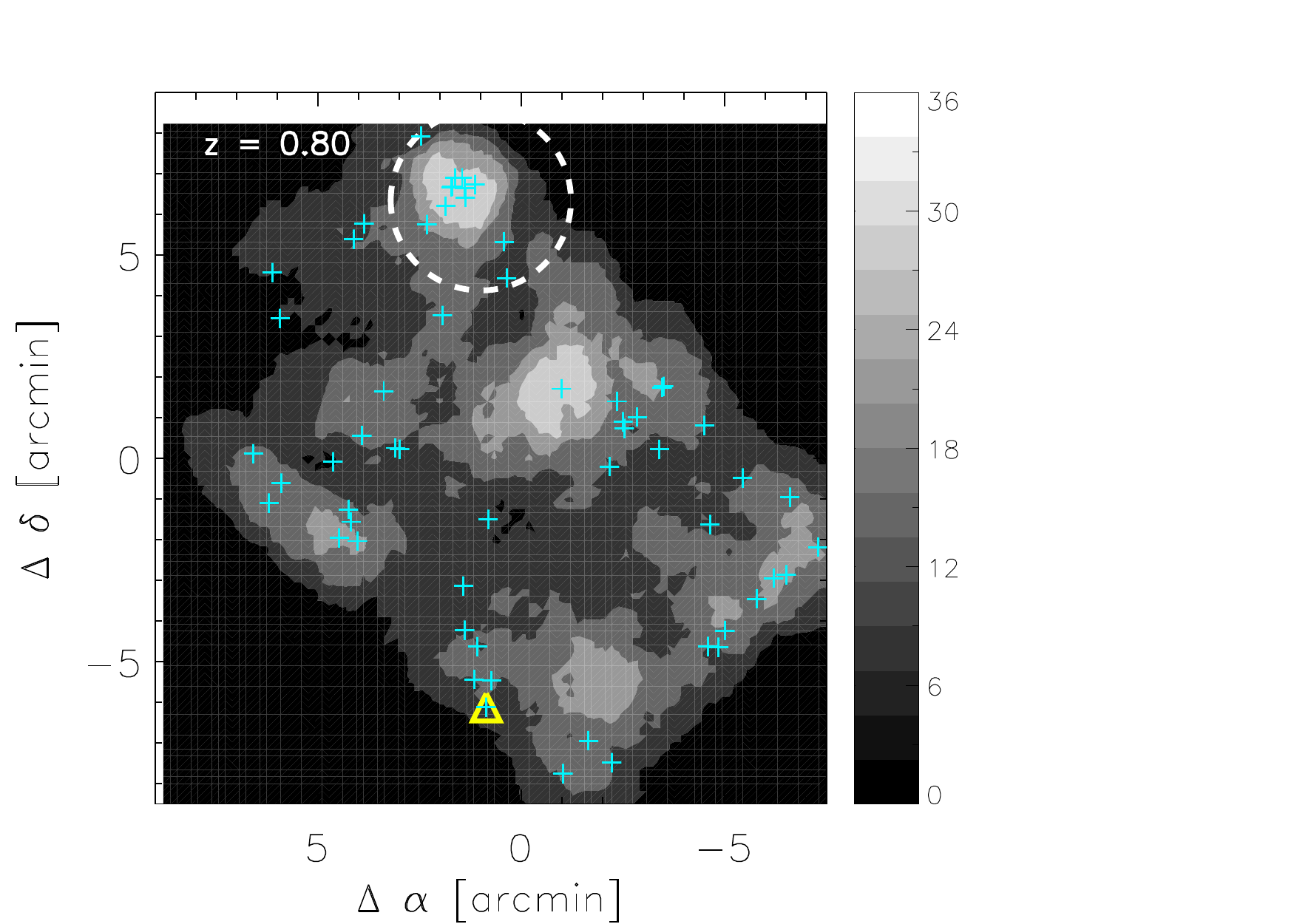}
       \hspace{-0.85in}
\includegraphics[width=81mm]{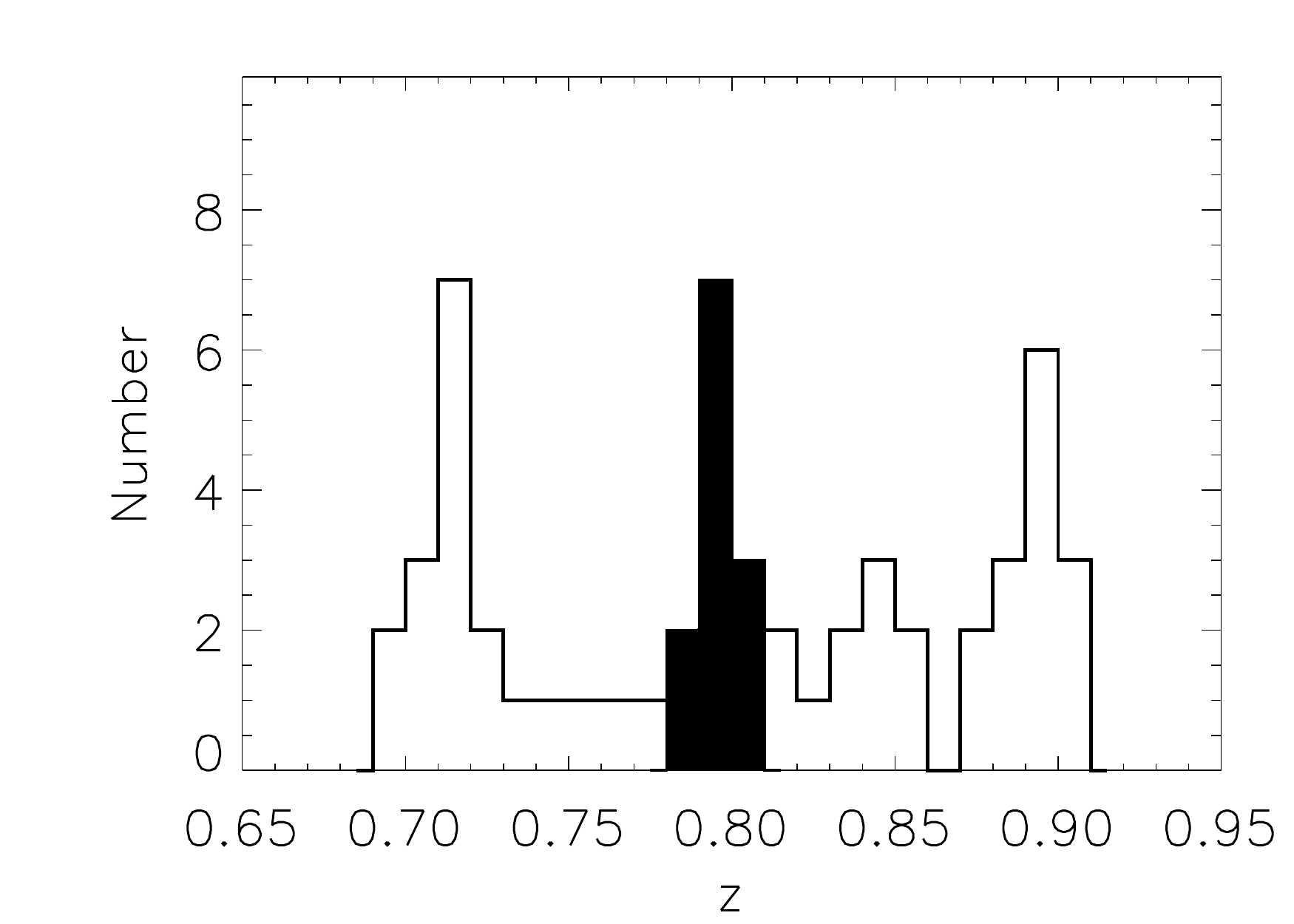}
\\
\includegraphics[width=81mm]{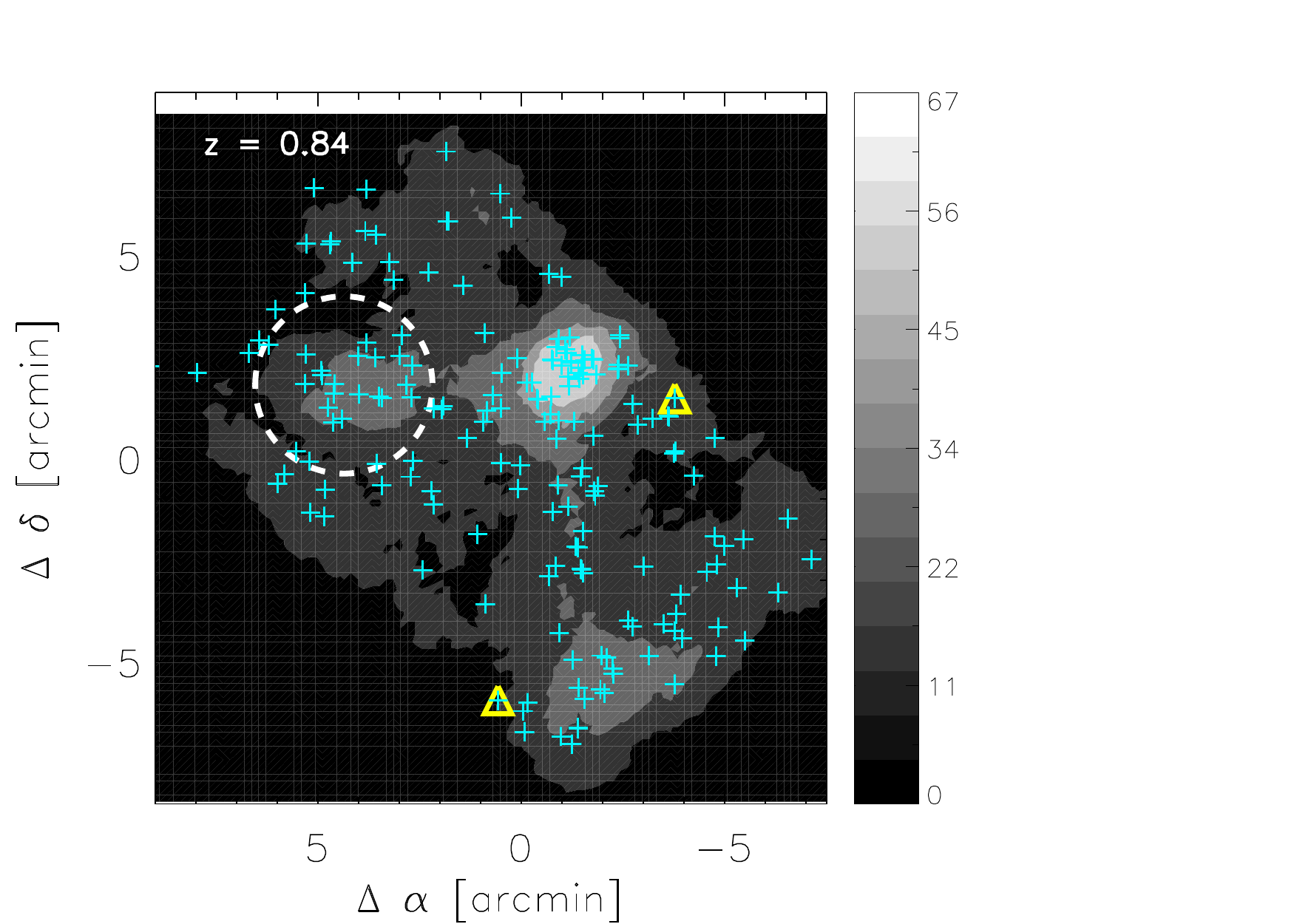}
       \hspace{-0.85in}
\includegraphics[width=81mm]{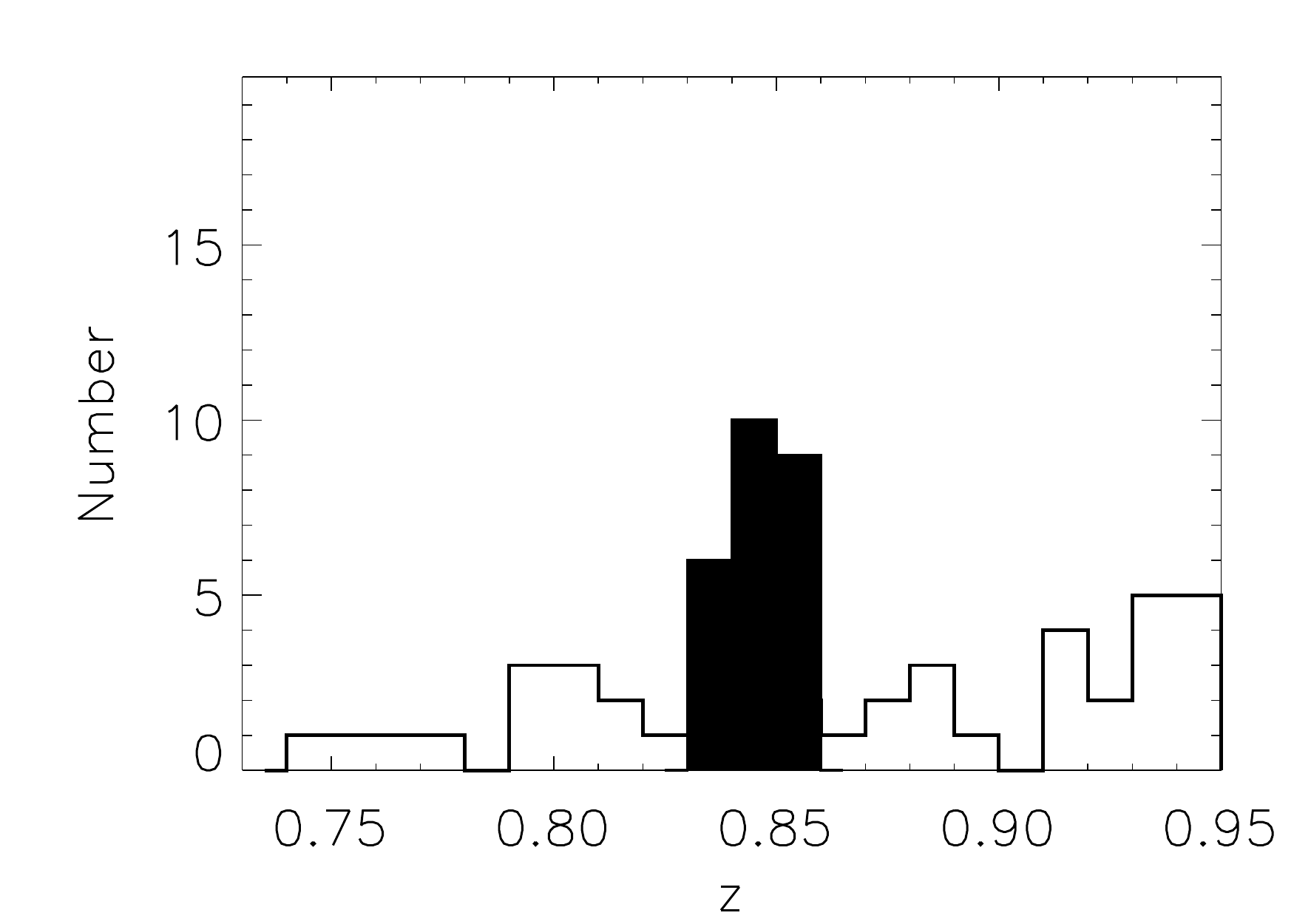}
\\
\includegraphics[width=81mm]{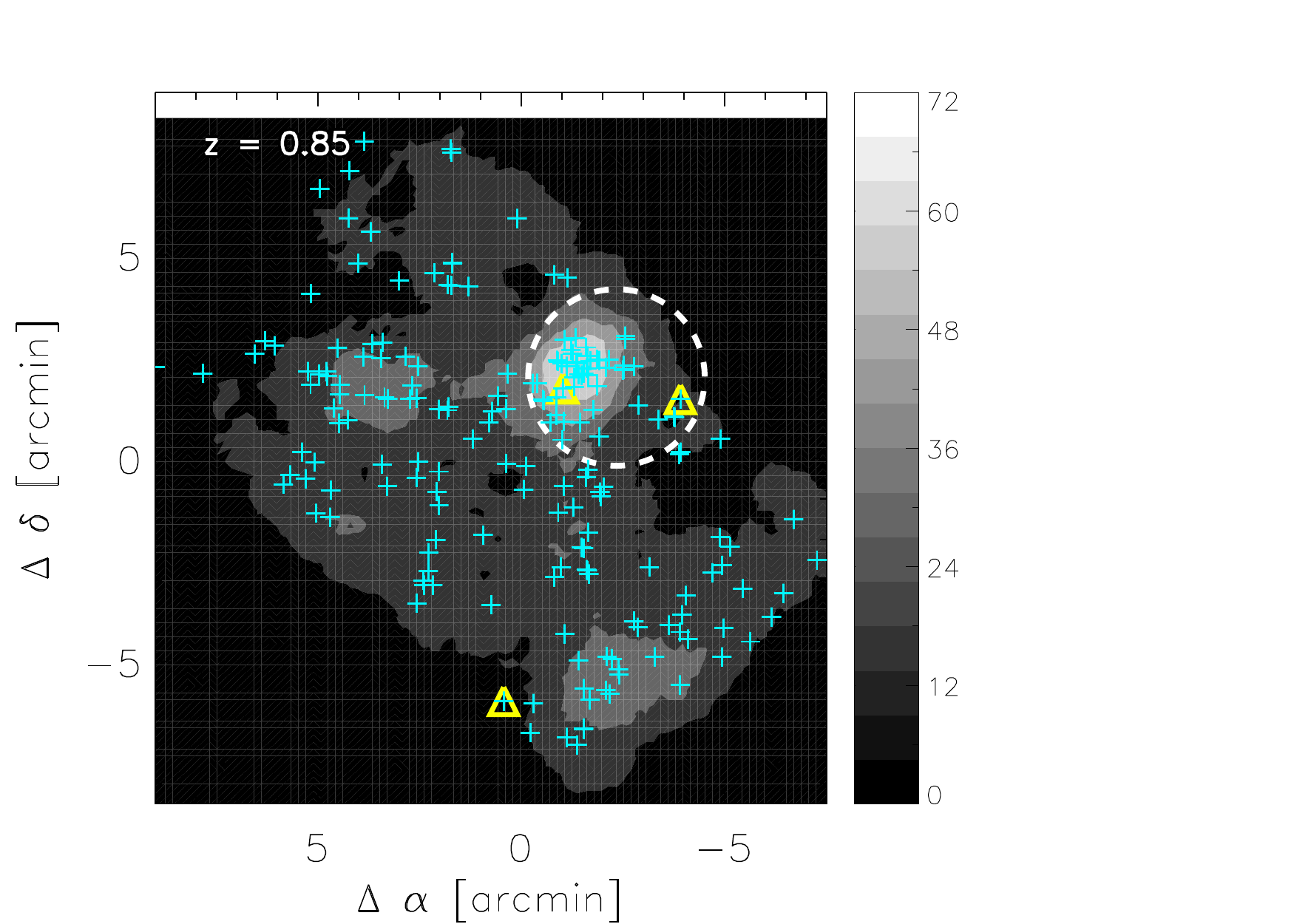}
       \hspace{-0.85in}
\includegraphics[width=81mm]{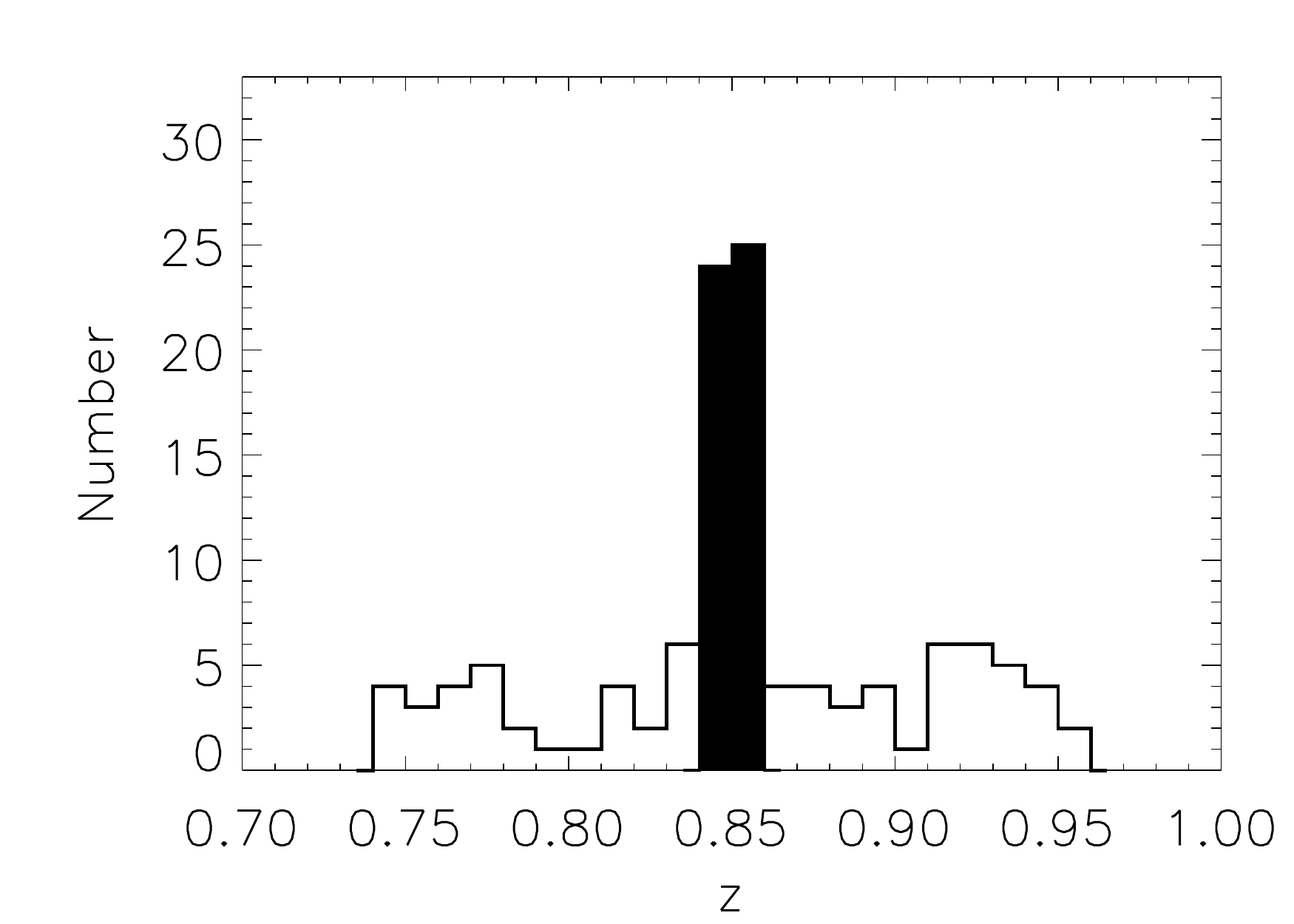}
\\
\includegraphics[width=81mm]{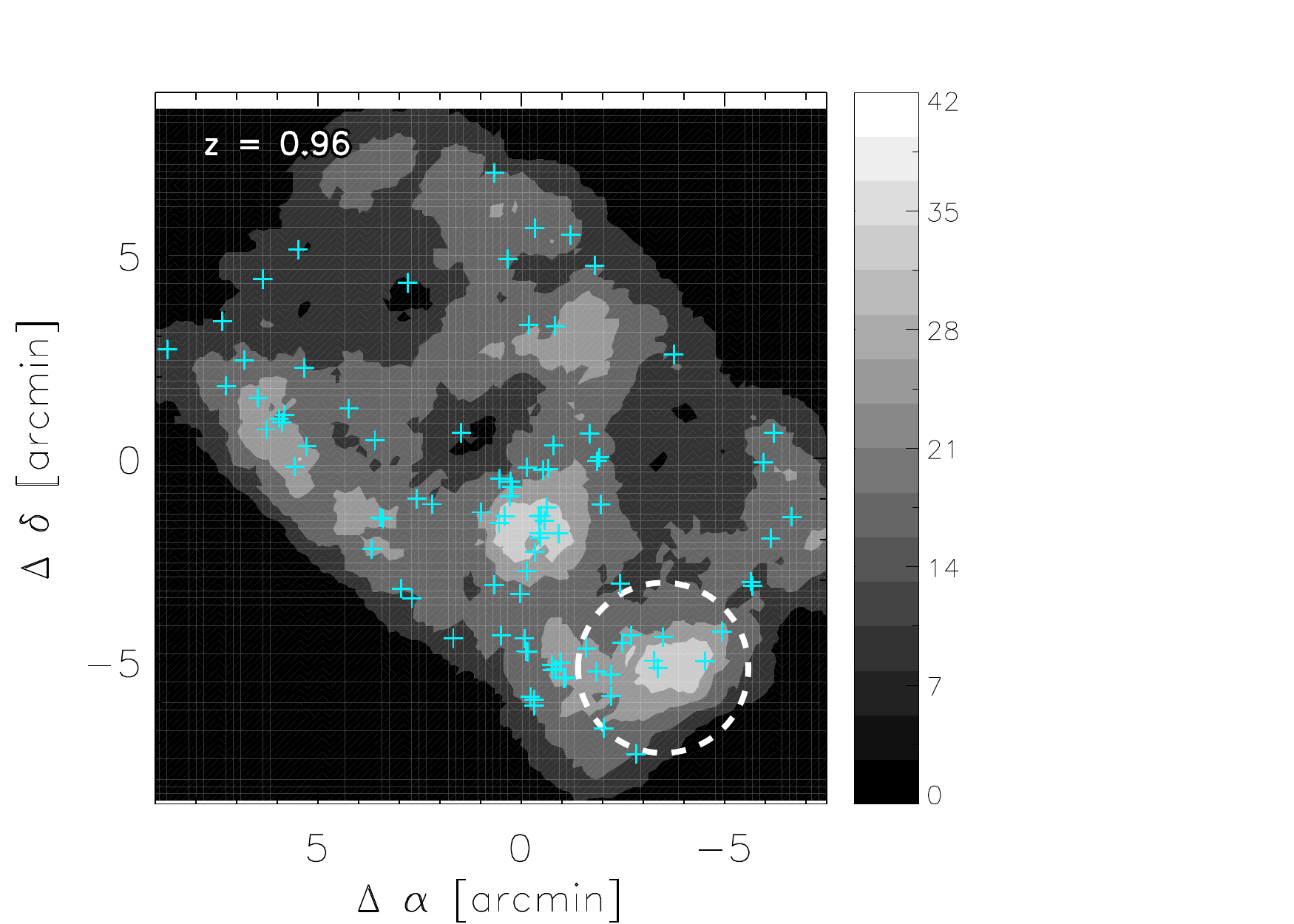}
       \hspace{-0.85in}
\includegraphics[width=81mm]{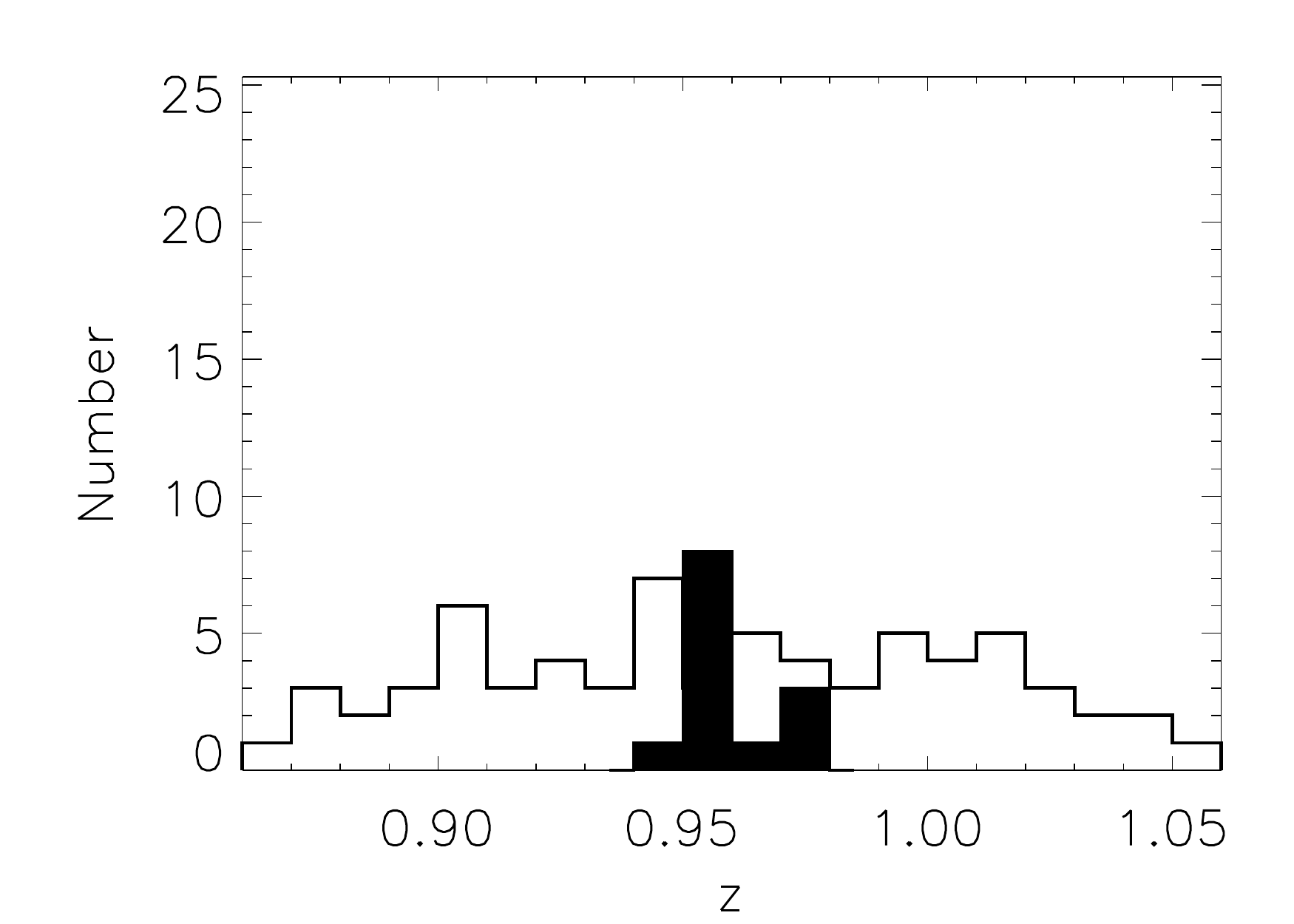}
\caption{(Continued) The yellow open triangles represent radio galaxies (\citealt{17}) within the same redshift interval.}
\label{f5b}
\end{figure*}


\begin{figure*}[tp!]
 \setcounter{figure}{6}
\centering
\includegraphics[width=81mm]{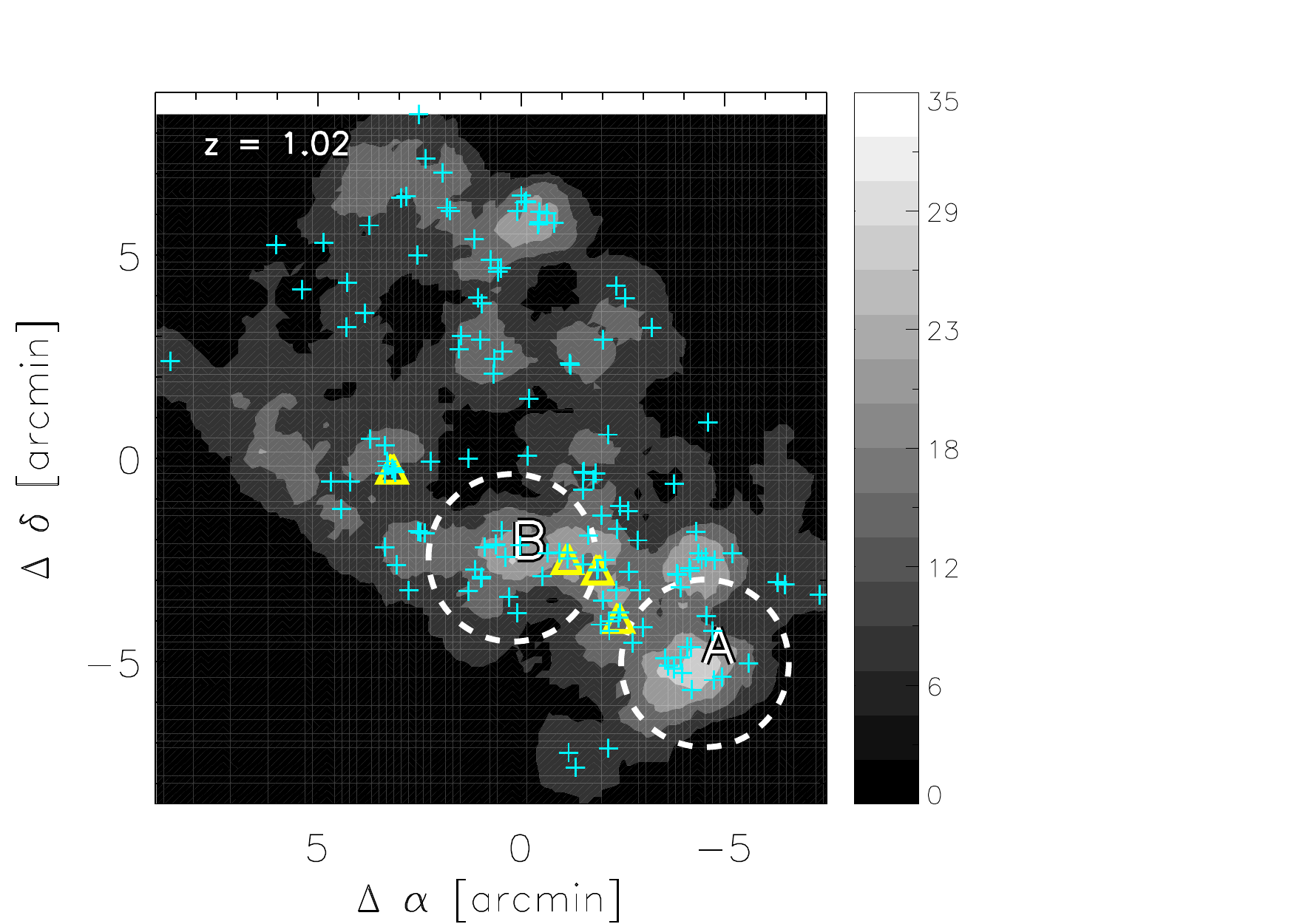}
       \hspace{-0.85in}
\includegraphics[width=81mm]{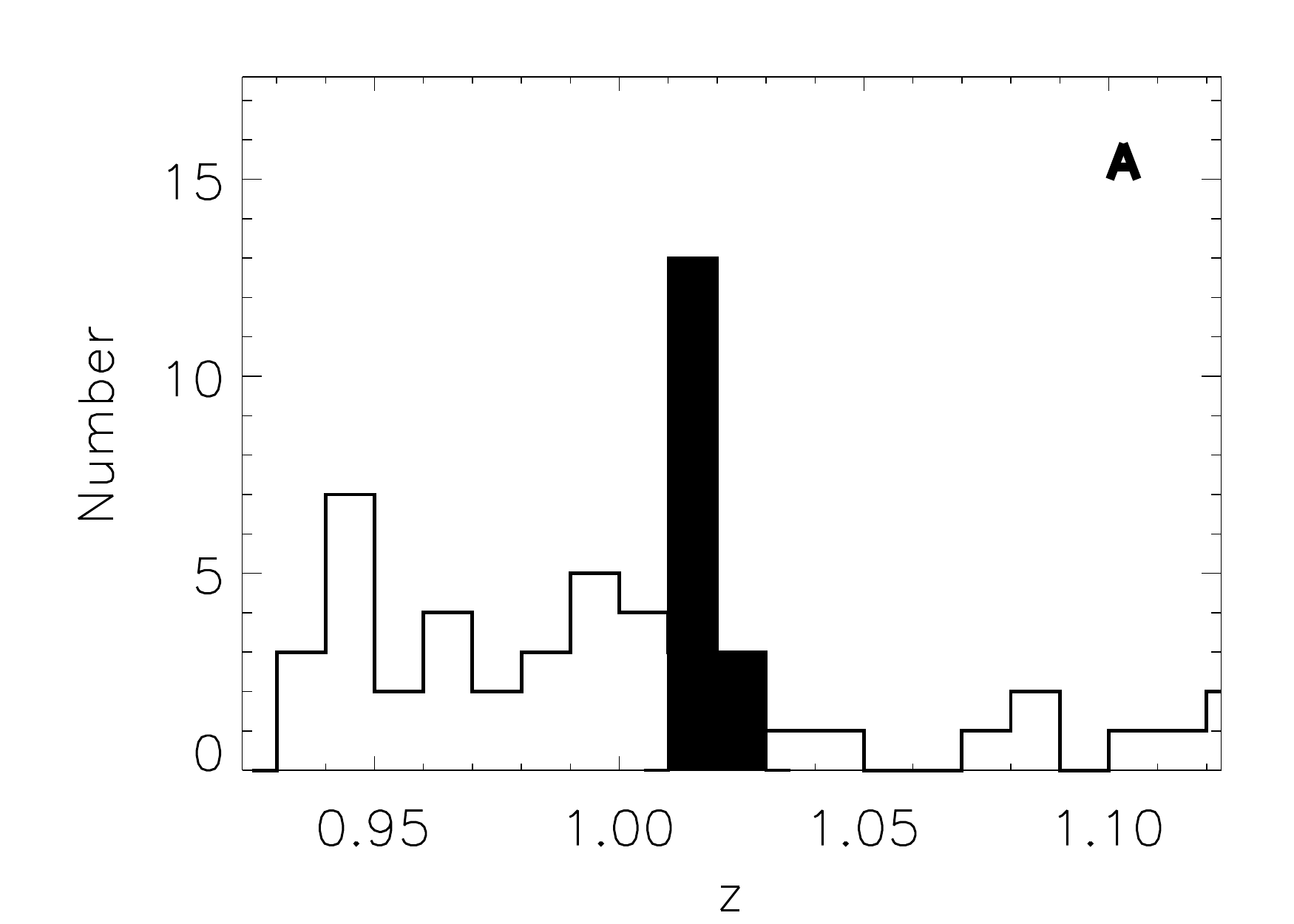} \\
\includegraphics[width=81mm]{f7h1-eps-converted-to.pdf}
       \hspace{-0.85in}
\includegraphics[width=81mm]{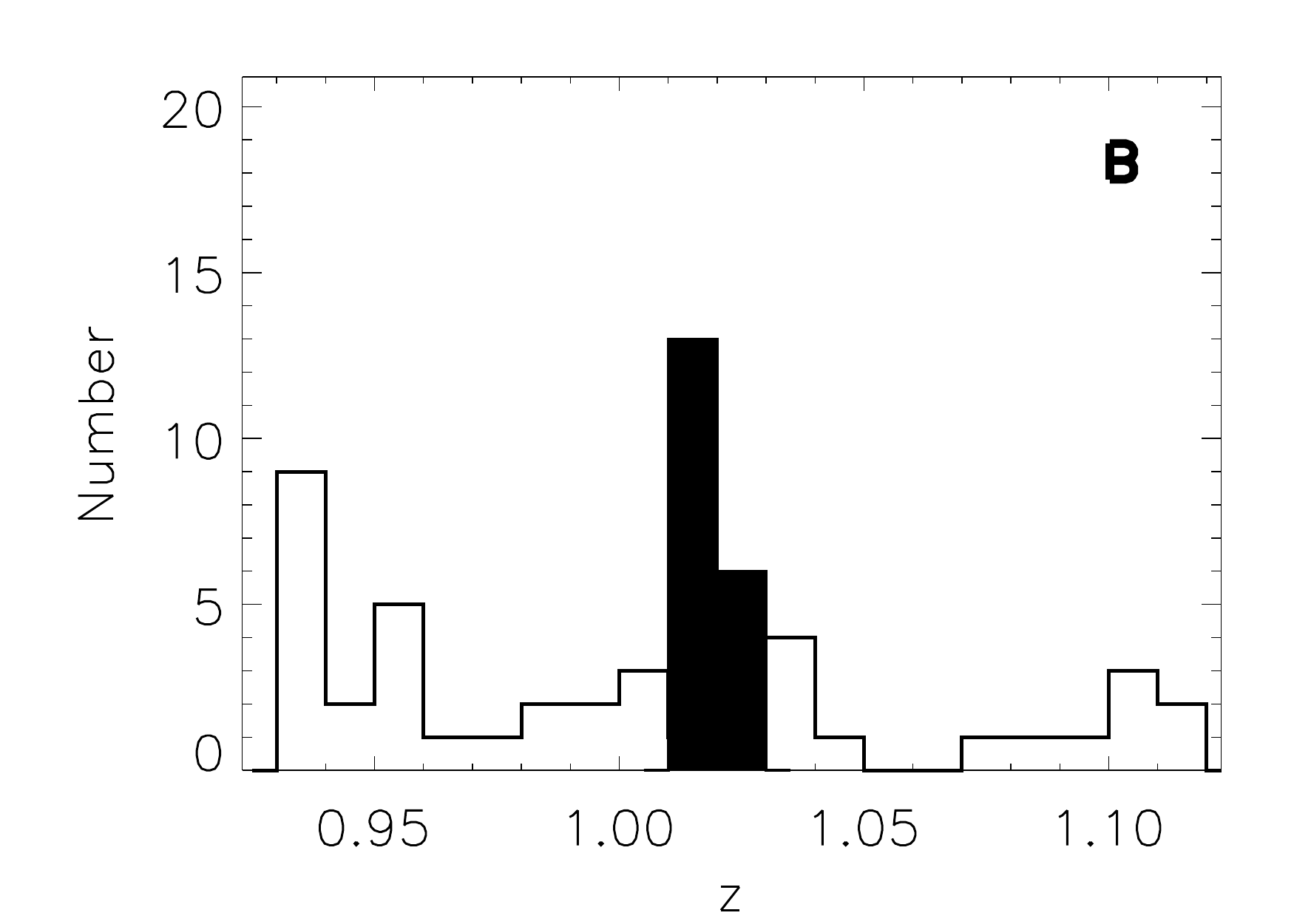}
\\
\includegraphics[width=81mm]{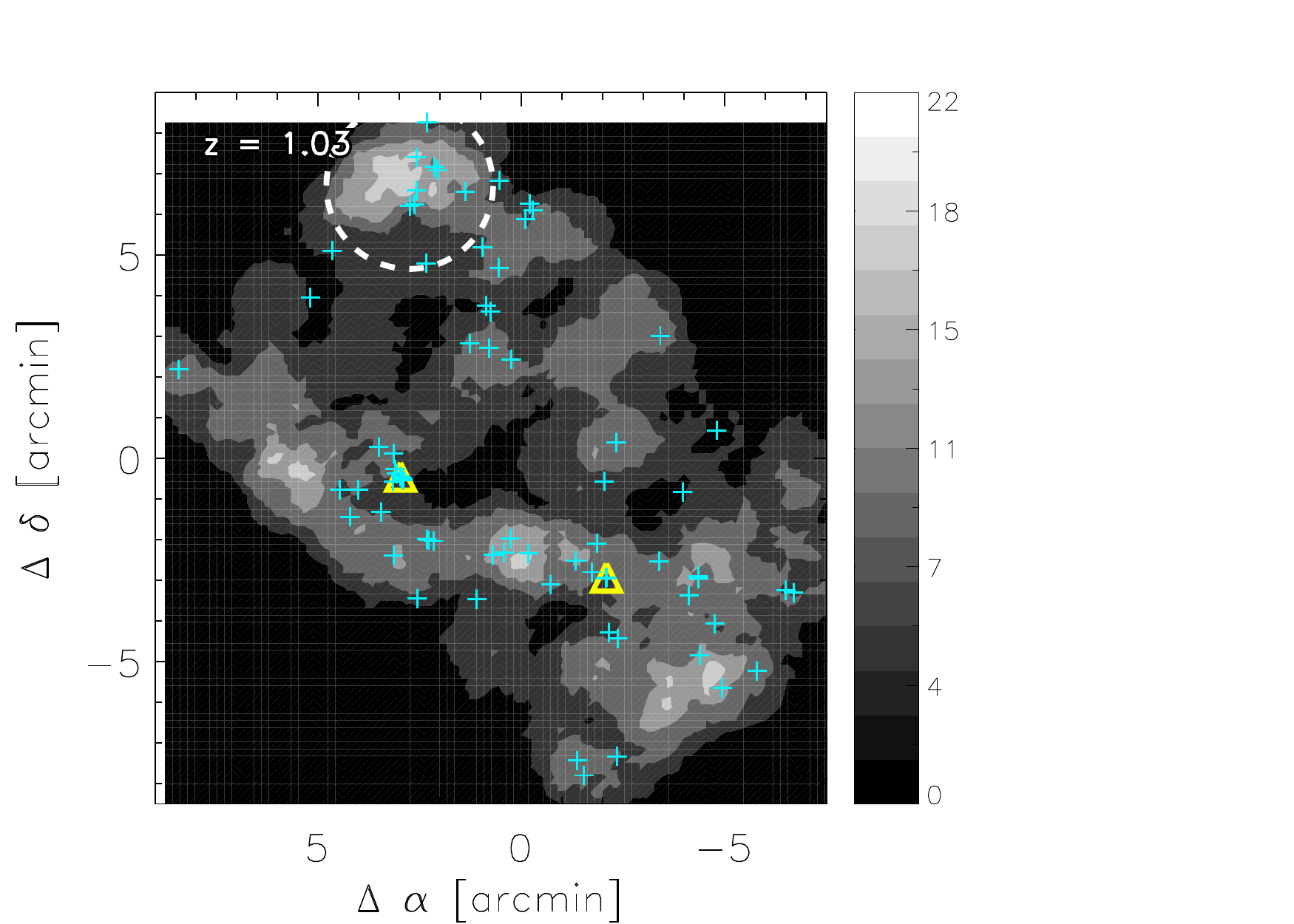}
       \hspace{-0.85in}
\includegraphics[width=81mm]{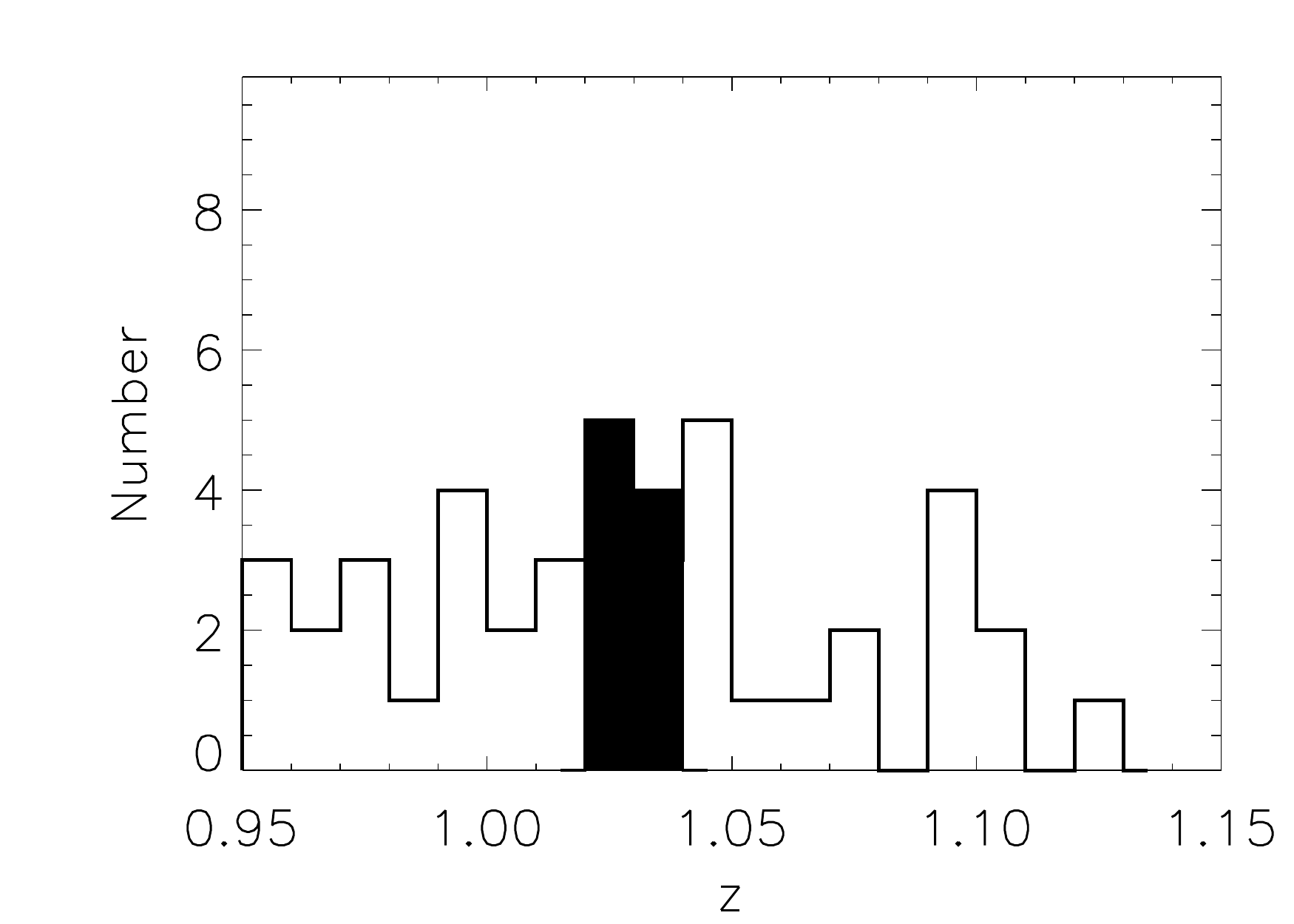}
\\
\includegraphics[width=81mm]{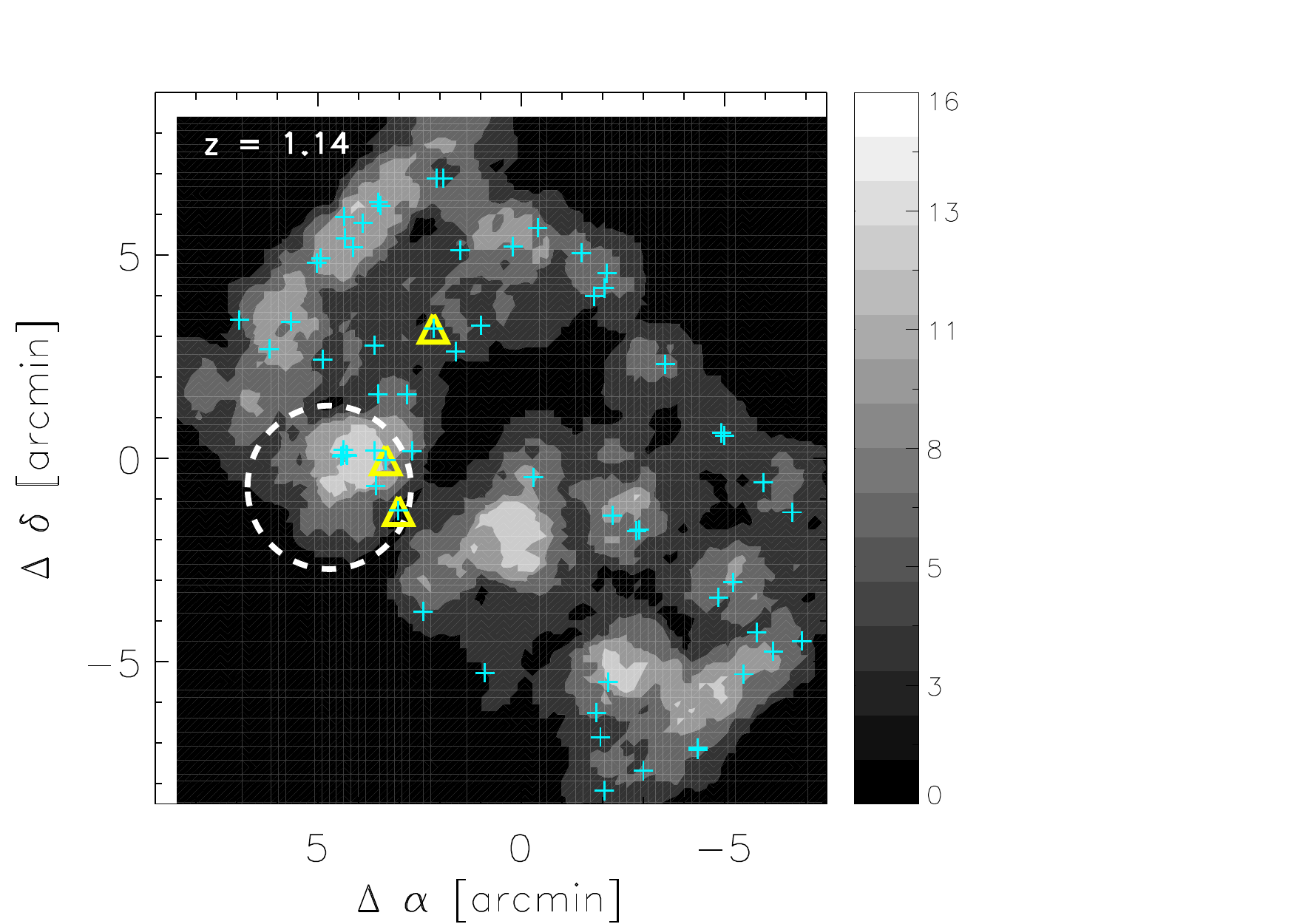}
       \hspace{-0.85in}
\includegraphics[width=81mm]{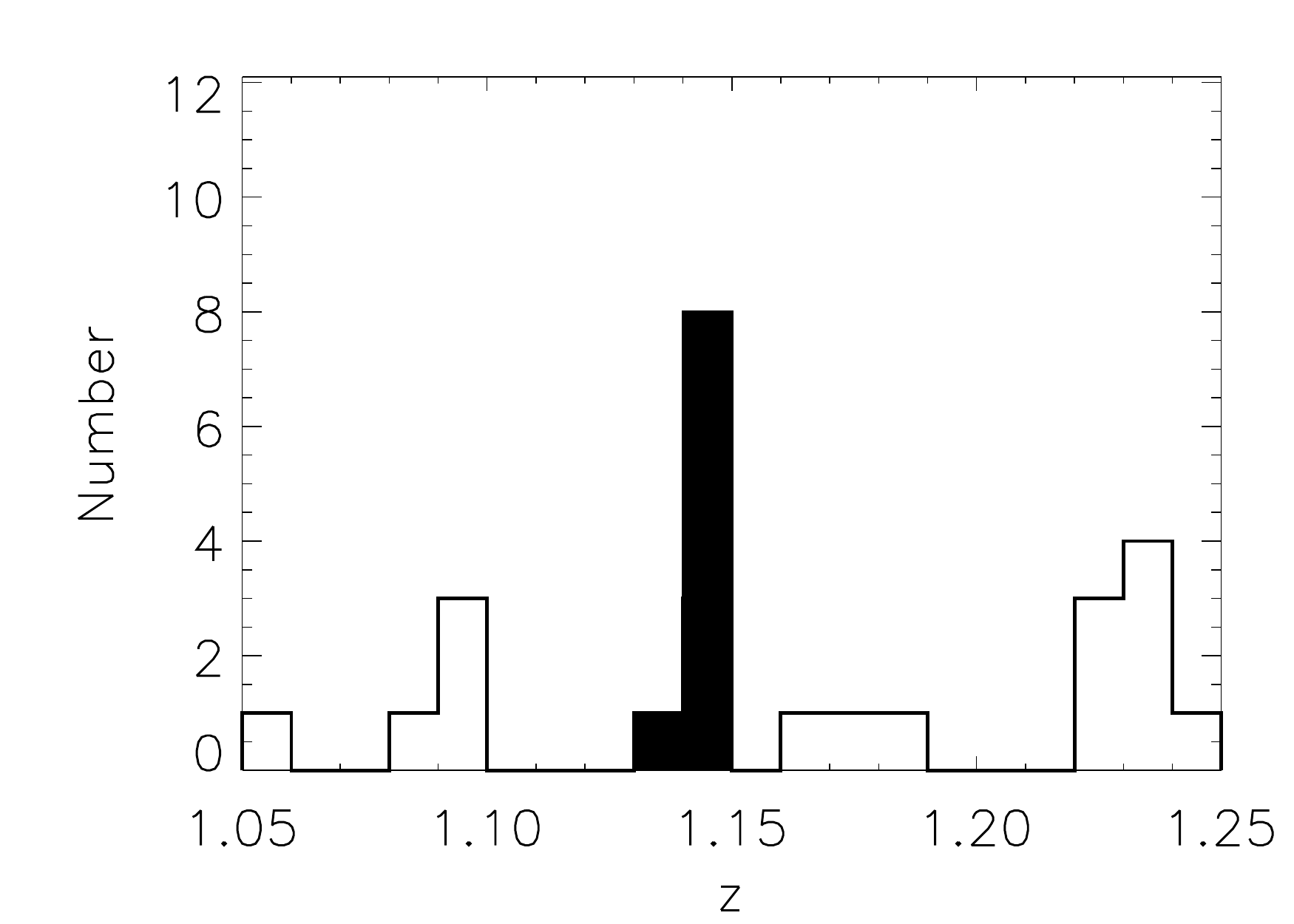}
\caption{(Continued)}
\label{f5c}
\end{figure*}


\begin{figure*}[tp!]
 \setcounter{figure}{6}
\centering
\includegraphics[width=81mm]{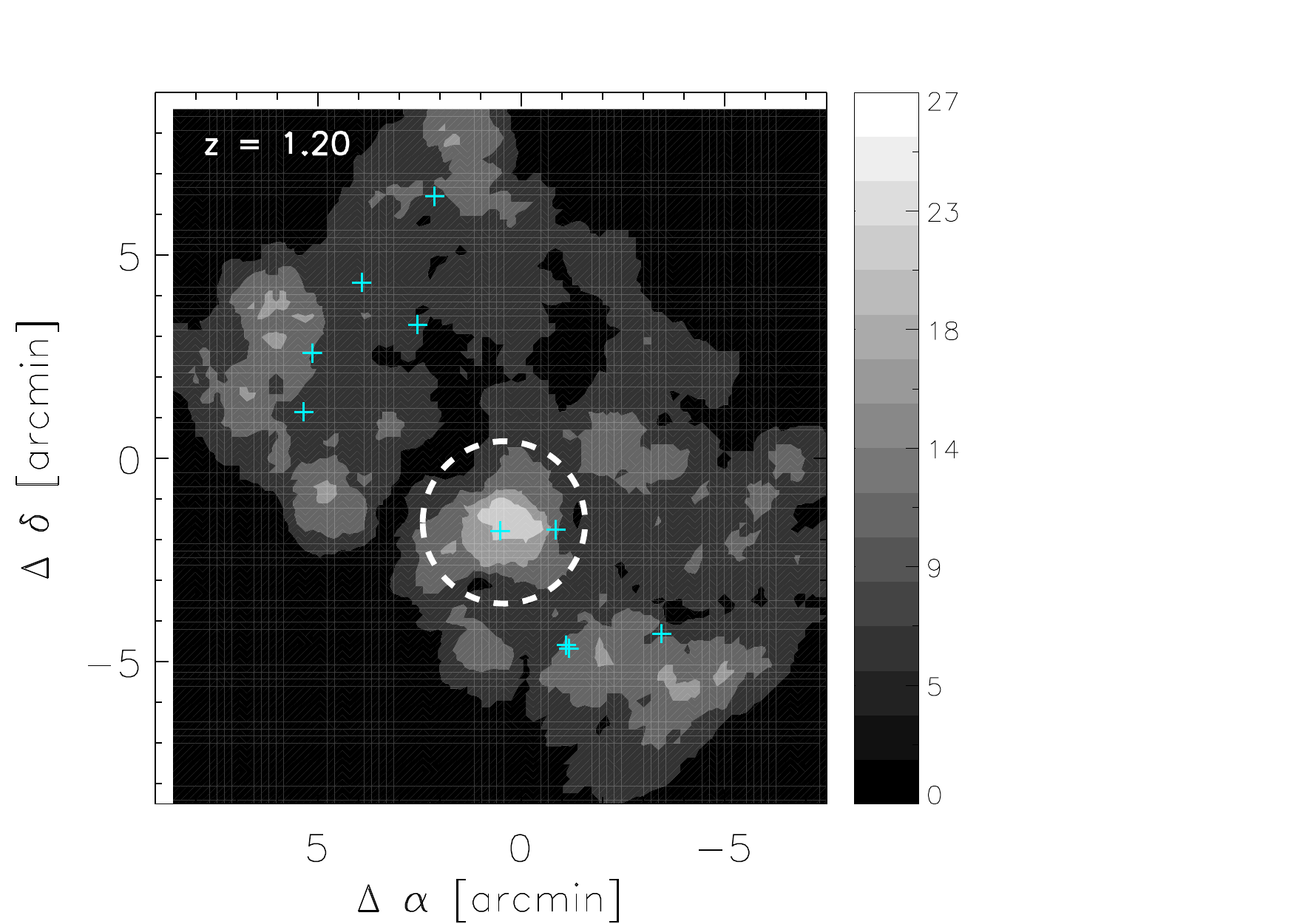}
       \hspace{-0.85in}
\includegraphics[width=81mm]{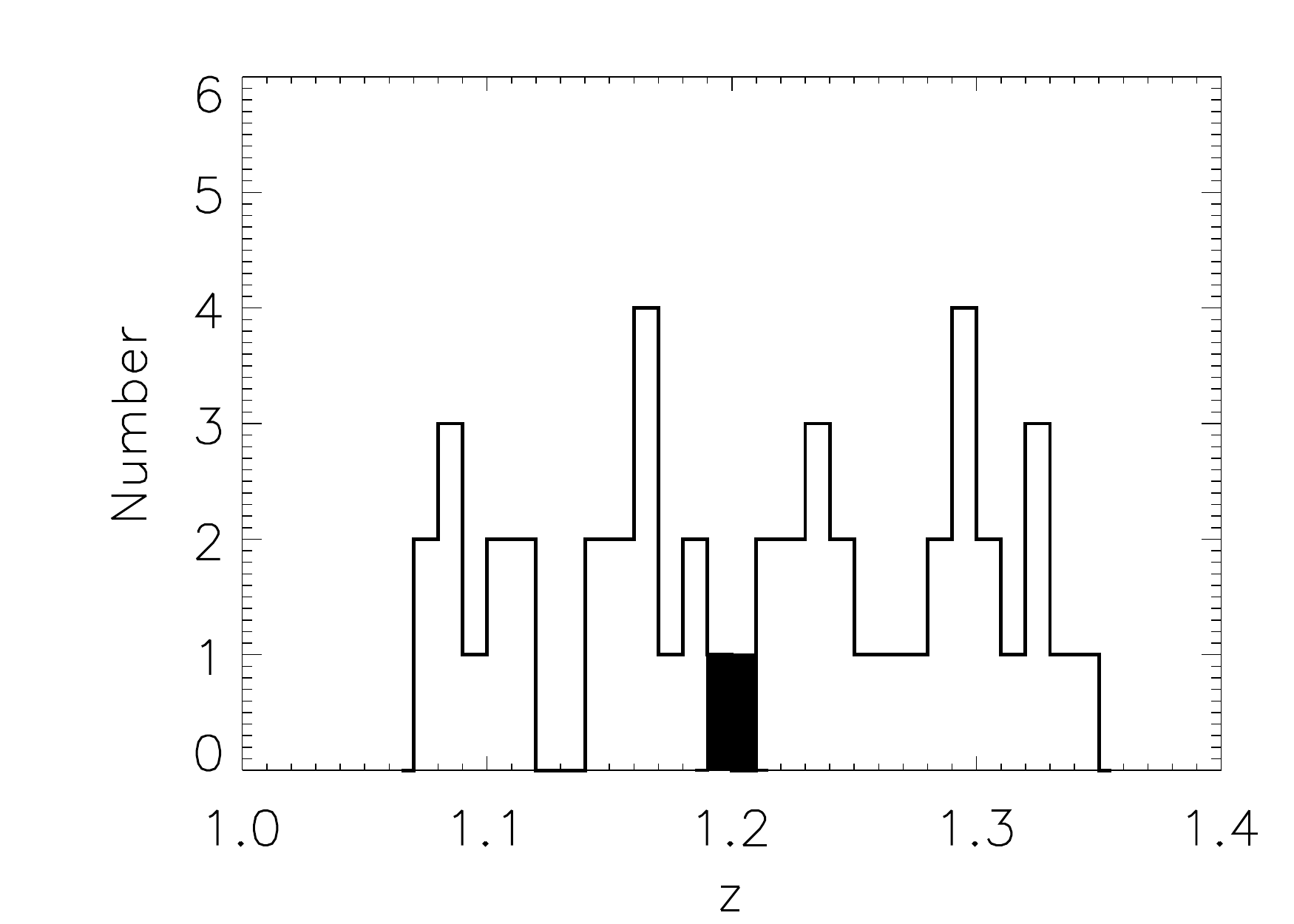}
\\
\includegraphics[width=81mm]{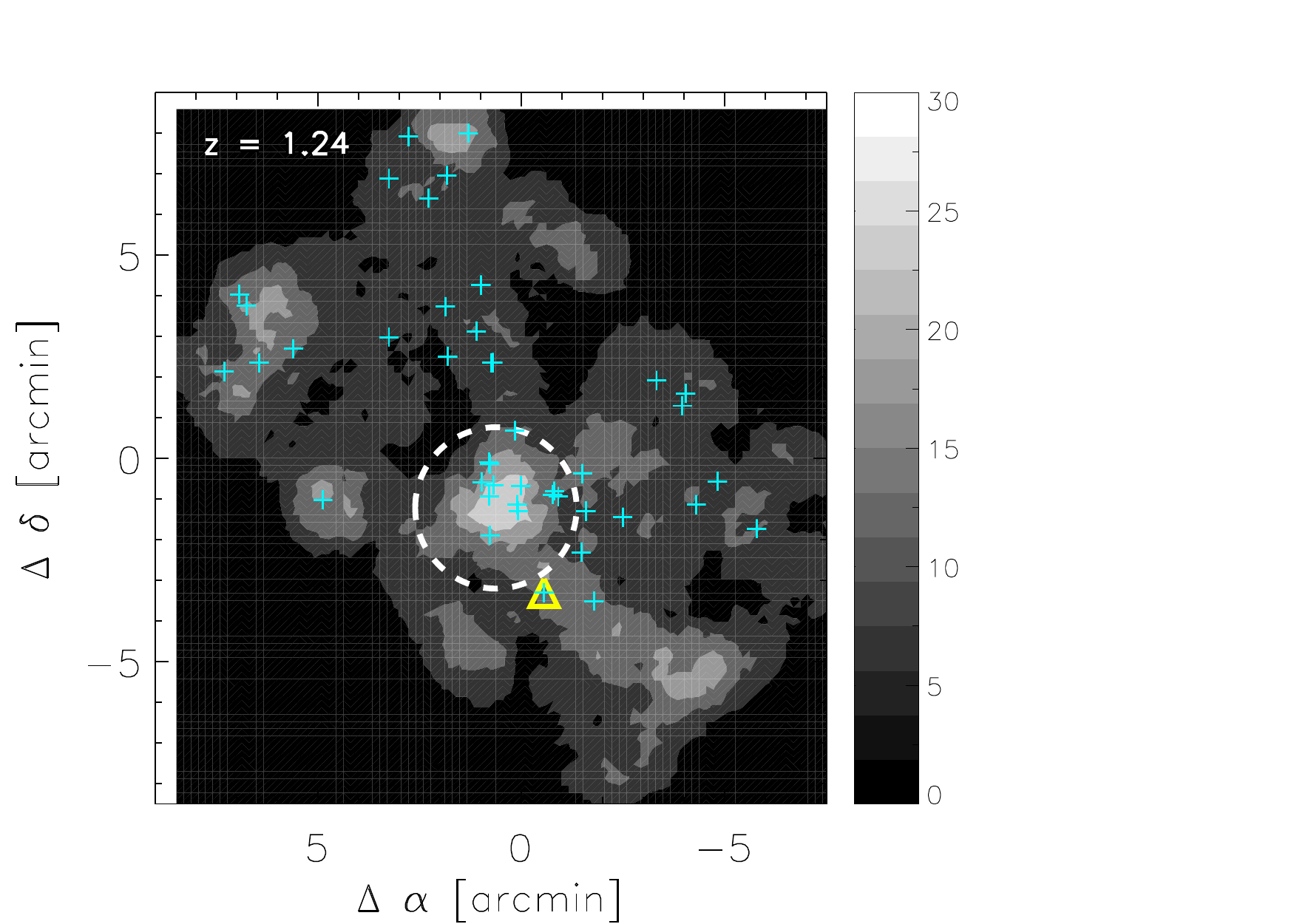}
       \hspace{-0.85in}
\includegraphics[width=81mm]{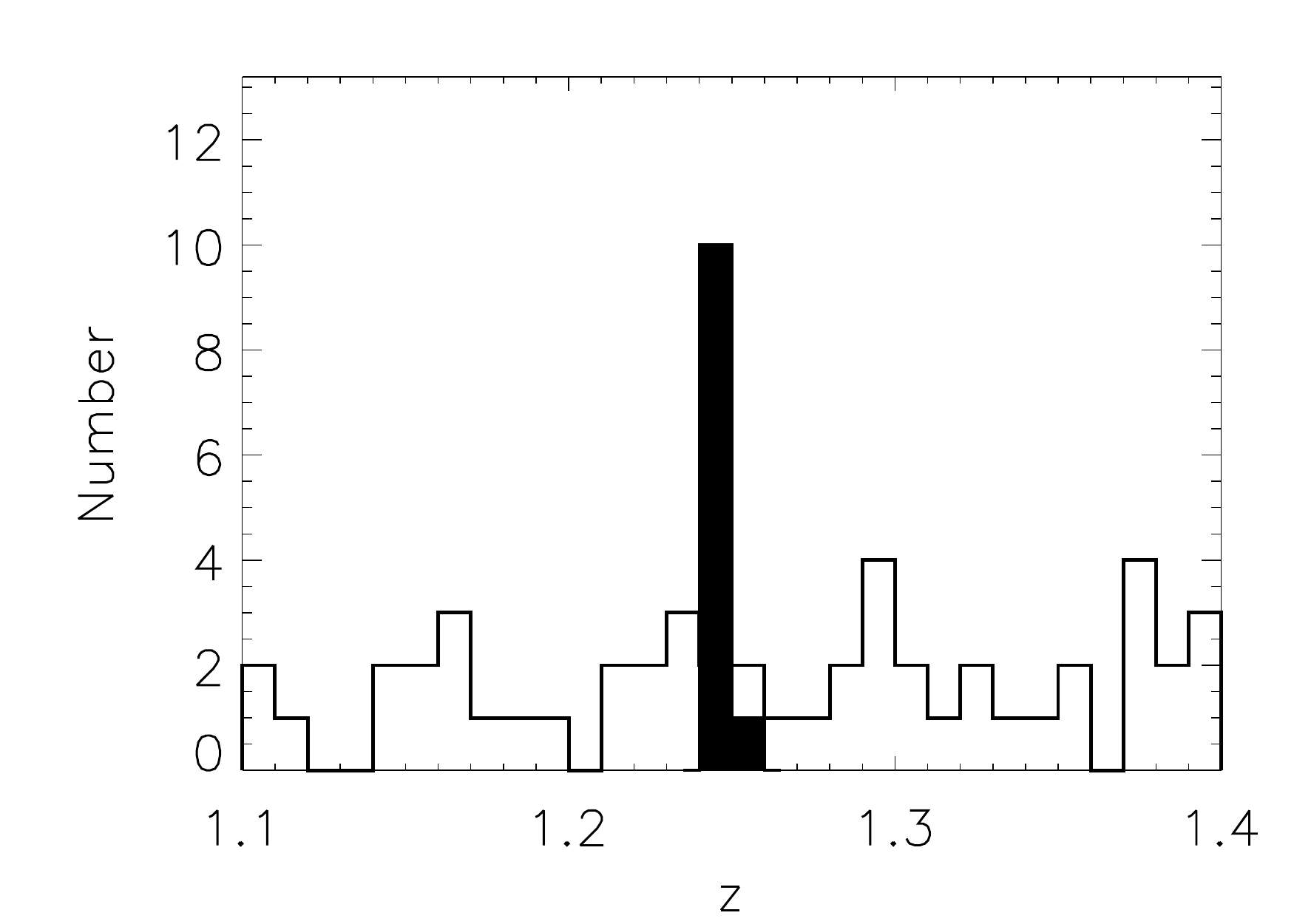}
\\
\includegraphics[width=81mm]{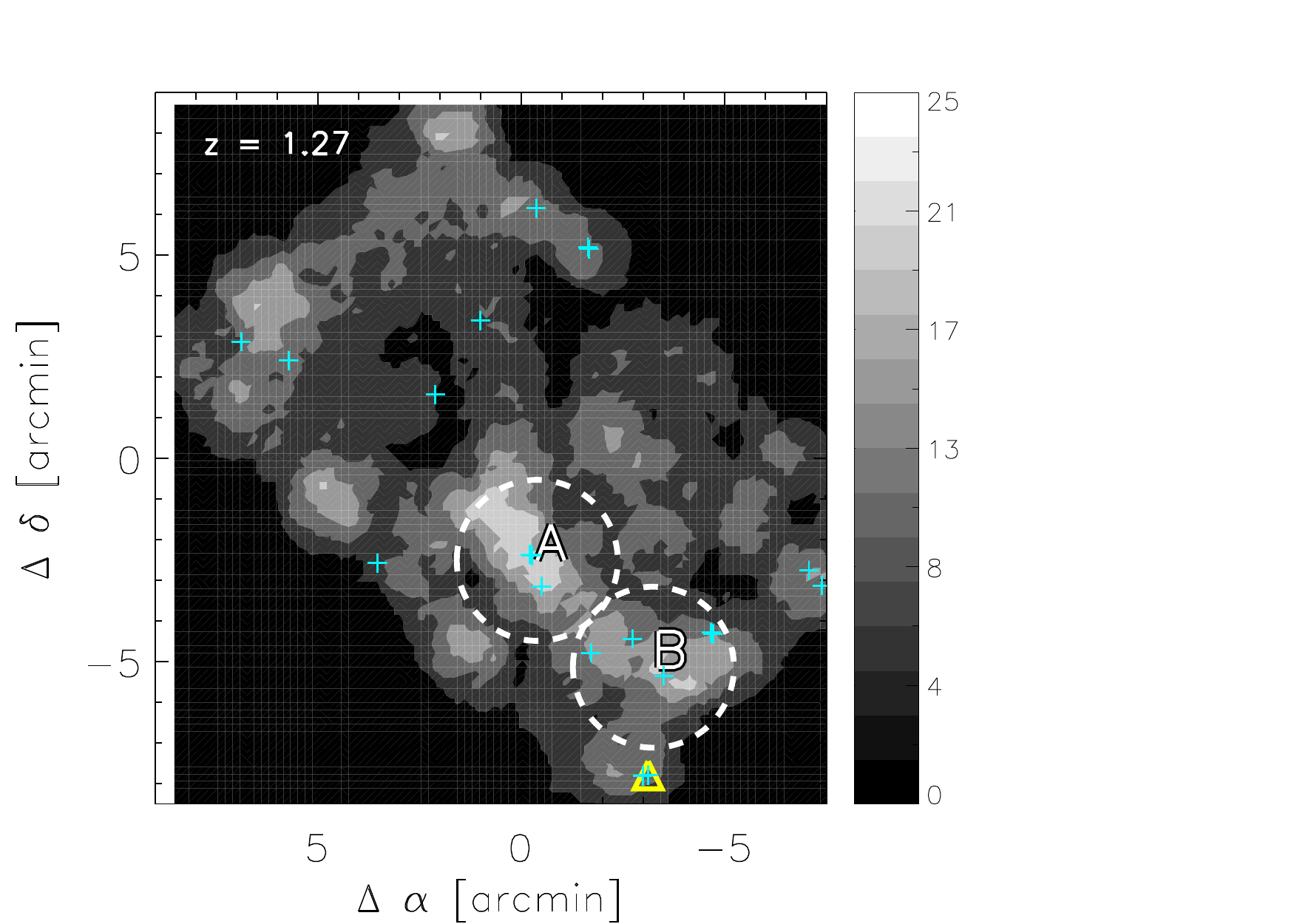}
       \hspace{-0.85in}
\includegraphics[width=81mm]{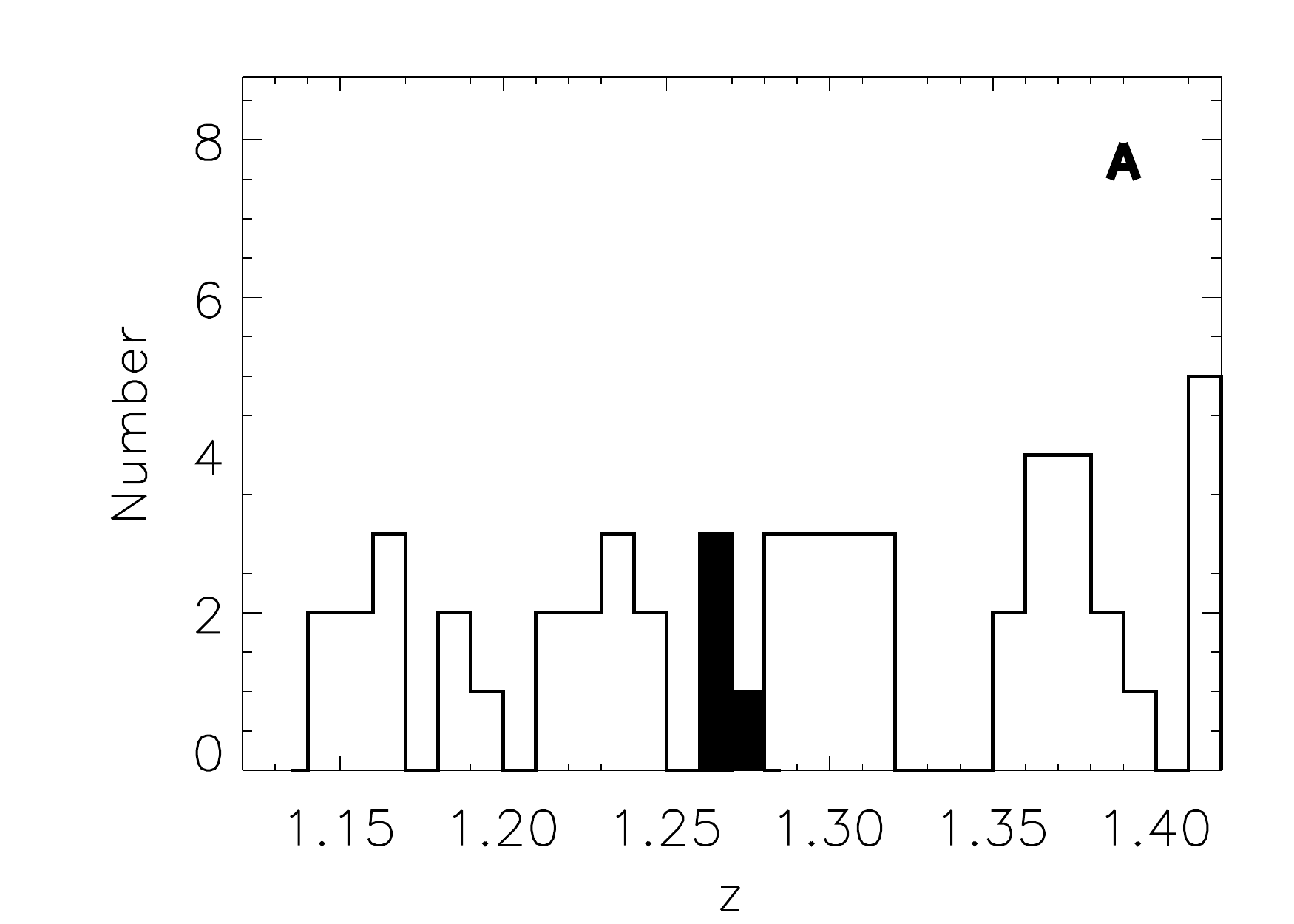} \\
\includegraphics[width=81mm]{f7m1-eps-converted-to.pdf}
       \hspace{-0.85in}
\includegraphics[width=81mm]{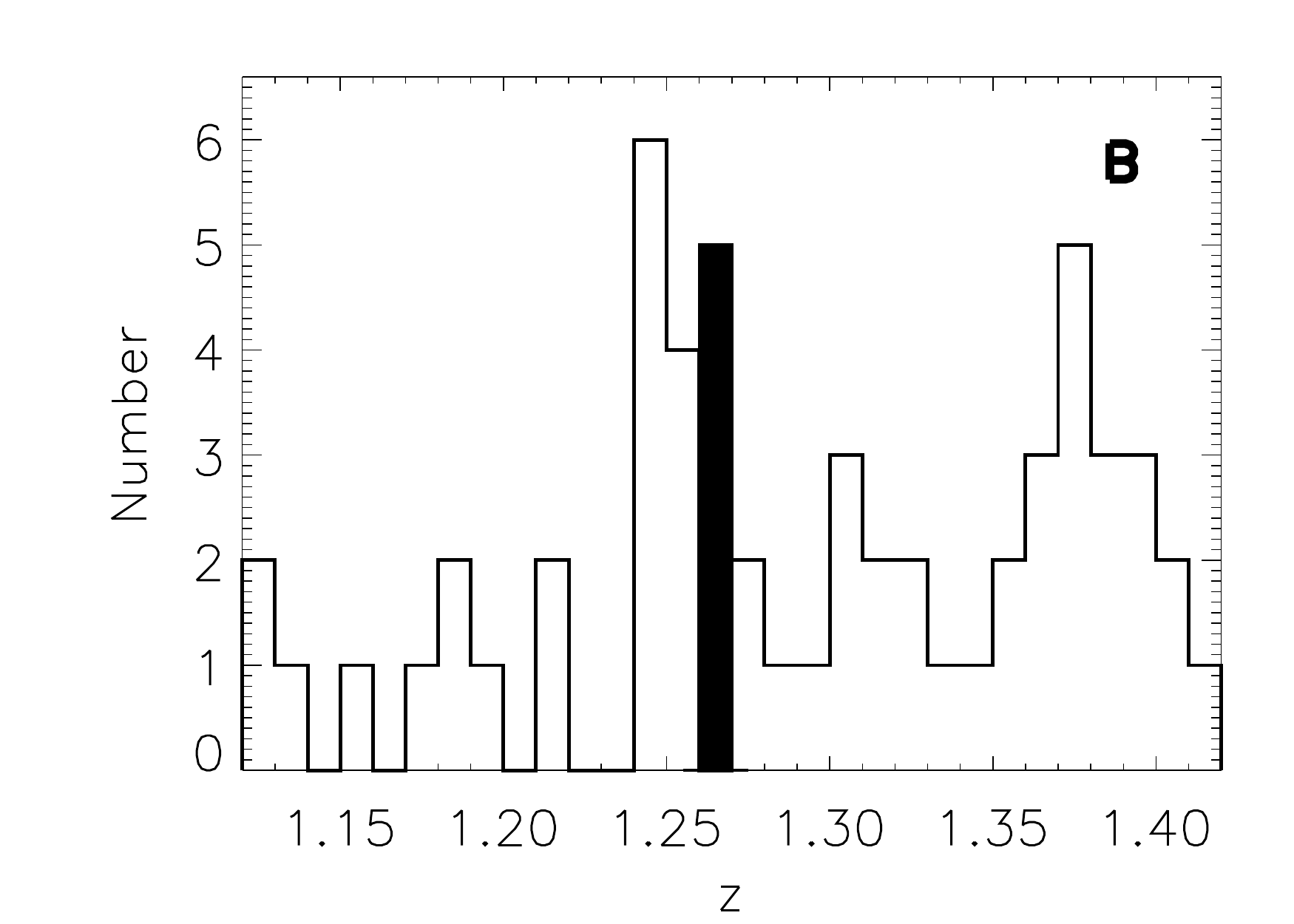}
\caption{(Continued)}
\label{f5d}
\end{figure*}


\begin{figure*}[tp!]
 \setcounter{figure}{6}
\centering
\includegraphics[width=81mm]{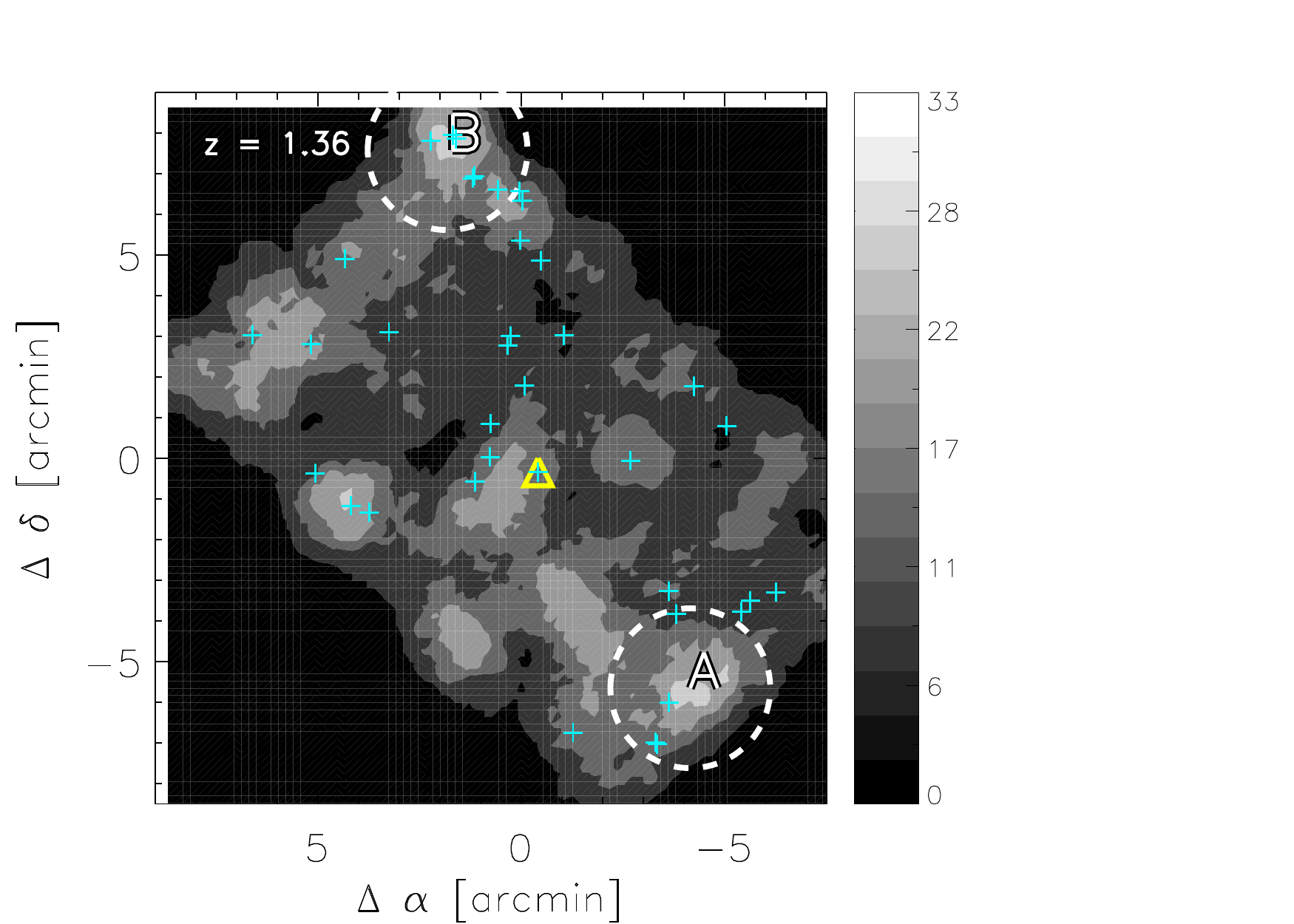}
       \hspace{-0.85in}
\includegraphics[width=81mm]{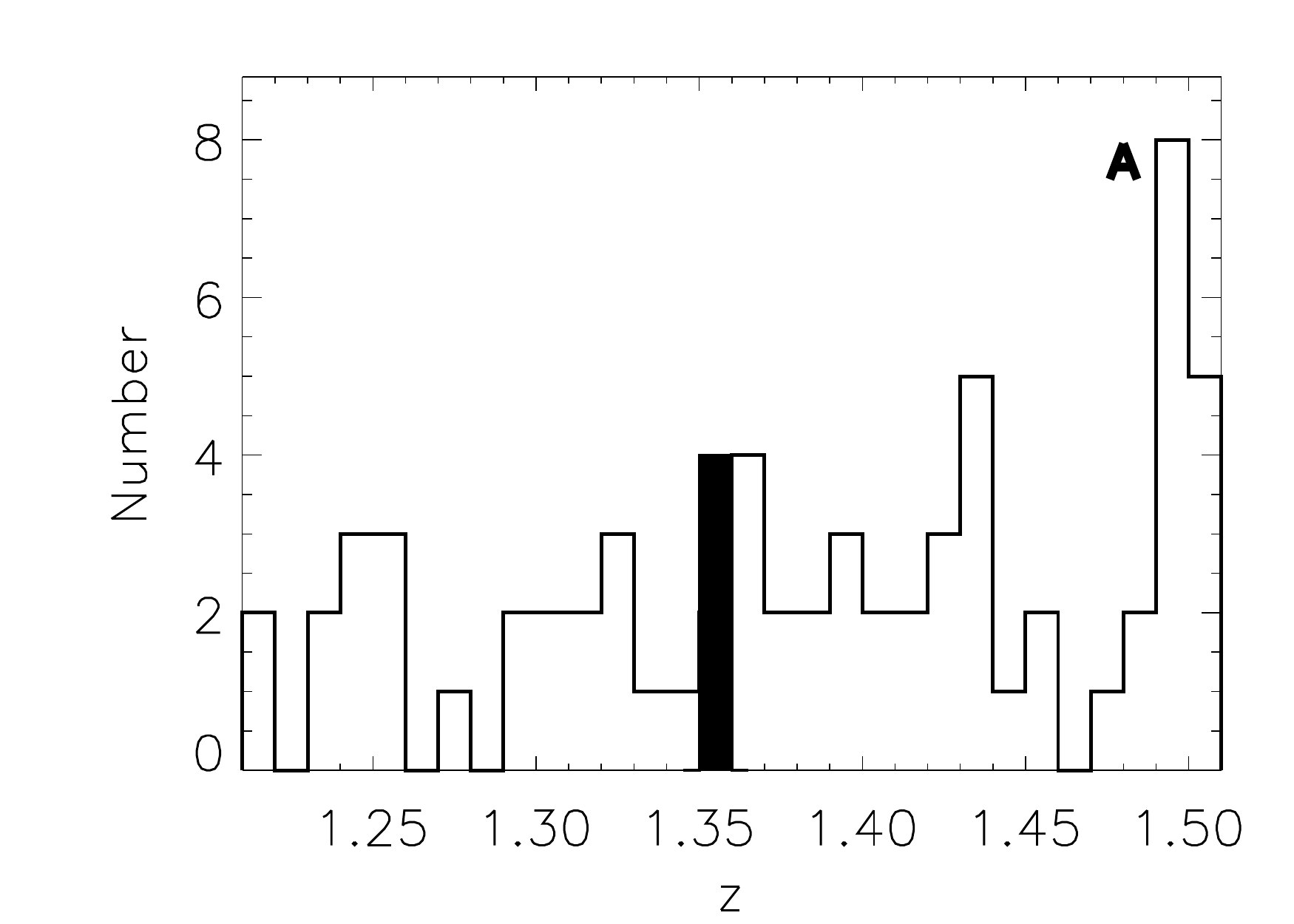} \\
\includegraphics[width=81mm]{f7n1-eps-converted-to.pdf}
       \hspace{-0.85in}
    \includegraphics[width=81mm]{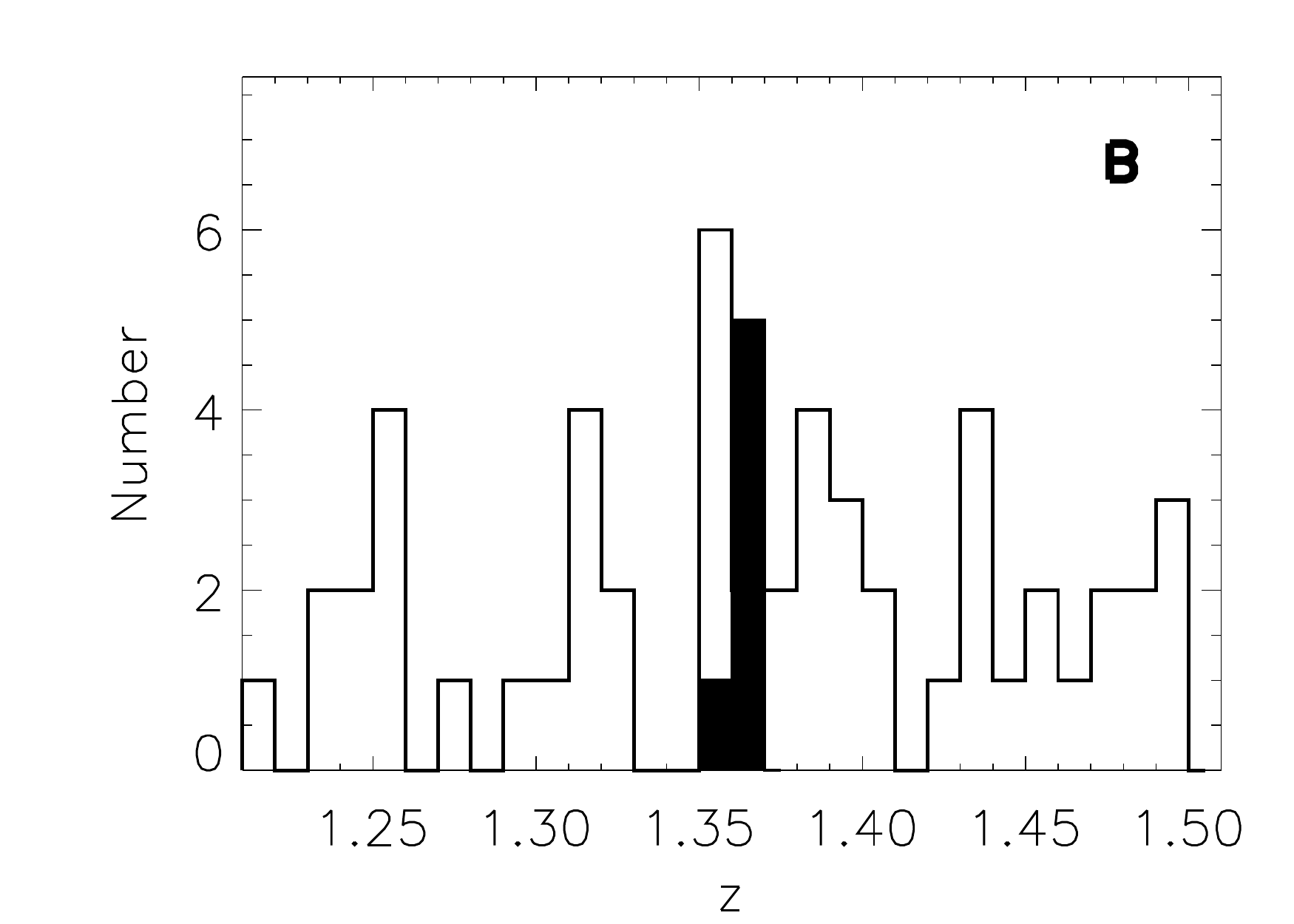}
\\
\includegraphics[width=81mm]{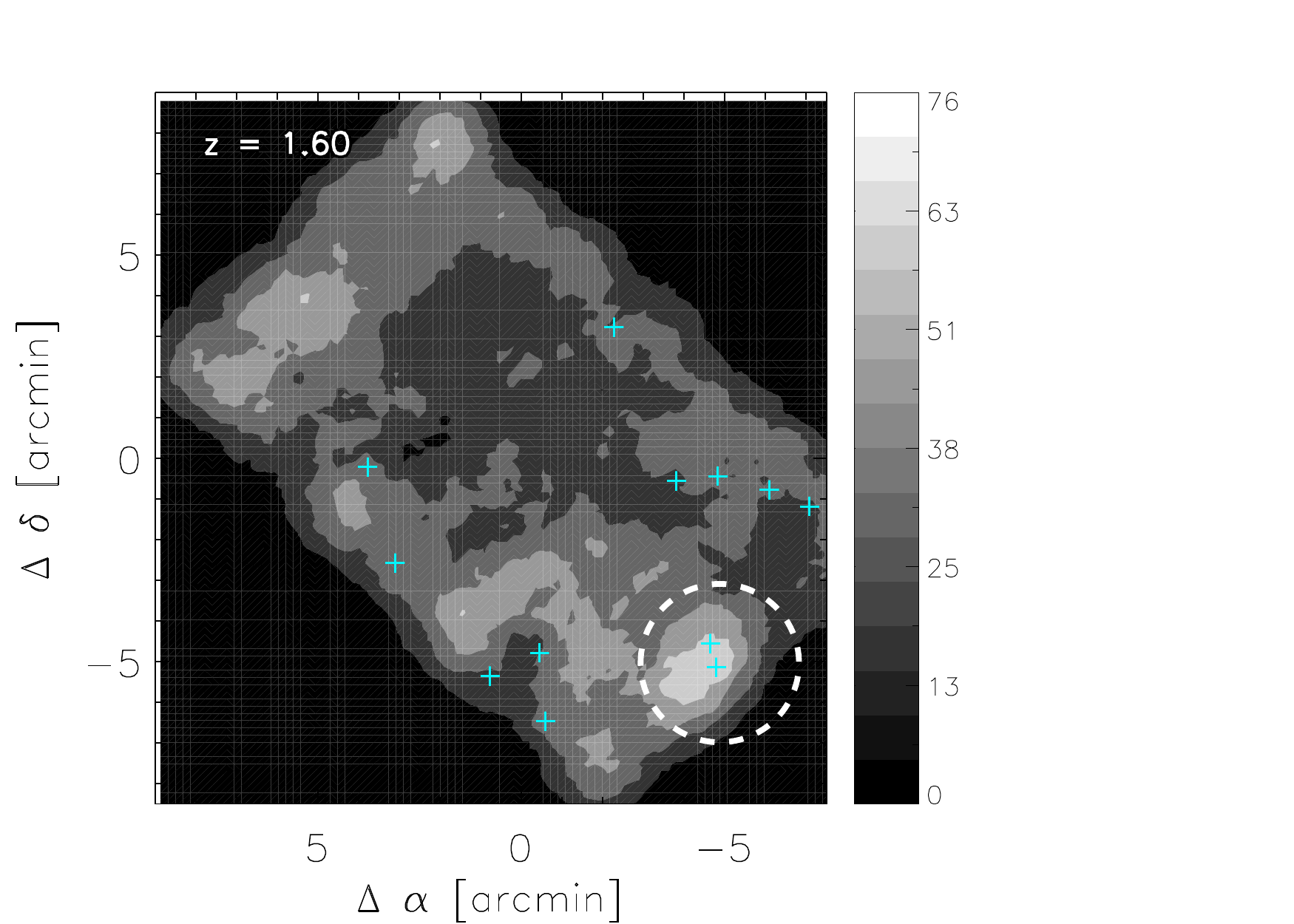}
       \hspace{-0.85in}
\includegraphics[width=81mm]{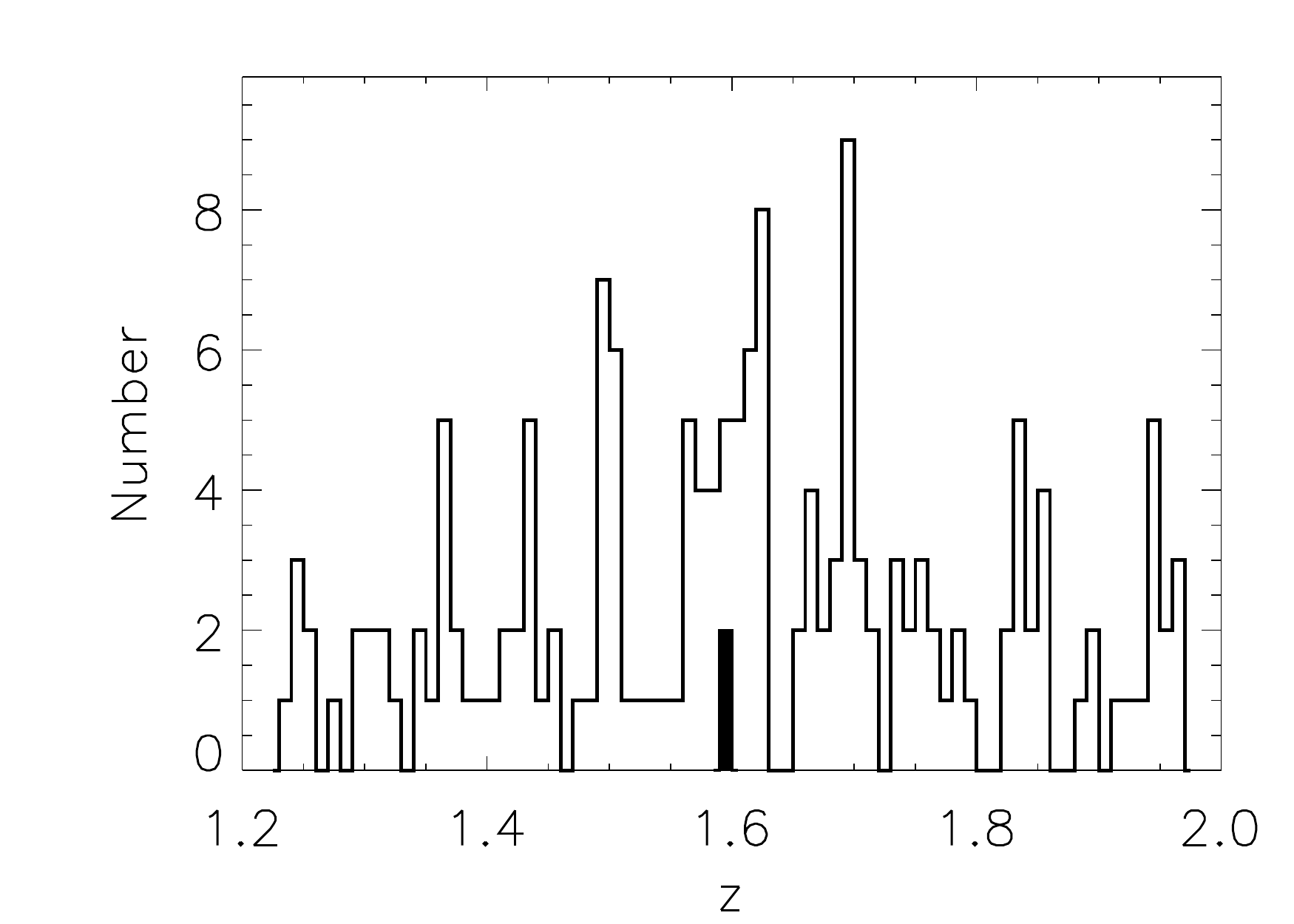}

\includegraphics[width=81mm]{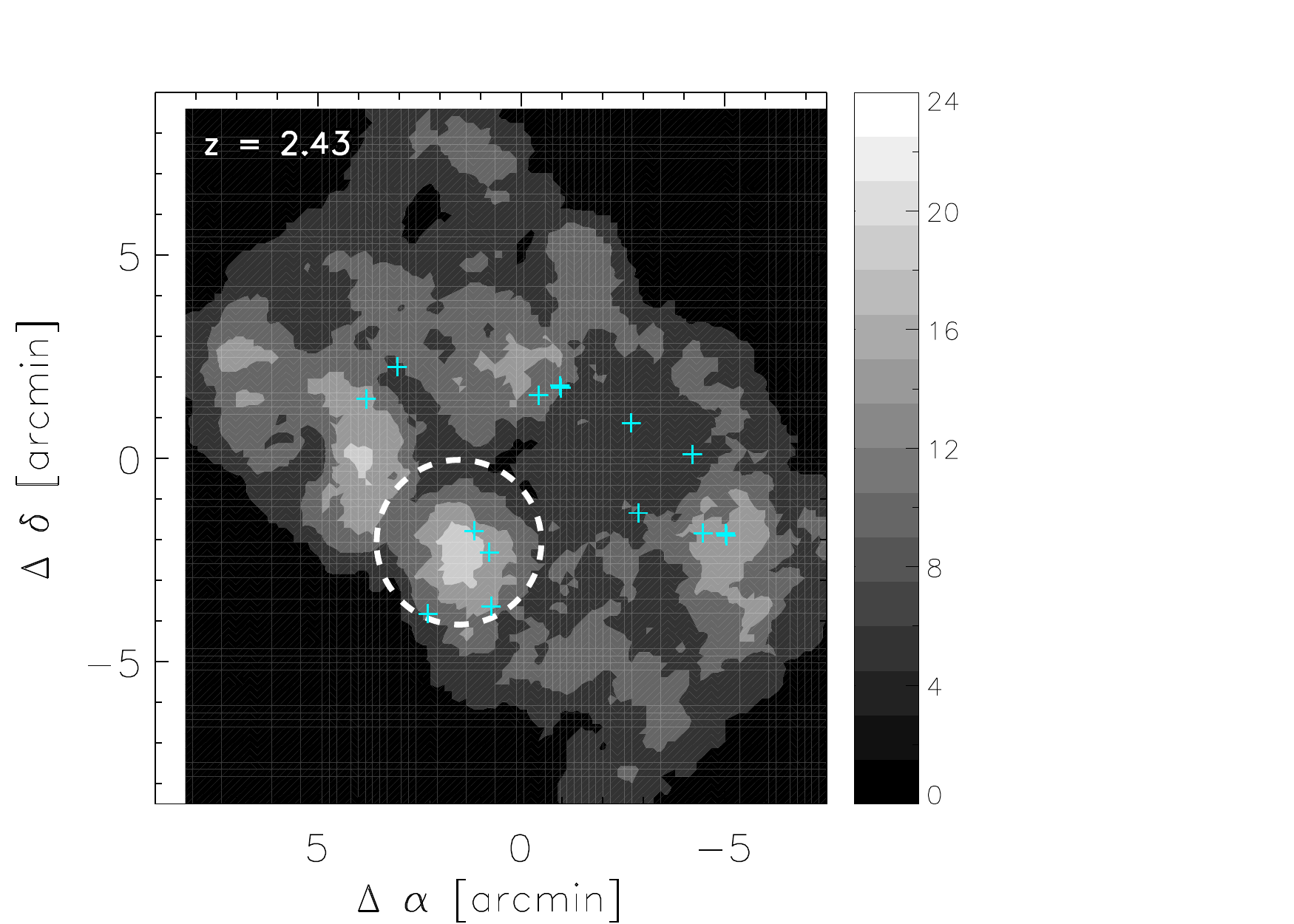}
       \hspace{-0.85in}
\includegraphics[width=81mm]{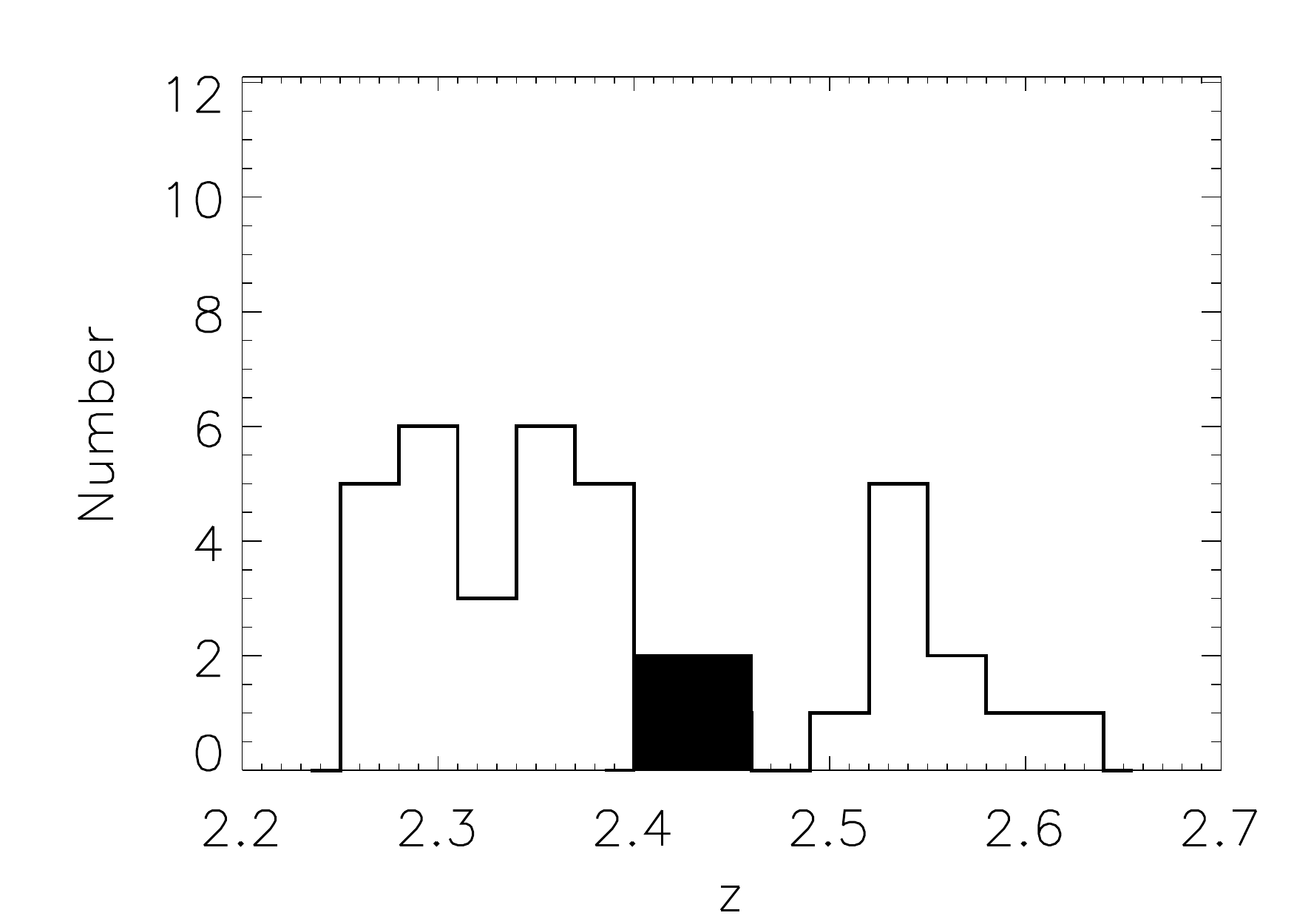}
\caption{(Continued)}
\label{f5e1}
\end{figure*}


\begin{figure*}[tp!]
 \setcounter{figure}{6}
\centering
\includegraphics[width=81mm]{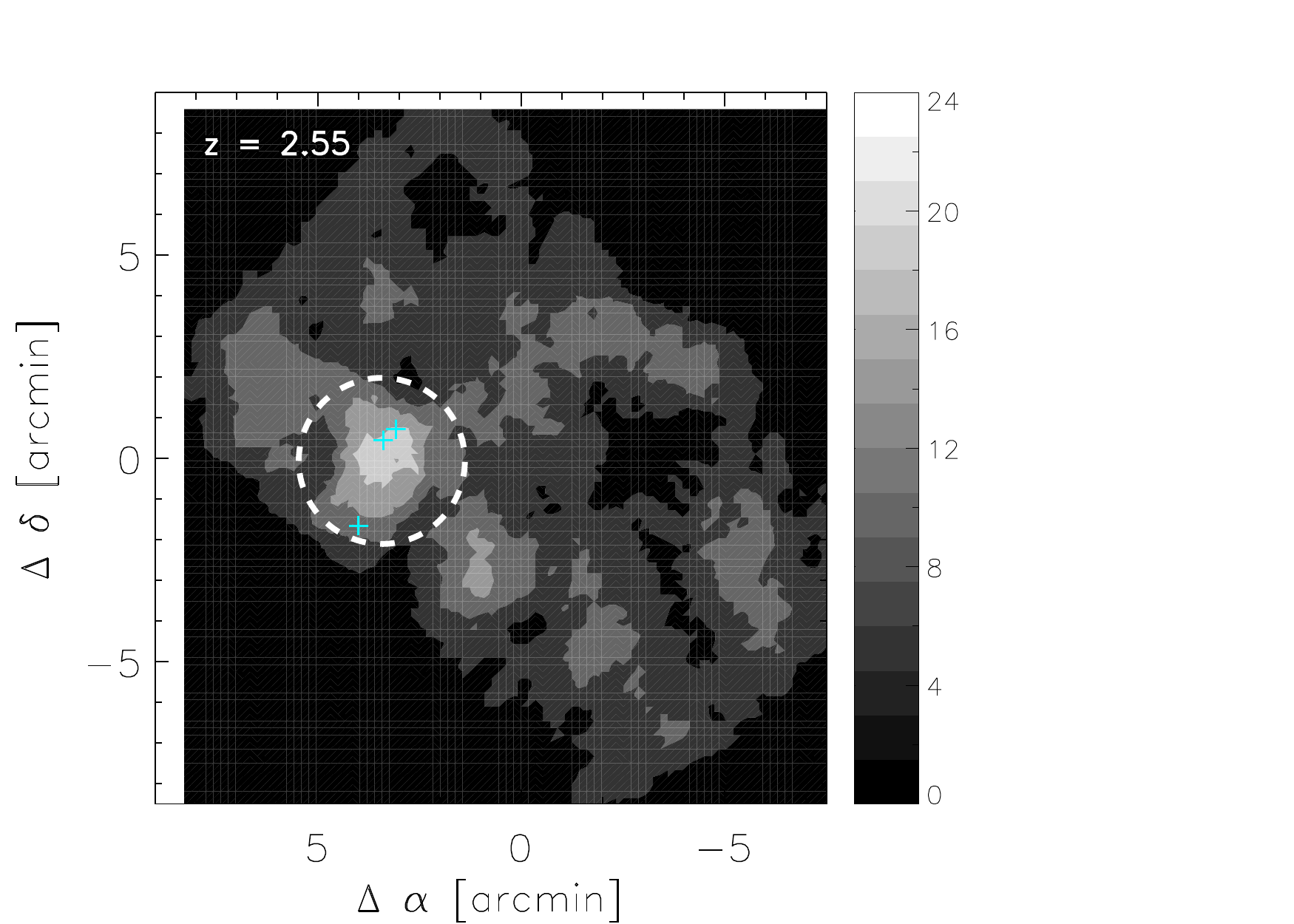}
       \hspace{-0.85in}
\includegraphics[width=81mm]{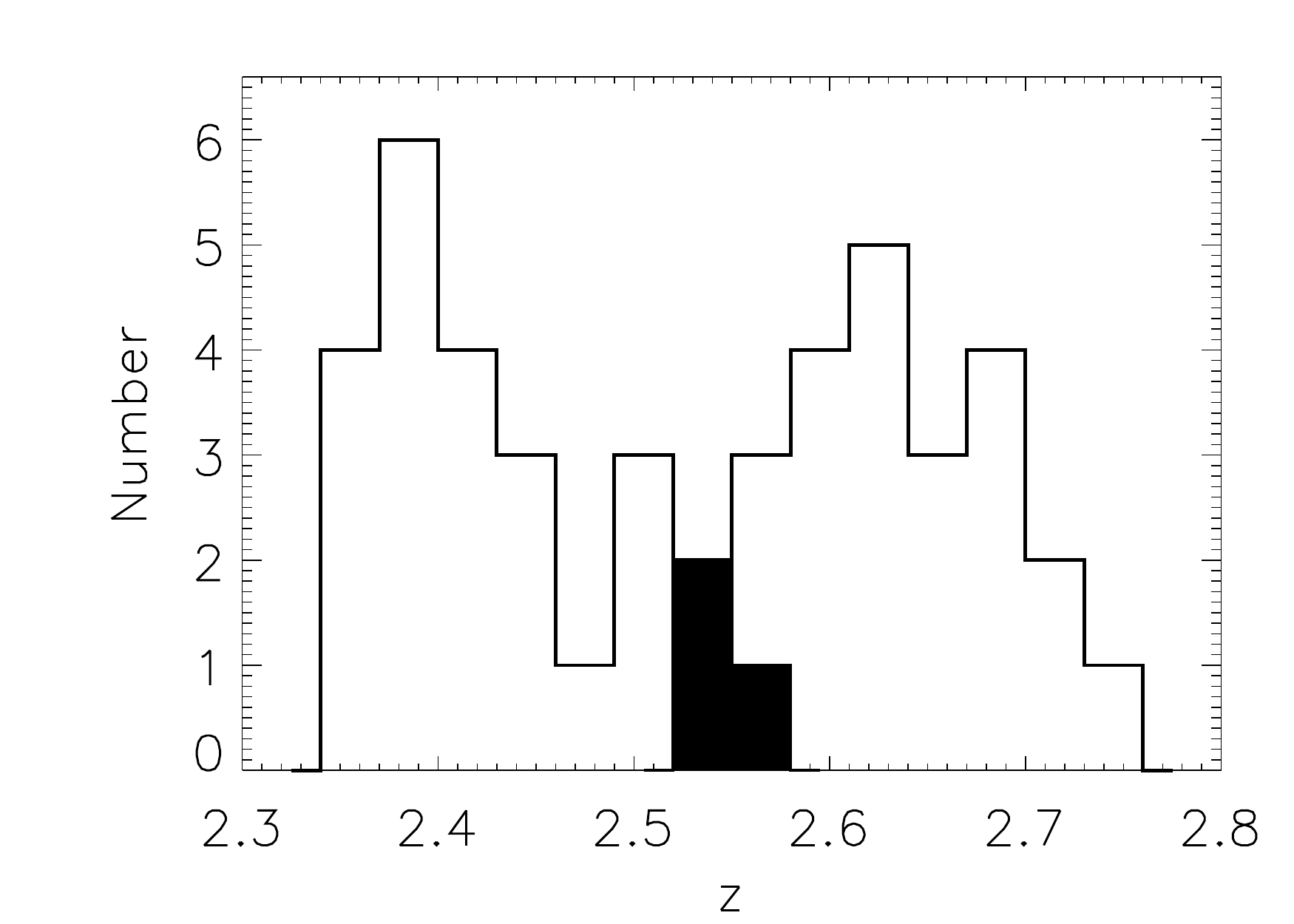}
\\
\includegraphics[width=81mm]{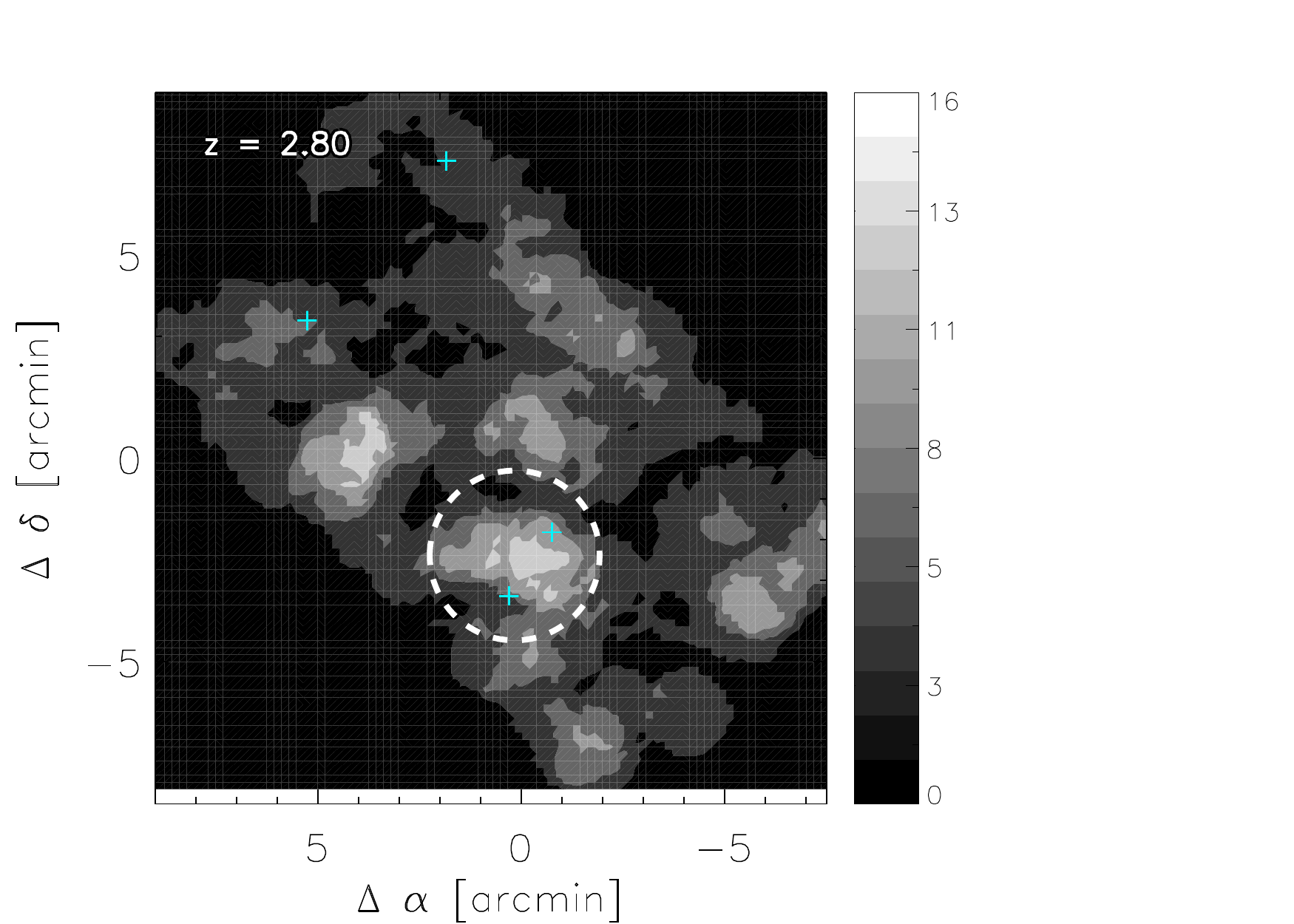}
       \hspace{-0.85in}
\includegraphics[width=81mm]{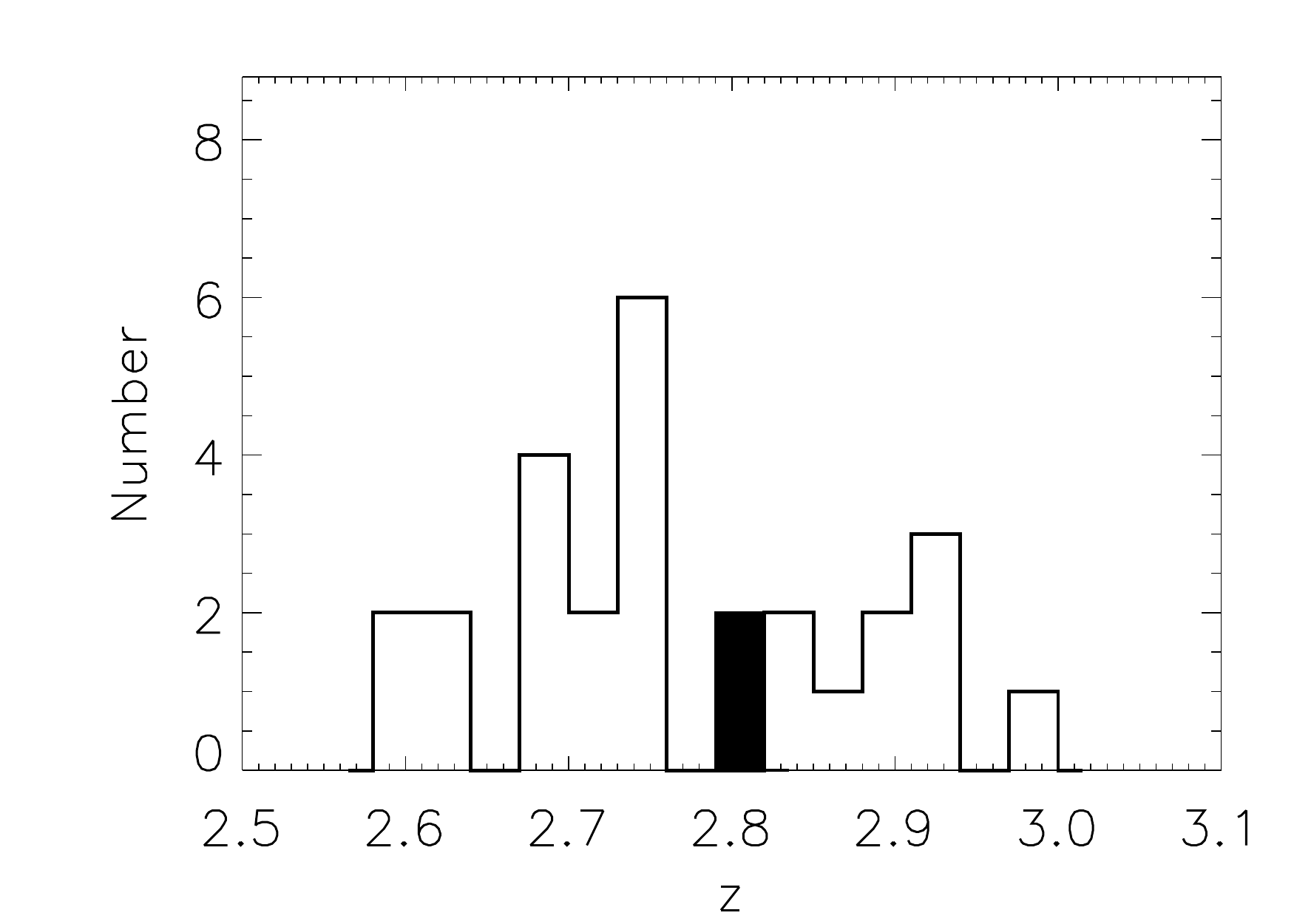}
\\
\includegraphics[width=81mm]{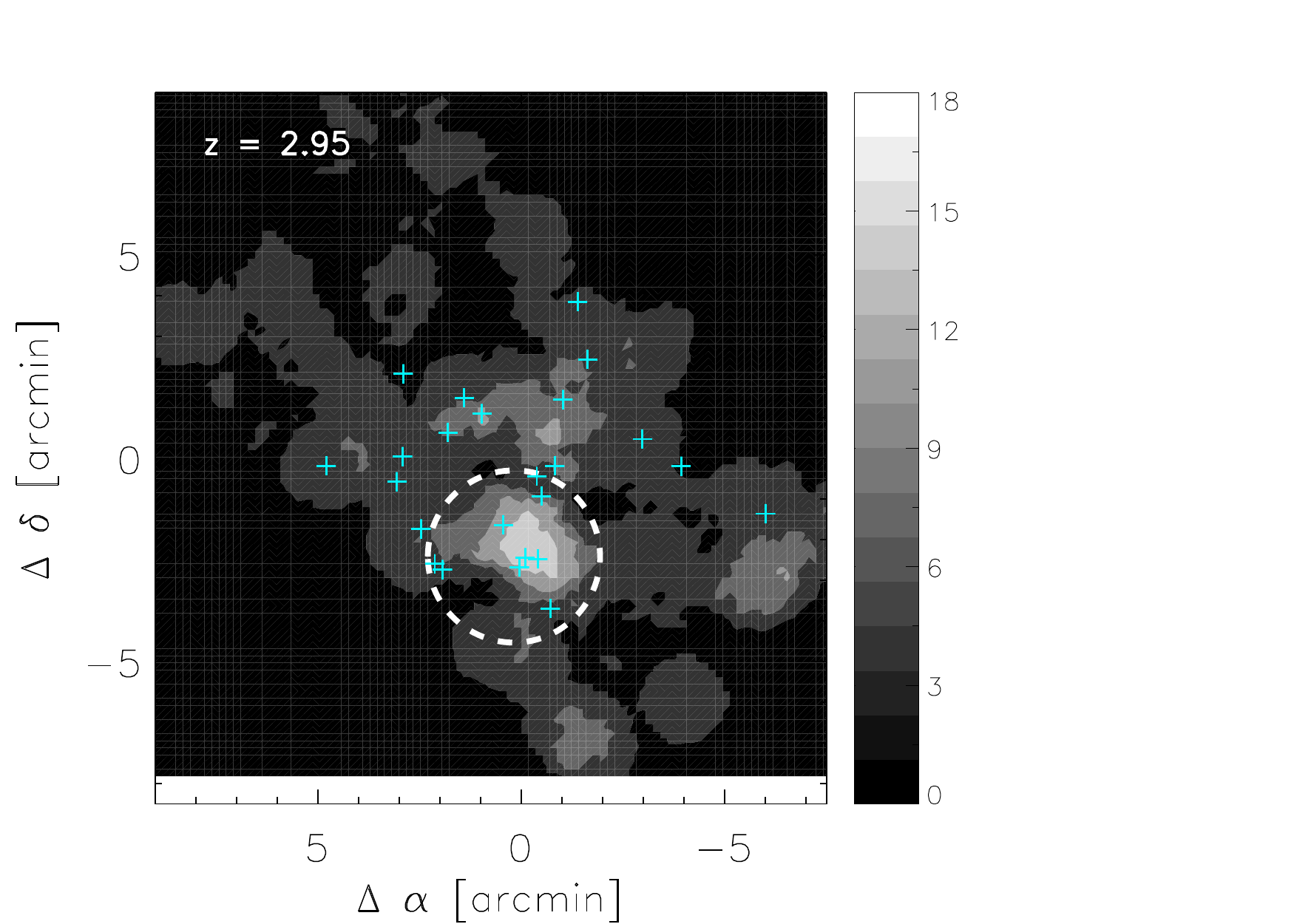}
       \hspace{-0.85in}
\includegraphics[width=81mm]{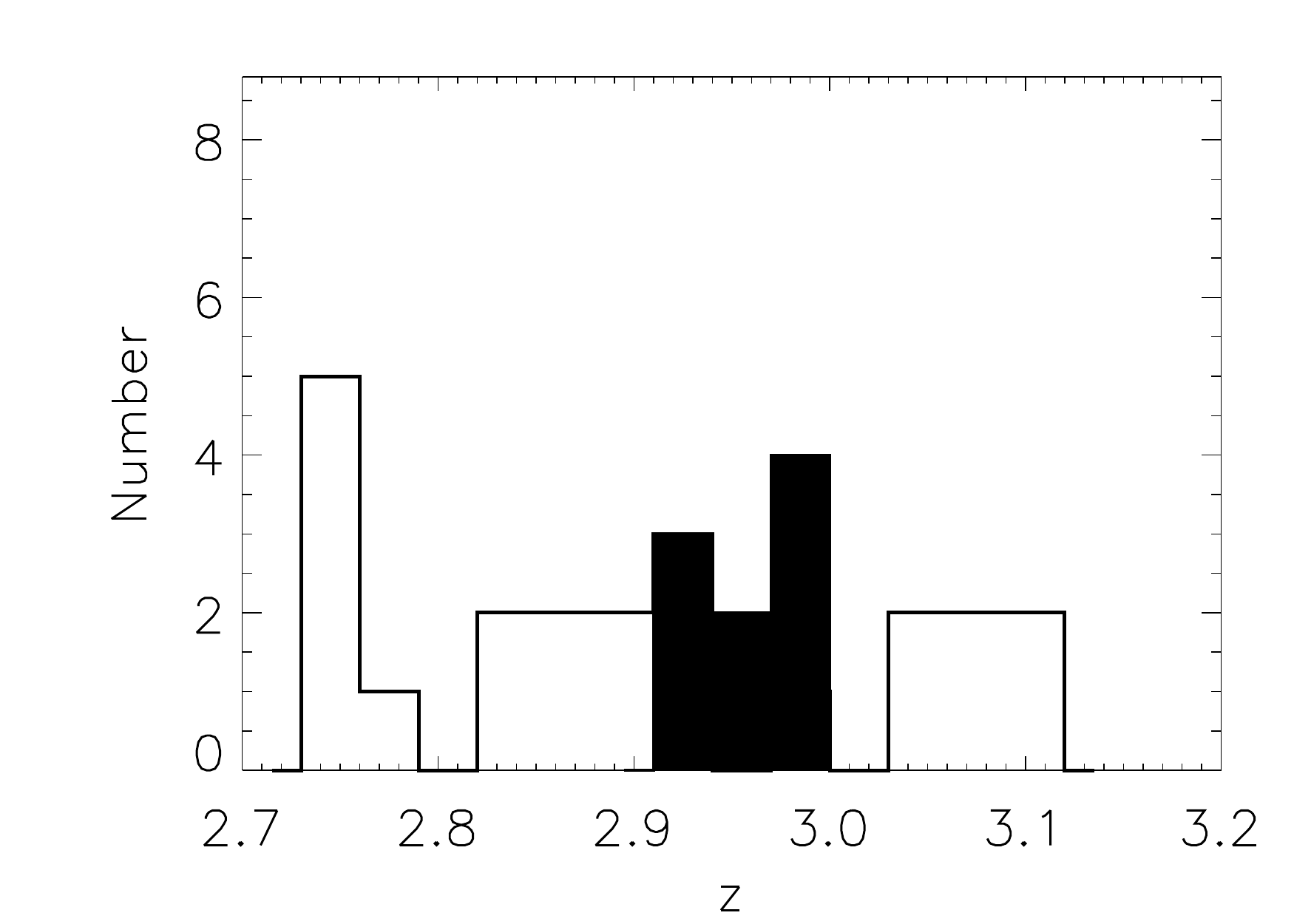}
\\
\includegraphics[width=81mm]{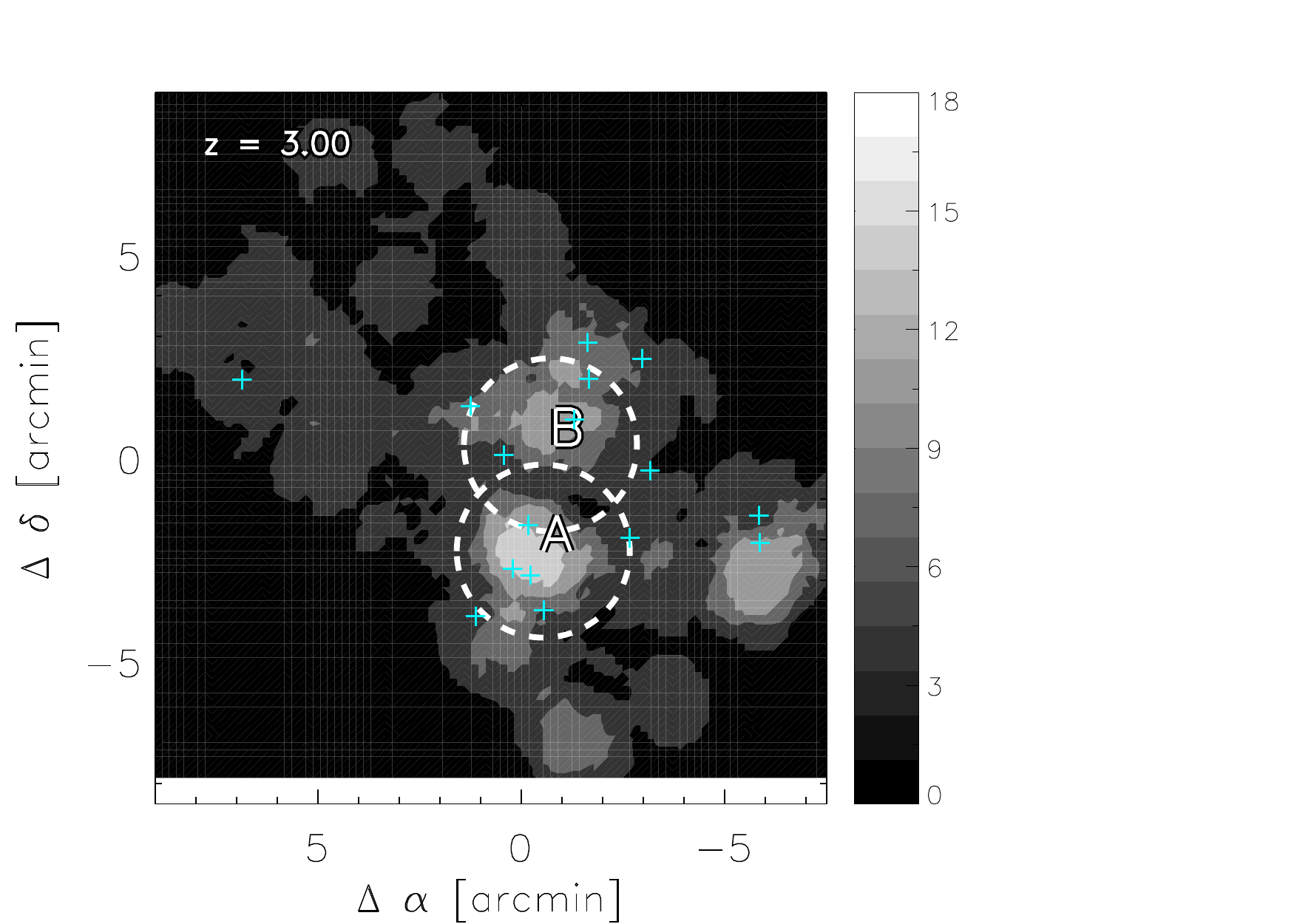}
       \hspace{-0.85in}
\includegraphics[width=81mm]{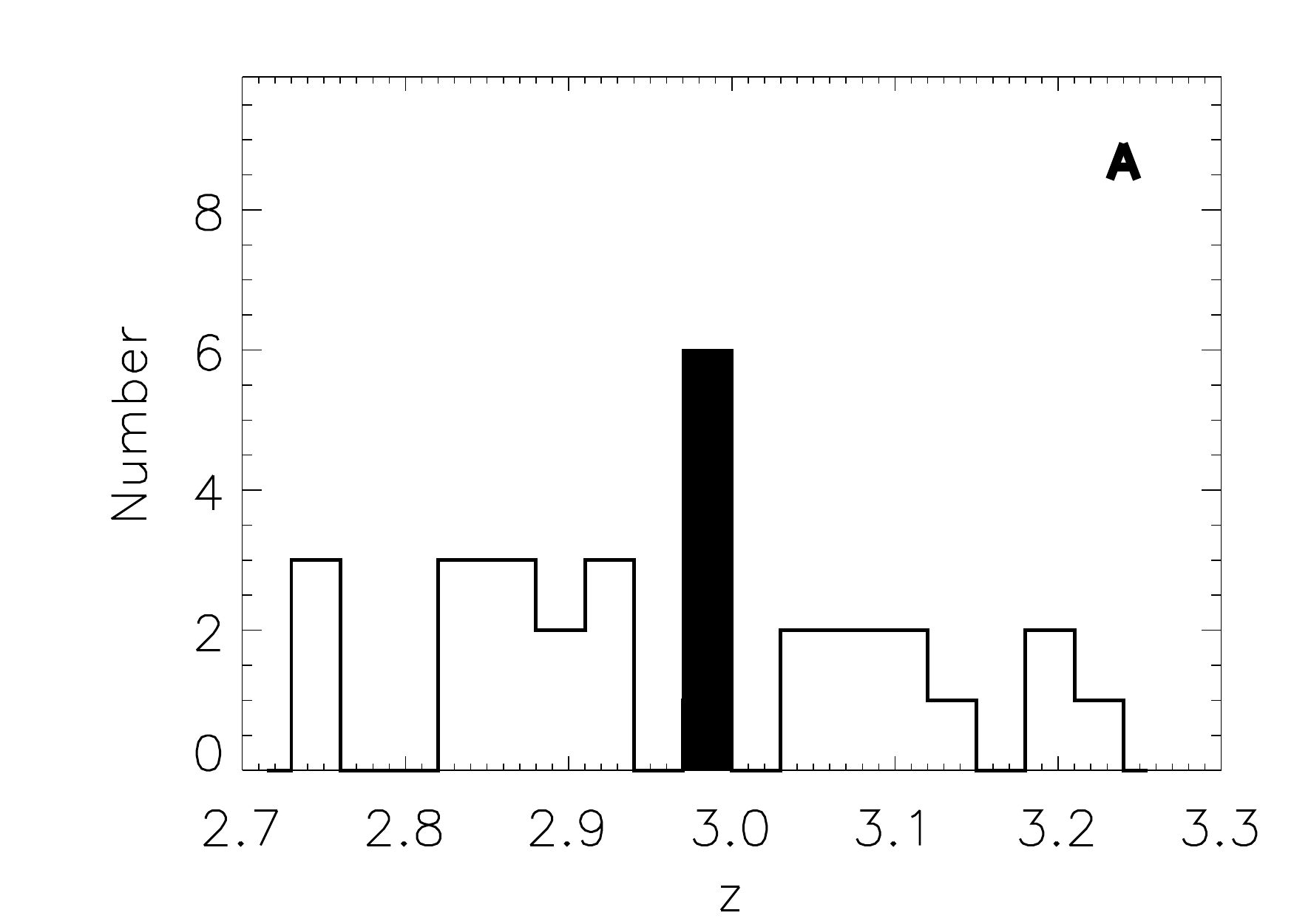}
\caption{(Continued)}
\label{f5e}
\end{figure*}


\begin{figure*}[tp!]
 \setcounter{figure}{6}
\centering
\includegraphics[width=81mm]{f7t1-eps-converted-to.pdf}
       \hspace{-0.85in}
\includegraphics[width=81mm]{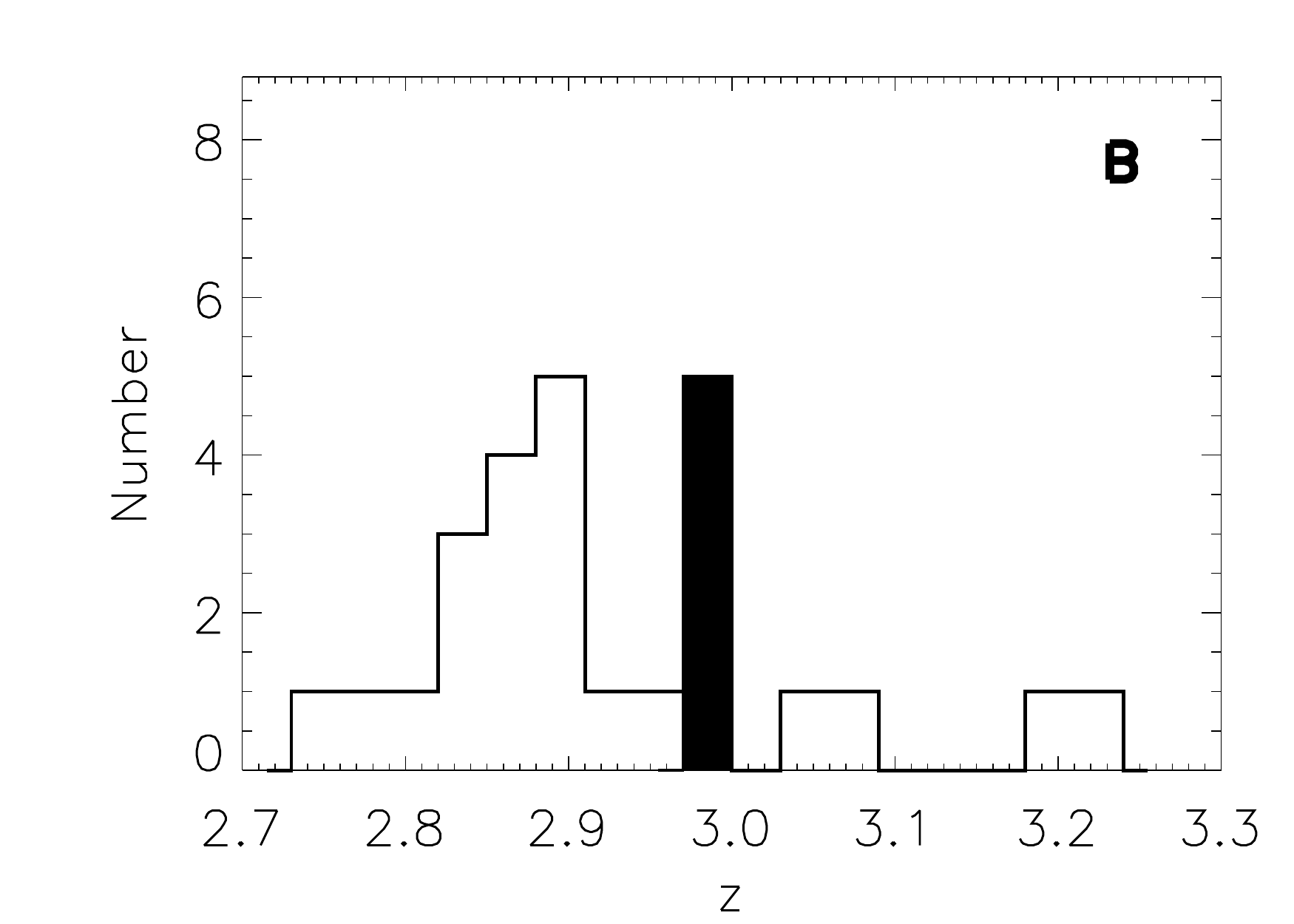}
\\
\includegraphics[width=81mm]{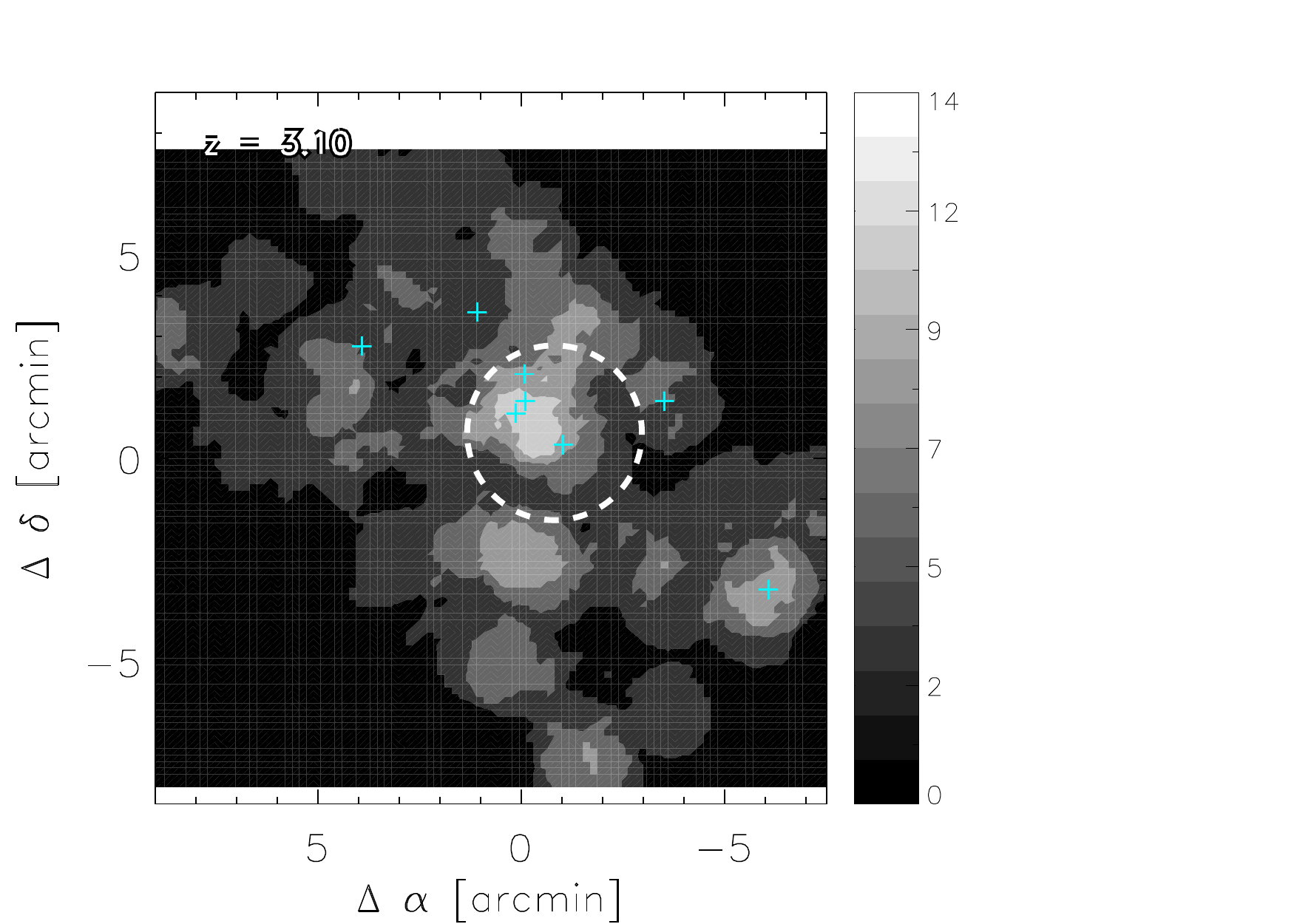}
       \hspace{-0.85in}
\includegraphics[width=81mm]{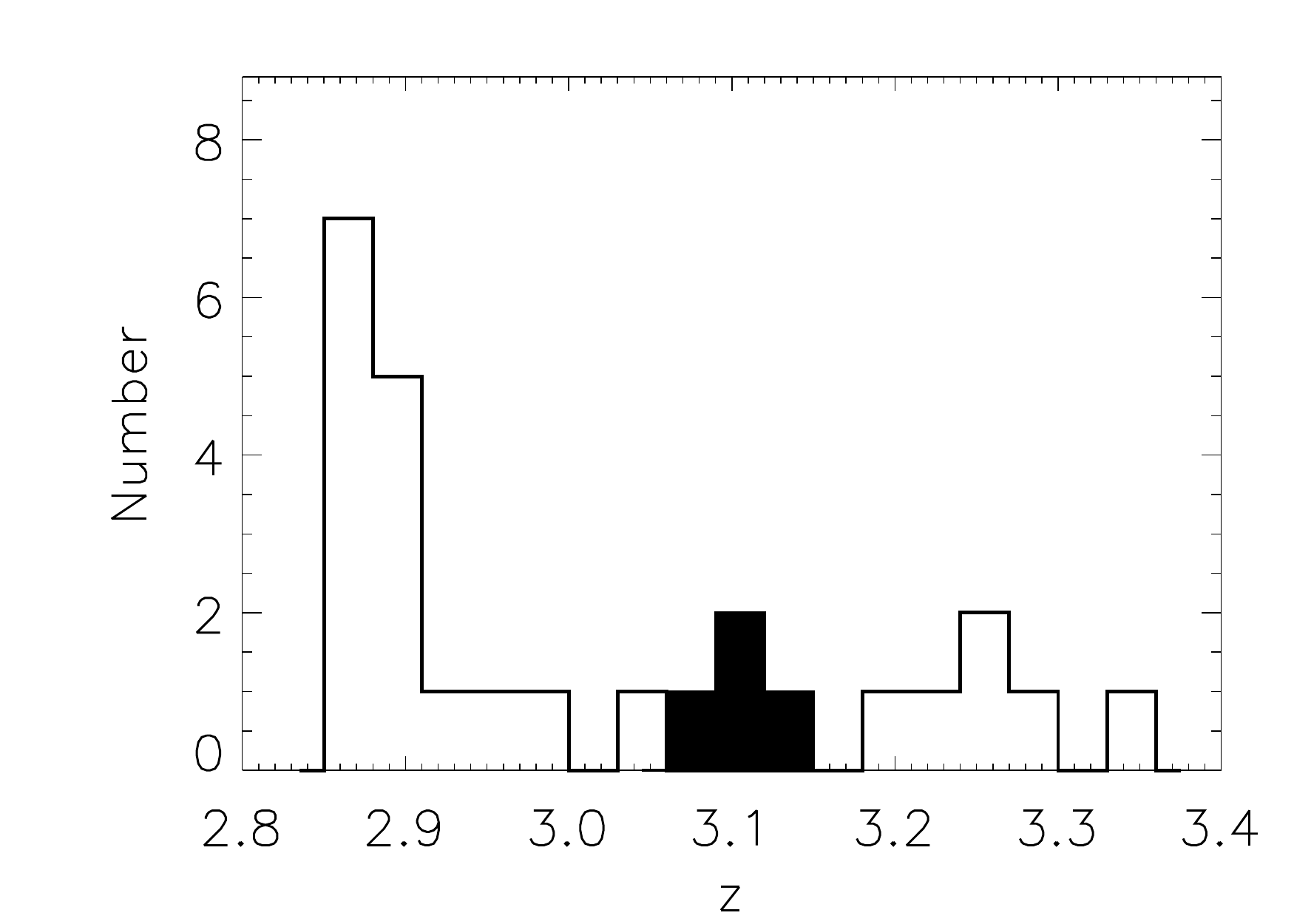}
\\
\includegraphics[width=81mm]{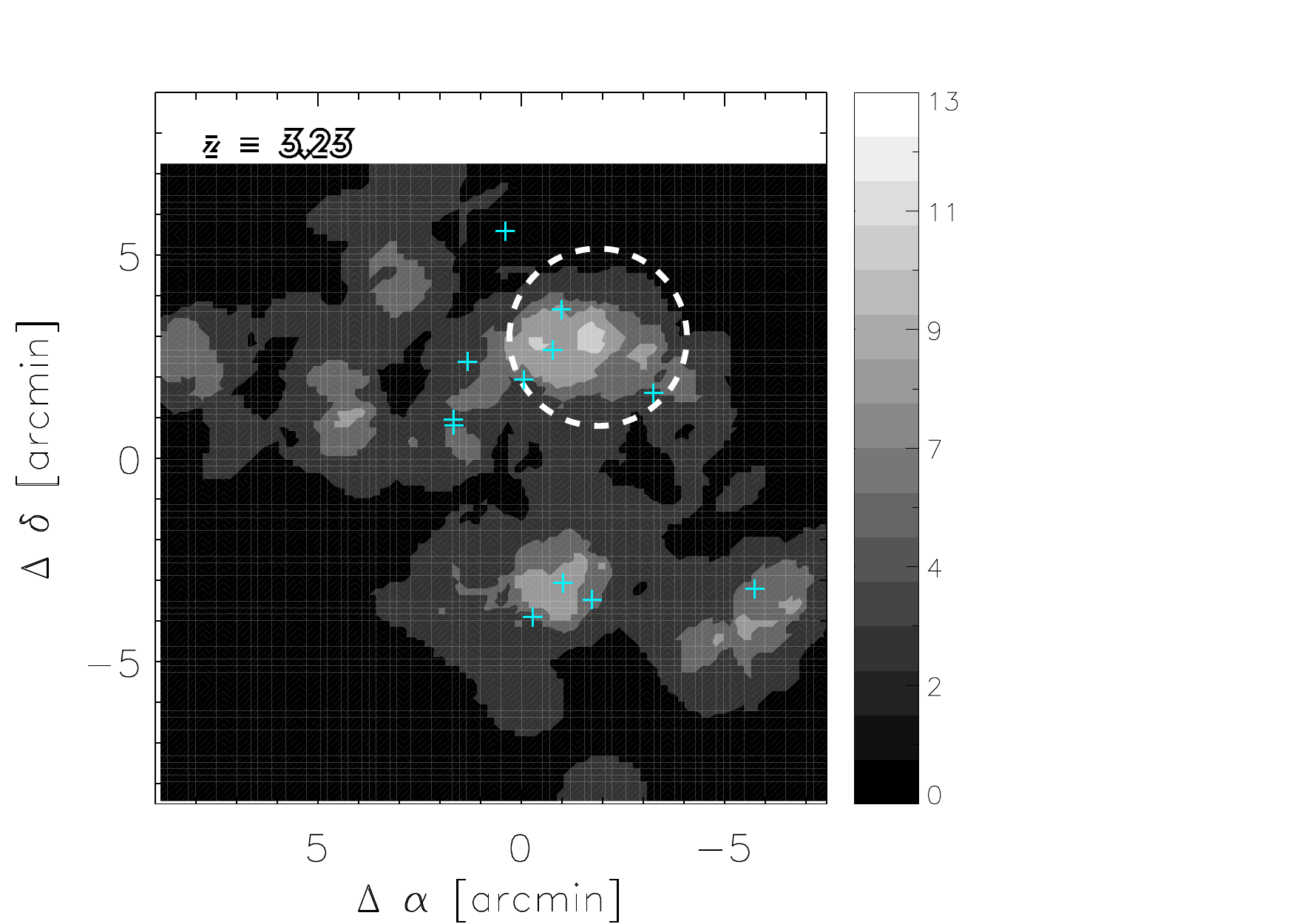}
       \hspace{-0.85in}
\includegraphics[width=81mm]{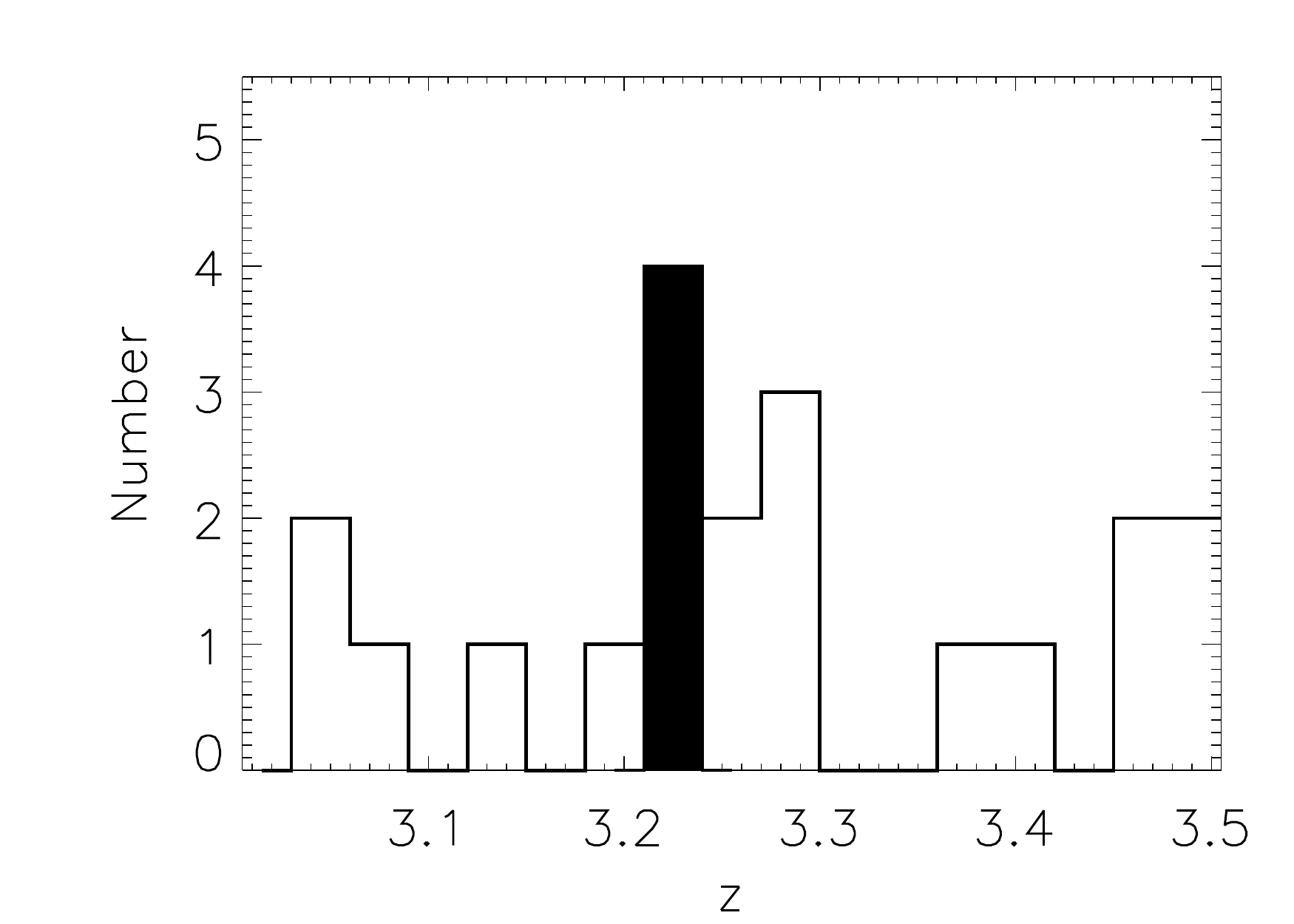}
\\
\includegraphics[width=81mm]{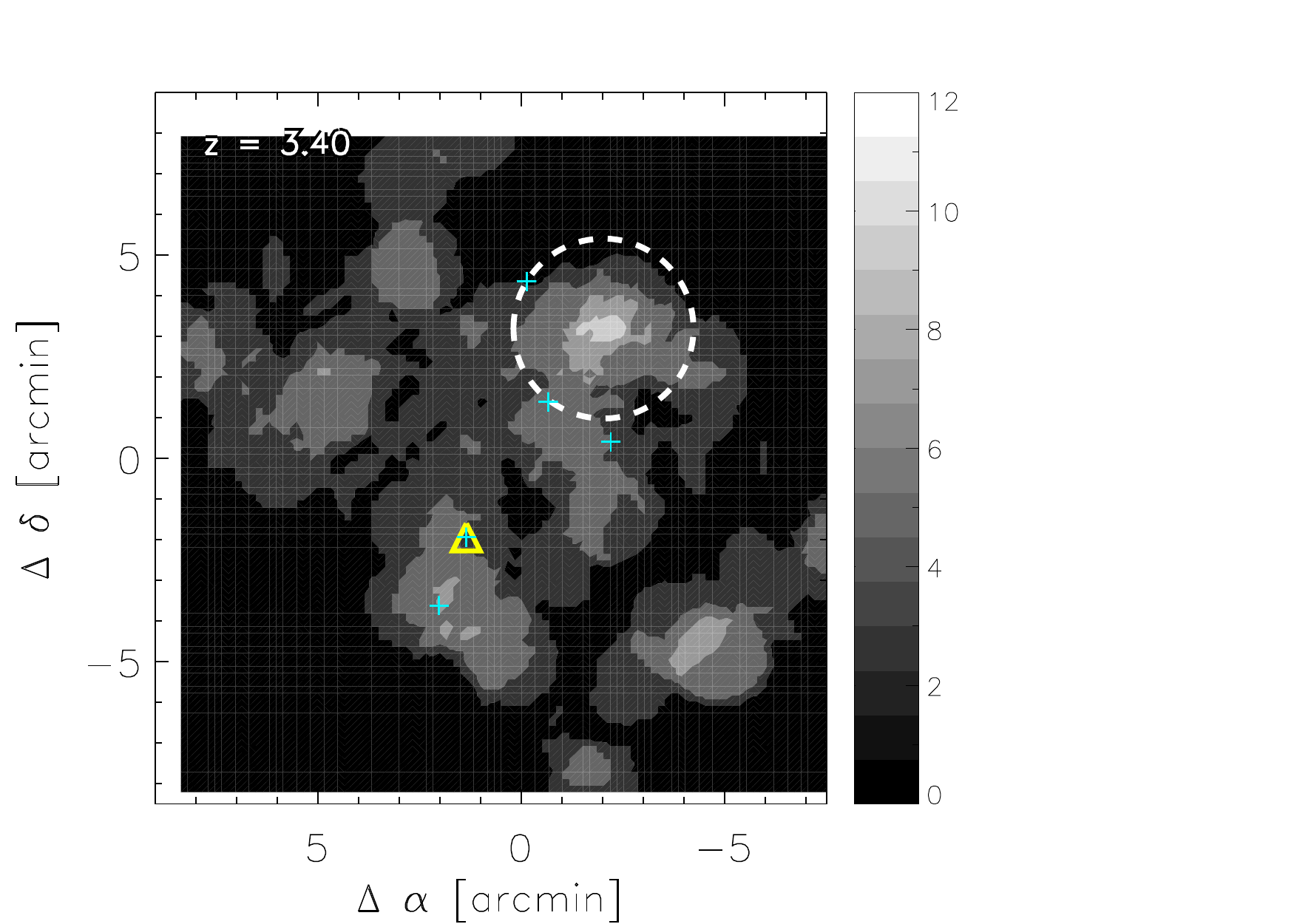}
       \hspace{-0.85in}
\includegraphics[width=81mm]{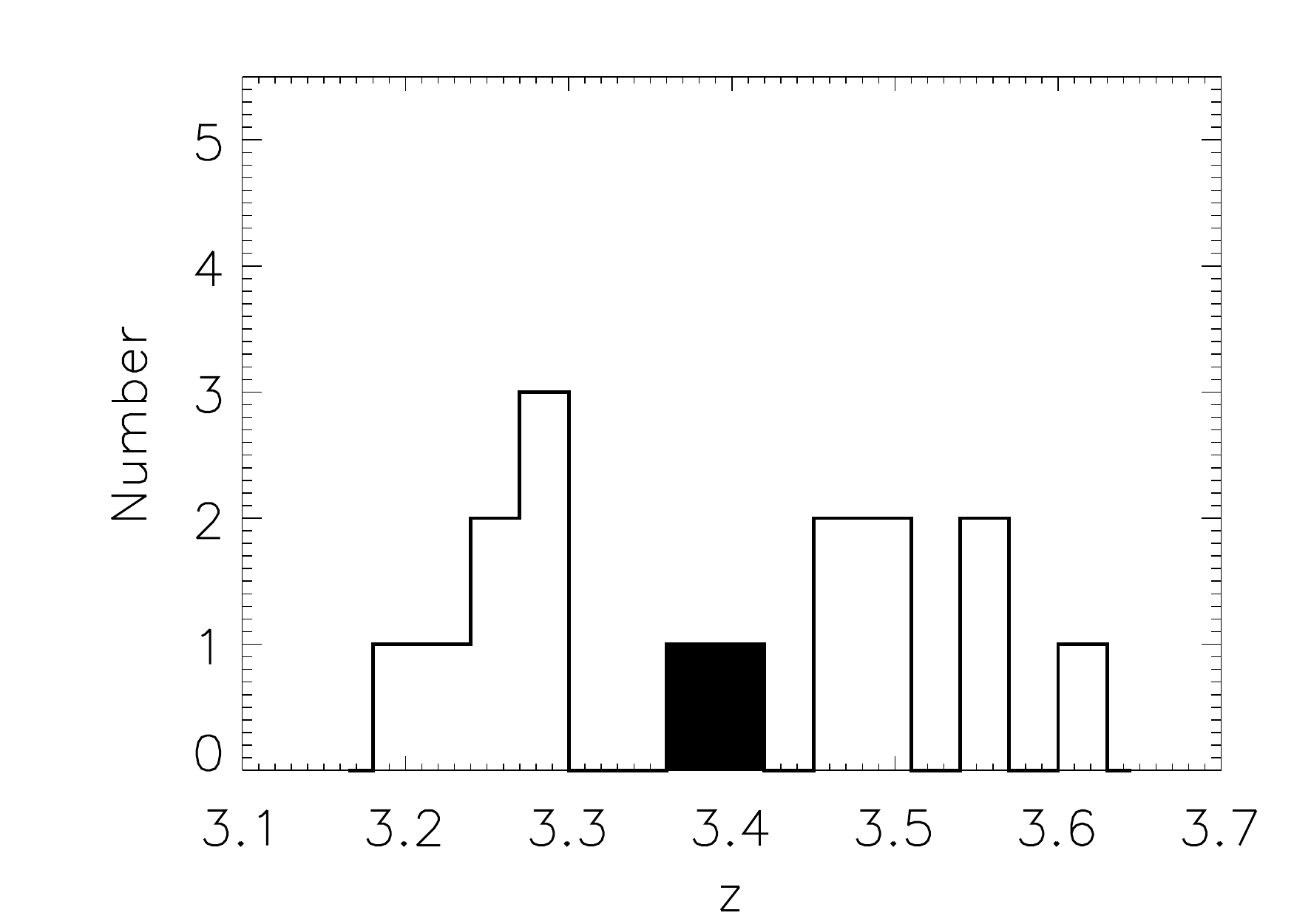}
\caption{(Continued)}
\label{f5f}
\end{figure*}


\begin{figure*}[tp!]
 \setcounter{figure}{6}
\centering
\includegraphics[width=80mm]{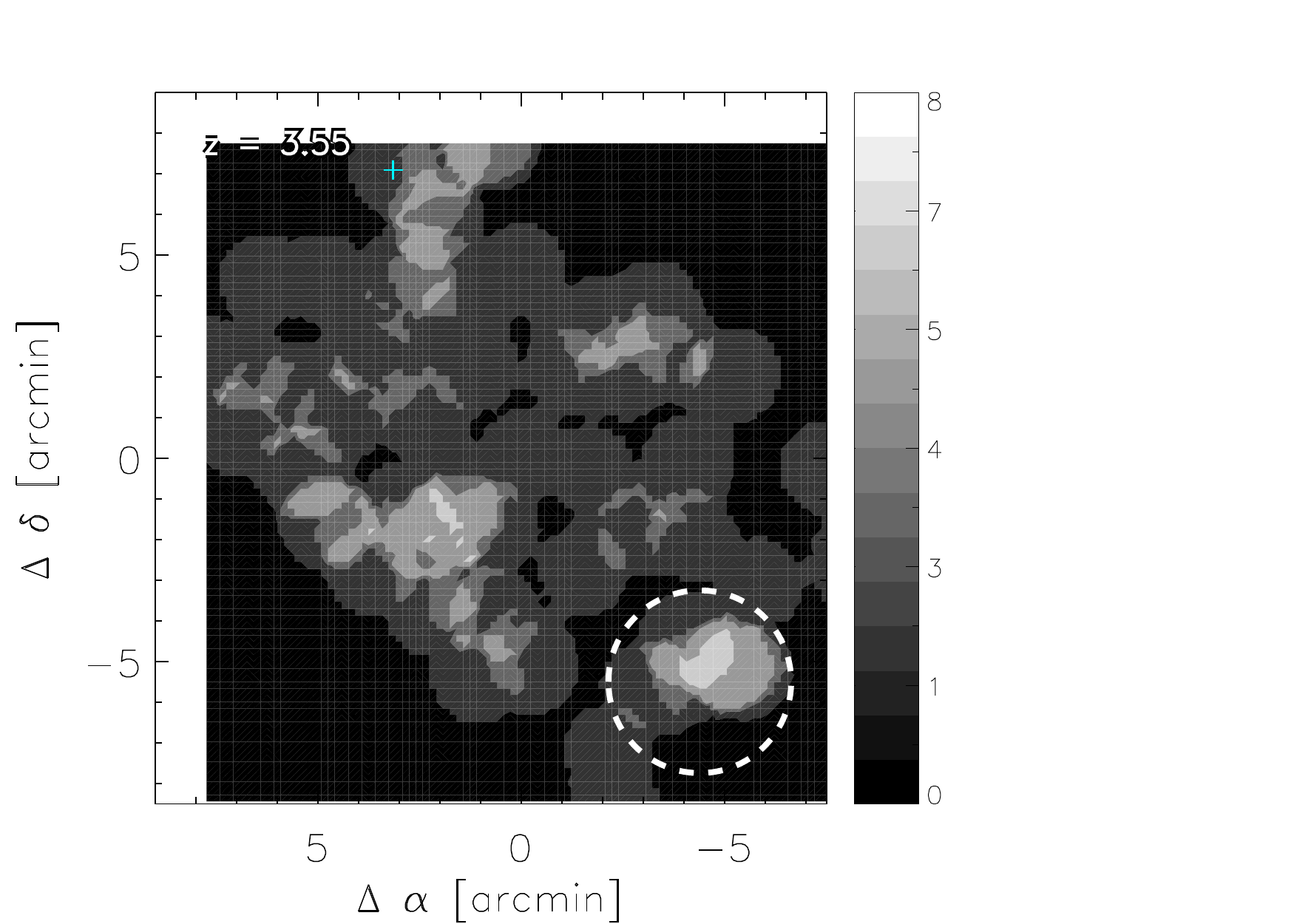}
       \hspace{-0.85in}
\includegraphics[width=80mm]{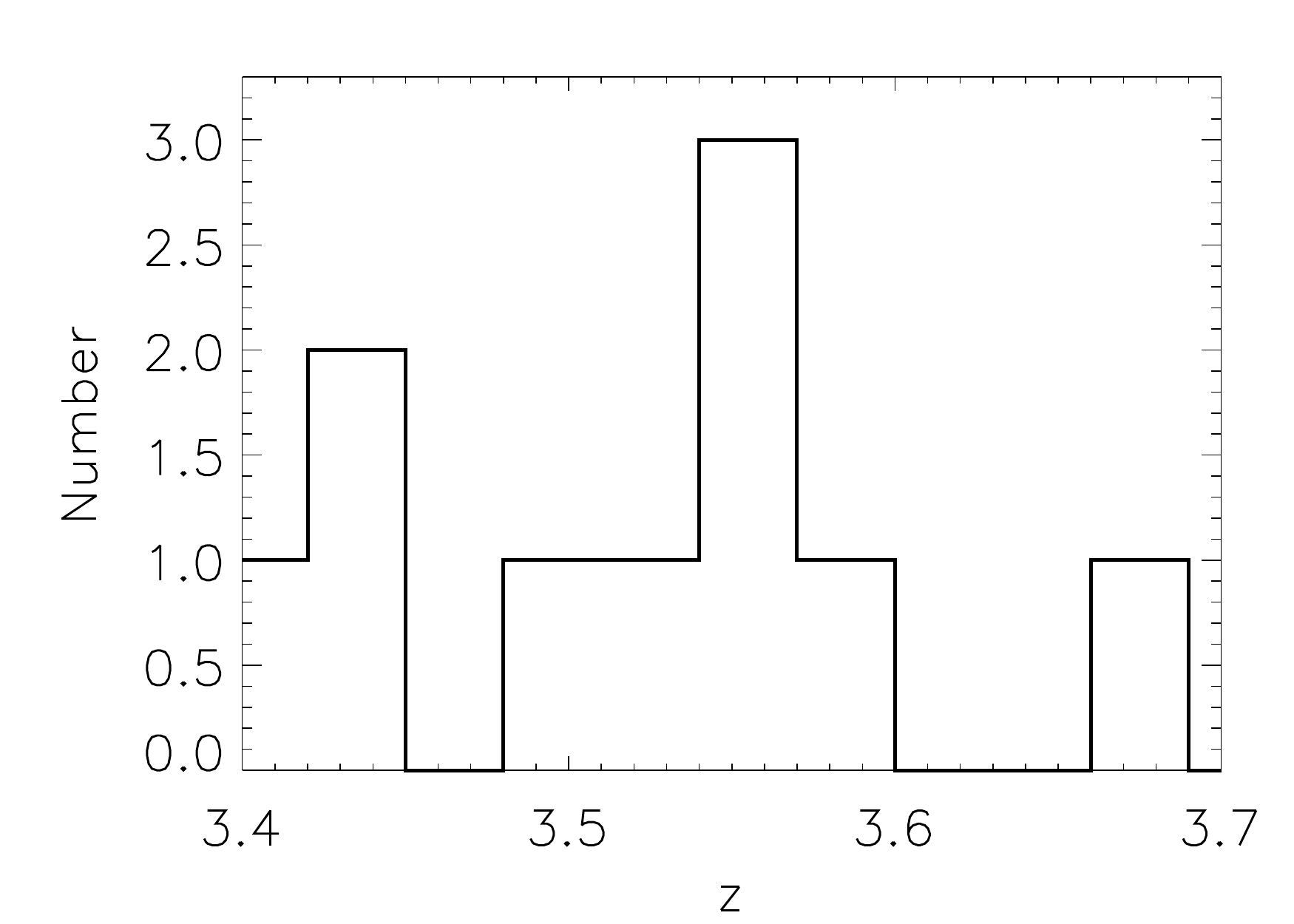}
\\
\includegraphics[width=80mm]{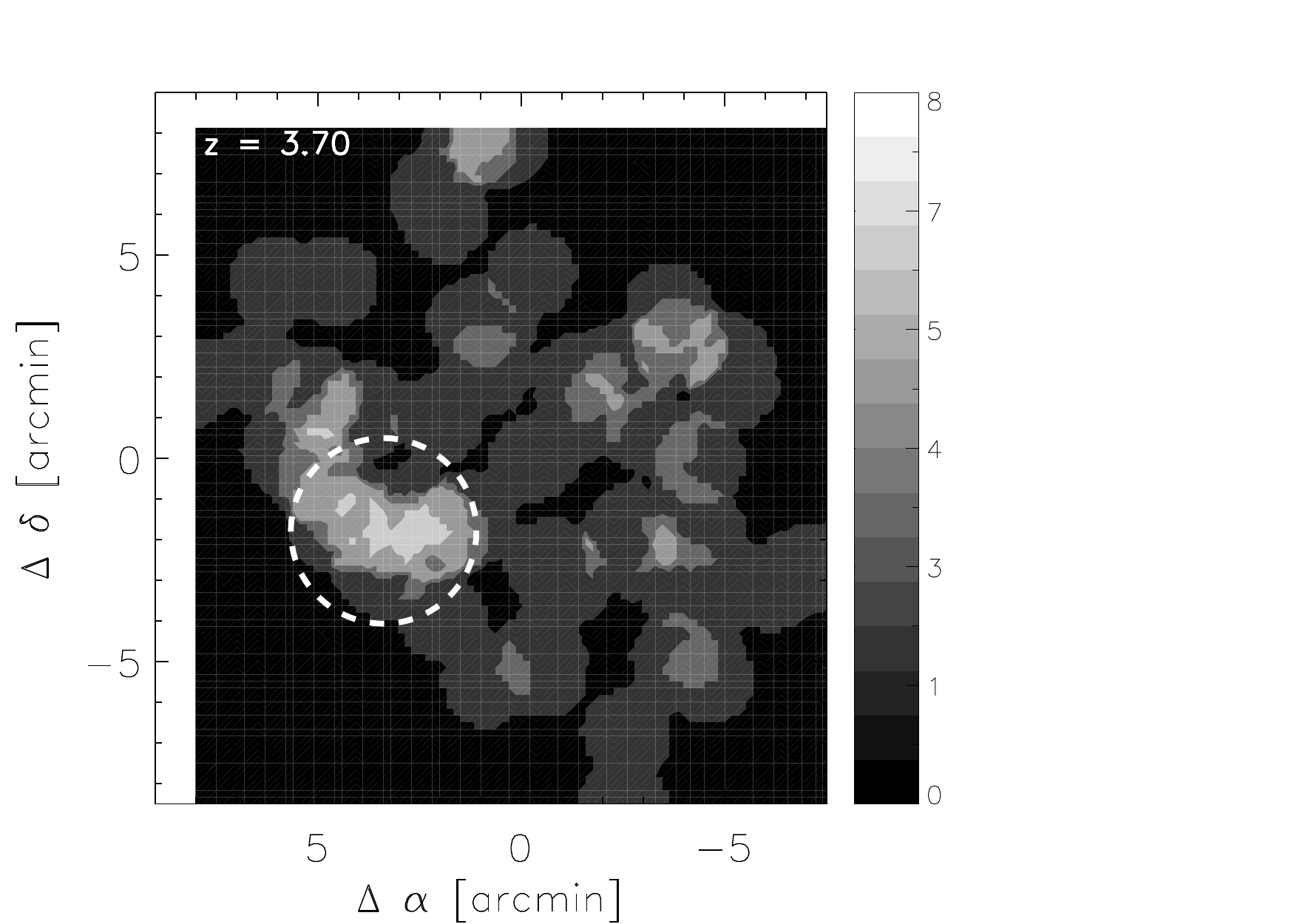}
       \hspace{-0.85in}
\includegraphics[width=80mm]{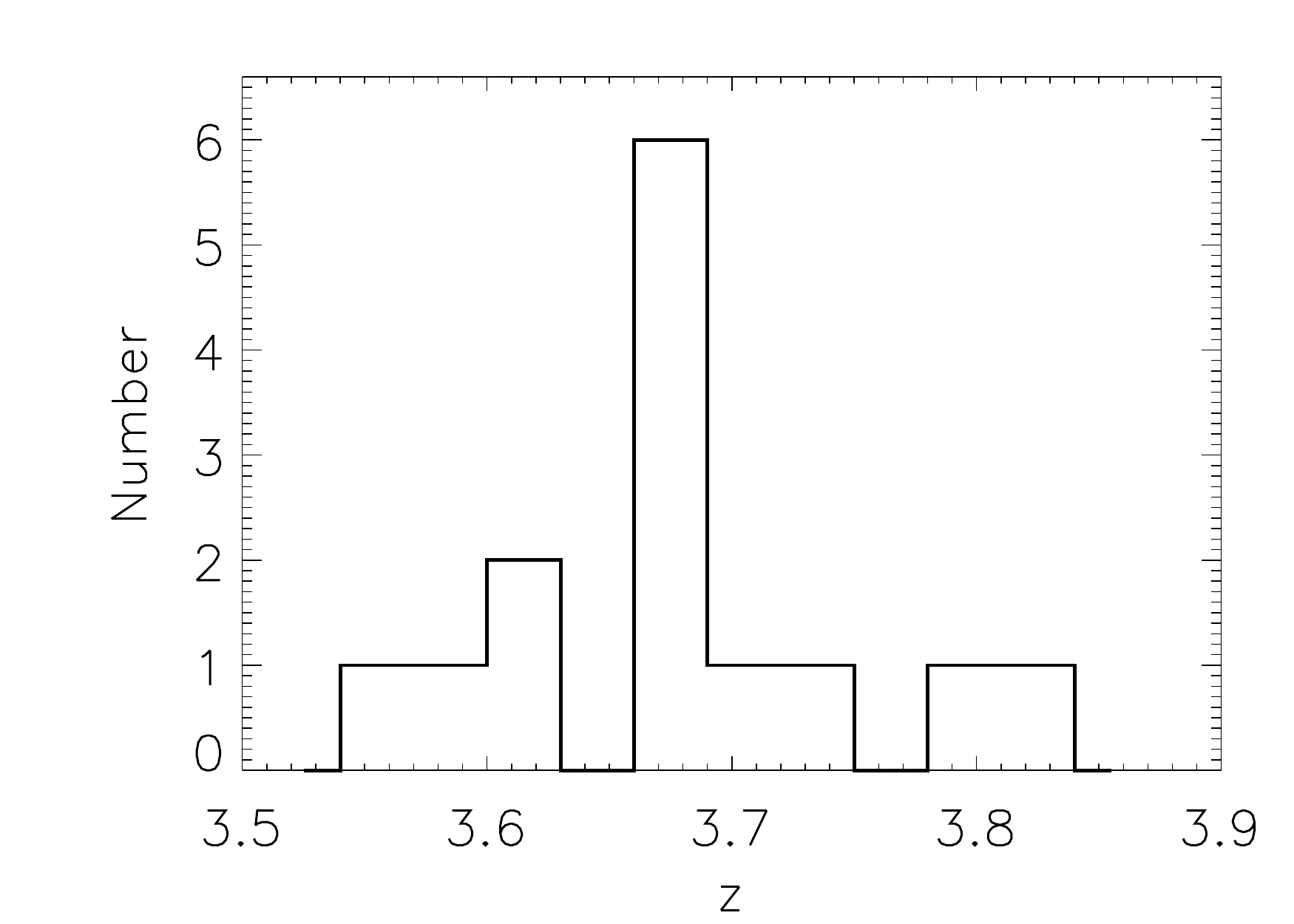}
\\
\includegraphics[width=80mm]{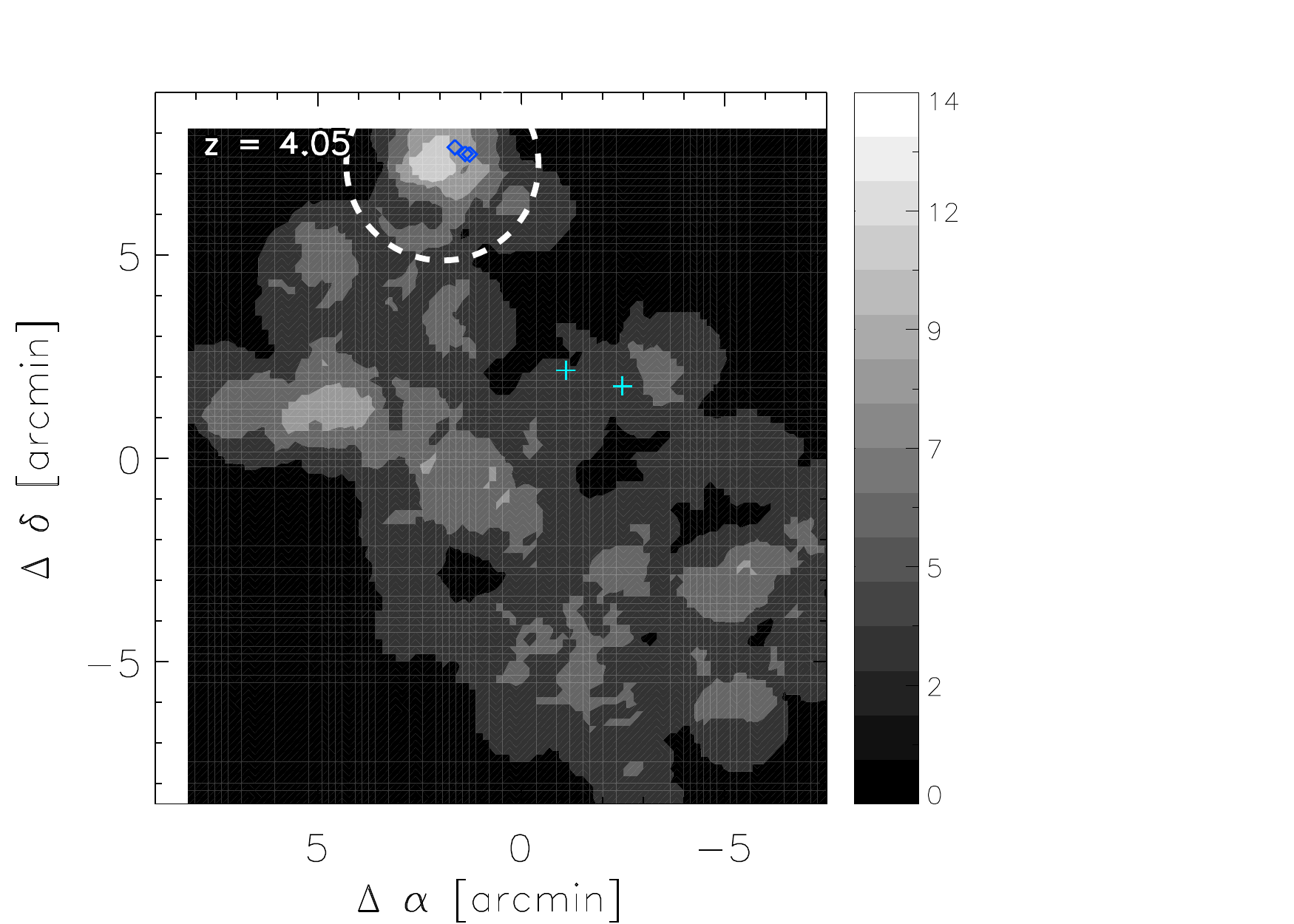}
       \hspace{-0.85in}
\includegraphics[width=80mm]{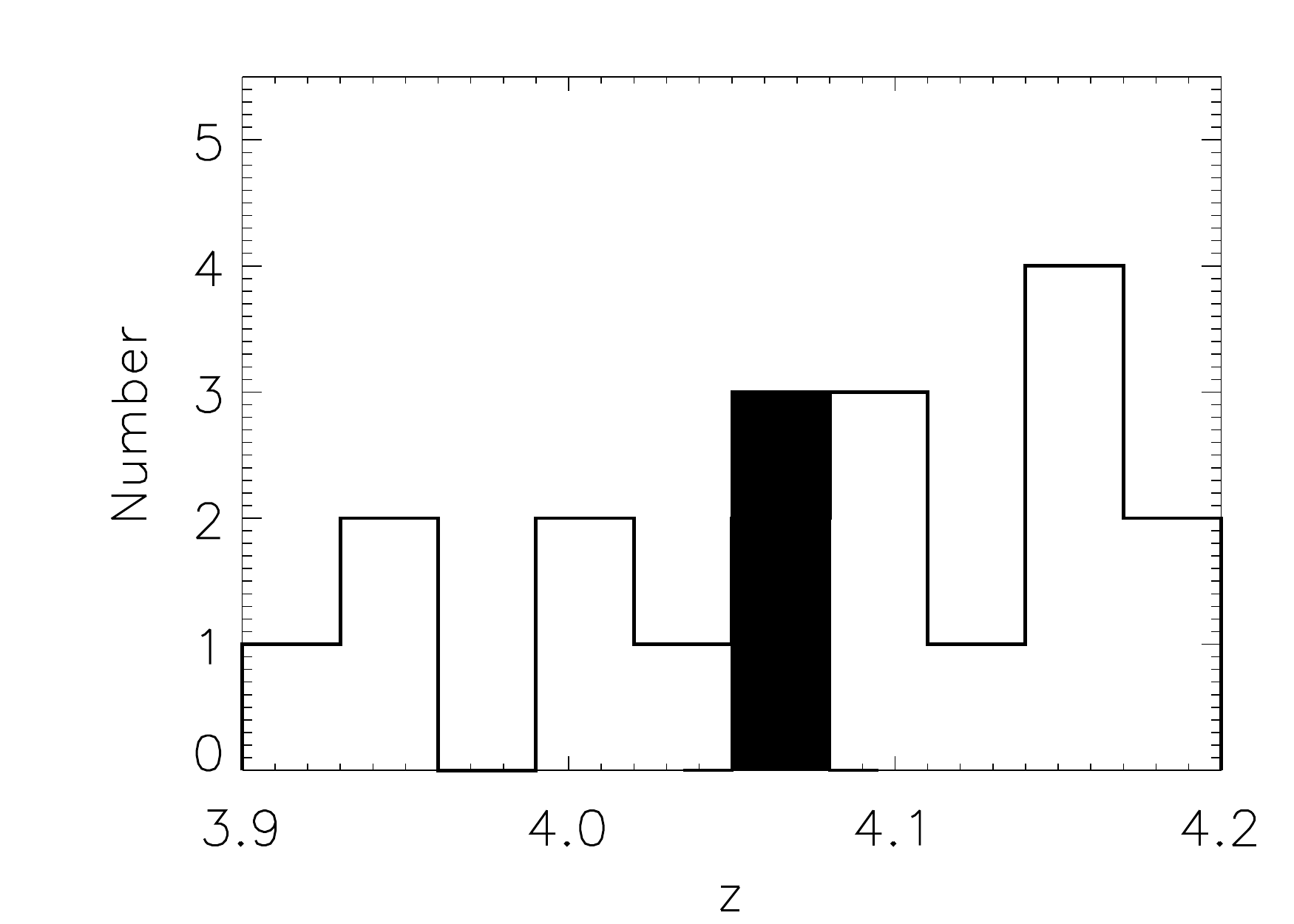}
\\
\includegraphics[width=80mm]{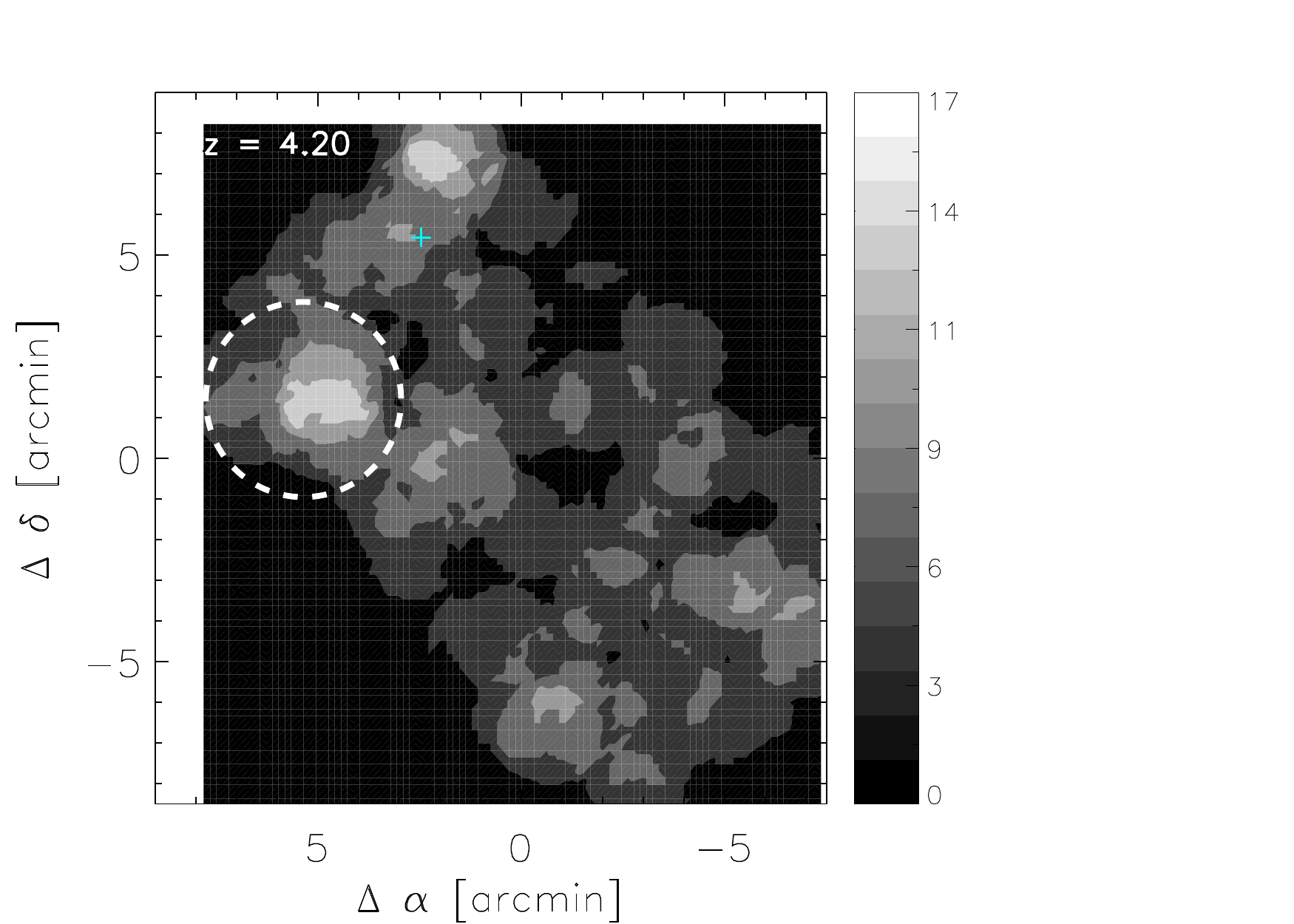}
       \hspace{-0.85in}
\includegraphics[width=80mm]{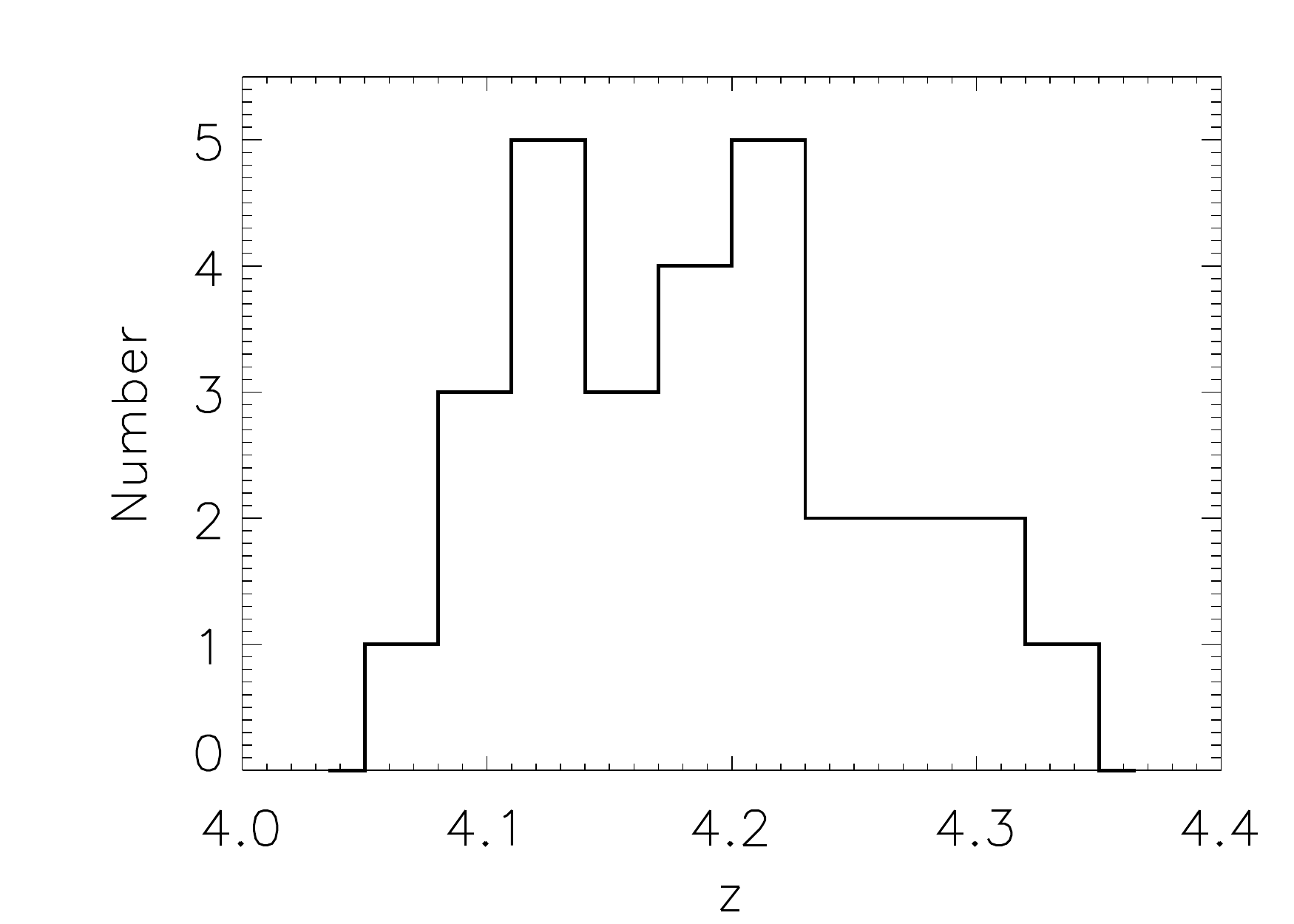}
\caption{(Continued) Note that the blue diamonds in the surface density map of MSG at $z\sim4.05$ indicate the three submm
 galaxies at $z=4.05$ discussed in Daddi et al. (2009).
}
\label{f5g}
\end{figure*}


\begin{figure*}[tp!]
 \setcounter{figure}{7}
\centering
\includegraphics[width=80mm]{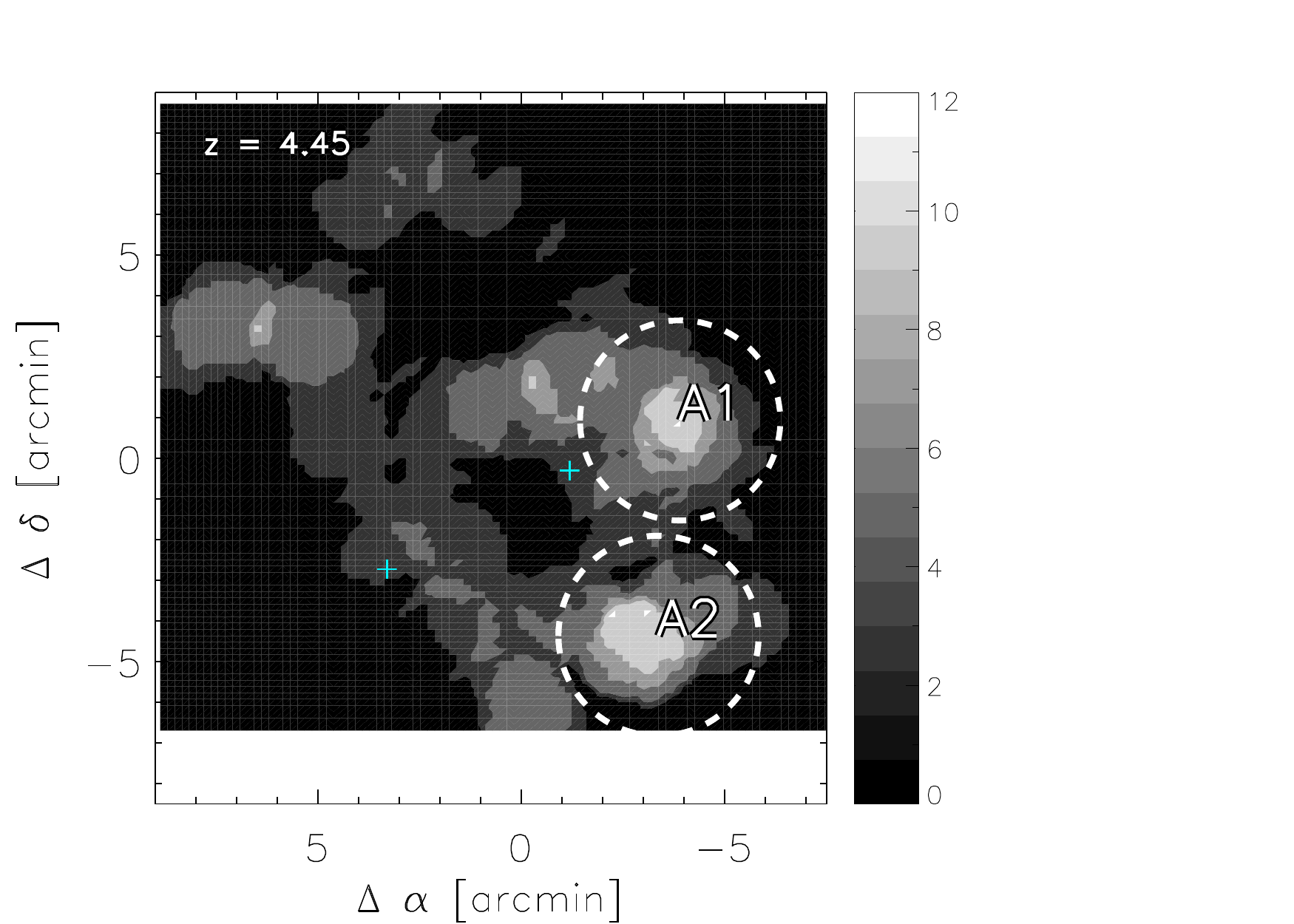}
       \hspace{-0.85in}
\includegraphics[width=80mm]{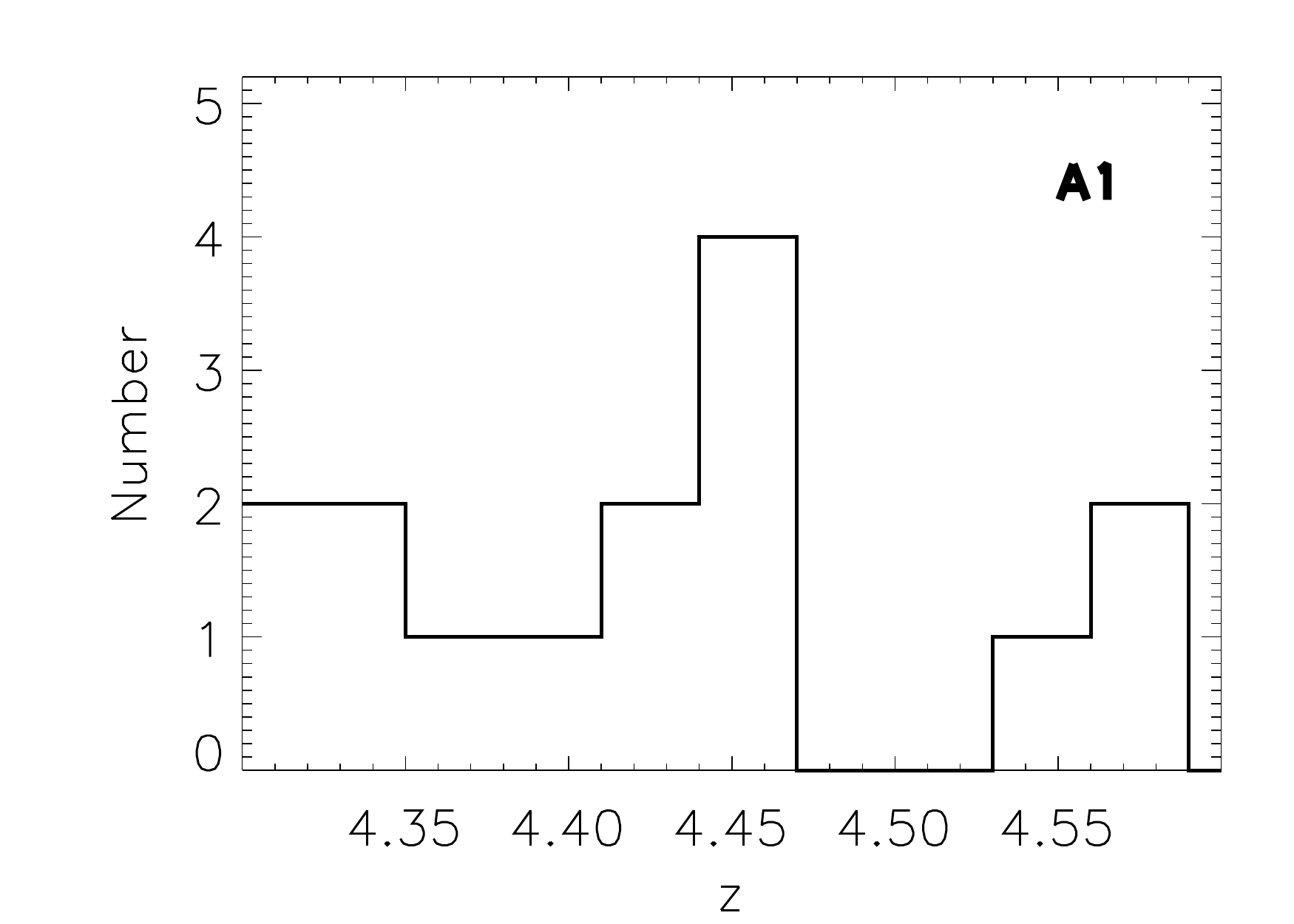}
\\
\includegraphics[width=80mm]{f7z5-eps-converted-to.pdf}
       \hspace{-0.85in}
\includegraphics[width=80mm]{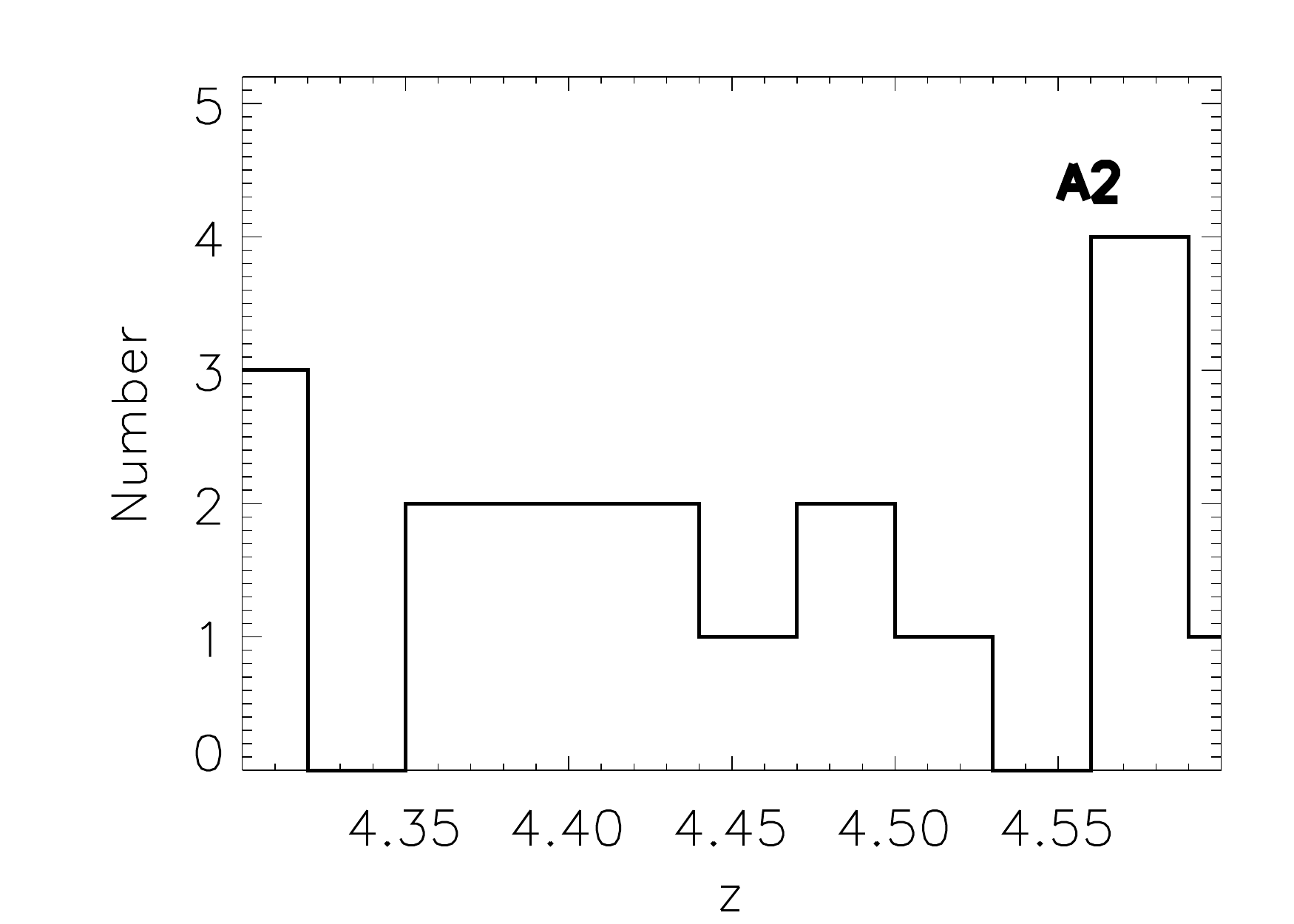}
\includegraphics[width=80mm]{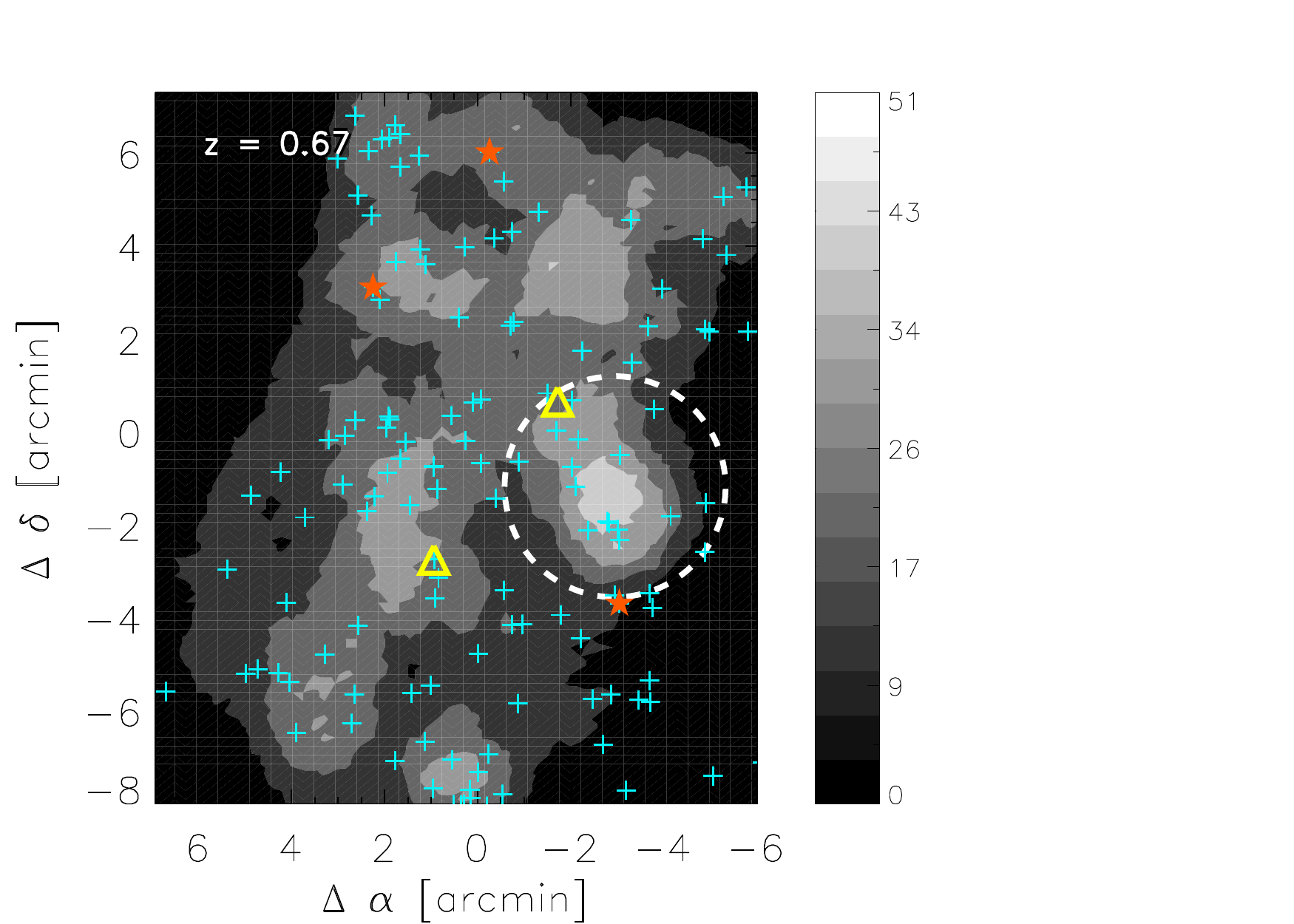}
       \hspace{-0.85in}
       \includegraphics[width=80mm]{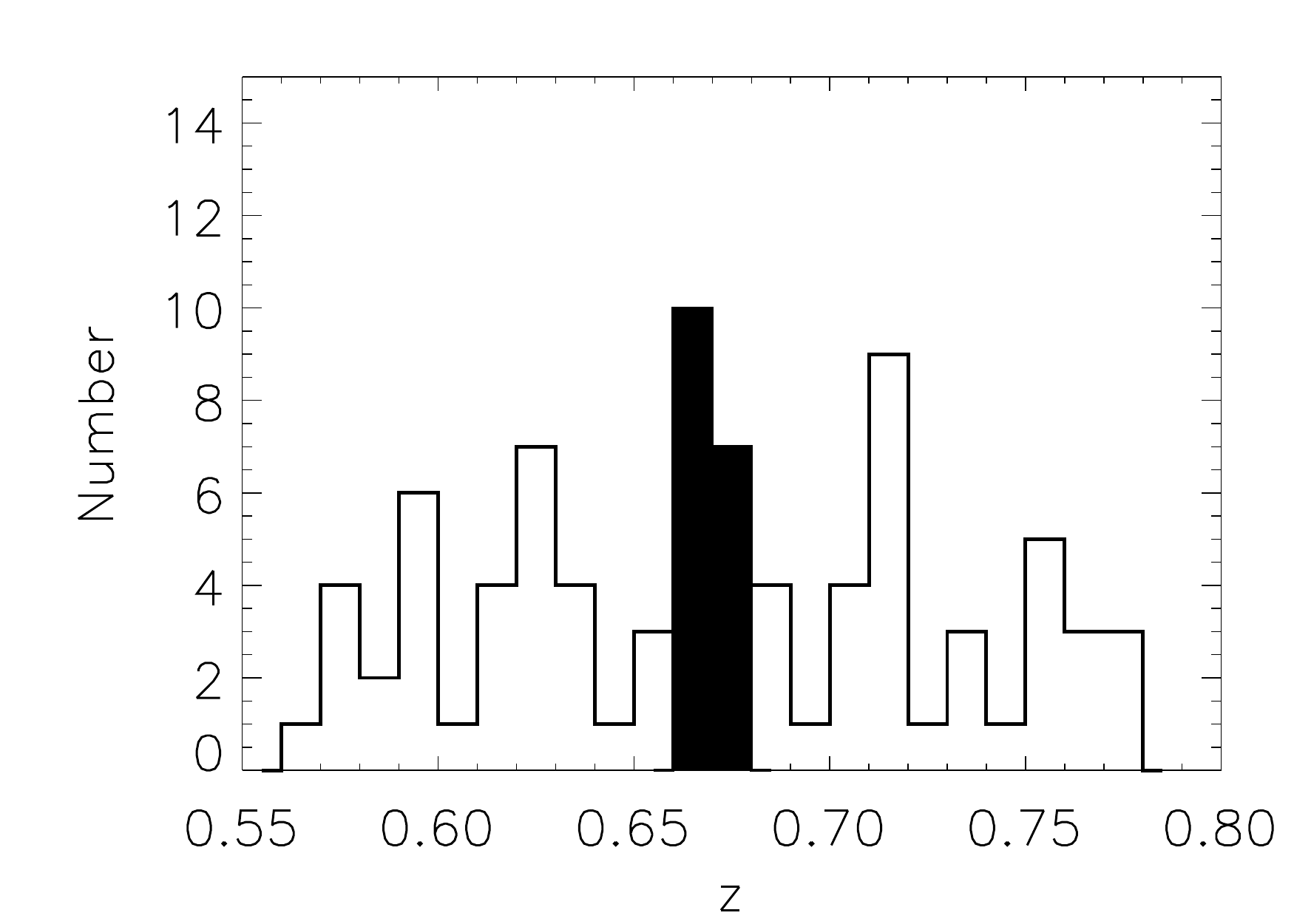}
\\
\includegraphics[width=80mm]{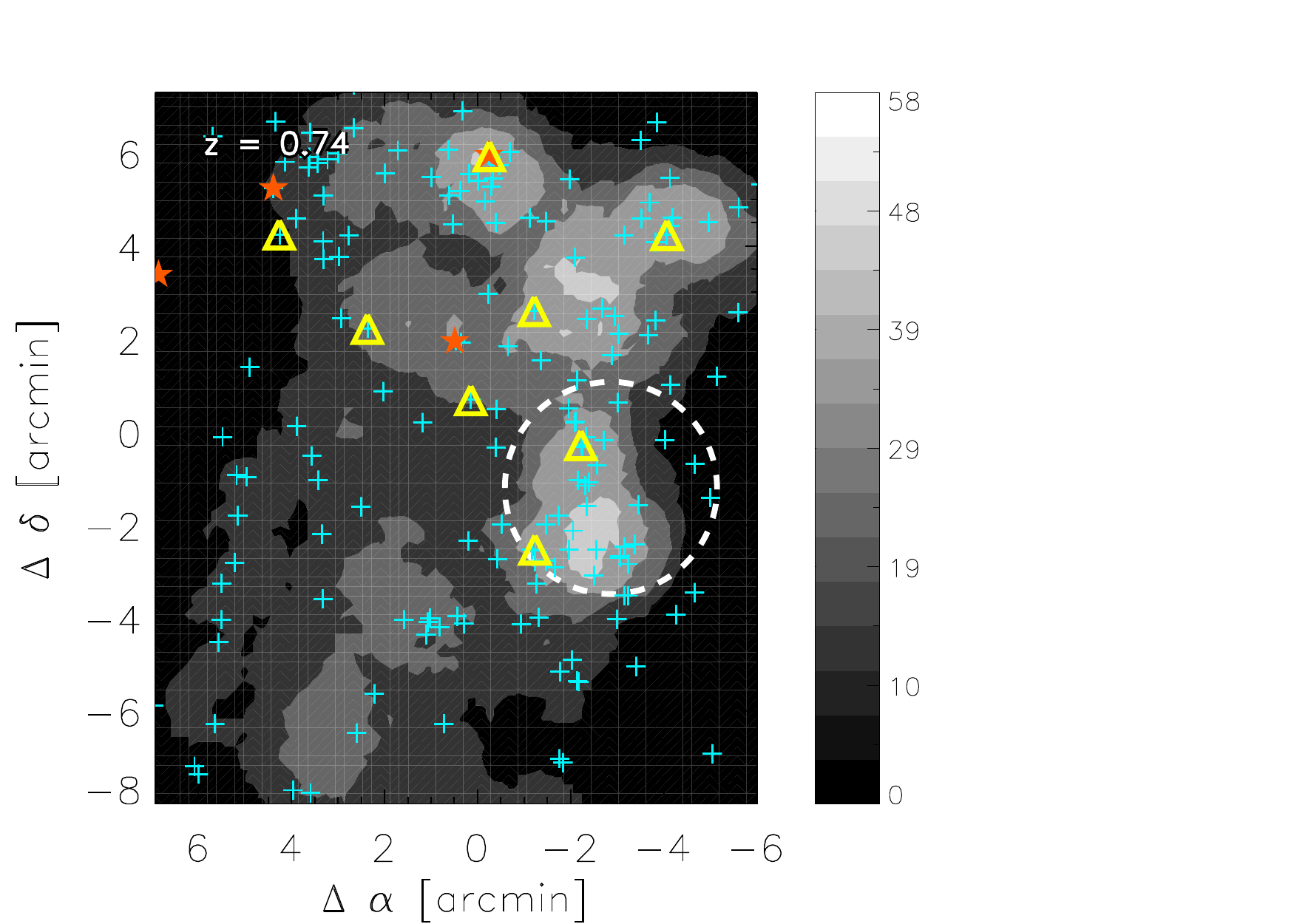}
       \hspace{-0.85in}
\includegraphics[width=80mm]{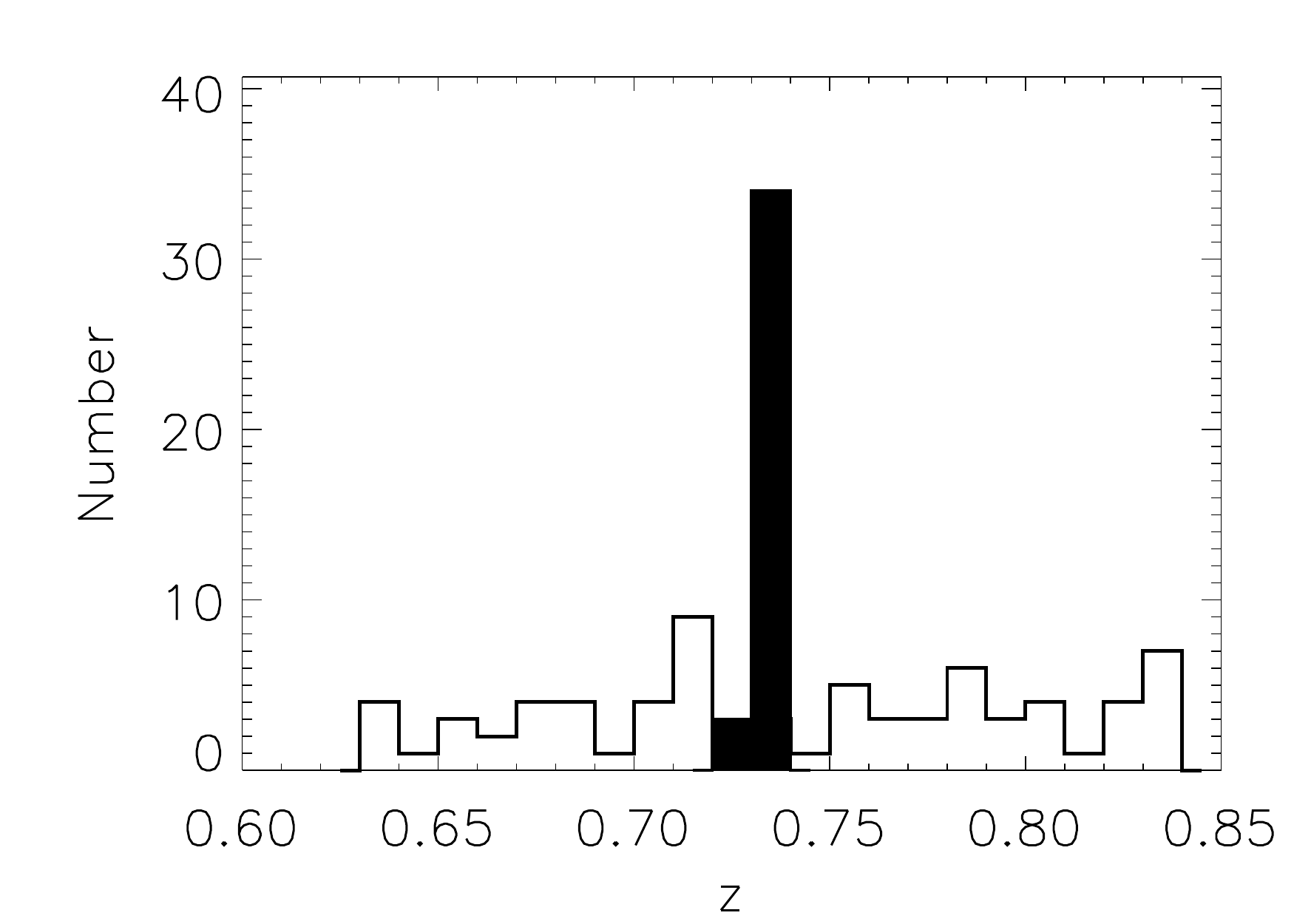}
\caption[The spatial distribution and redshift distributions of MSGs
 in the GOODS-South.]
{ Same as Figure 7, but the GOODS-South field starts from the third panel in this figure.
  The filled red stars represent AGNs (\citealt{18}) and the yellow open triangles
 represent radio galaxies (\citealt{19}) within the same redshift interval.
}
\label{f6a}
\end{figure*}


\begin{figure*}[tp!]
 \setcounter{figure}{7}
\centering
\includegraphics[width=81mm]{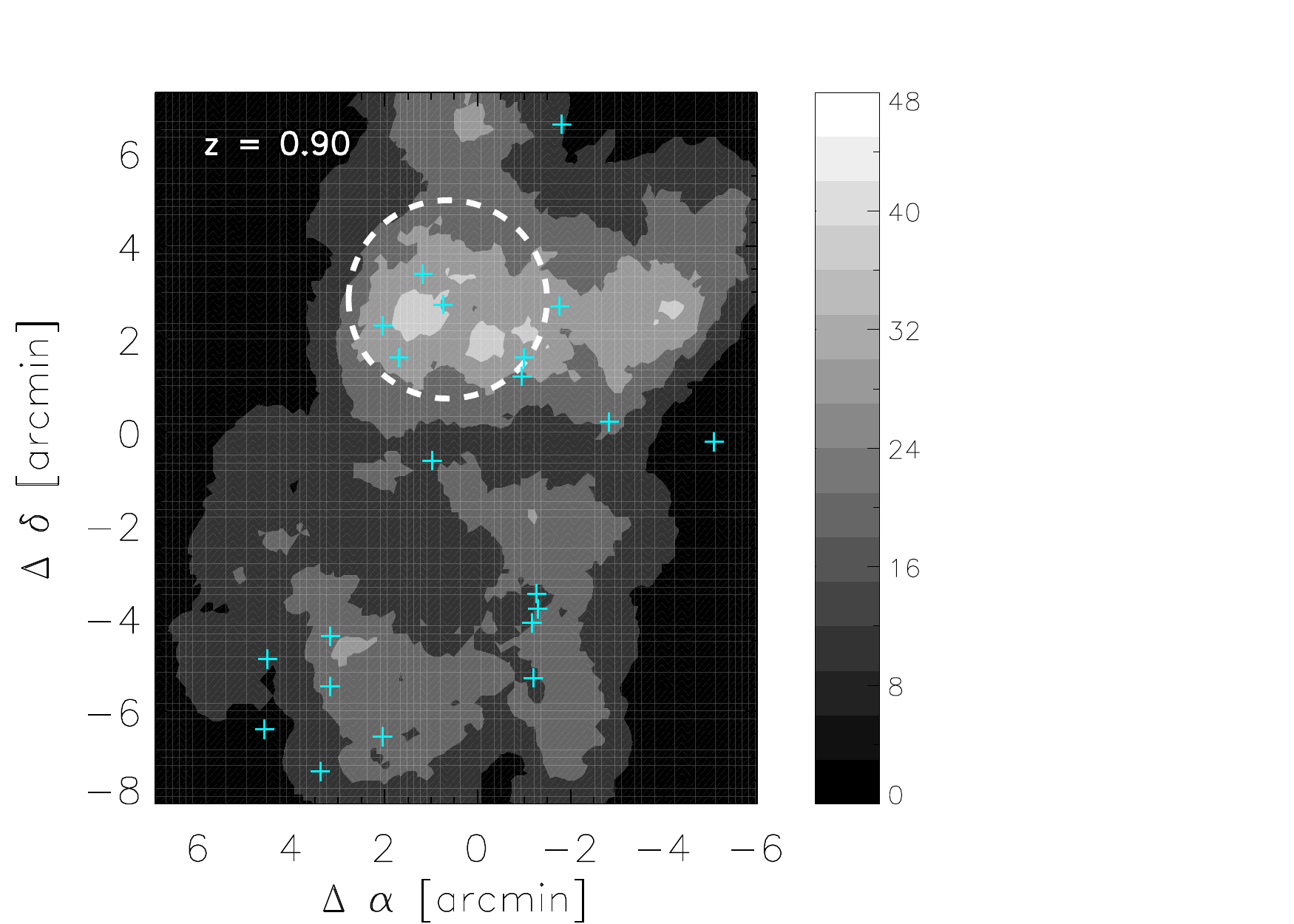}
       \hspace{-0.85in}
\includegraphics[width=81mm]{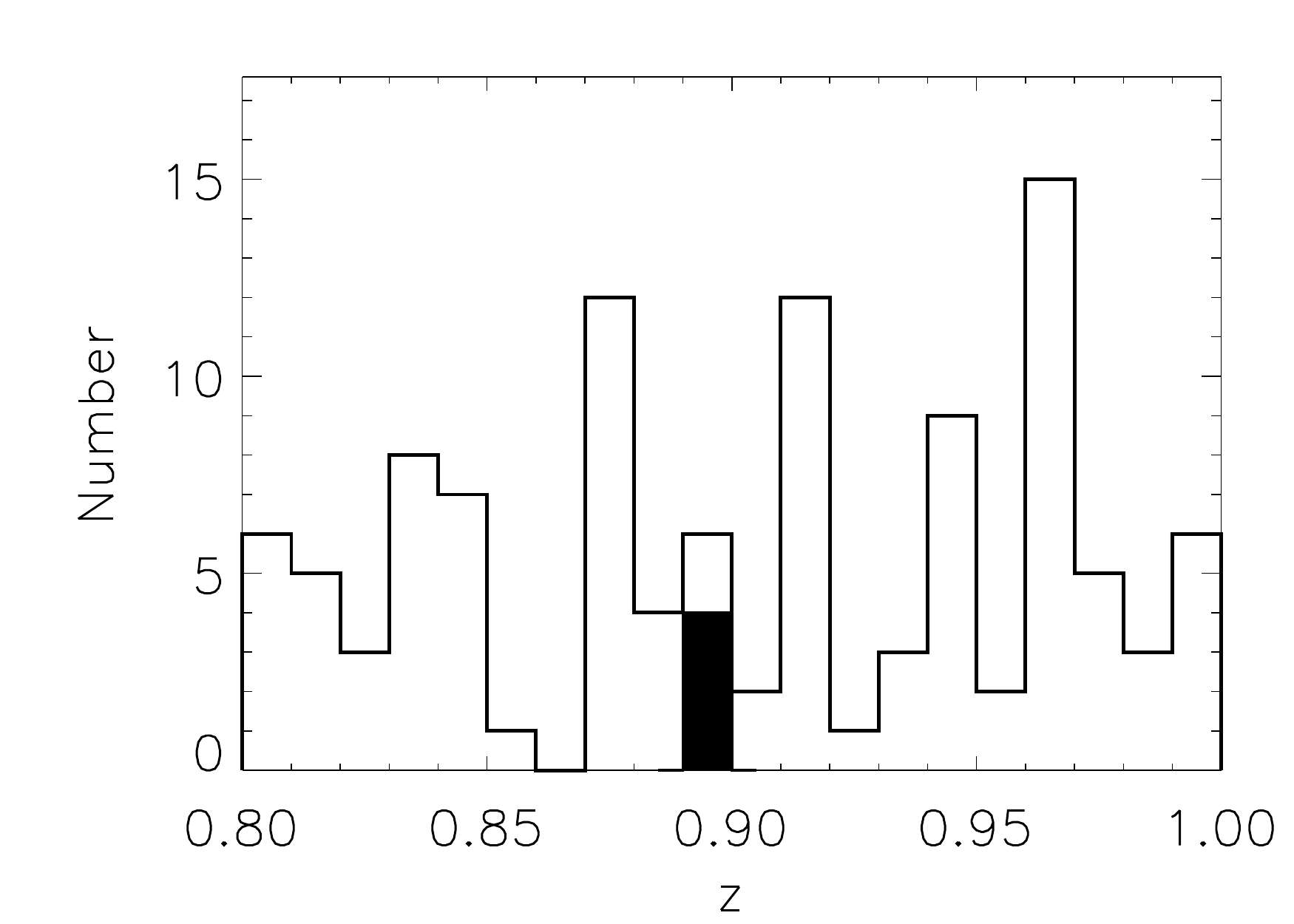}
\\
\includegraphics[width=81mm]{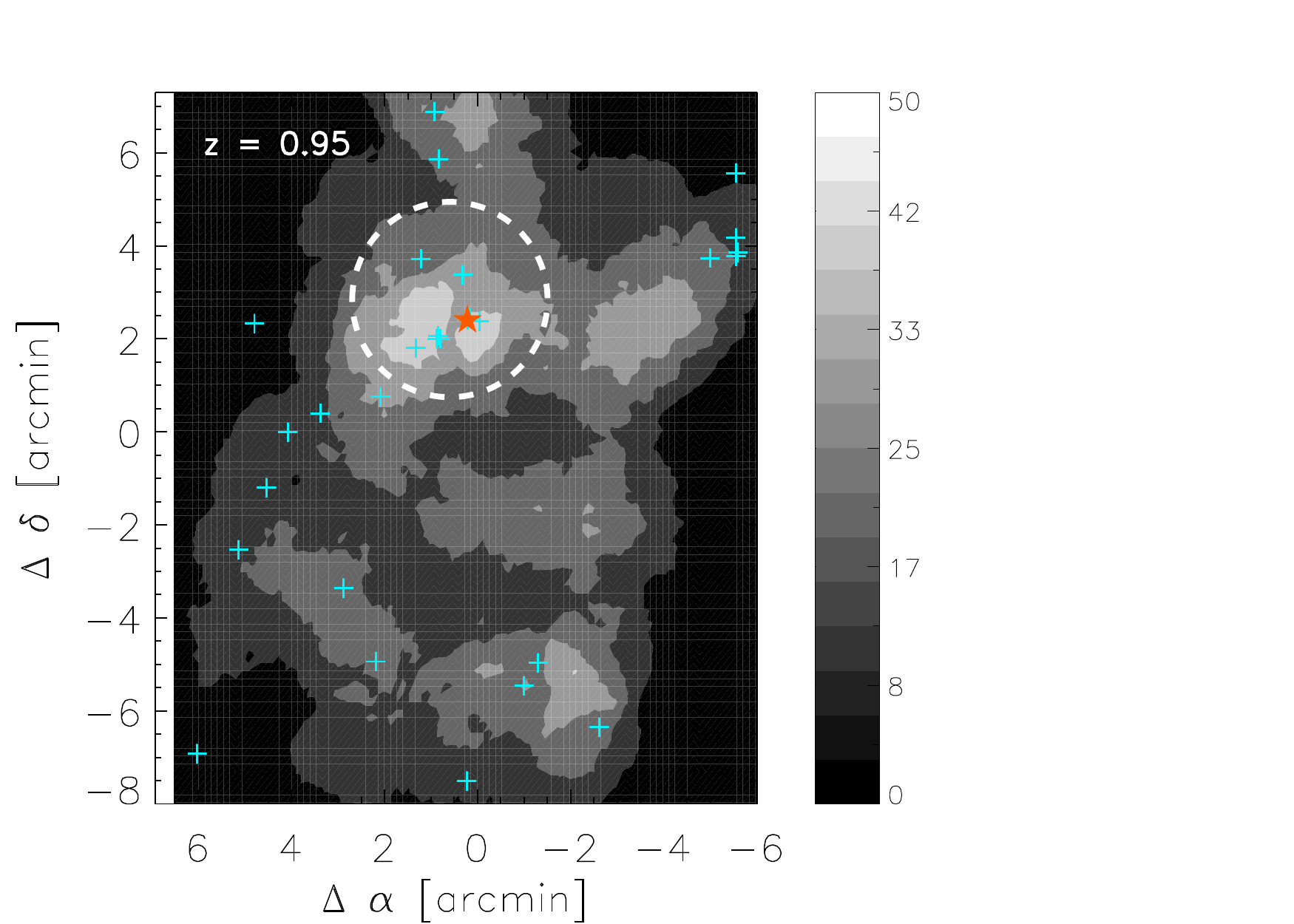}
       \hspace{-0.85in}
\includegraphics[width=81mm]{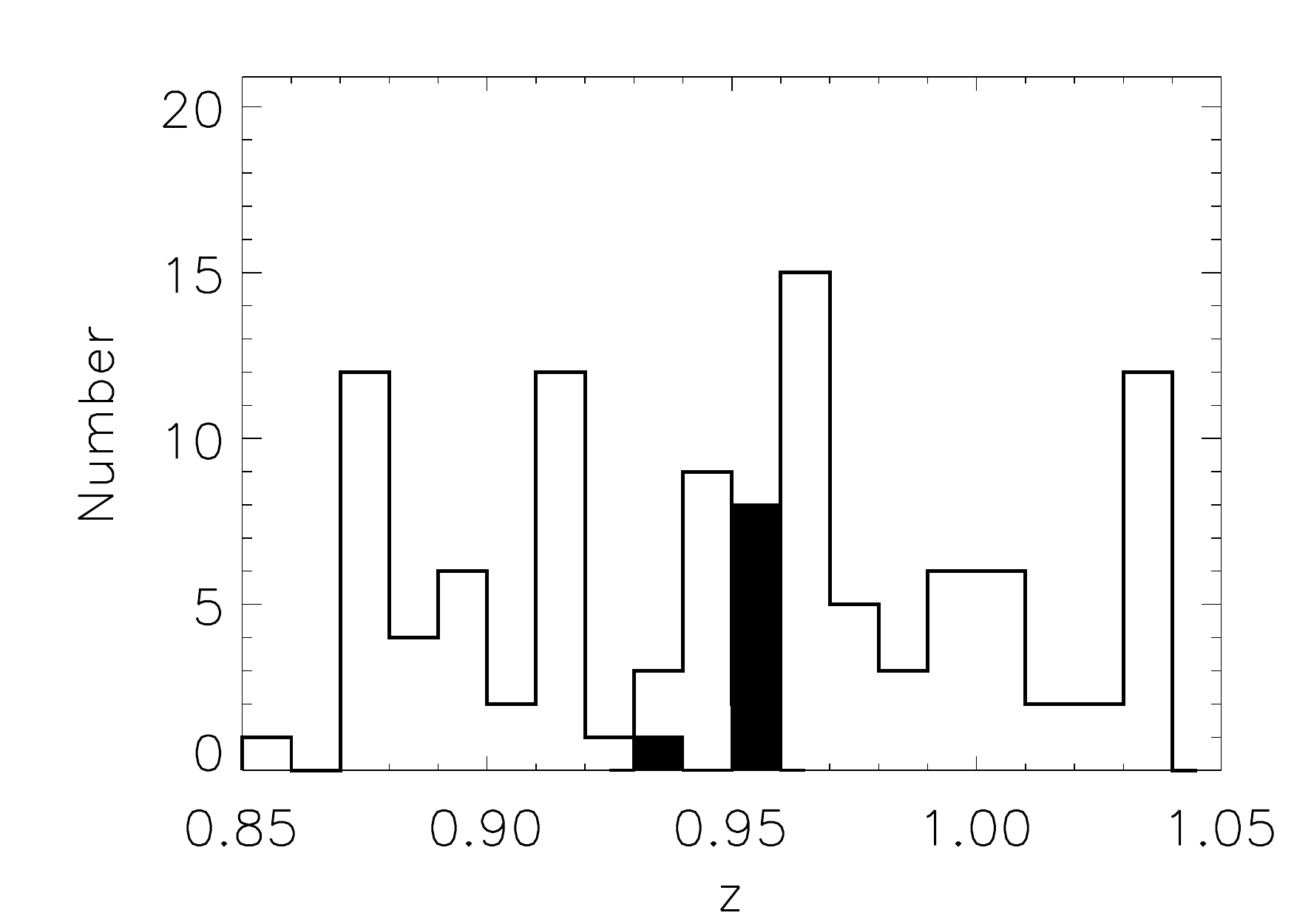}
\\
\includegraphics[width=81mm]{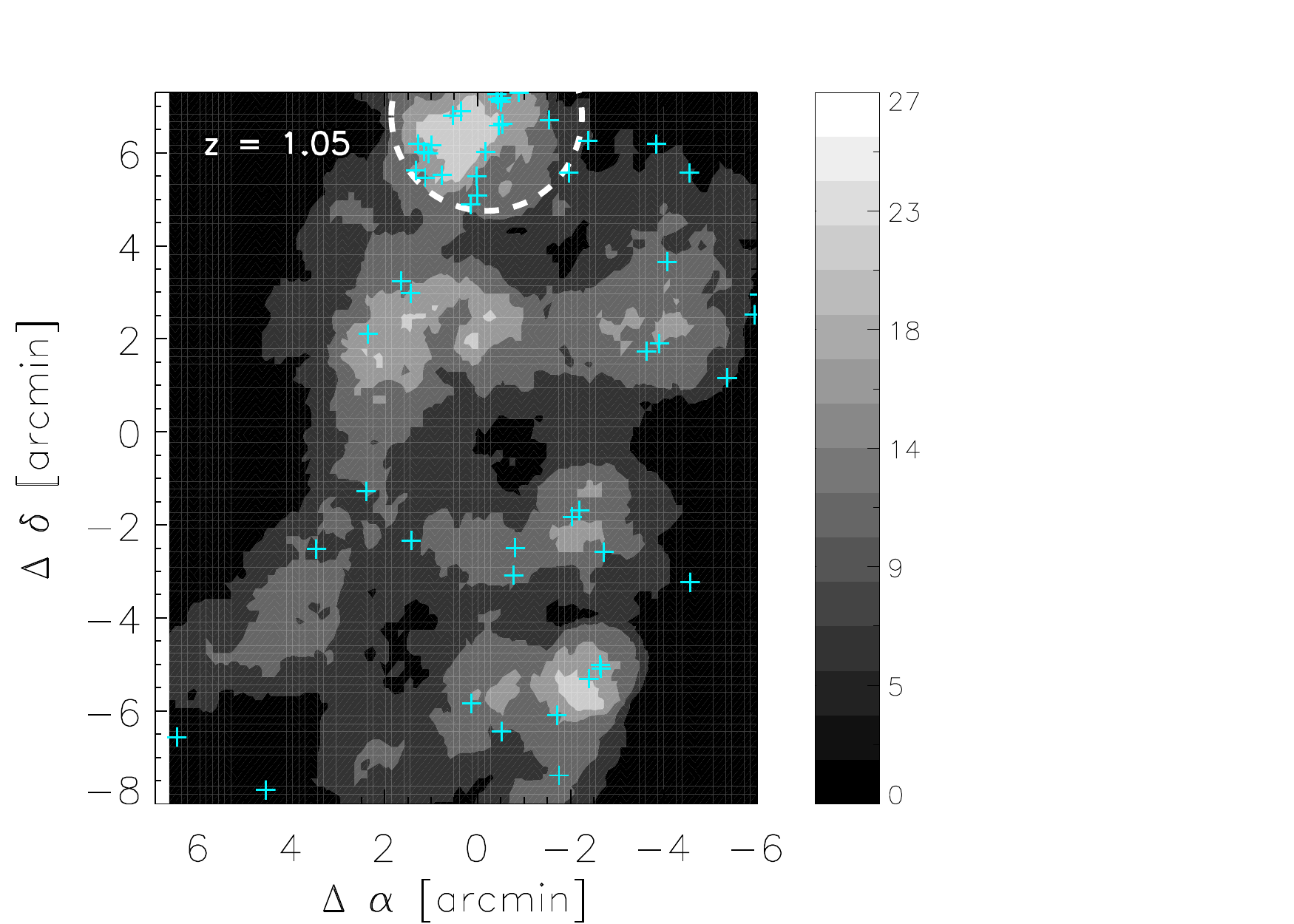}
       \hspace{-0.85in}
\includegraphics[width=81mm]{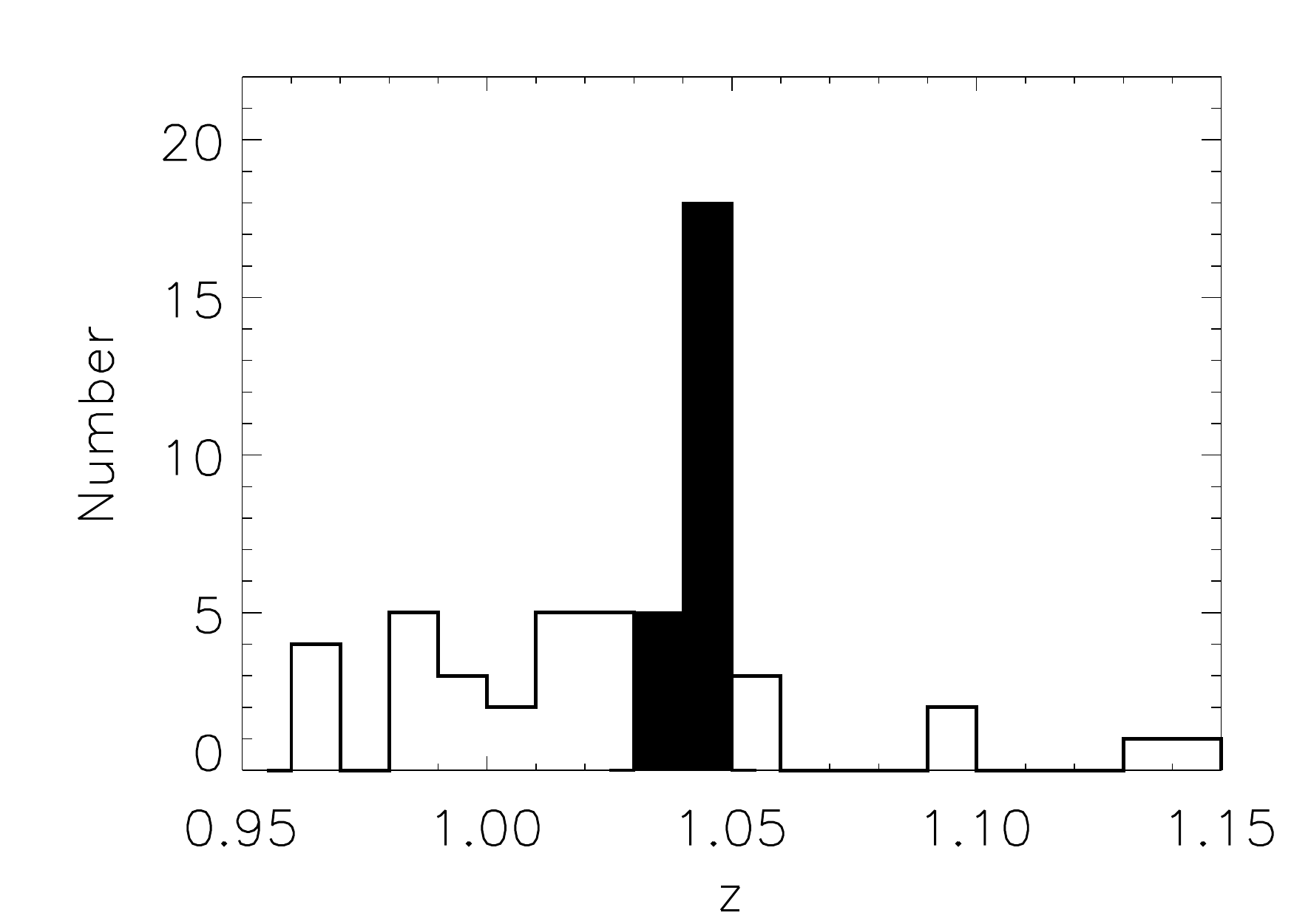}
\\
\includegraphics[width=81mm]{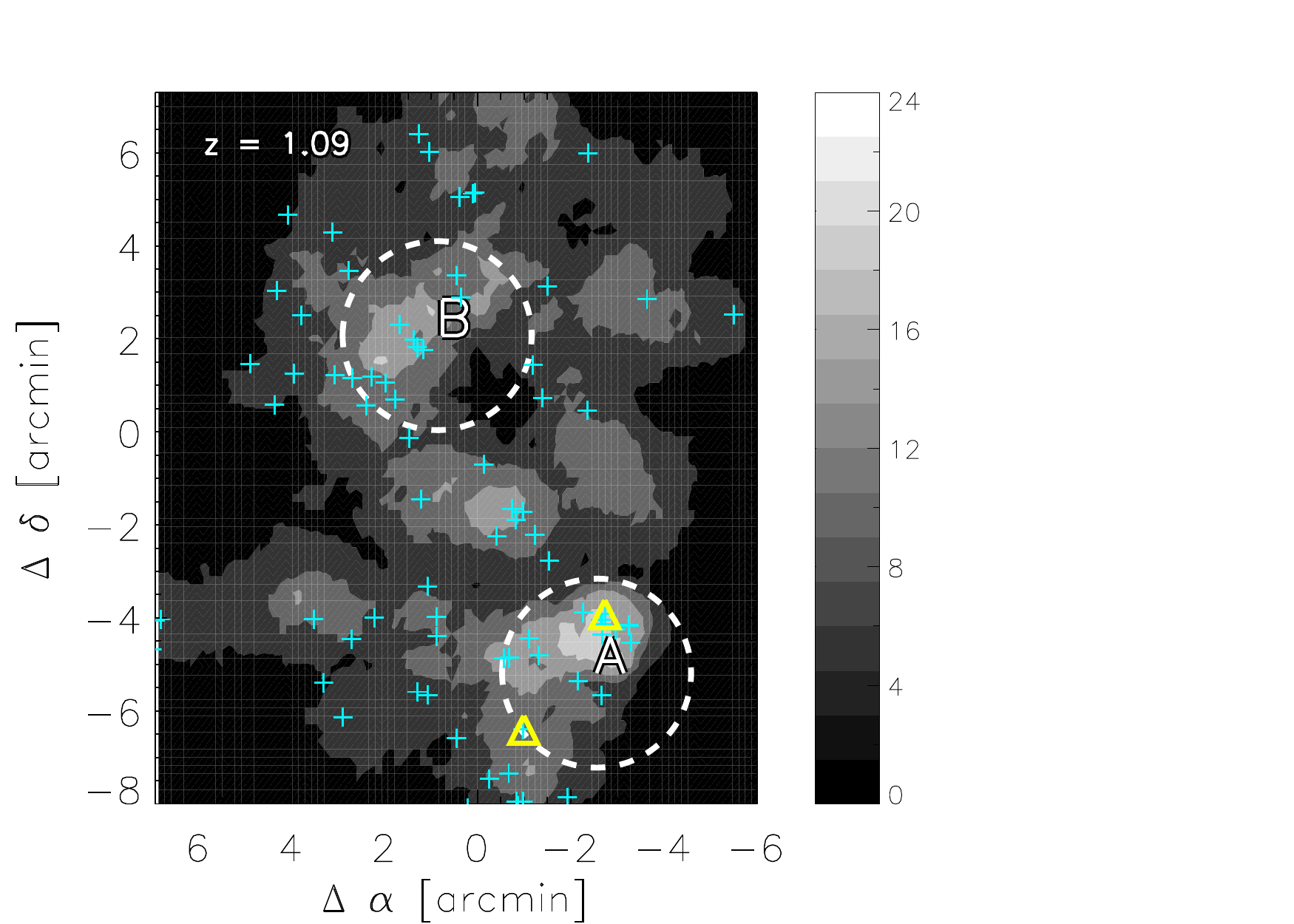}
       \hspace{-0.85in}
\includegraphics[width=81mm]{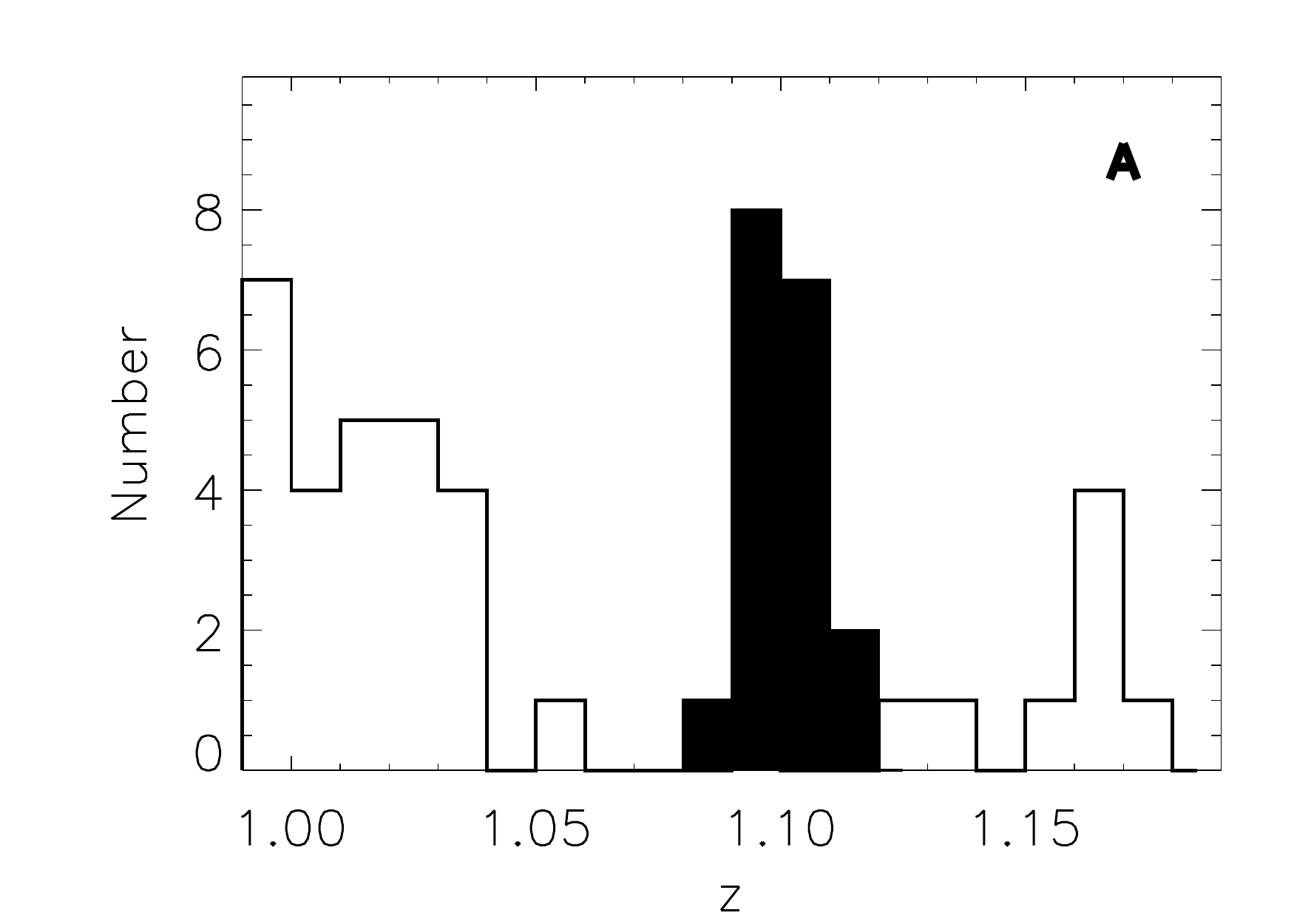}
\caption{(Continued)}
\label{f6b}
\end{figure*}


\begin{figure*}[tp!]
 \setcounter{figure}{7}
\centering
\includegraphics[width=81mm]{f8f1-eps-converted-to.pdf}
       \hspace{-0.85in}
\includegraphics[width=81mm]{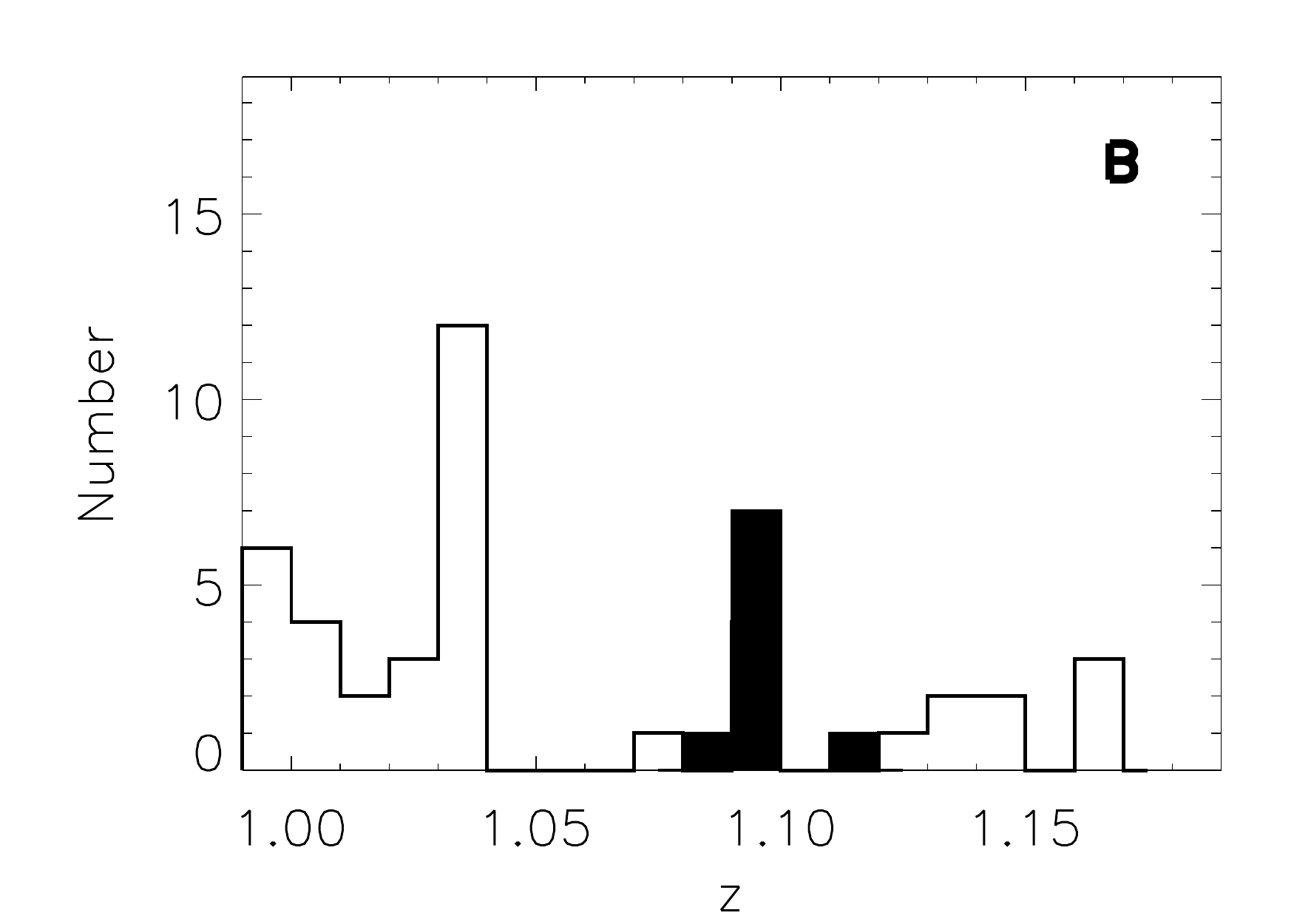}
\\
\includegraphics[width=81mm]{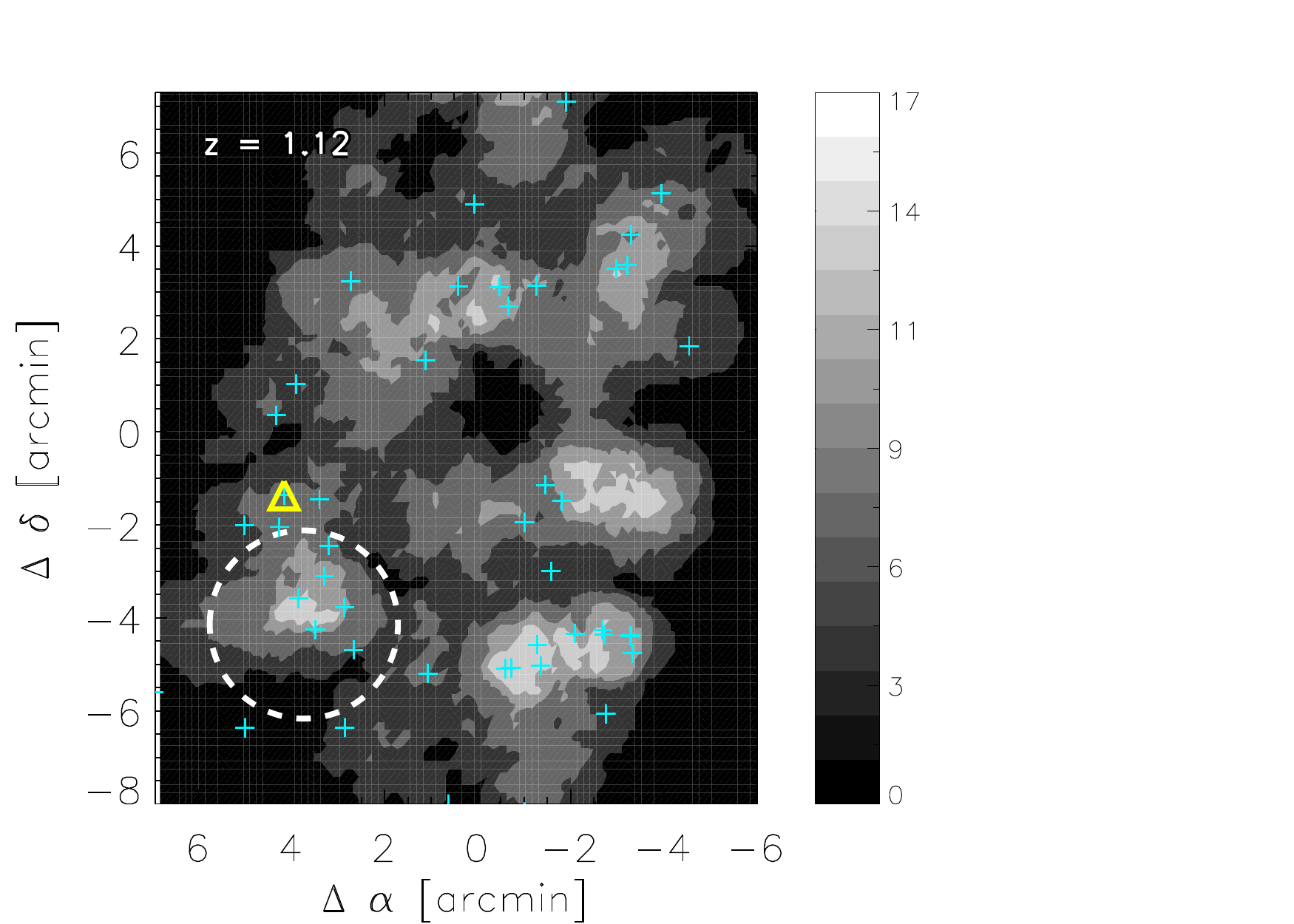}
       \hspace{-0.75in}
\includegraphics[width=81mm]{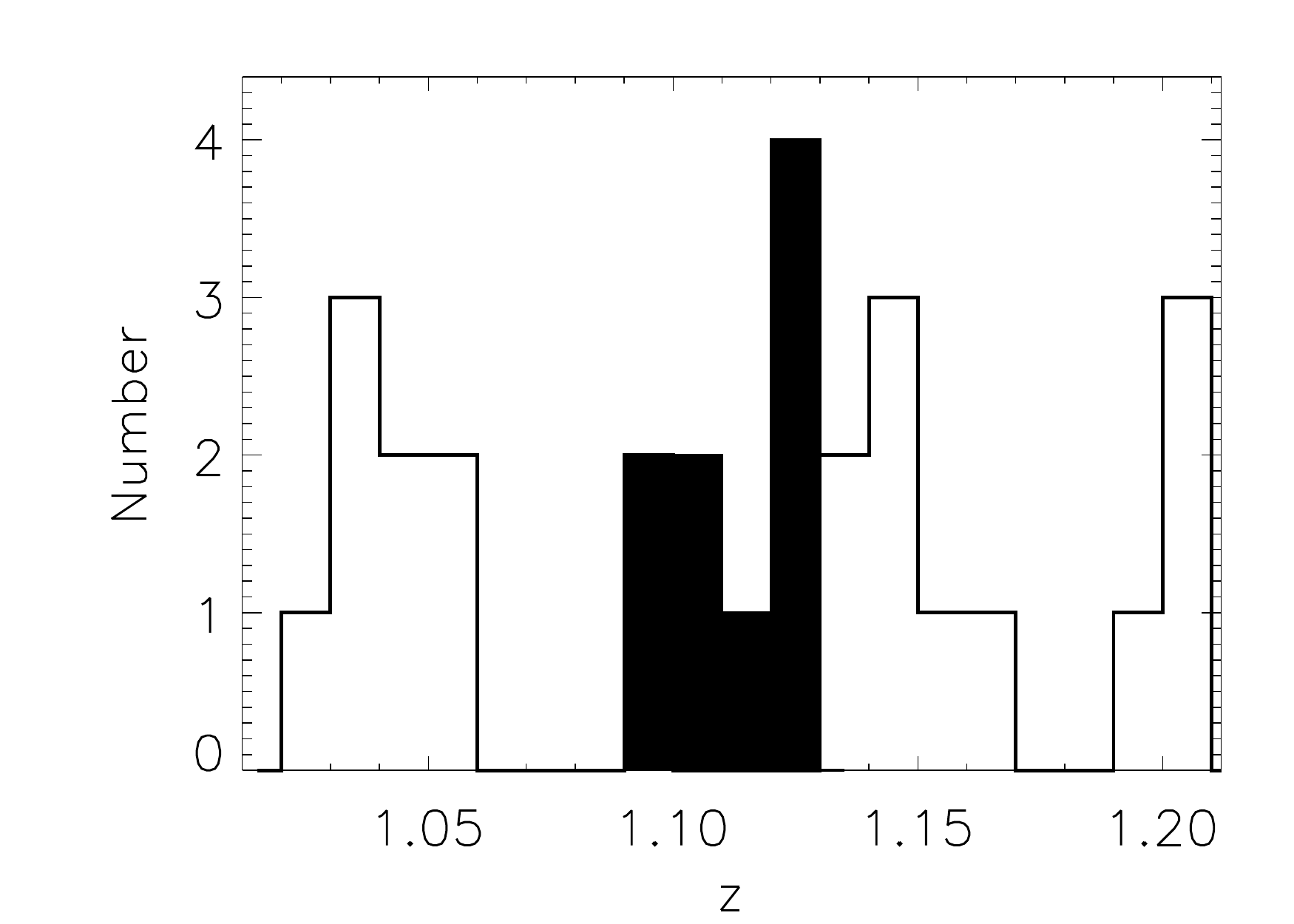}
\\
\includegraphics[width=81mm]{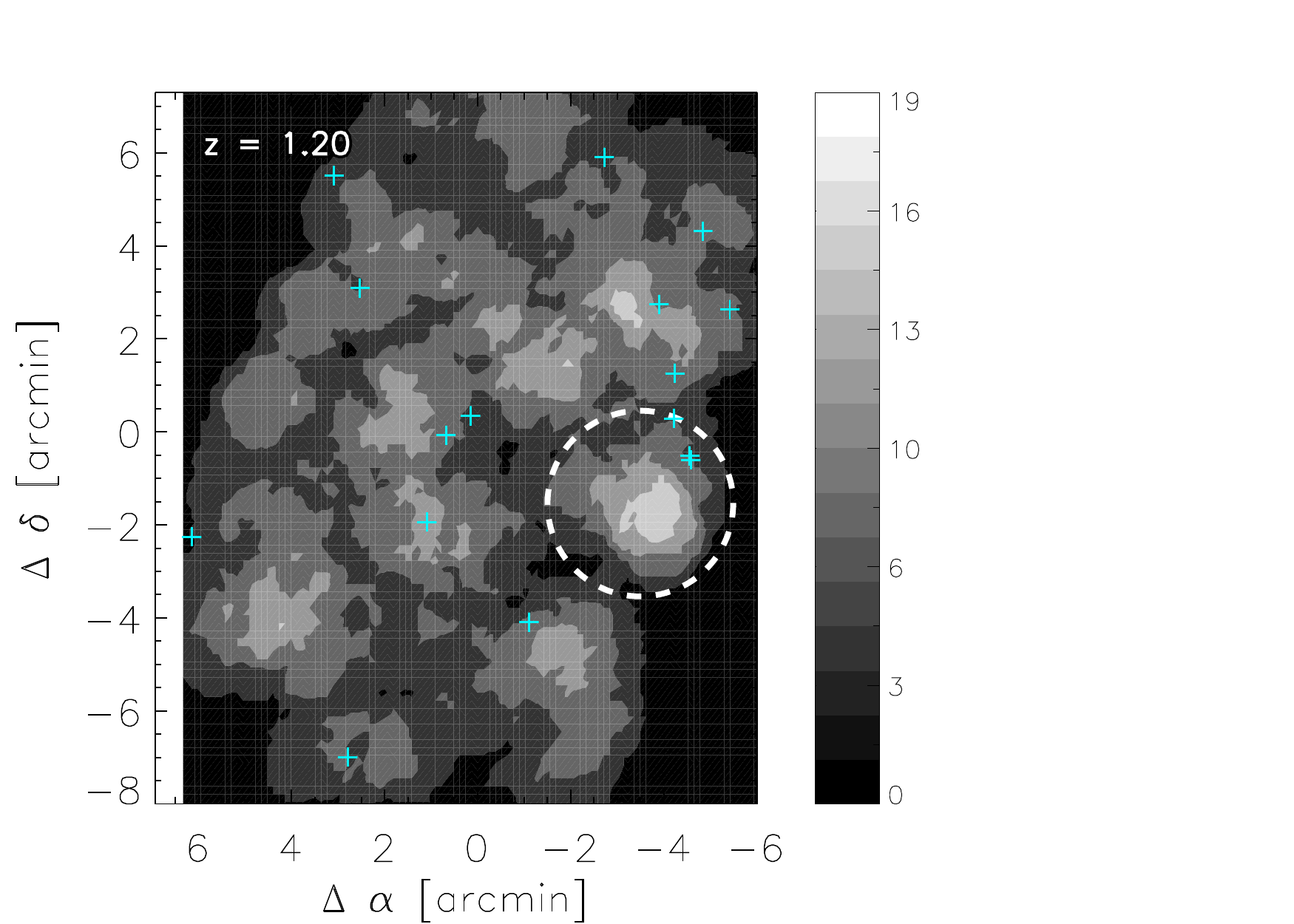}
       \hspace{-0.85in}
\includegraphics[width=81mm]{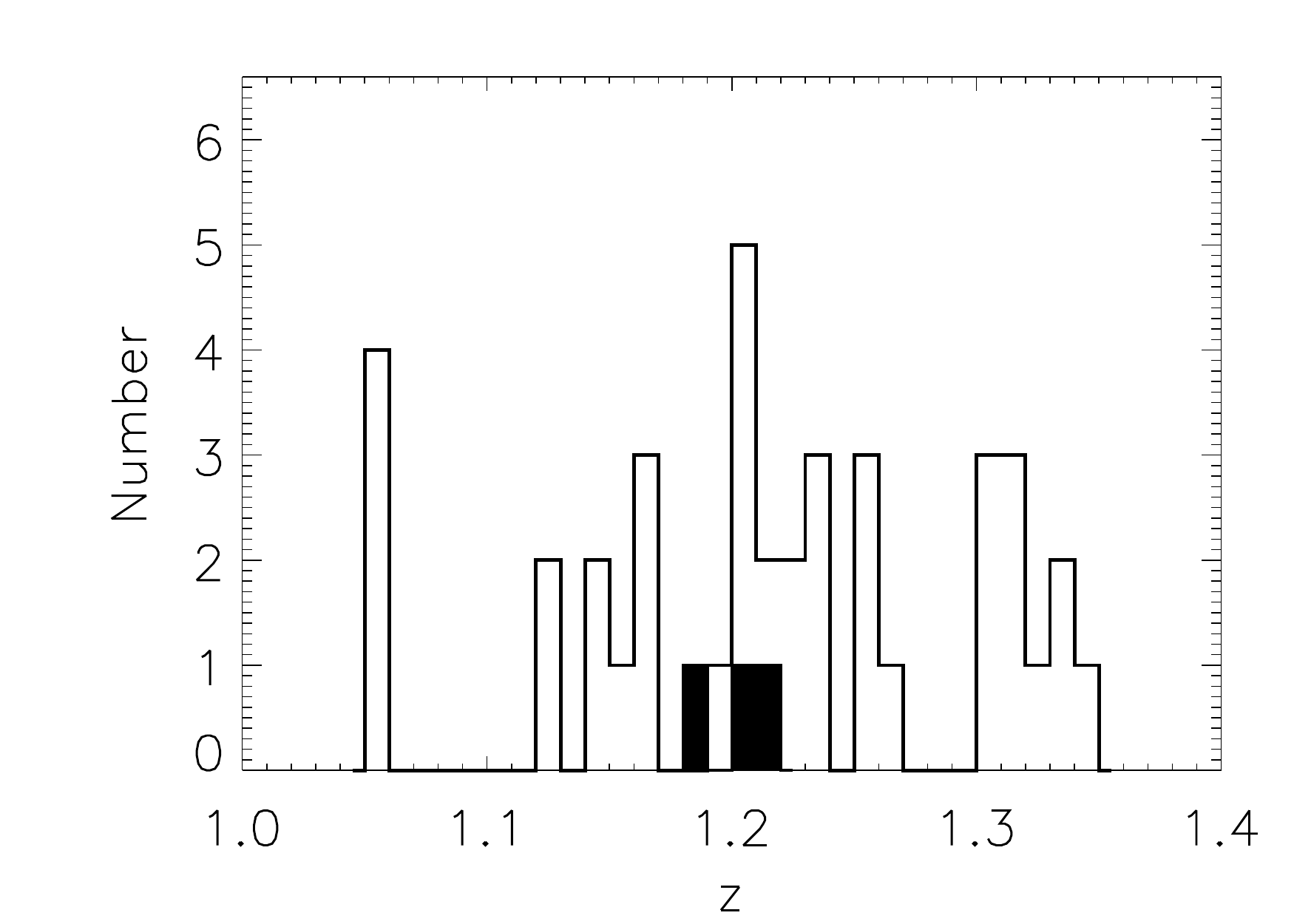}
\\
\includegraphics[width=81mm]{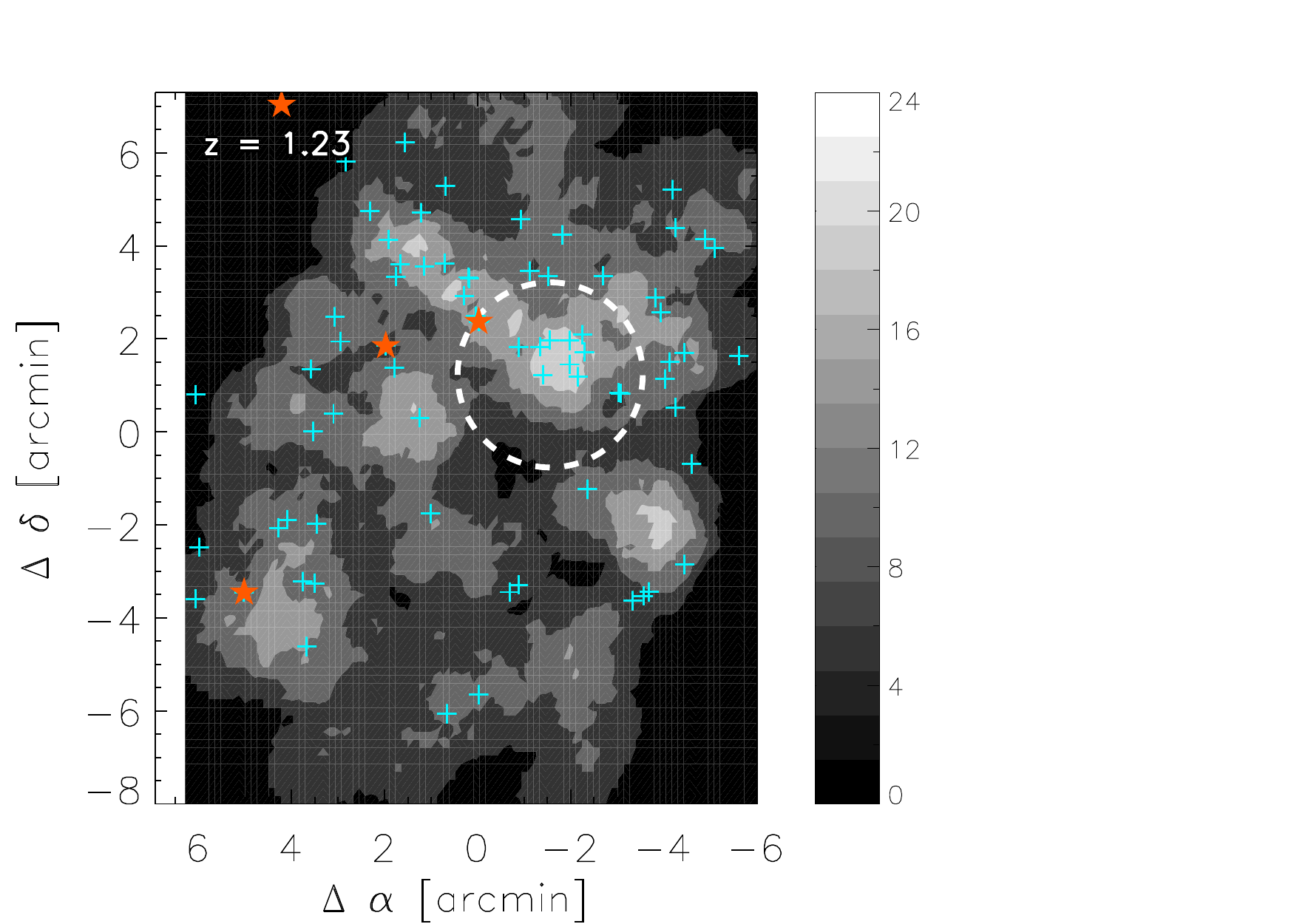}
       \hspace{-0.75in}
\includegraphics[width=81mm]{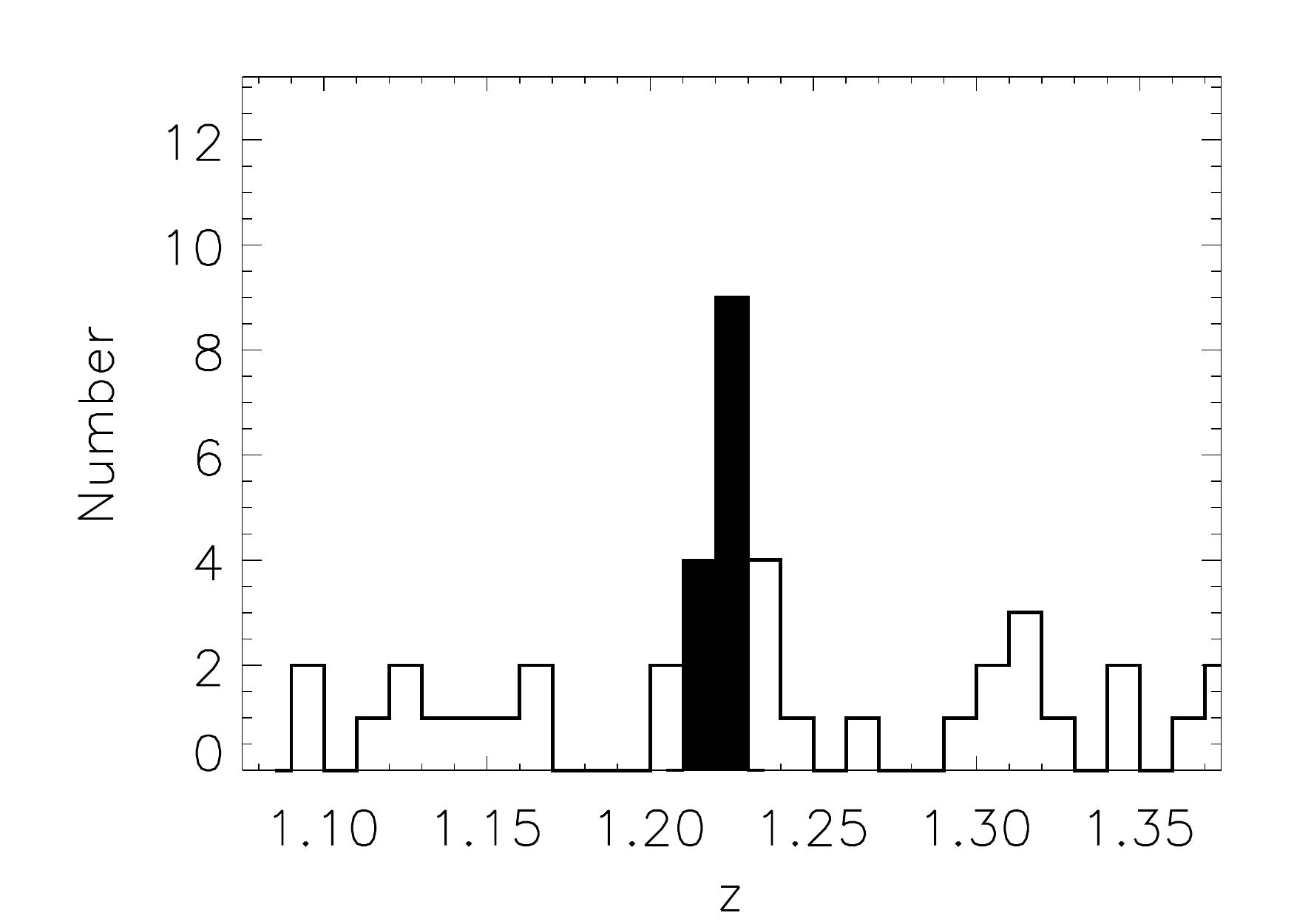}
\caption{(Continued)}
\label{f6c}
\end{figure*}


\begin{figure*}[tp!]
 \setcounter{figure}{7}
\centering
\includegraphics[width=81mm]{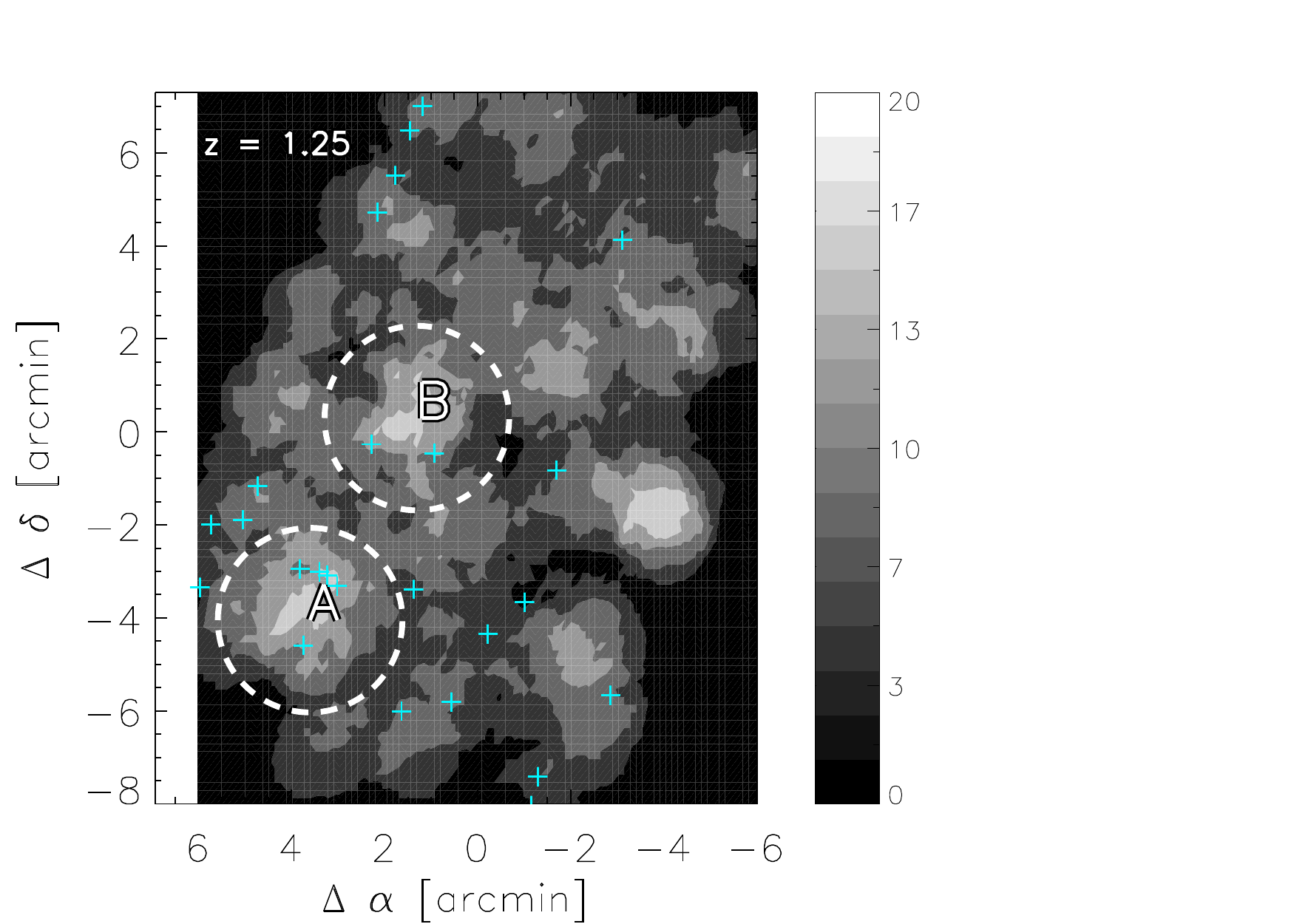}
       \hspace{-0.85in}
\includegraphics[width=81mm]{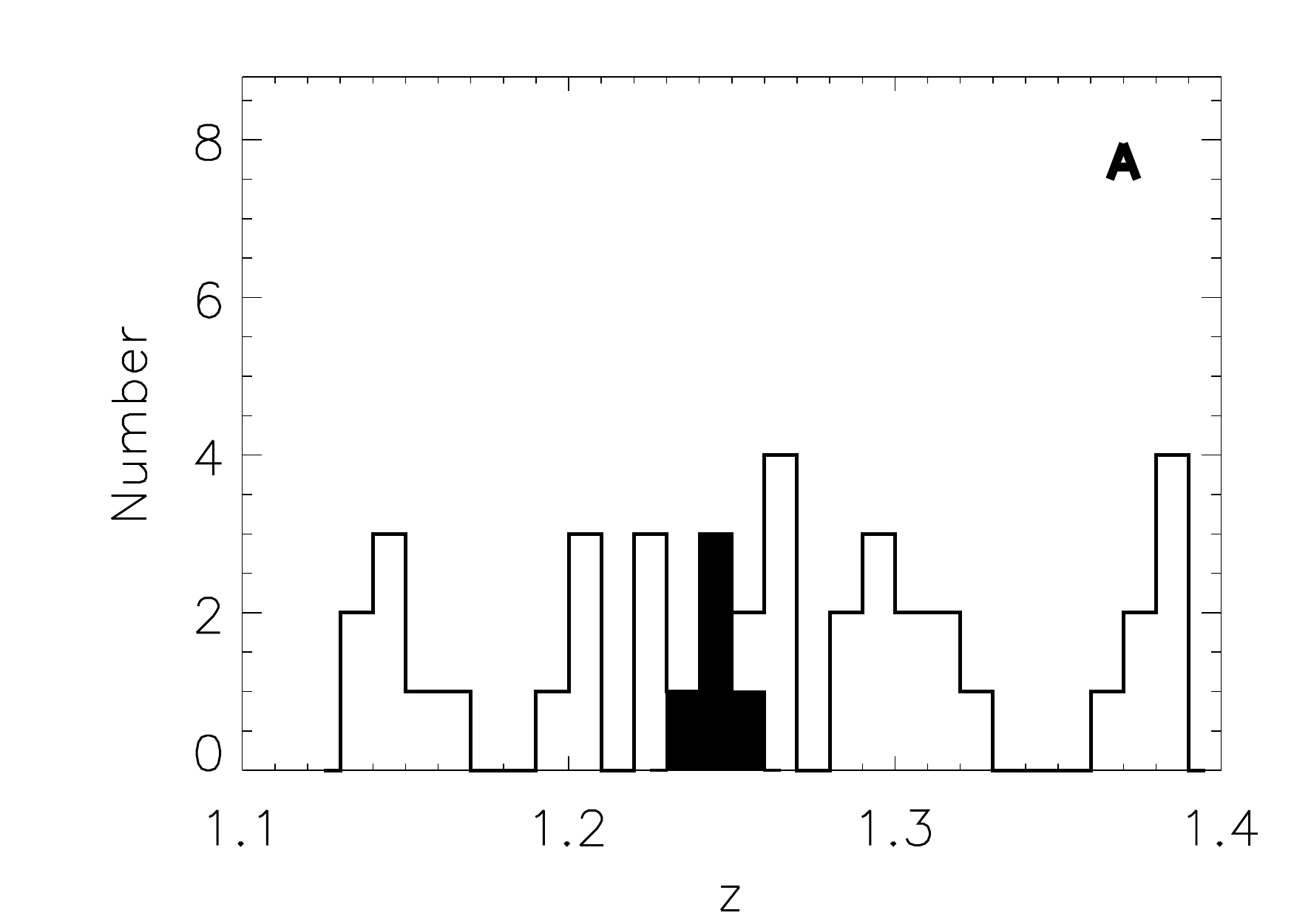}
\\
\includegraphics[width=81mm]{f8j1-eps-converted-to.pdf}
       \hspace{-0.85in}
\includegraphics[width=81mm]{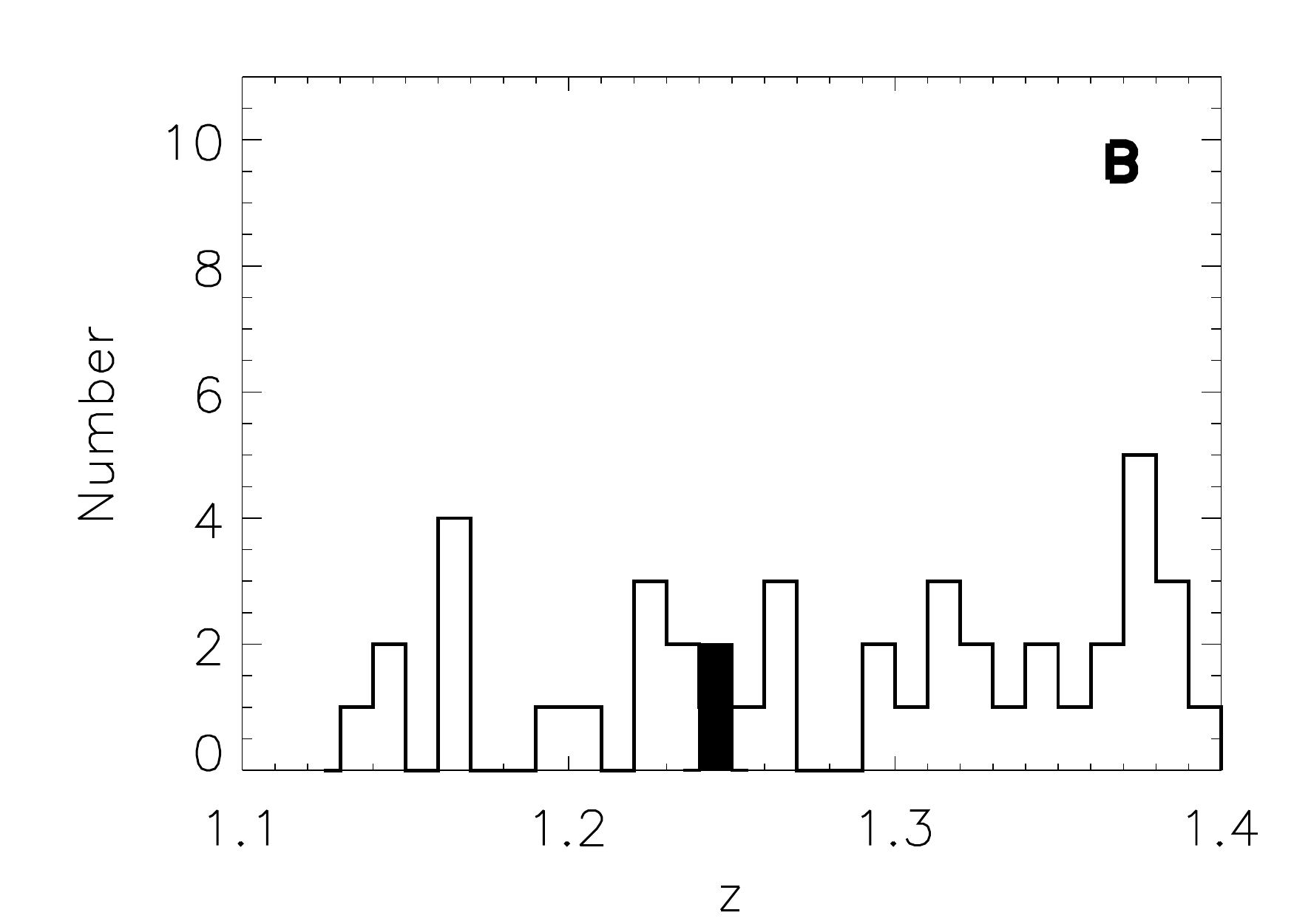}
\\
\includegraphics[width=81mm]{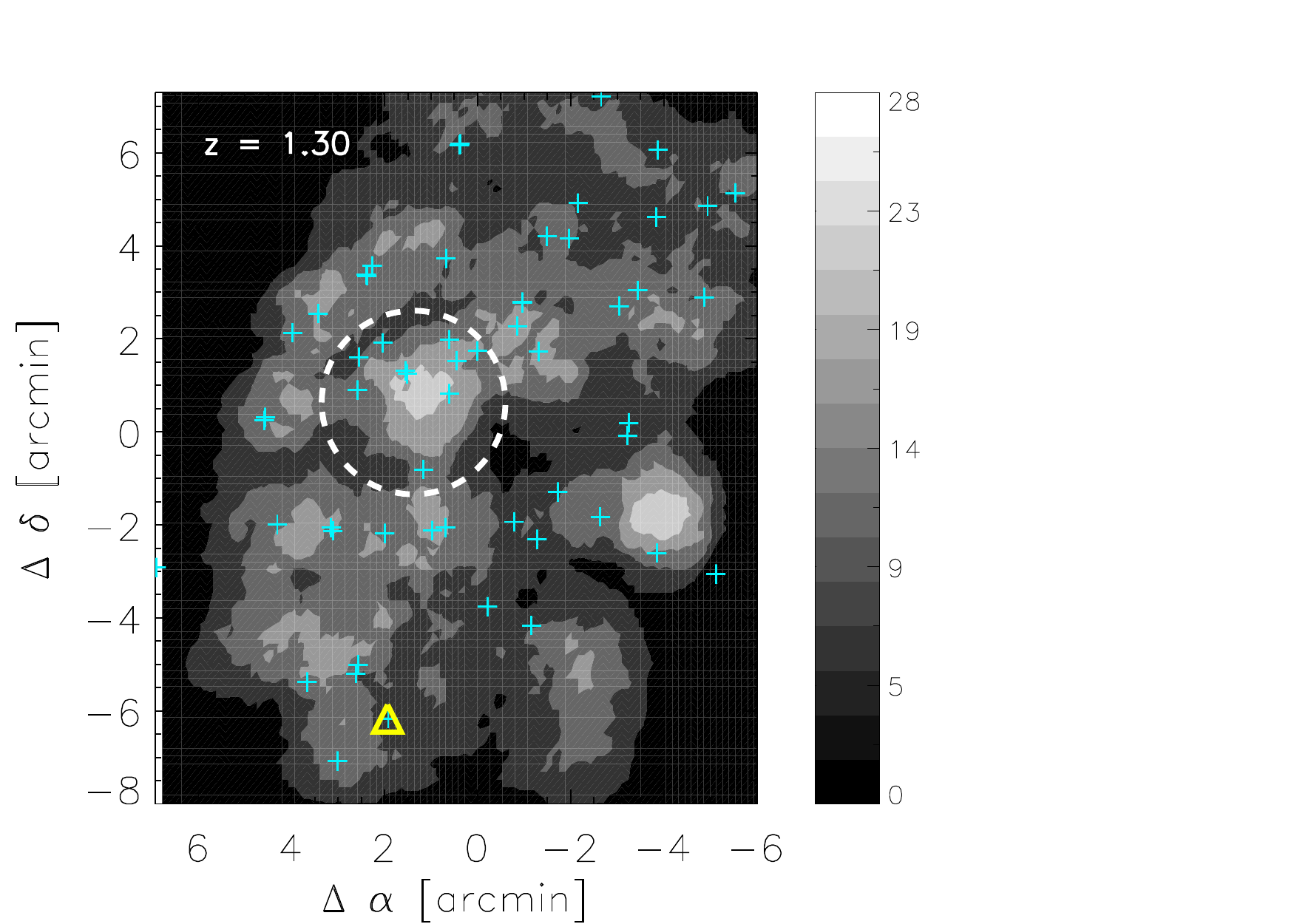}
       \hspace{-0.85in}
\includegraphics[width=81mm]{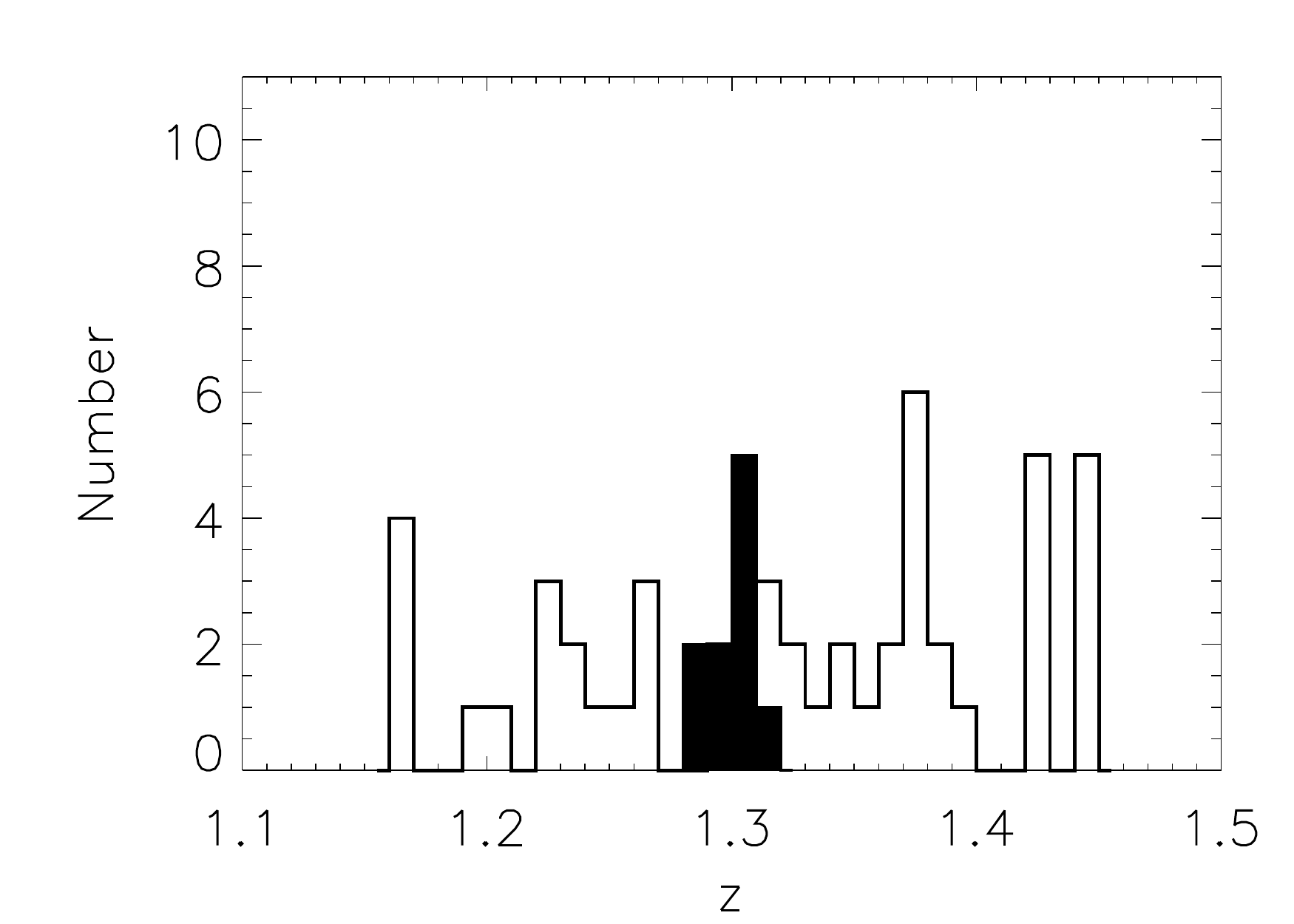}
\\
\includegraphics[width=81mm]{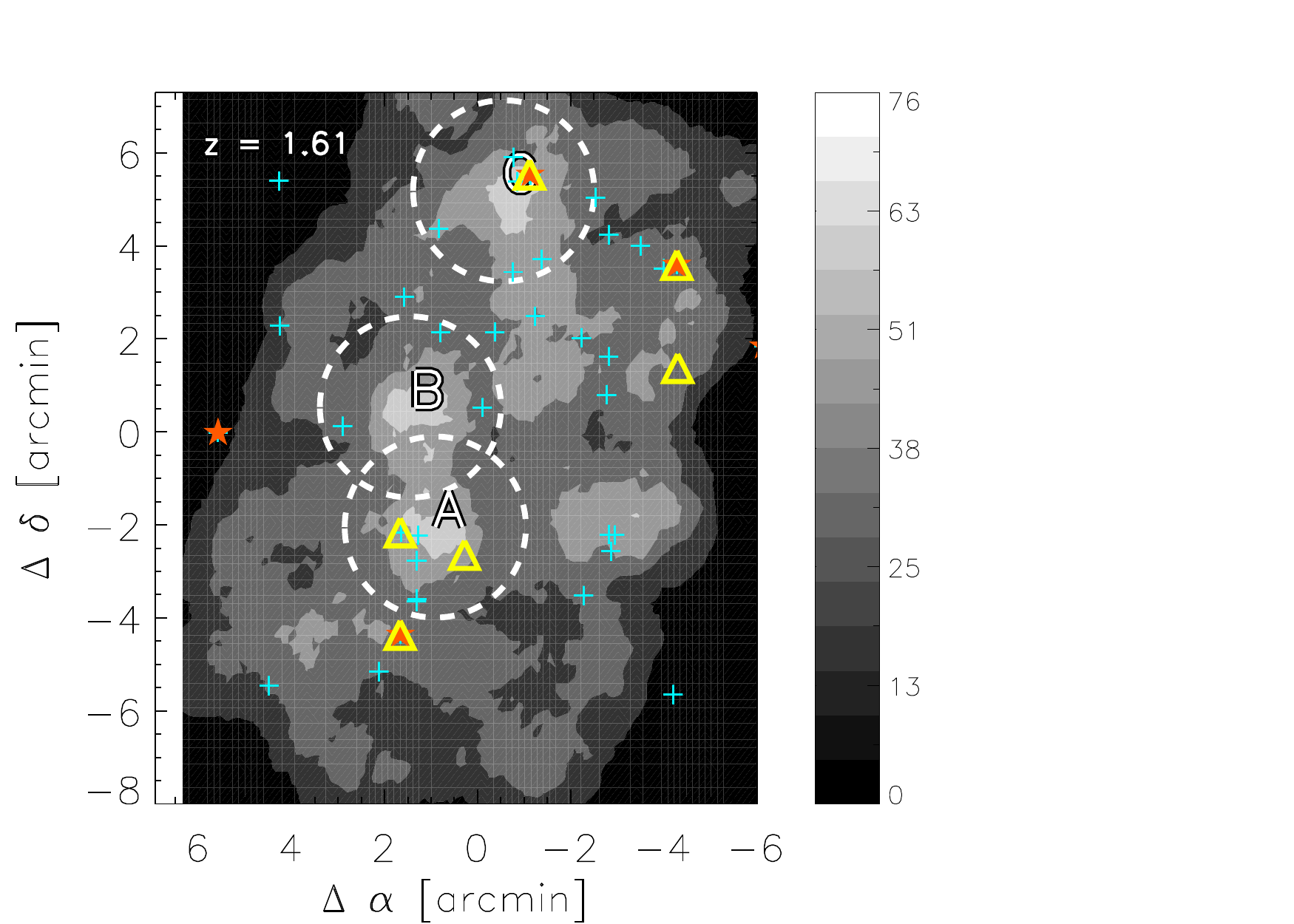}
       \hspace{-0.85in}
\includegraphics[width=81mm]{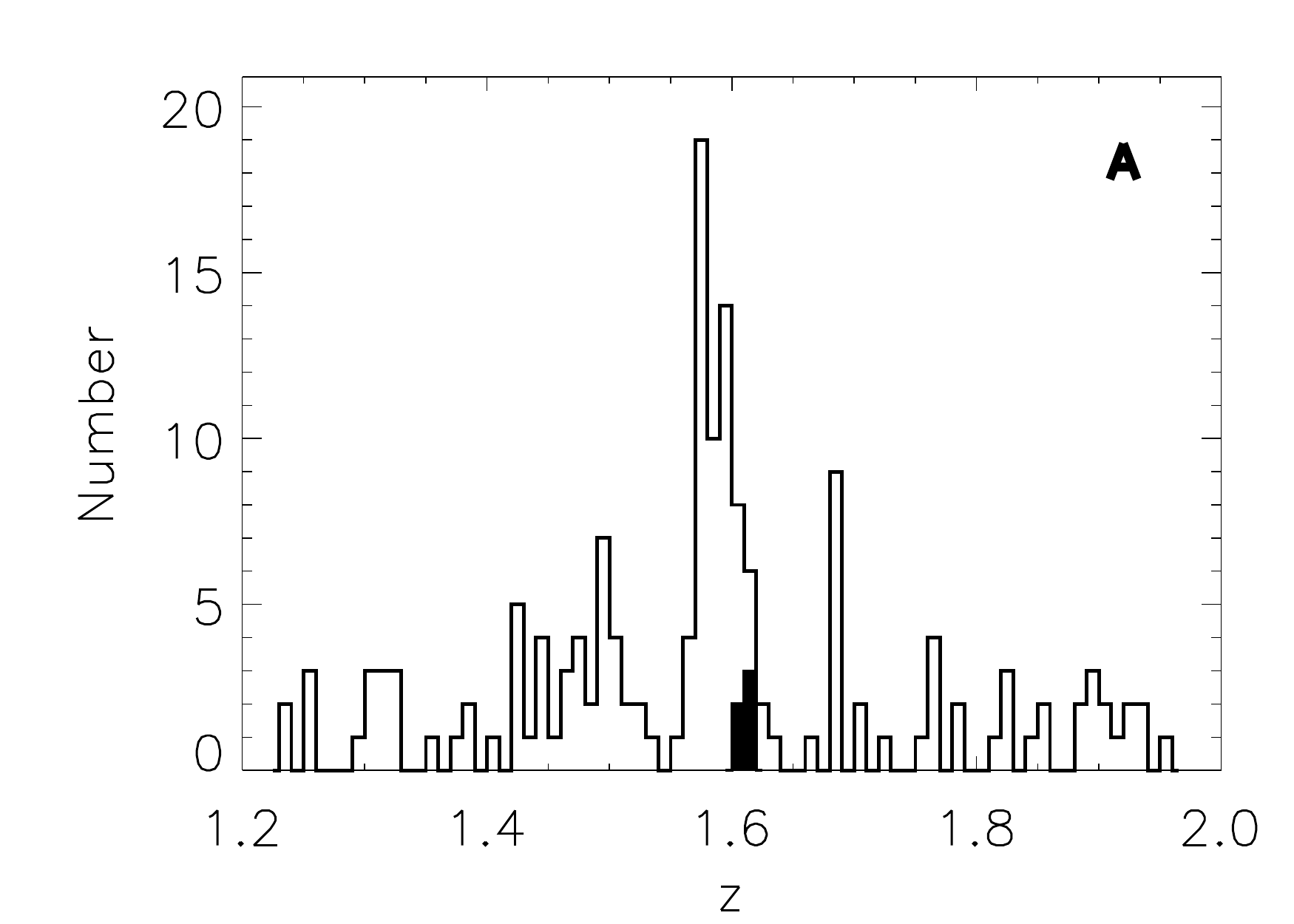}
\caption{(Continued)}
\label{f6c2}
\end{figure*}


\begin{figure*}[tp!]
 \setcounter{figure}{7}
\centering
\includegraphics[width=81mm]{f8l1-eps-converted-to.pdf}
       \hspace{-0.85in}
\includegraphics[width=81mm]{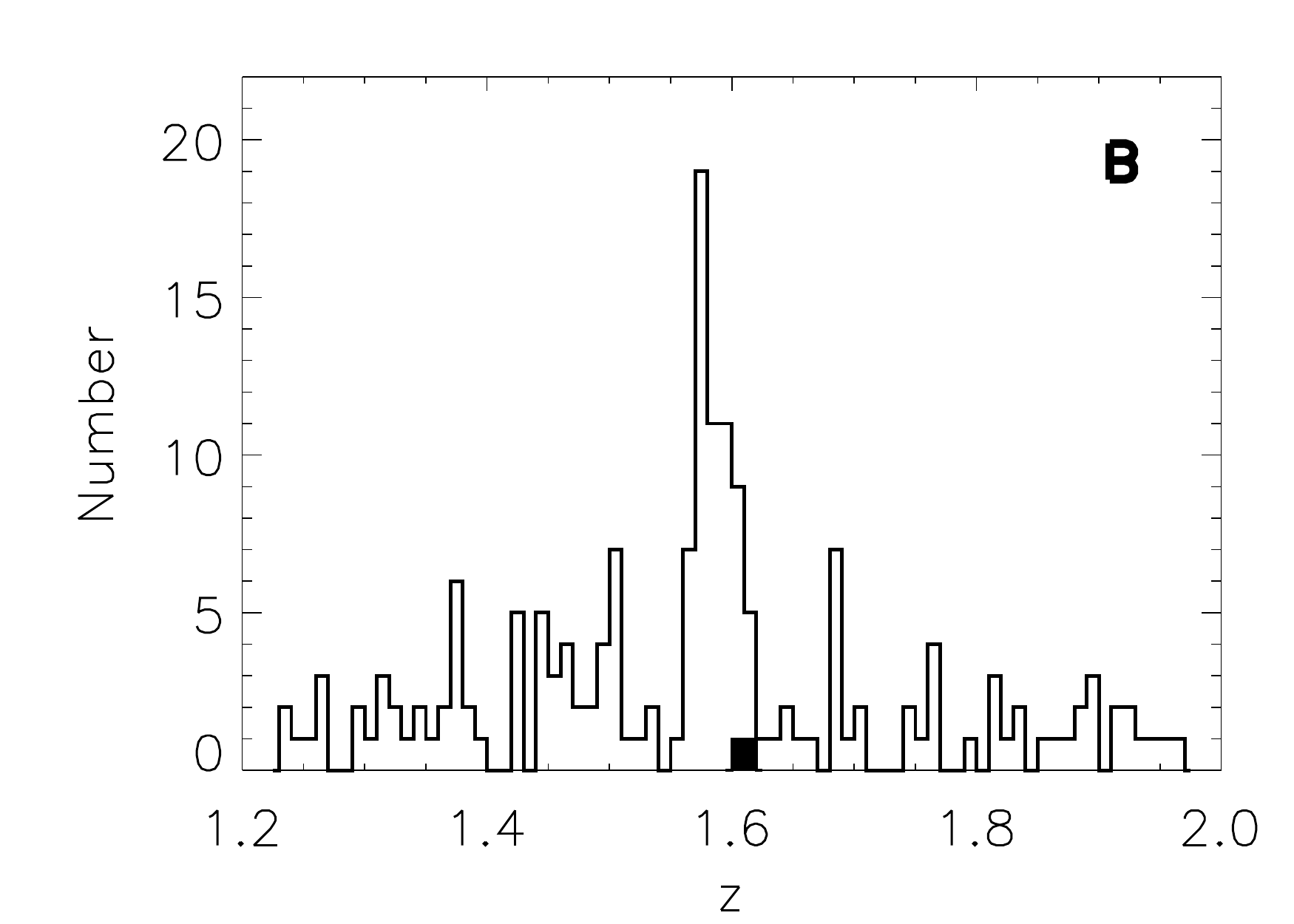} \\
\includegraphics[width=81mm]{f8l1-eps-converted-to.pdf}
       \hspace{-0.85in}
\includegraphics[width=81mm]{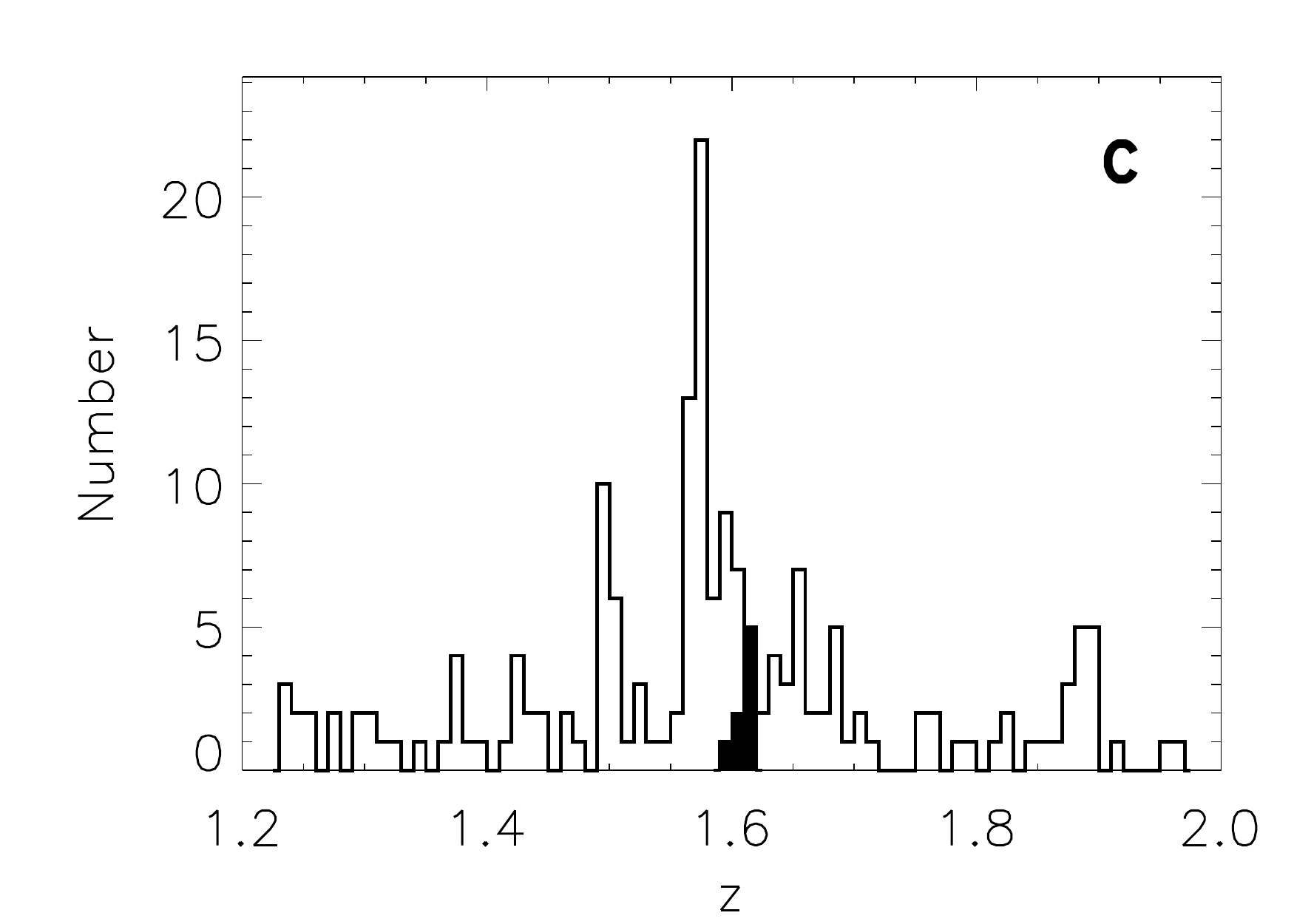}
\\
\includegraphics[width=81mm]{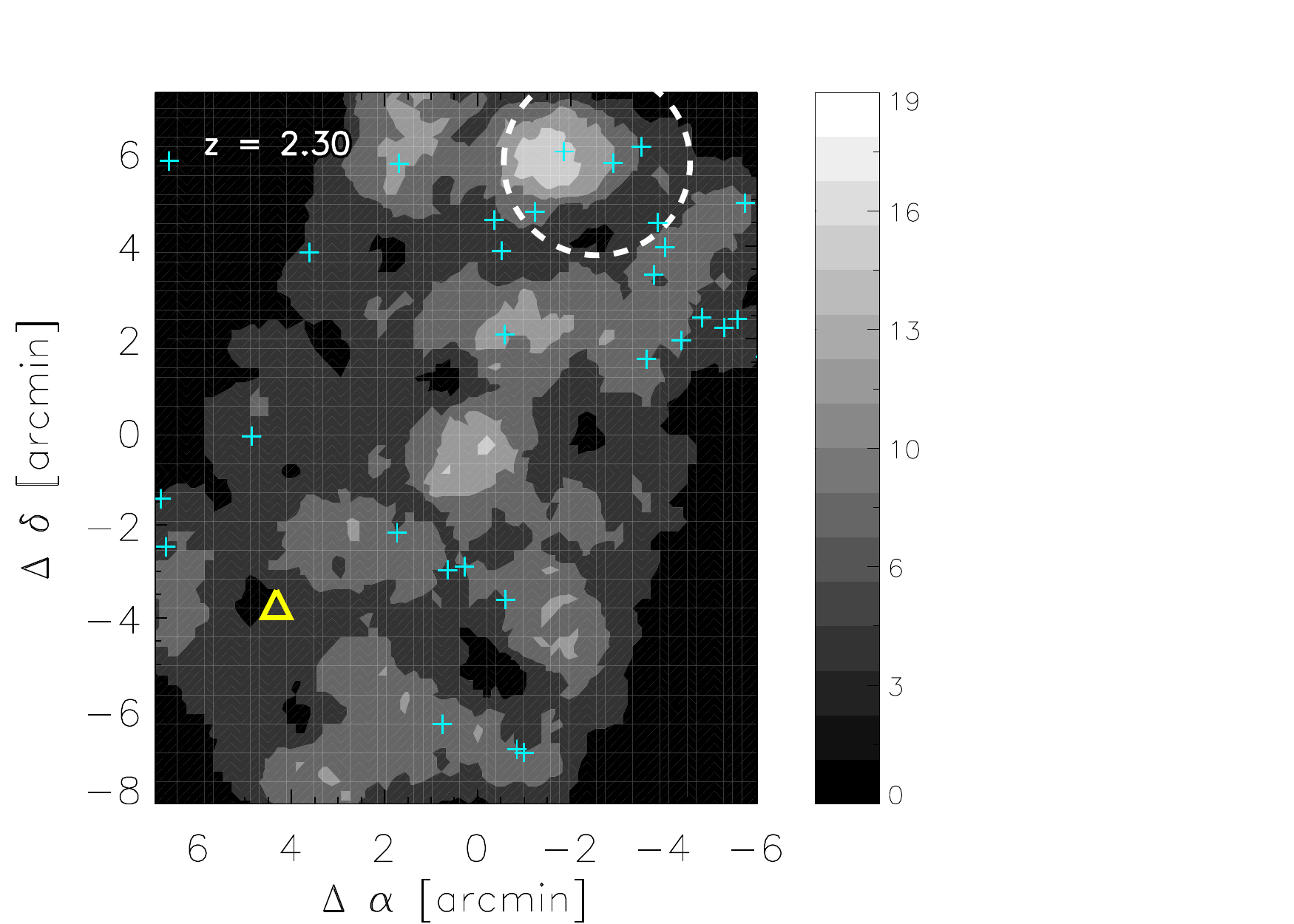}
       \hspace{-0.85in}
\includegraphics[width=81mm]{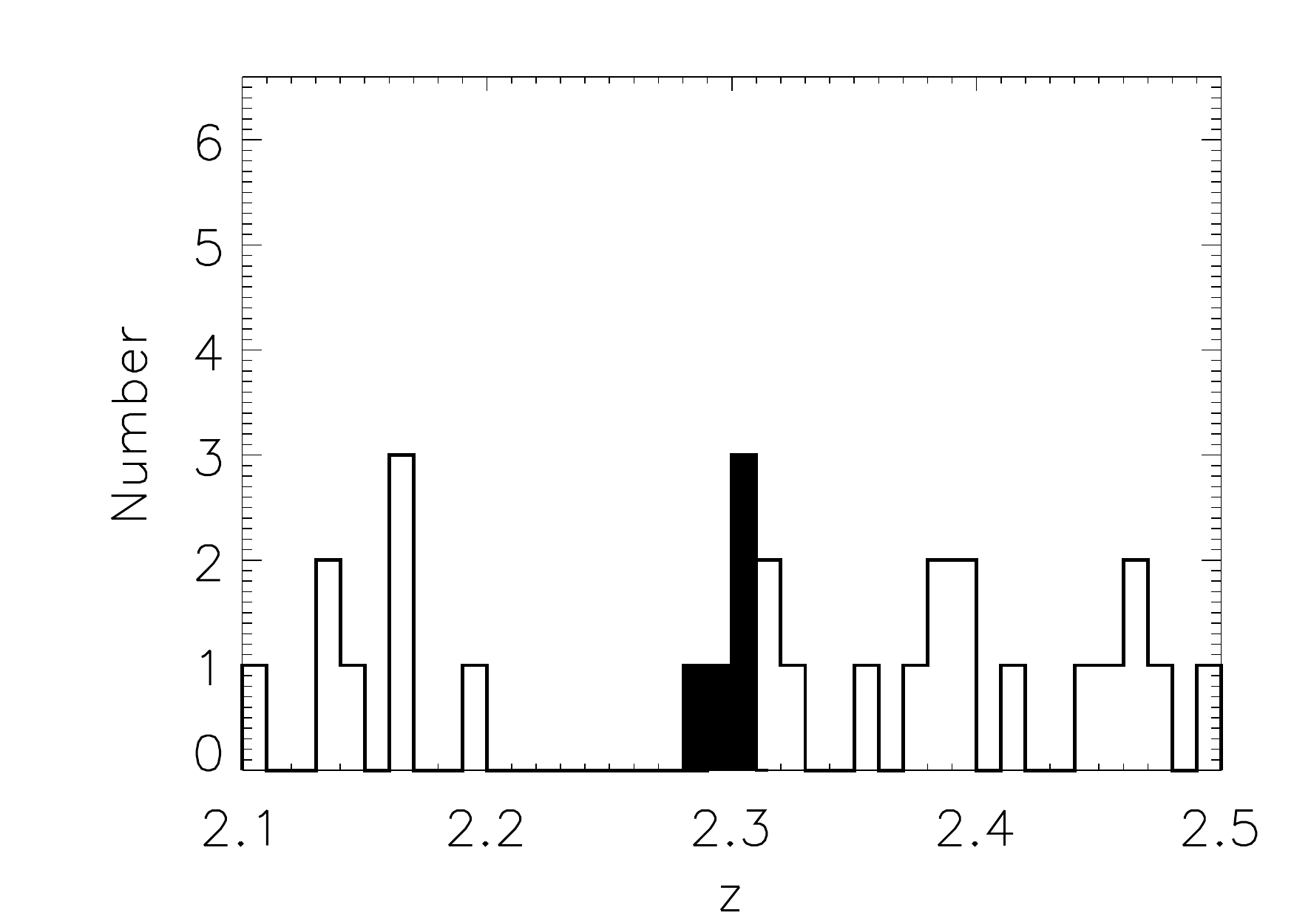}
\\
\includegraphics[width=81mm]{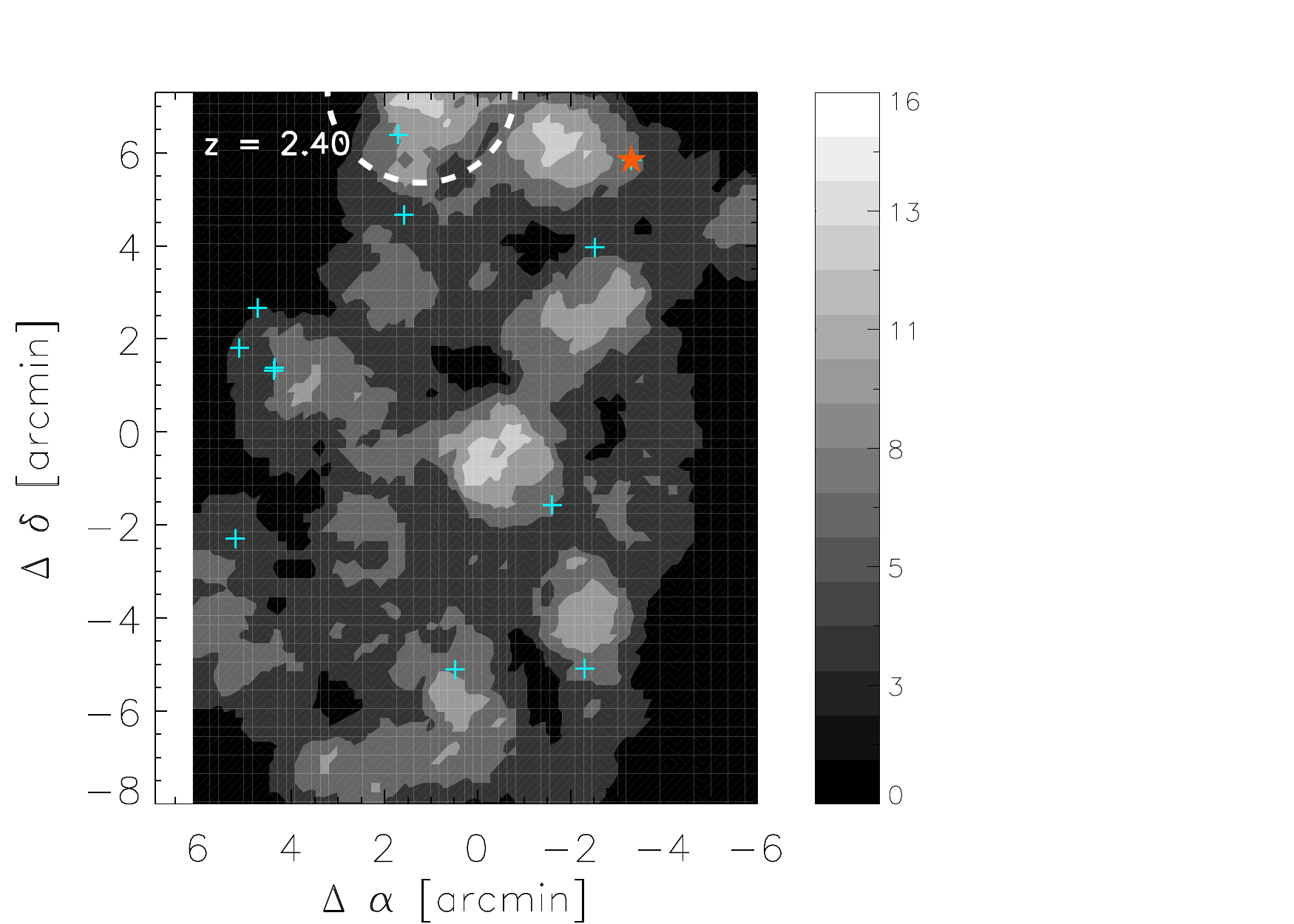}
       \hspace{-0.85in}
\includegraphics[width=81mm]{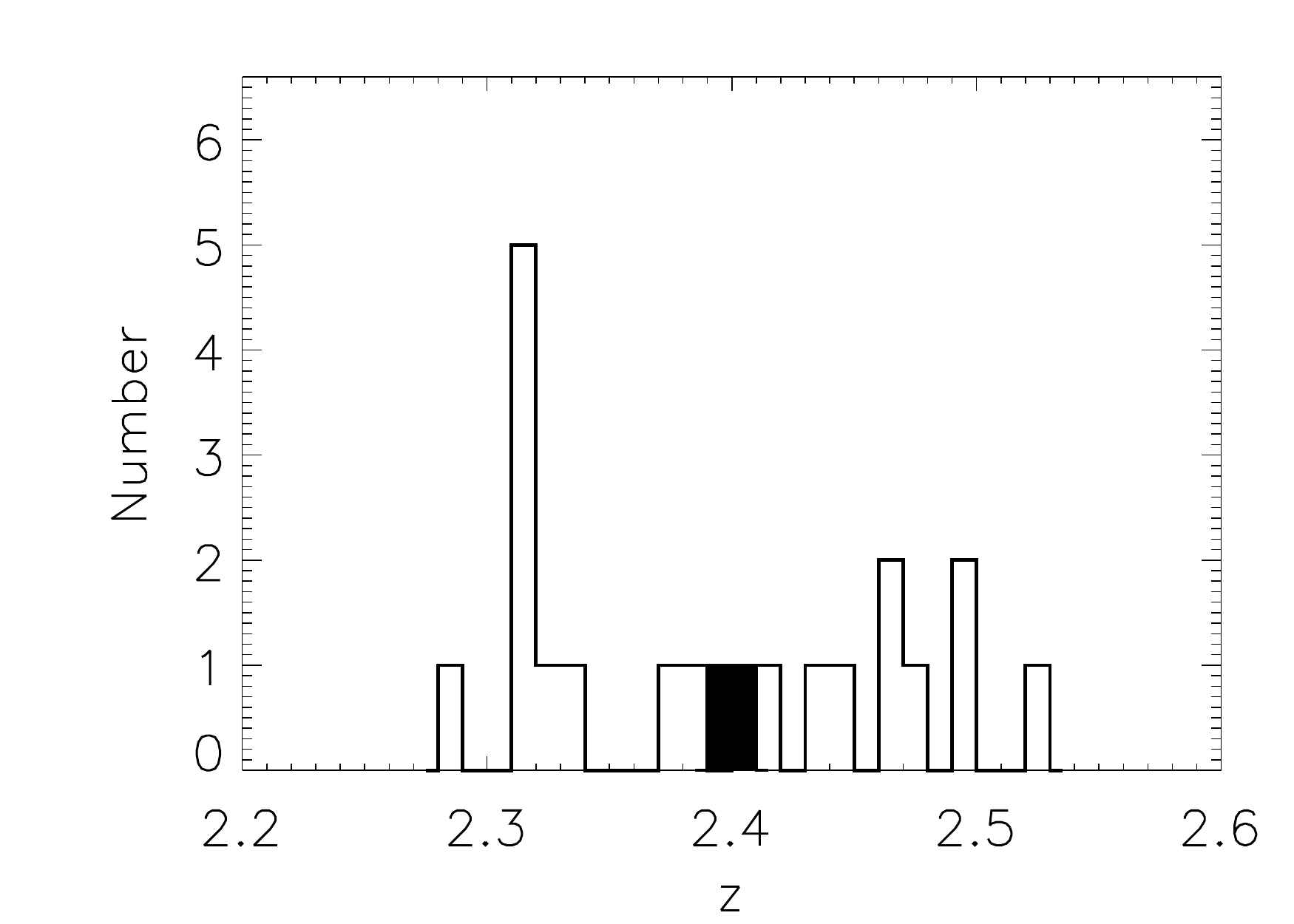}
\caption{(Continued)}
\label{f6d}
\end{figure*}


\begin{figure*}[tp!]
 \setcounter{figure}{7}
\centering
\includegraphics[width=81mm]{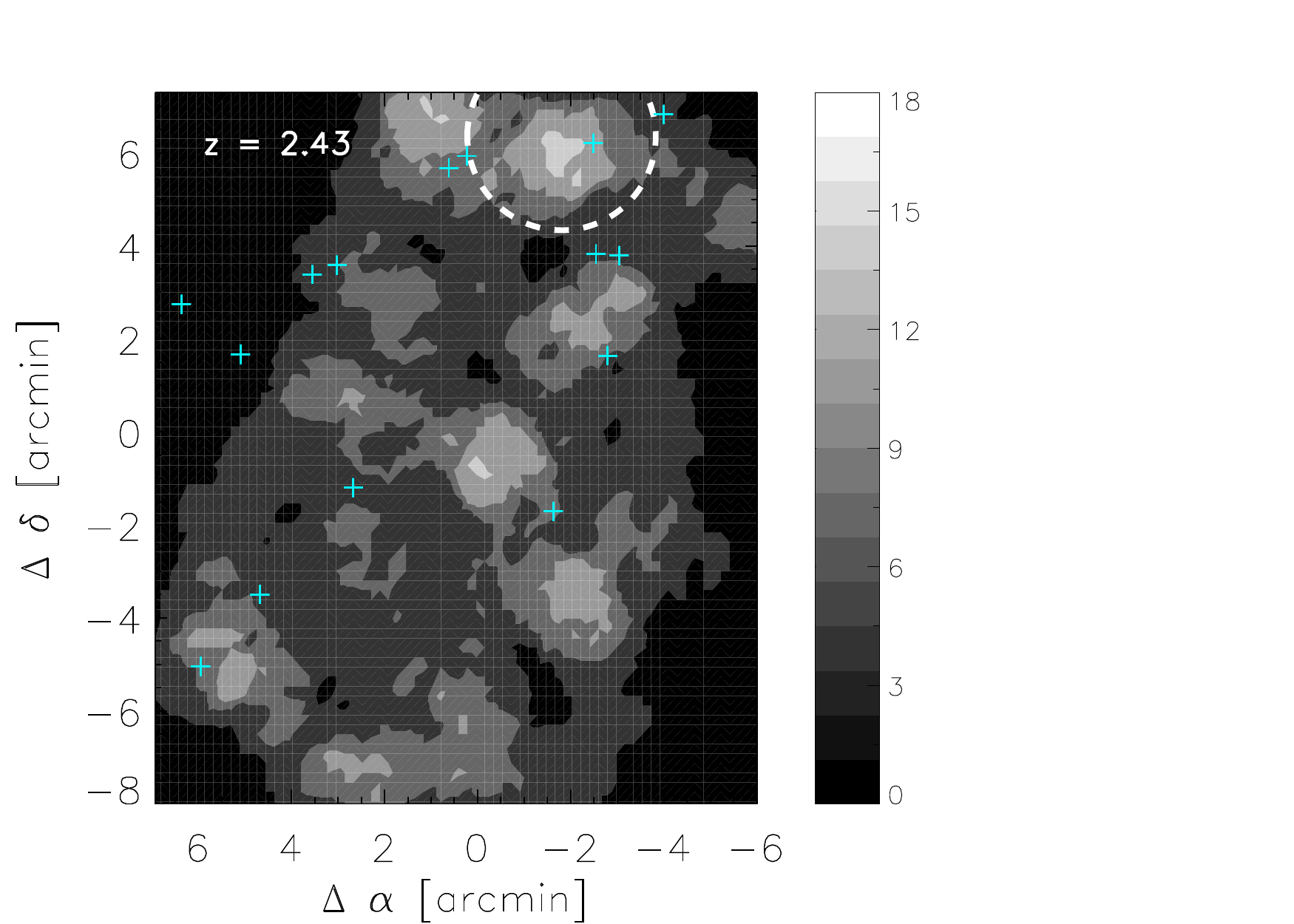}
        \hspace{-0.85in}
\includegraphics[width=81mm]{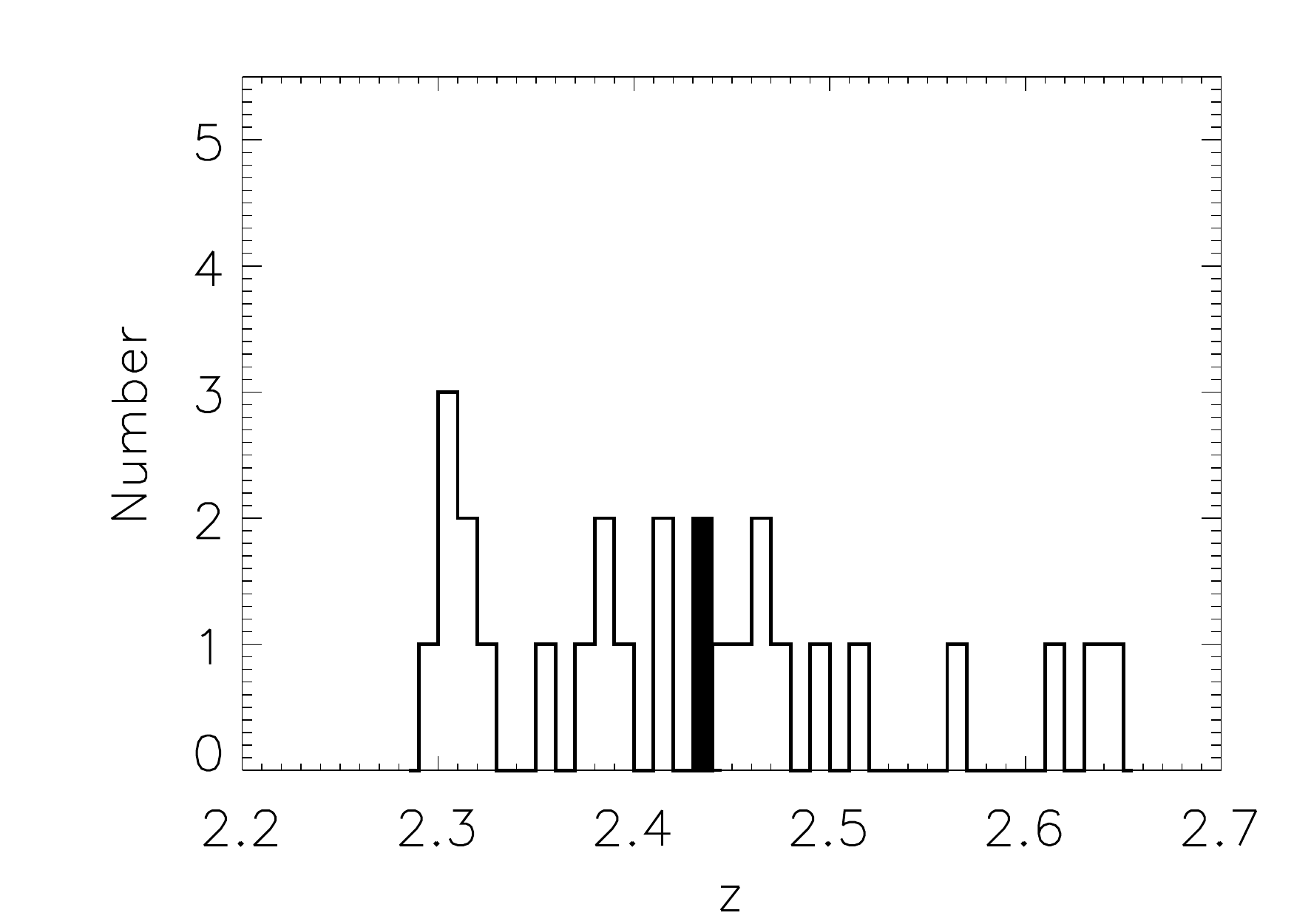}
\\
\includegraphics[width=81mm]{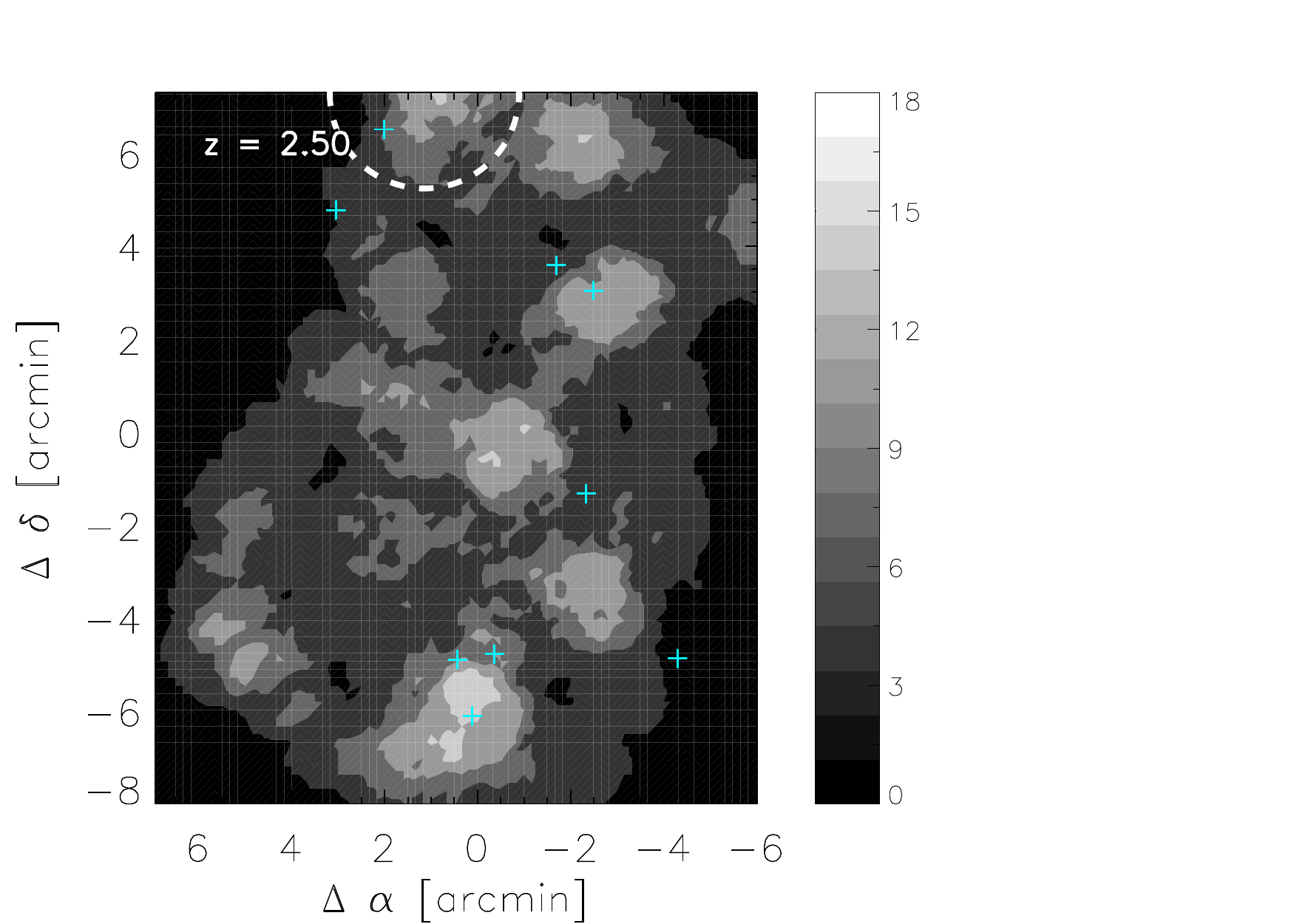}
       \hspace{-0.85in}
\includegraphics[width=81mm]{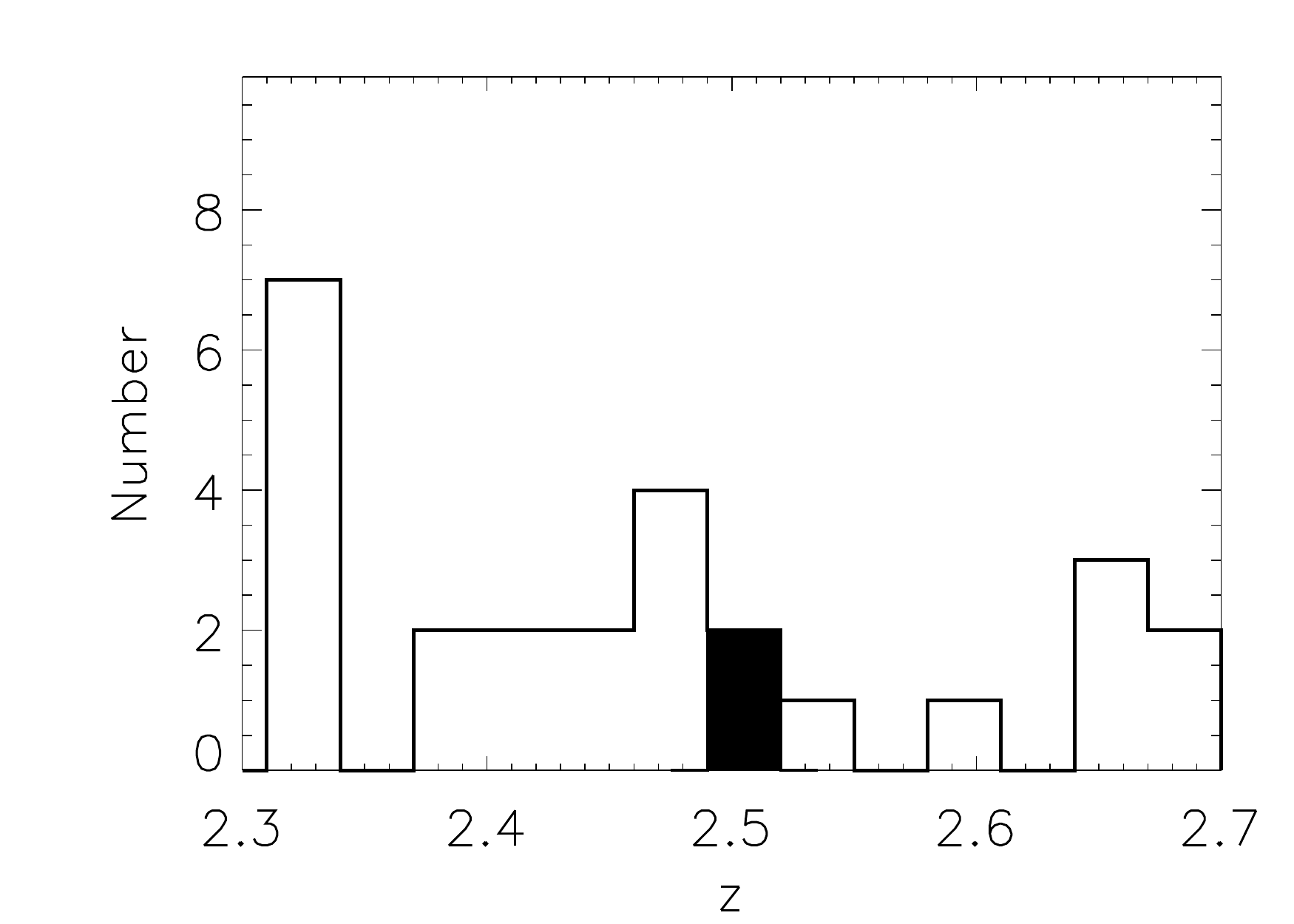}
\\
\includegraphics[width=81mm]{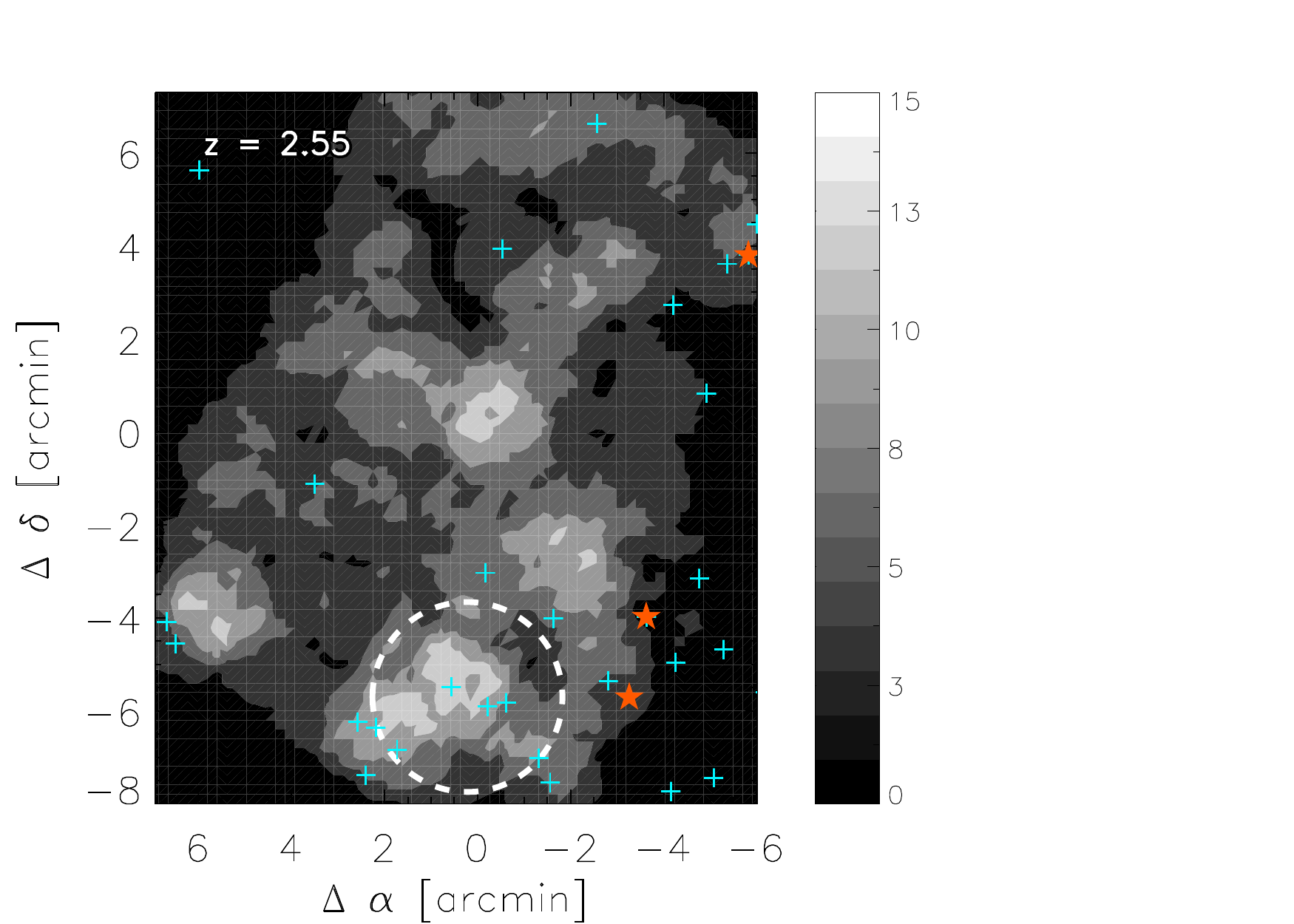}
       \hspace{-0.85in}
\includegraphics[width=81mm]{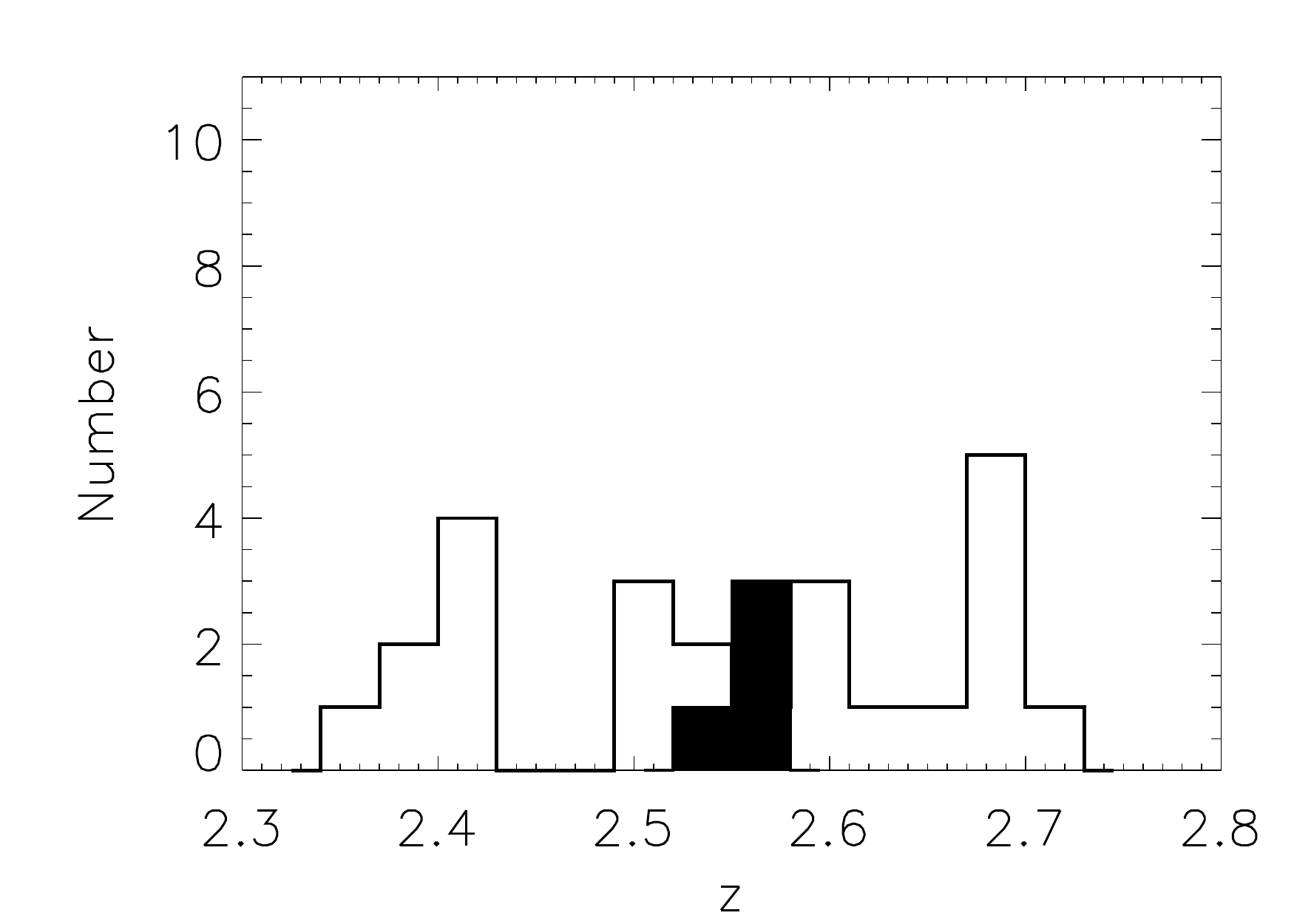}
\\
\includegraphics[width=81mm]{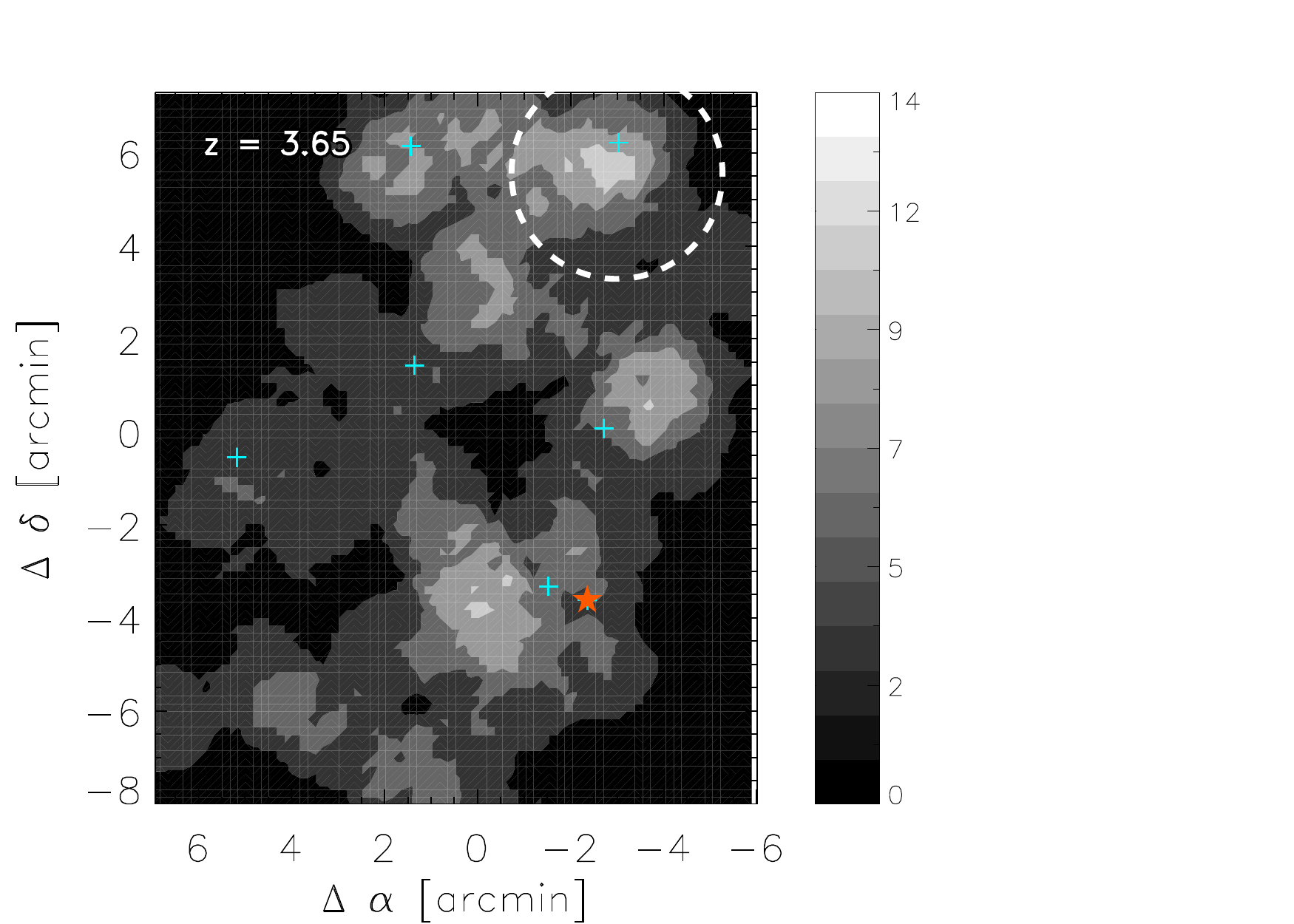}
       \hspace{-0.75in}
\includegraphics[width=81mm]{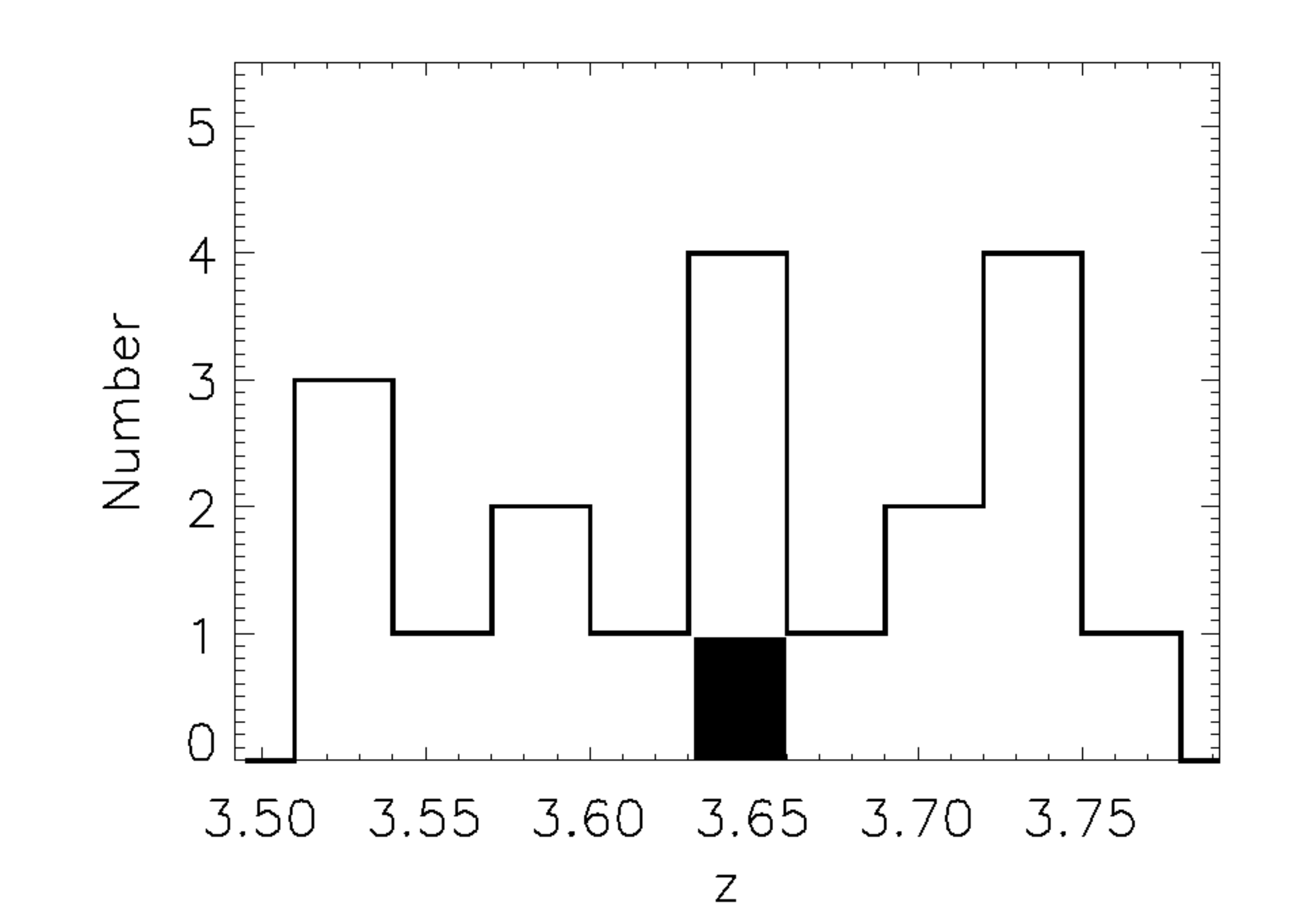}
\caption{(Continued)}
\label{f6e}
\end{figure*}


\begin{figure*}[t!]
 \setcounter{figure}{7}
\centering
\includegraphics[width=81mm]{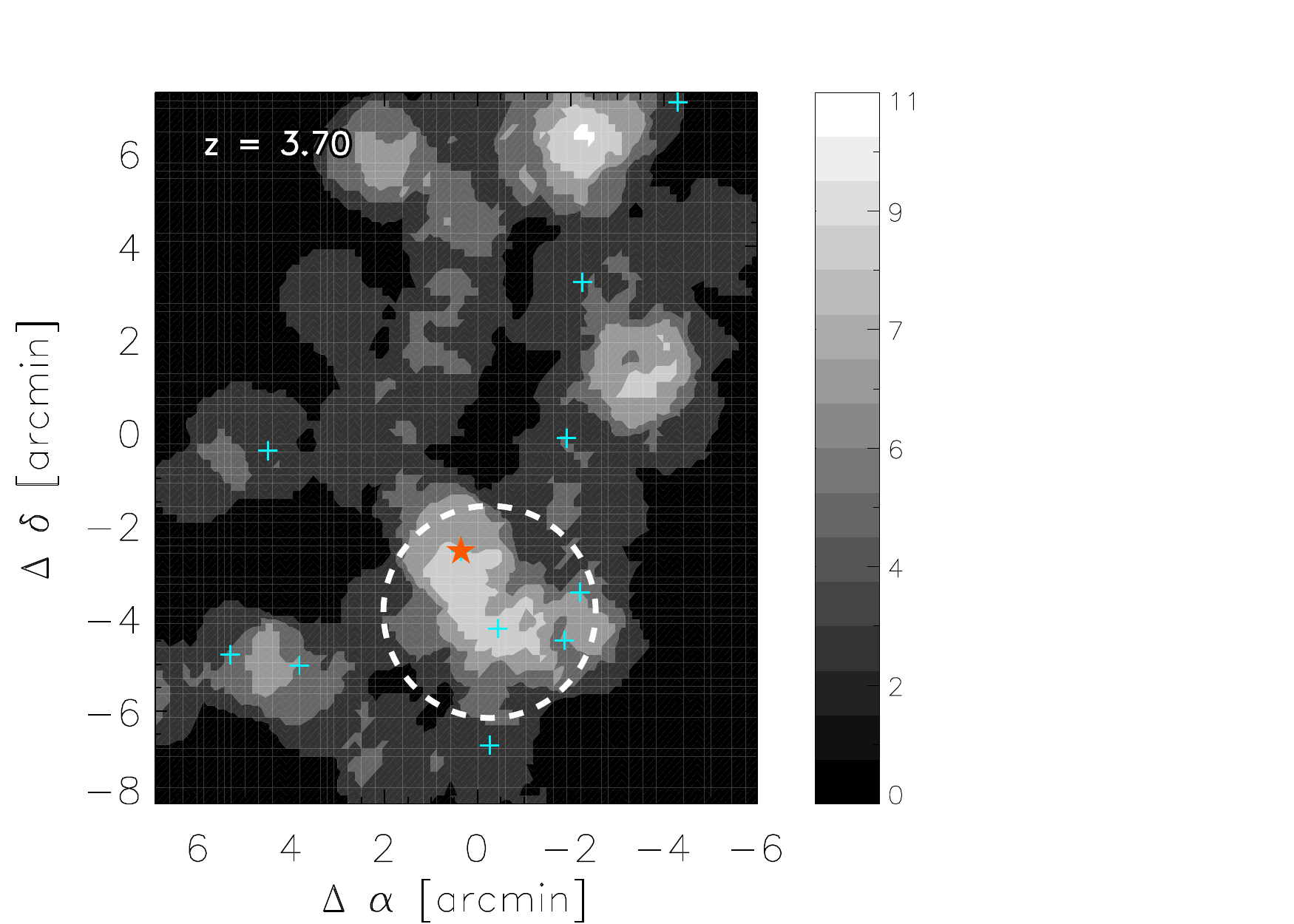}
       \hspace{-0.85in}
\includegraphics[width=81mm]{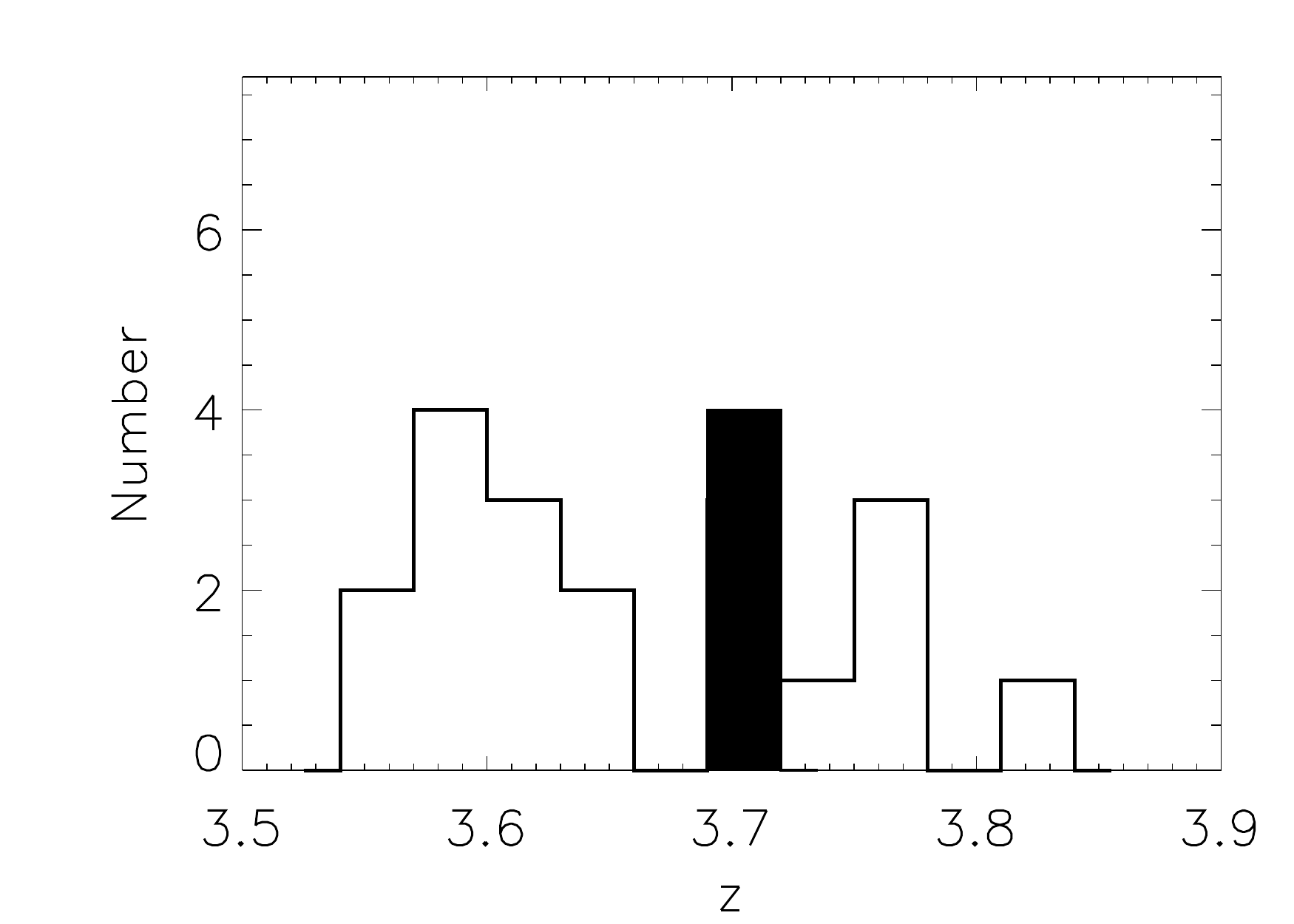}
\\
\includegraphics[width=81mm]{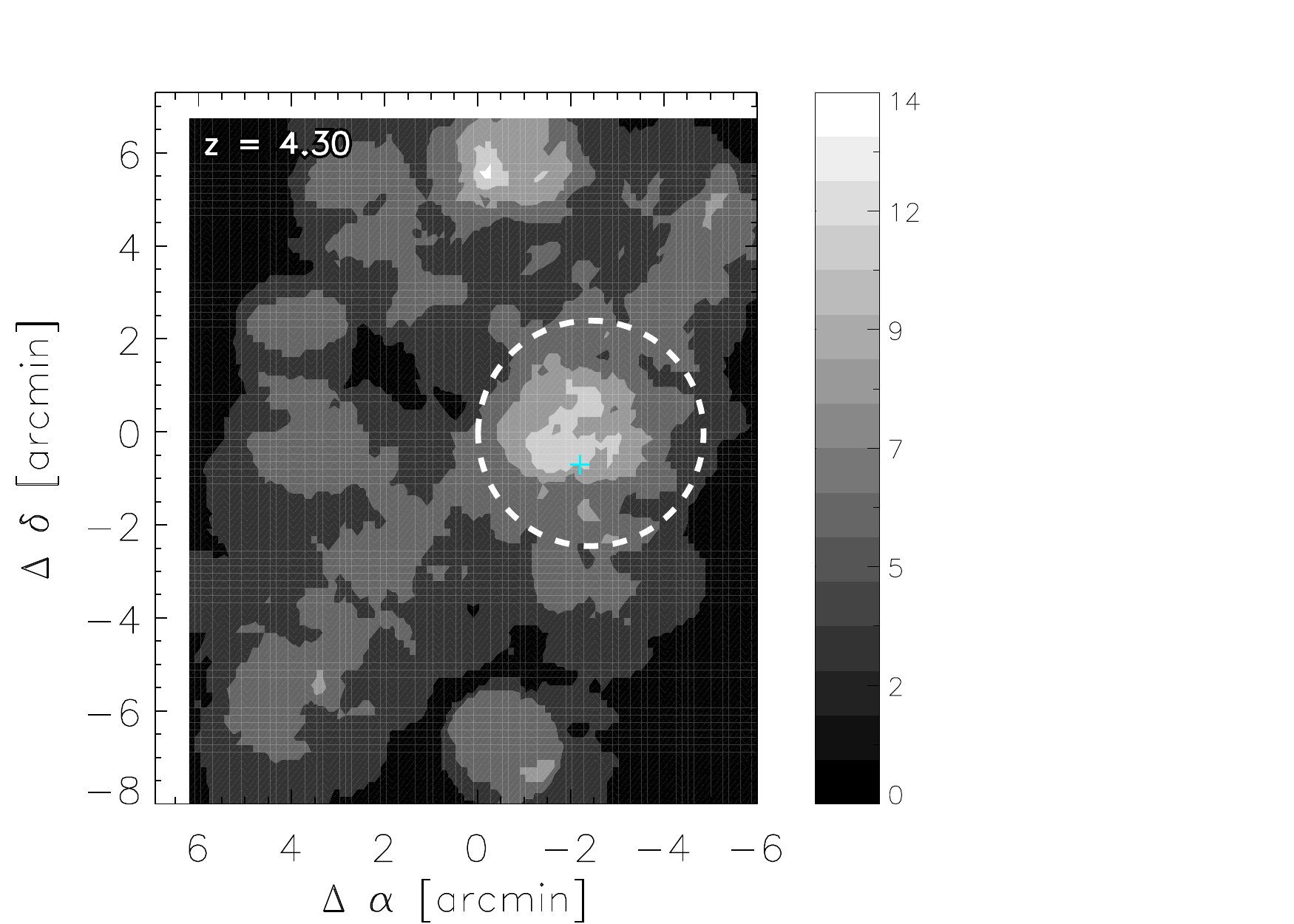}
       \hspace{-0.85in}
\includegraphics[width=81mm]{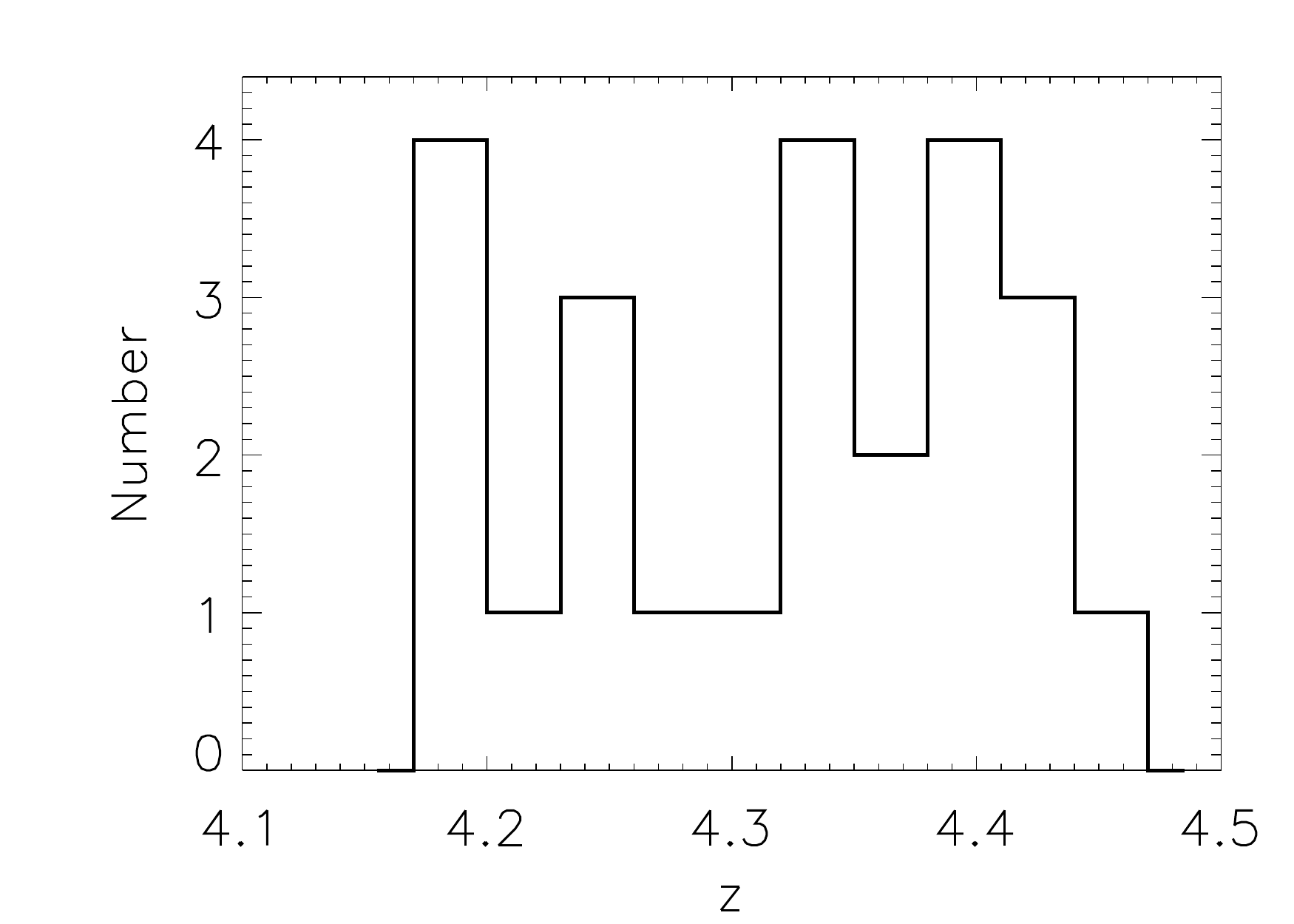}
\\
\includegraphics[width=81mm]{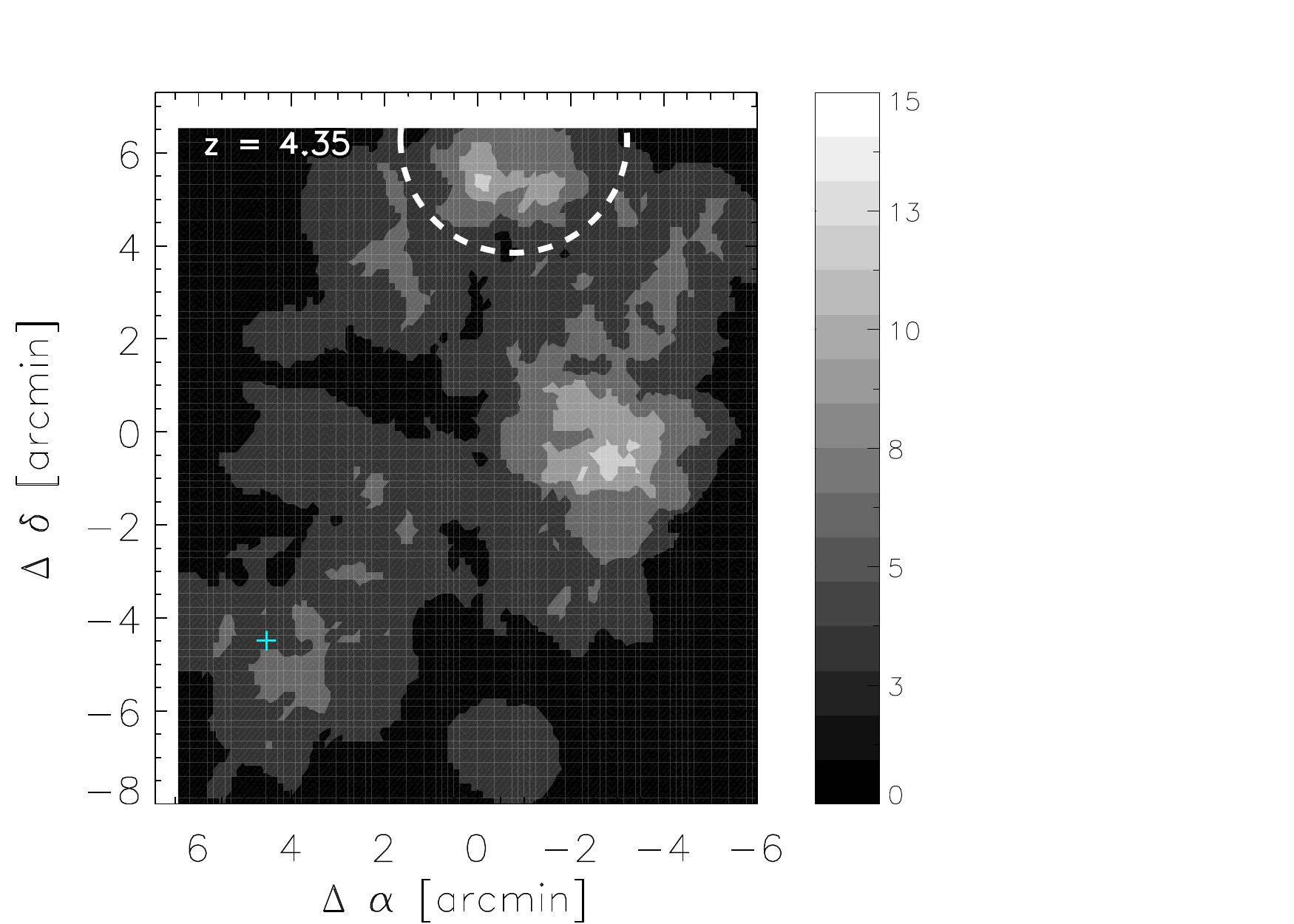}
       \hspace{-0.85in}
\includegraphics[width=81mm]{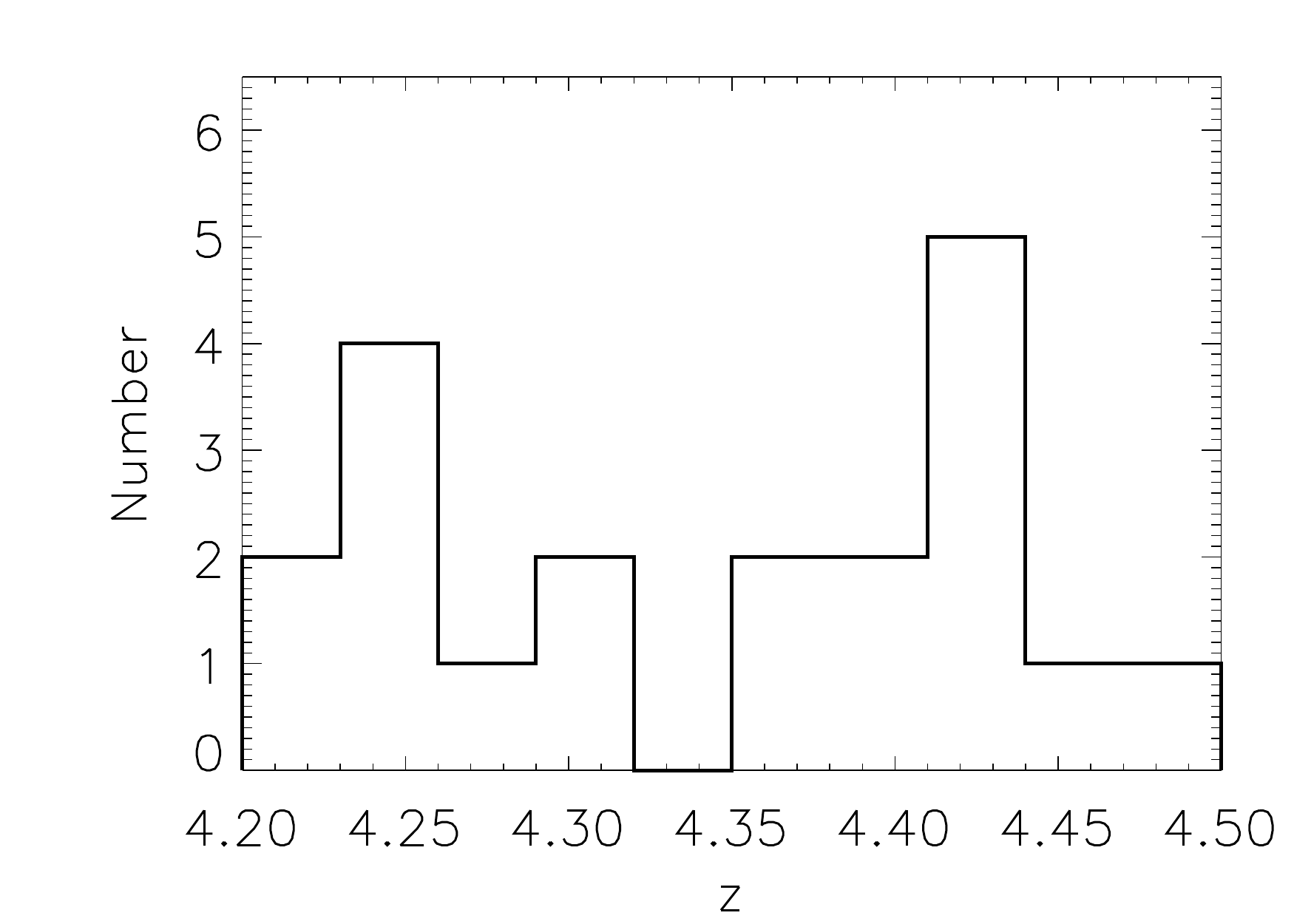}
\caption{(Continued)}
\label{f6f}
\end{figure*}

\subsection{Mass of MSGs}

  We estimate MSG masses (including dark matter) by adding up the stellar mass of galaxies within $1.5\times$
 the projected harmonic mean radius ($r_{p}$)
  and converting it to the total mass assuming a M$_{*}$-to-M$_{\rm halo}$ ratio of galaxies at $z < 1$.
 The projected harmonic mean radius $r_{p}$ may not be the best measure of cluster radius, but it is an effective way to measure the size of gravitationally bound systems with small number of member galaxies (\citealt{ca1}),
 independently of the center definition.
 We use 1.5 $\times$ $r_{p}$ since this corresponds roughly to the distance from the MSG peaks where the number density approaches the 1-$\sigma$ value of the surface number density
 distribution.


 We determine $r_{p}$ using galaxies within 1 Mpc radius from the surface number density peak as follows:
\begin{equation}
   r_{p}= \left[\Pi_{i = 1}^{N}r_i\right]^{1/N} ~ .
\end{equation}
 Here, N is the number of objects considered, and $r_{i}$ is the distance to the
{\it i}th galaxy from the surface density peak measured in physical scale.
 This gives a range of projected harmonic radii $r_{p} = 0.3-0.7$ Mpc for the MSGs.

  The stellar mass of each member galaxy within $1.5\times r_{p}$ is determined by performing
  an SED fit using the same templates as for the photometric redshift estimates.
 For sources with non-detections in the NIR or IRAC bands, we use upper limits for the SED  fitting and the stellar mass estimation.
 Note that our templates assume the Salpeter IMF, therefore, the values can shift by $0.2-0.24$ dex if different IMFs (Kroupa IMF, Chabrier IMF) are used.
  Next, these stellar masses are summed, and several corrections are made to the summed stellar mass of MSG in the following order: (1) the foreground or background galaxy contamination due to uncertainty in $z_{\rm phot}$; (2) incompleteness of the galaxy stellar mass function in the magnitude-limited sample; and (3) conversion of the stellar mass to the total MSG mass.

\begin{figure}[t!]
\centering
\includegraphics[trim=12mm 0mm 4mm 4mm, clip, width=82mm]{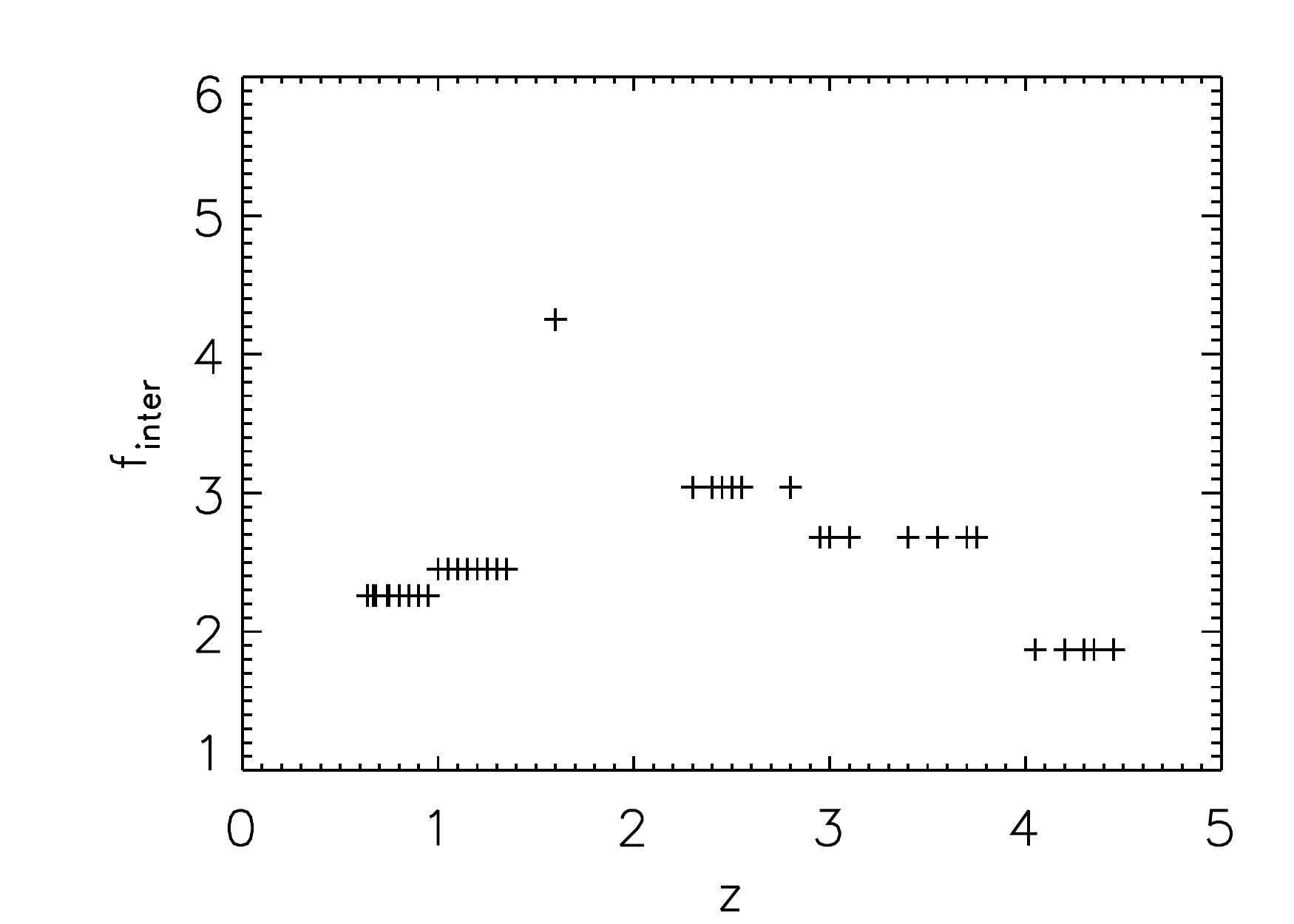}
\caption{The correction factor $f_{\rm inter}$ as a function of redshift. $f_{\rm inter}$ is
 derived from simulation data and corrects for the foreground or background galaxy contamination in the $z_{\rm phot}$ sample.}
\label{f7}
\end{figure}

 First, the summed stellar mass is corrected for foreground or background galaxies in the photometric redshift sample. The level of contamination is determined from the simulation (See Section 5 for details), and defined as the mass of MSG member candidates within the 1.5 $\times$ $r_{p}$ radius versus the mass of the true members of MSGs. The interloper fraction, $f_{\rm inter}$, is about a factor of 2 - 3 for most redshift intervals, except for $1.5 < z < 2$. At $1.5 < z < 2$, the large uncertainty in $z_{\rm phot}$ and the paucity of $z_{\rm spec}$ leads to an increased level of contamination by interlopers ($\times 5$, Figure \ref{f7}).

\begin{figure}[t!]
\centering
\includegraphics[trim=12mm 0mm 4mm 4mm, clip, width=82mm]{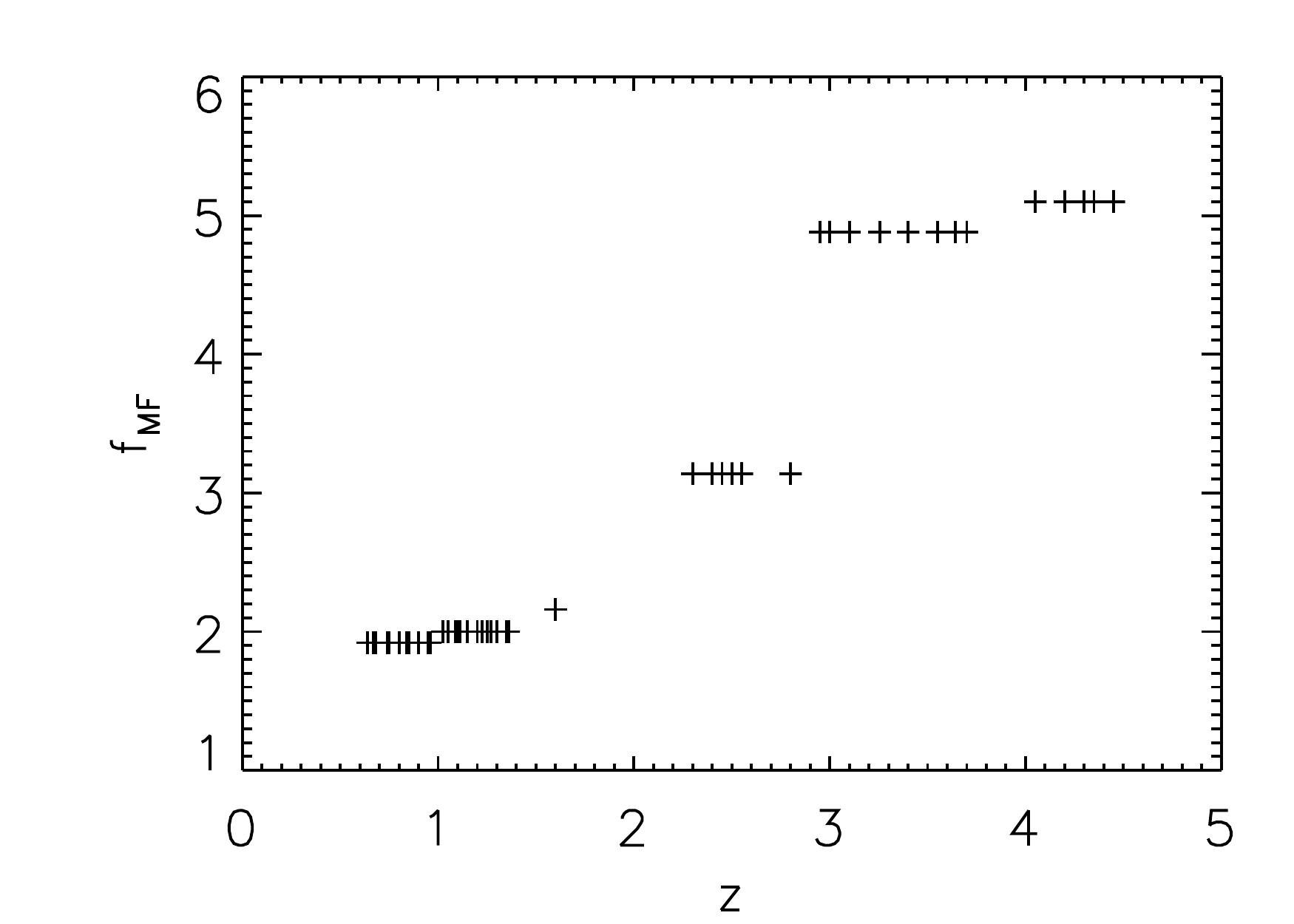}
\caption{The correction factor $f_{\rm MF}$ as a function of redshift. It is derived from
 simulation data and corrects
 for the incompleteness in the stellar mass function in a magnitude-limited sample.}
\label{f8}
\end{figure}

 Then, we make a correction for the incompleteness in the coverage of the faint (less massive) end of the luminosity (mass) function, since the MSG selection is based on a magnitude-limited sample. This correction factor, $f_{\rm MF}$, is determined from the simulation data
 (Section 5).
 We estimate this value using the ratio of the total stellar mass of all galaxies that
 belong to an MSG to the total stellar mass of the galaxies that are brighter than the limiting magnitude.
  Figure \ref{f8} shows $f_{\rm MF}$ as a function of redshift.

 With the corrections of $f_{\rm inter}$ and $f_{\rm MF}$, we obtain the total stellar mass of an MSG.
  In the next step, $M_{*,\rm total}$ is converted to the total MSG mass, including dark-matter,
 by adopting the stellar mass to the observed halo mass relation of galaxies at $0 < z < 1$ (\citealt{21, 22}). According to these and other relevant works, it is known that the ratio of the stellar mass to the cluster or halo mass changes with the halo mass. This factor, $f_{*,\rm MSG}$, is (\citealt{21}):
%
%
\begin{equation}
  1/f_{*,\rm MSG} = \frac{M_{*,\rm total}}{M_{\rm halo}} = 0.05 \times \left(\frac{M_{\rm halo}}{5 \times 10^{13} M_{\odot}}\right)^{-0.26\pm0.09}.
\end{equation}

   If $M_{*,\rm total} = 10^{13.7}$, then $1/f_{*,\rm MSG} = 0.05$ is adopted.
 Here, the derivation of the stellar mass assumes the Salpeter IMF.
This relation is suggested to change little as a function of redshift out to $z=4$
(\citealt{23, 24}), therefore we apply the same relation to galaxies at $z > 1$.

We note that some of the MSGs at higher redshifts are suspected to be a collection of small mass halos. In such a case, the procedure described above could underestimate the MSG mass since the smaller halos have smaller $f_{*,\rm MSG}$ values. If an MSG is made of three halos with an equal mass, this could lead to an underestimation of $M_{\rm MSG}$ by 25\%. This amount is small compared to the uncertainties related to the correction procedures (a factor of a few), therefore it can be neglected. We also note from the simulation that most of the MSGs are the dominant density peaks within the search redshift range.
Finally, $M_{\rm MSG}$ is derived as
\begin{equation}
M_{\rm MSG} = \frac{f_{*,\rm MSG}\,f_{\rm MF}}{f_{\rm inter}} \sum_i m_{*,i},
\end{equation}
where $m_{*,i}$ is the stellar mass of the {\it i}th galaxy identified as a MSG member in the magnitude-limited sample.
 Note that the systematic error of the stellar mass estimate due to
the assumed IMF gets cancels due to the multiplication of
$f_{*,\rm MSG}$ by $m_{*,i}$, as long as
an identical IMF is assumed for deriving both the $f_{*, \rm MSG}$ and $m_{*,i}$ values.
  The derived $M_{\rm MSG}$ and $M_{*,\rm total}$ are listed in Table \ref{t3}.

 Errors in the MSG mass estimates are dominated by dispersion in $f_{\rm inter}$,
$f_{\rm MF}$, and $f_{*, \rm MSG}$, and errors in the stellar mass estimates. The dispersions in
$f_{\rm inter}$ and $f_{\rm MF}$ are inferred from the simulation as described in the next section, and they amount to 0.16 dex and 0.1 dex respectively. The dispersion in $f_{*, \rm MSG}$ is based on Eq. (2), and is about 0.1 dex. Therefore, the combined dispersion of the three factors is 0.21 dex.
Typical random errors in galaxy stellar mass estimates are a factor of 2 - 3 (\citealt{pe1}), but the random errors
cancel out with each other when stellar masses of many galaxies are added up. More worrisome is  the systematic error in stellar mass estimates which could be as large as a factor of 2 (Kannappan \& Gawiser 2007; Lee et al. 2009).
 As we shall see in the next section, the ratio between our $M_{\rm MSG}$ estimates and $M_{\rm MSG}$ estimates from other groups shows a dispersion of about 0.4 dex $-$ 0.5 dex, with a systematic difference up to 0.2 dex. If two independent MSG estimates have equal uncertainties, this implies that the error in $M_{\rm MSG}$ estimates is 0.25 dex - 0.35 dex.
 With these considerations, we conclude that the errors in $M_{\rm MSG}$ are somewhere between 0.21 dex - 0.35 dex. As a fiducial error of $M_{\rm MSG}$, we take 0.28 dex which is the mid-point of 0.21 dex to 0.35 dex. The choice of the $M_{\rm MSG}$ error becomes important when determining the expected number of MSGs, and we will investigate how the result changes for different values of   the $M_{\rm MSG}$ error.

\subsection{Comparison with Previously Identified (Proto-) Clusters in GOODS Fields}

 Previous works have identified large scale structures and clusters/proto-clusters in the GOODS fields.
 Most of them are included as MSGs in our study,
and they are identified in Table \ref{t3} by indicating their mass in the literature in column 9.
 Among 16 MSGs at $z < 2$ in GOODS-South, 9 were previously identified
 (\citealt{ 7, 29, 31,20, 27, 30, 11}).
 For the GOODS-North field, only a few MSGs were previously identified as large scale structures (\citealt{ 28, 26}).

  At high redshifts, we note that an MSG at $z=4.05$ in the GOODS-North field was previously identified, and MSGs at $z=2.55$ and 3.70 were previously identified in the GOODS-South field (\citealt{32,18, 20, 26}).
 However, many of the MSGs found in this work are new identifications.

\begin{figure}[t!]
\centering
\includegraphics[trim=14mm 2mm 6mm 4mm, clip, width=83mm]{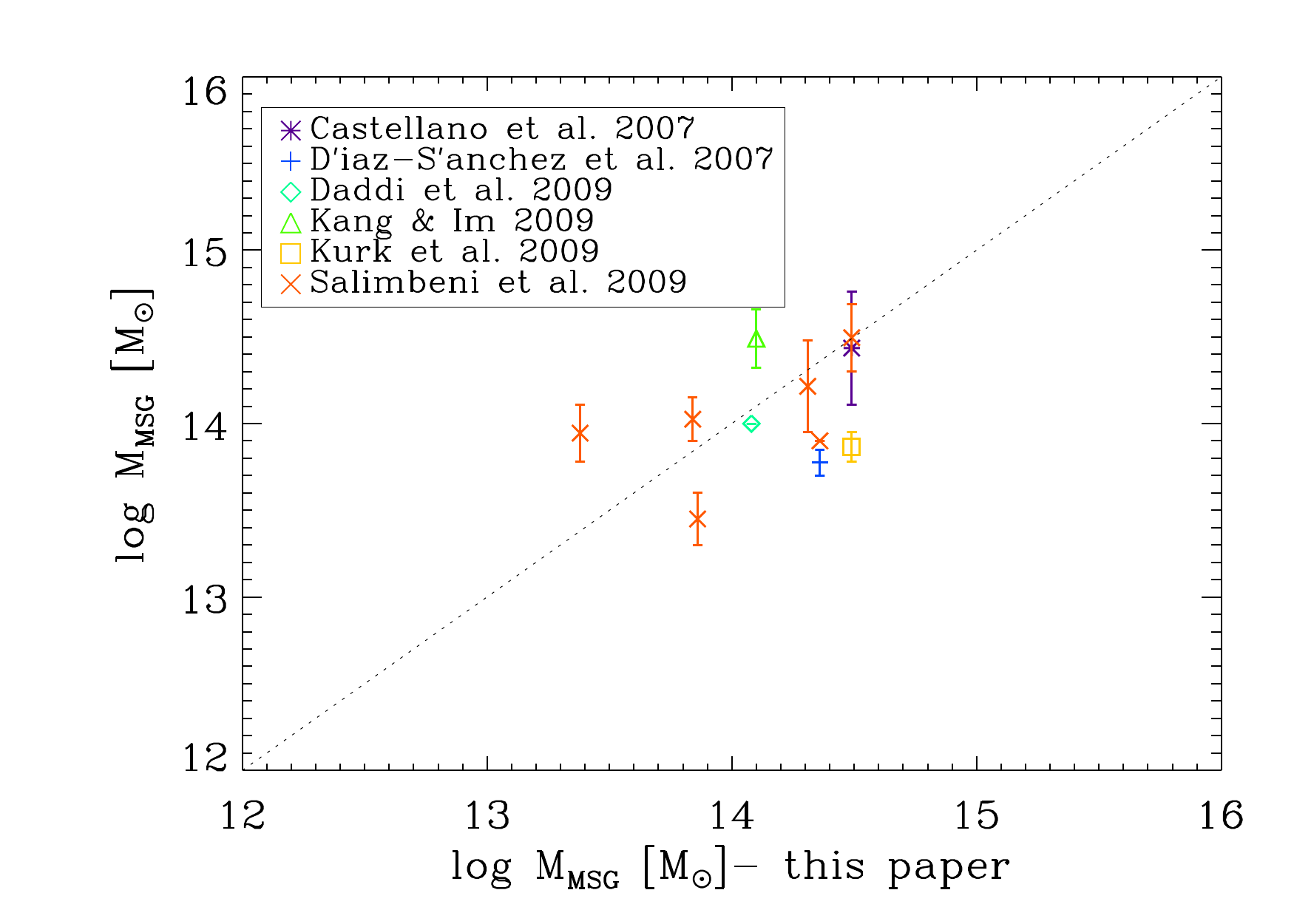}
\caption{The comparison of MSG masses from this paper with those obtained by other groups
 and available in the literature.
}
\label{f9}
\end{figure}

  In Figure \ref{f9}, we compare the MSG masses in the literature and our MSG mass estimates.
 The comparison shows that our $M_{\rm MSG}$ estimates correlates reasonably well with the literature values.
 The mean offset between our mass estimates and the literature values is between -0.03 dex and 0.21 dex ($\langle {\rm log}_{10}[M_{\rm MSG}({\rm ours})/M_{\rm MSG}({\rm literature})] \rangle$), depending on the largest values and the smallest values taken from the literature respectively. The rms dispersion of the relation ranges between $0.4-0.5$ dex, meaning that MSG mass estimates from different methods and groups are consistent with each other within a factor of a few.

\section{Analysis of Millennium Simulation Data}

 Since the identification and mass estimates of MSGs involve several steps that cannot be directly compared to theoretical predictions of massive halos, we make use of the simulation data. We compare the observed results with the theoretical predictions by mimicking the observation
and our analysis method of the GOODS data
as much as possible. To do so, we adopt the mock catalogs constructed by De Lucia \& Blaizot (2007) and Kitzbichler \& White (2007) which are based on the Millennium Simulation.
 The Millennium Simulation data (Springel et al. 2005)
follows trajectories $1 \times 10^{10}$ particles of mass $8.6 \times 10^{8}\,h^{-1}$
$M_{\odot}$ in a periodic box with 500 $h^{-1}$ Mpc on a side.
   The mock catalog identifies galaxies and assigns luminosity values in the filter bands that overlap with the GOODS data and therefore it is well suited for this purpose. The mock catalog includes information such as redshift (plus peculiar velocity), galaxy luminosity, galaxy stellar mass, halo mass where a galaxy belongs to, and the spatial information of galaxies.
 To perform a realistic analysis of the simulation data,
we assign photometric redshifts for some galaxies in the simulation, and use the photometric redshift information rather than their true redshift in the mock catalog during the analysis. This procedure is necessary since the GOODS data include a fair amount of photometric redshift information, and the uncertainties in the photometric redshift values diffuse MSGs in the radial direction and make the identification of MSG members less unambiguous.
 The quantification of what fraction of galaxies should be given photometric redshifts is done following the spectroscopic redshift completeness in the observed data as a function of the apparent $K_s$-band magnitude and redshift (Figure \ref{f02}).
 Following this spectroscopic redshift completeness, we assign either spectroscopic redshifts
  ($z_{\rm spec,sim}$) or photometric redshifts ($z_{\rm phot,sim}$) to galaxies in the simulation within a certain apparent magnitude and redshift bins as
\begin{equation}
z_{\rm phot,sim} = z_{\rm spec,sim} + \delta z,
\end{equation}
where $\delta z$ is a value randomly picked from a Gaussian distribution
with a mean of zero and a standard deviation of $\sigma_{z_{\rm phot}}$ as in Figure \ref{f2}.

  We name the mock catalog that implements the photometric redshifts as the ``modified" mock catalog, to distinguish it from the ``original" mock catalog (De Lucia \& Blaizot 2007; Kitzbichler \& White 2007).
  MSGs are searched in the modified mock catalog using the same method as we used for the analysis of the GOODS data. For the MSGs identified in the simulation, we have the true redshift distribution of galaxies and the mass of halos associated with these galaxies and MSGs. These information is then used for deriving the correction factors that are used for the analysis of the observed data.

\begin{figure}[t!]
\centering
\includegraphics[trim=1mm 2mm 4mm 2mm, clip, width=73mm]{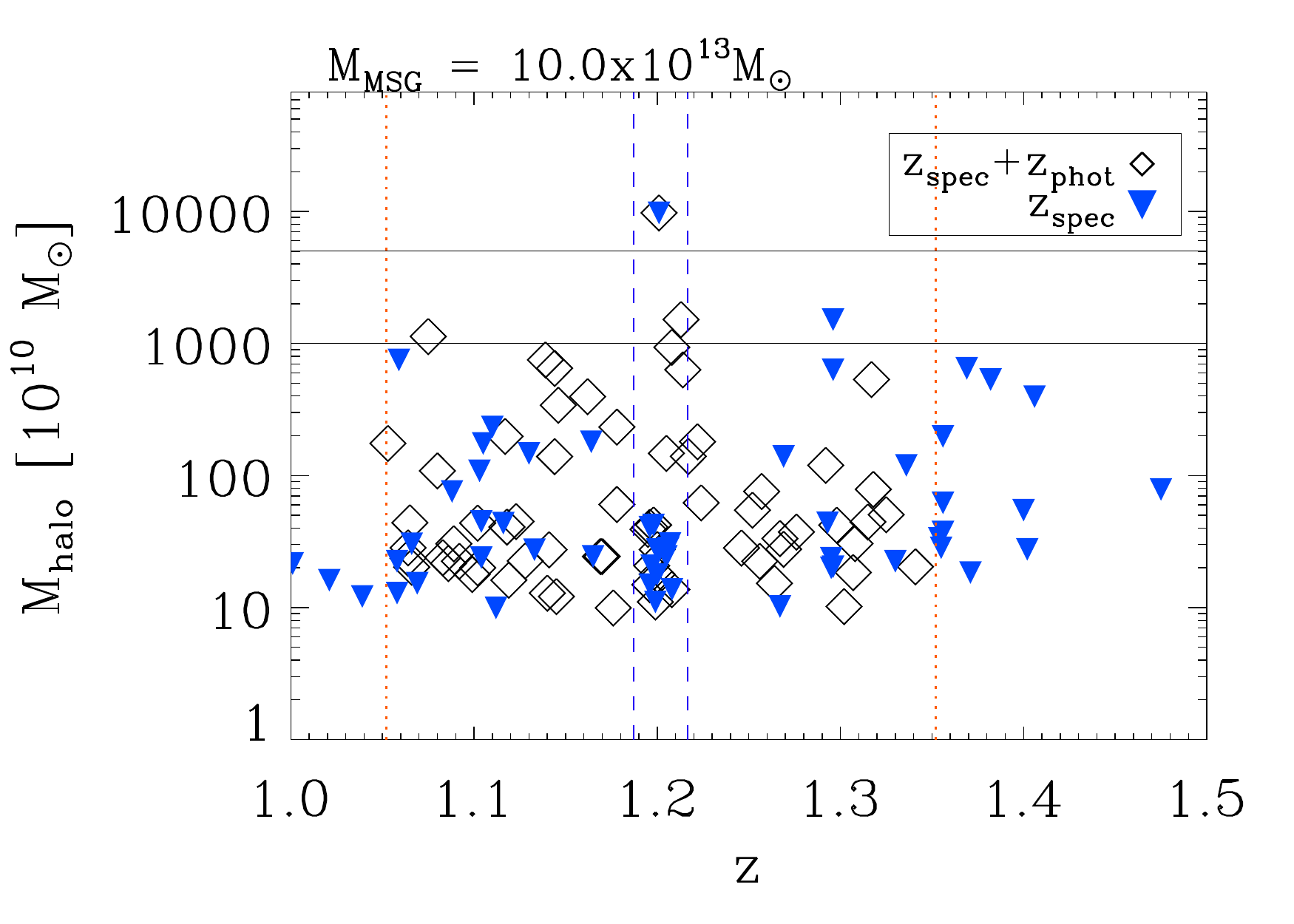} \\
\includegraphics[trim=1mm 2mm 4mm 2mm, clip, width=73mm]{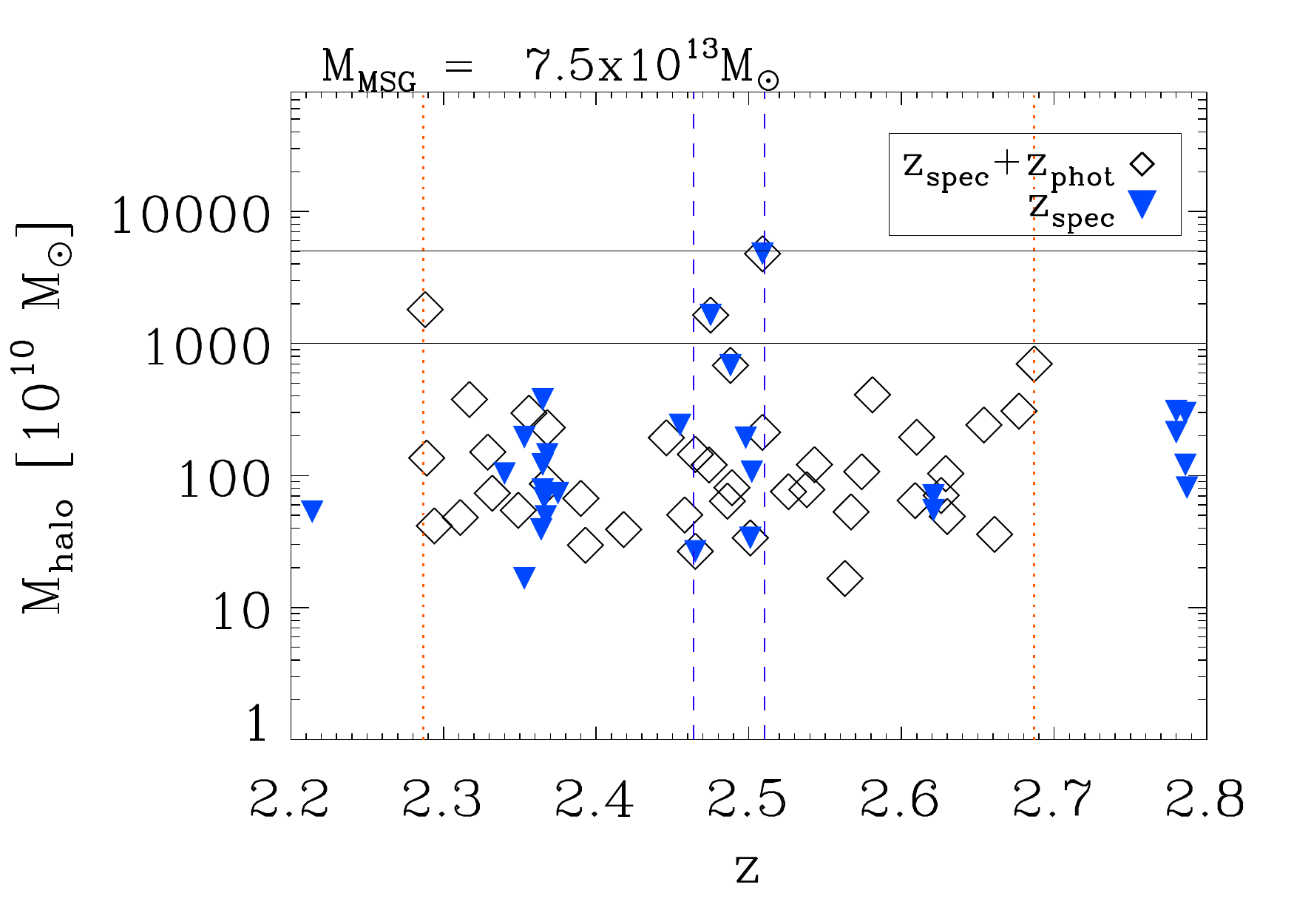} \\
\includegraphics[trim=1mm 2mm 4mm 2mm, clip, width=73mm]{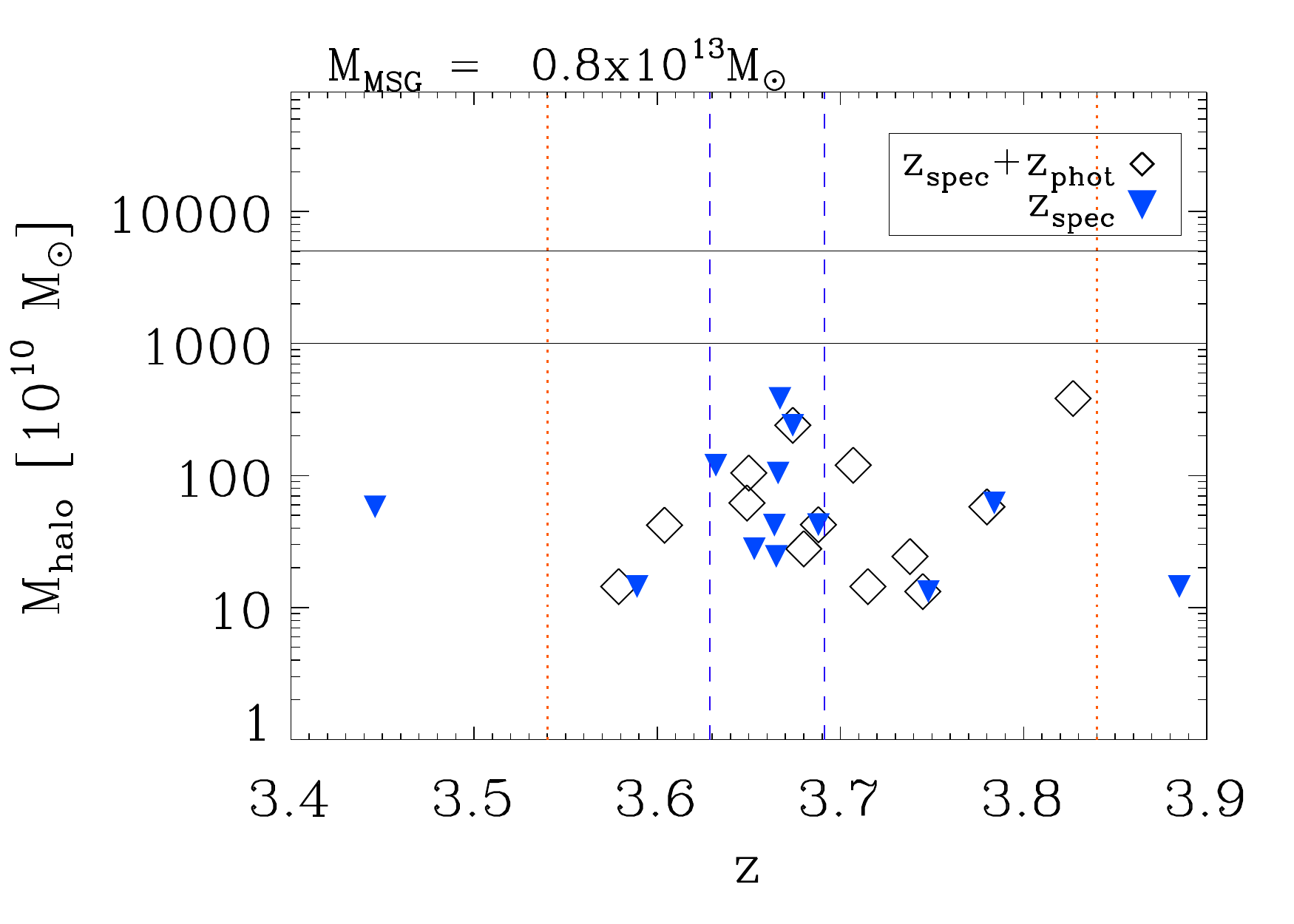}
\caption{
Examples showing how MSG masses are determined in the simulation. Plotted are the redshifts and masses of halos of which at least one galaxy is identified to belong to an MSG during the MSG search. The diamonds are the points from the modified mock catalog, and the filled, inverted triangles are the points from the true redshift catalog.
  The vertical dashed lines indicate the redshifts at $\Delta v = \pm 2000 \,km/s$ from the redshift of the corresponding MSG and the
 vertical dotted lines indicate the median photometric redshift error (Figure \ref{f2}) at the search redshift of each respective MSG.
 The horizontal lines indicate the halo mass of $10^{13}M_\odot$ and $5\times10^{13}M_\odot$, respectively, to help readers judge the mass of subhalos.
 The examples are for MSGs at $z = 1.20, 2.49$ and $3.67$. We also note the
 true mass of the identified MSGs on top of each panel.
 To derive the true MSG mass, we add up the masses of the halos with true redshifts (blue points between the dashed lines). For the observed MSG mass before $f_{\rm inter}$ correction, halo masses represented by the diamonds between the red-dotted lines are added.
}
\label{f11}
\end{figure}

  Below, we describe how the correction factor $f_{\rm inter}$ that corrects for foreground or background interlopers is derived from the simulation.
  The factor $f_{\rm inter}$ corresponds to the ratio between the MSG mass derived from the true redshift values given in the simulation and the MSG mass estimate based on the modified mock catalog before the interloper correction. Since the photometric redshifts have large errors compared to spectroscopic redshifts, interloper correction is important for a sample with many photometric redshifts.
 Figure \ref{f11} shows examples of the procedure for determining MSG masses and $f_{\rm inter}$ at three different redshift bins. The figure shows the redshift versus the halo mass ($M_{\rm halo}$) of member galaxies that are selected as possible MSG members in the modified mock catalog
  by imposing the same observational selection constraints as the MSG search done in the GOODS fields.
 Two different sets of points are plotted in the figure. In one set, true redshifts in the original mock catalog are used (the inverted blue triangles). The other set uses a combination of true and photometric redshifts as in our modified mock catalog. When a halo includes multiple galaxies and the true redshifts are available for some of these galaxies, the true redshift is used to indicate the redshift of the halo.
 The two vertical dotted lines indicate the redshift interval over which the MSG search was performed and
 the two vertical dashed lines indicate the rest-frame redshift interval of $\Delta v = \pm 2000 \,km/s$ centered on the redshift of the MSG found from the modified mock catalog. Some of the blue triangles spill over the boundary defined by the two vertical dotted lines. They are galaxies that have $z_{\rm phot}$ values in the modified mock catalog, and their $z_{\rm phot}$'s fall
 within the boundary of the dotted lines although their true redshifts are outside the boundary.
 The ``true" MSG mass without interlopers is determined by adding up the masses of halos whose true redshifts (the inverted triangles) lie within the redshift interval of  $\Delta v < 2000 \, $km s$^{-1}$ (the vertical dashed lines) and within 1.5$\times \, r_{p}$ of the peak of the projected number density.  By doing so, we exclude less significant peaks in the foreground and the background.
 Another MSG mass is estimated by summing up the masses of  halos in the search redshift range (the open diamonds), which we call, ``$z_{\rm phot}$-diluted" MSG mass. Then, $f_{\rm inter}$ is the ratio of the ``true" MSG mass to the ``$z_{\rm phot}$-diluted" MSG mass. Figure \ref{f7} shows $f_{\rm inter}$ values as a function of redshift. The figure indicates that MSG masses determined from the modified mock catalog are overestimated by a factor of 2 to 3 on average, if we do not take into account $f_{\rm inter}$. The situation becomes worse ($> 4\times$ mass overestimate) if both the accuracy of $z_{\rm phot}$ and the fraction of $z_{\rm spec}$ sample is low such as at the redshift interval of $1.5 < z < 2$.

 The factor $f_{\rm MF}$ is estimated from the simulation too. We estimate this value as the ratio of the total stellar mass of all galaxies that belong to an MSG to the stellar mass of galaxies associated with the MSG that are brighter than the limiting magnitude used in the observations.

 Finally, we derive the dispersion in $f_{\rm inter}$ and $f_{\rm MF}$ values, a measure of how large the errors in MSG mass estimates are in the GOODS data. The rms dispersions are found to be 0.16 dex and 0.1 dex, respectively.

 For the MSGs found in the simulation, we add a random error of 0.28 dex to the derived MSG mass in order to mimic the observation. Adding the dispersion is important since we are studying the massive end of the halo mass function, and the number of very massive MSGs is sensitive to the measurement errors. The number of MSGs can increase significantly if the errors are large, since the errors are more likely to move less massive MSGs to the more massive regime than the other way around.

\section{Number Density of MSGs}

\begin{figure}[t!]
\centering
\includegraphics[trim=2mm 1mm 5mm 2mm, clip, width=83mm]{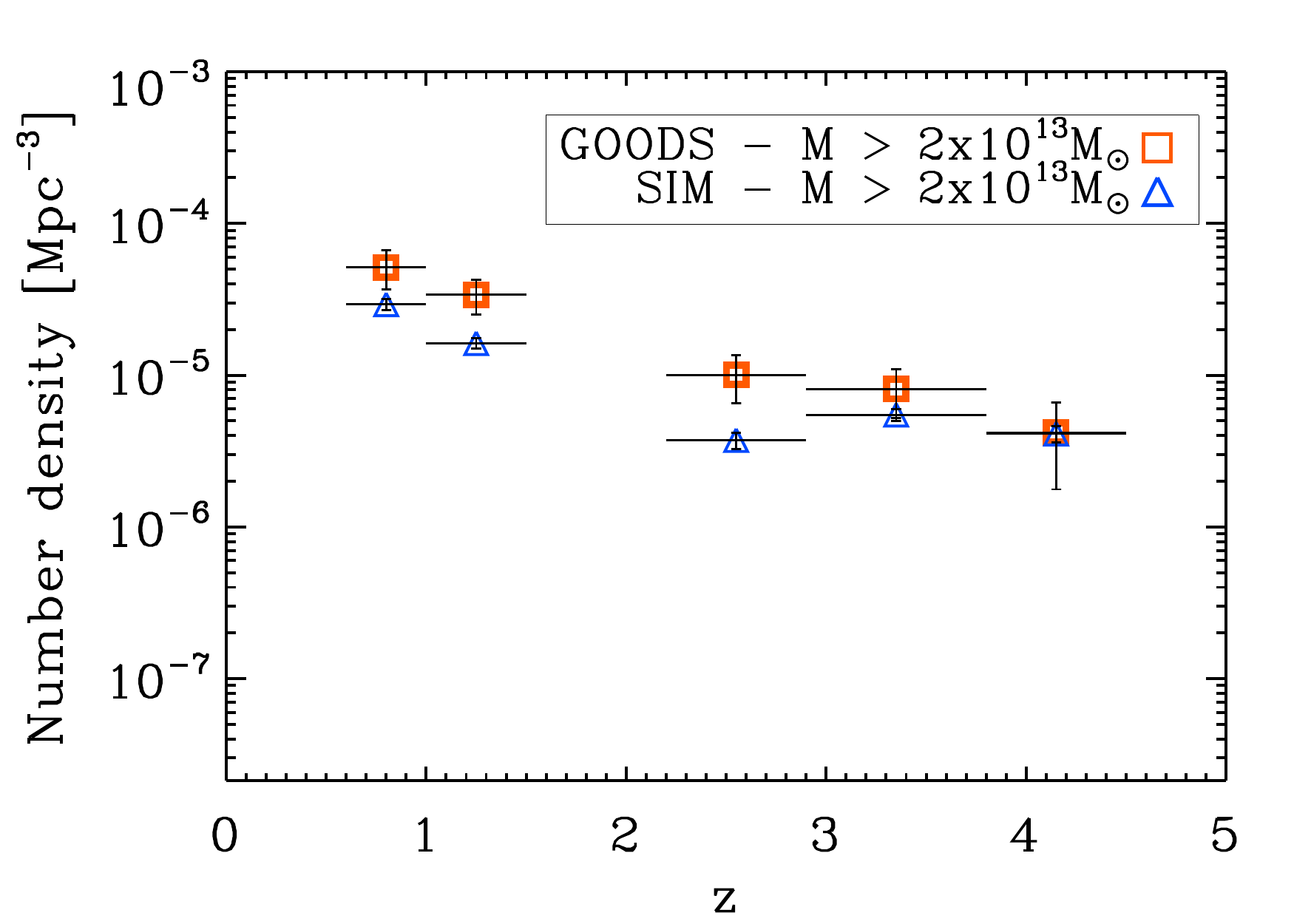}
\includegraphics[trim=2mm 1mm 5mm 2mm, clip, width=83mm]{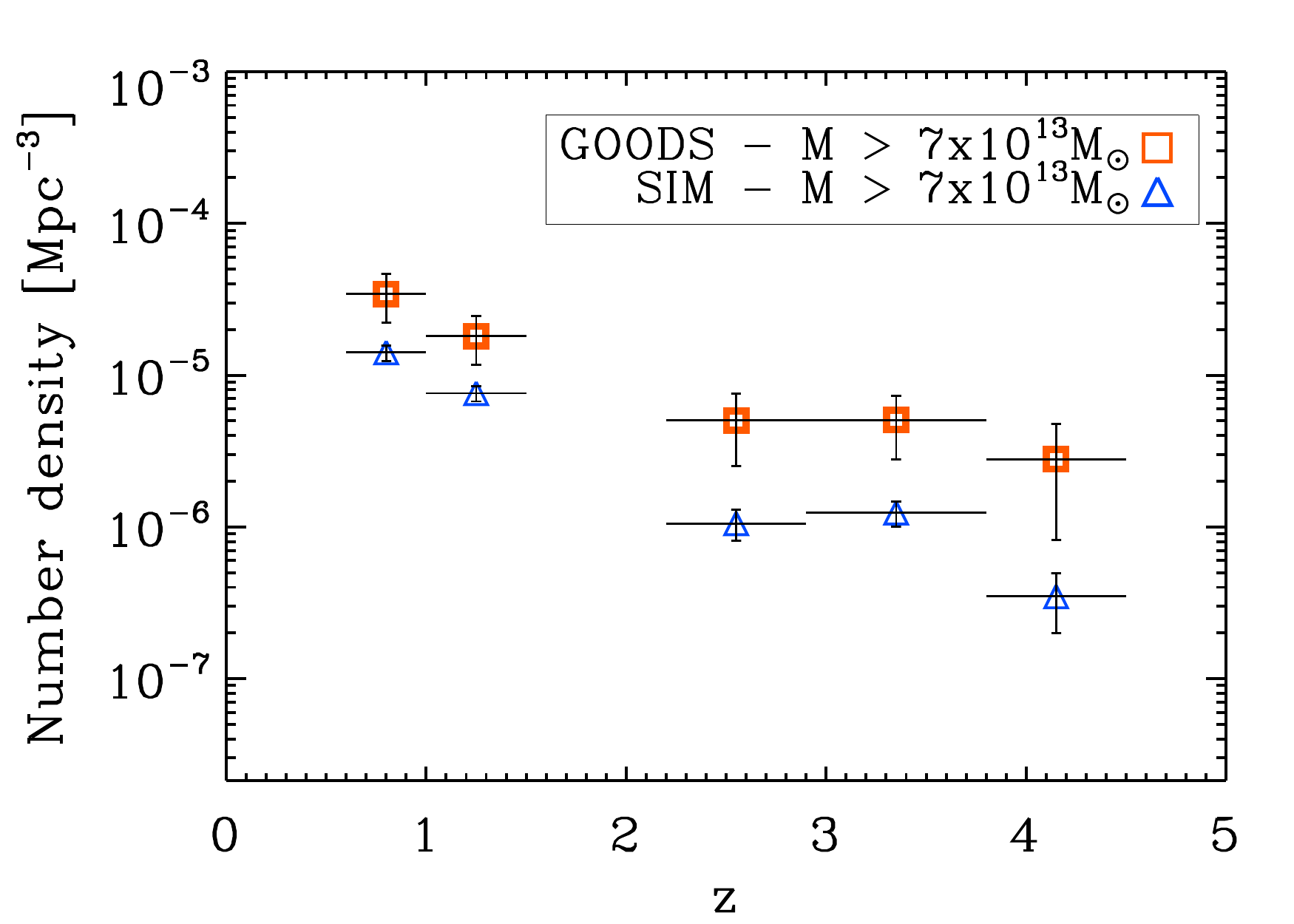}
\caption{ The observed number densities of MSGs as a function of redshift (red squares), compared with the simulation result (blue triangles). The upper panel shows the number density of MSGs with $M > 2 \times 10^{13}\, M_{\odot}$, while the bottom panels show the number density of MSGs with $M > 7 \times 10^{13}\, M_{\odot}$. The error bars represent the Poisson errors. When constructing the simulation result, a random scatter of 0.28 dex is taken into account. For the case of the lower mass threshold, the observed MSG number densities are marginally consistent with the simulation. However, the number densities of massive MSGs are significantly greater than the simulation at $z > 2$ when the higher mass MSGs are considered.
}
\label{f12}
\end{figure}

 In order to see if the number density of MSGs in the GOODS fields agrees with the simulation
 data, we derive the number density evolution of MSGs as a function of redshift, and compare it with the simulation result.

 The number density of MSGs is derived by summing the number of MSGs within each redshift bin with a bin size of $\Delta z \simeq 0.2 \times (1+z)$,
 and then dividing the number by the surveyed comoving volume covered
by the redshift bin. We estimate the number densities only over the mass range
 where our MSG search scheme is effective at picking up MSGs (i.e., $M >$ a few $\times10^{13} \, M_{\odot}$).
 A certain amount of incompleteness is expected in the number density at low mass end (by missing low mass MSGs), but we do not adopt any incompleteness correction since we expect the same level of incompleteness to be present for both the observations and the simulation data.
 Note that MSGs at $ 1.5 < z < 2$ are excluded in this analysis due to their large mass
 uncertainties that arise from inaccuracy in photometric redshifts and a small fraction of galaxies with $z_{\rm spec}$.

 Figure \ref{f12} shows the number density of MSGs at different redshifts
with two mass thresholds (top: $2 \times 10^{13} \, M_{\odot}$, bottom: $7 \times 10^{13} \, M_{\odot}$). The red thick squares indicate the observed number densities from the GOODS fields, and the blue triangles show the numbers derived from the analysis of the simulation data.
When deriving the number densities from the simulation, a random error of 0.28 dex is added to MSG mass values as described in the previous section.
  At the lower mass cut of $M > 2 \times 10^{13}\,M_{\odot}$ (top panel of Figure 13), the MSG number densities from $z=0.6$ to $z=4.5$ are marginally consistent with the simulation within a factor of two, with more MSGs in the GOODS fields than in the simulation. The discrepancy between the observations and the simulation becomes more prominent for a higher mass cut of $M > 7 \times 10^{13}\,M_{\odot}$ (bottom panel of Figure 13). Here, we find that the number densities of MSGs from the GOODS data are significantly higher than those from the simulation at $z > 2$ by a factor of $\sim$5.
  There are too many MSGs in the observed data at $z > 2$, compared to the simulation.

\begin{figure*}[tp!]
\centering
\includegraphics[width=78mm]{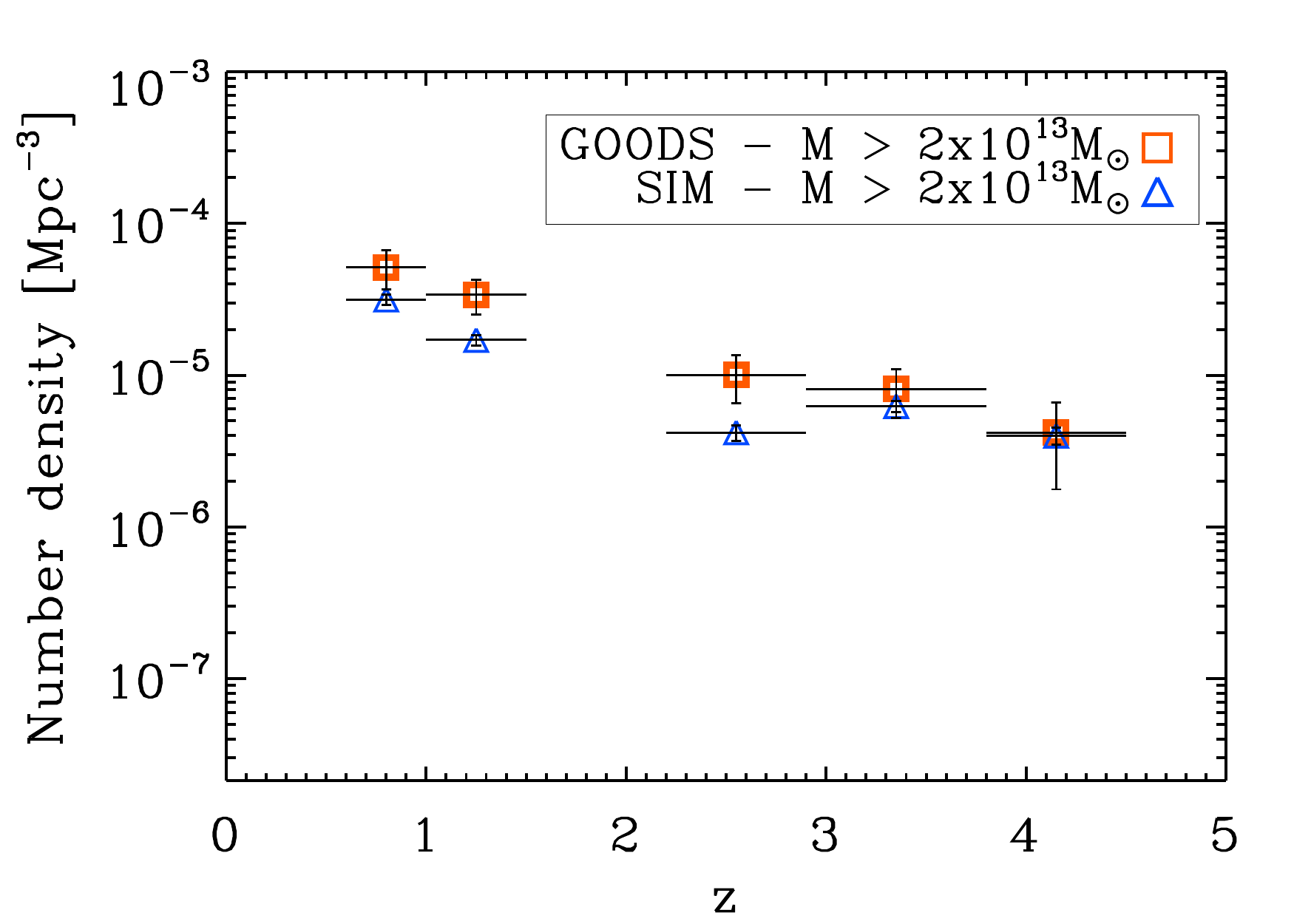} \includegraphics[width=78mm]{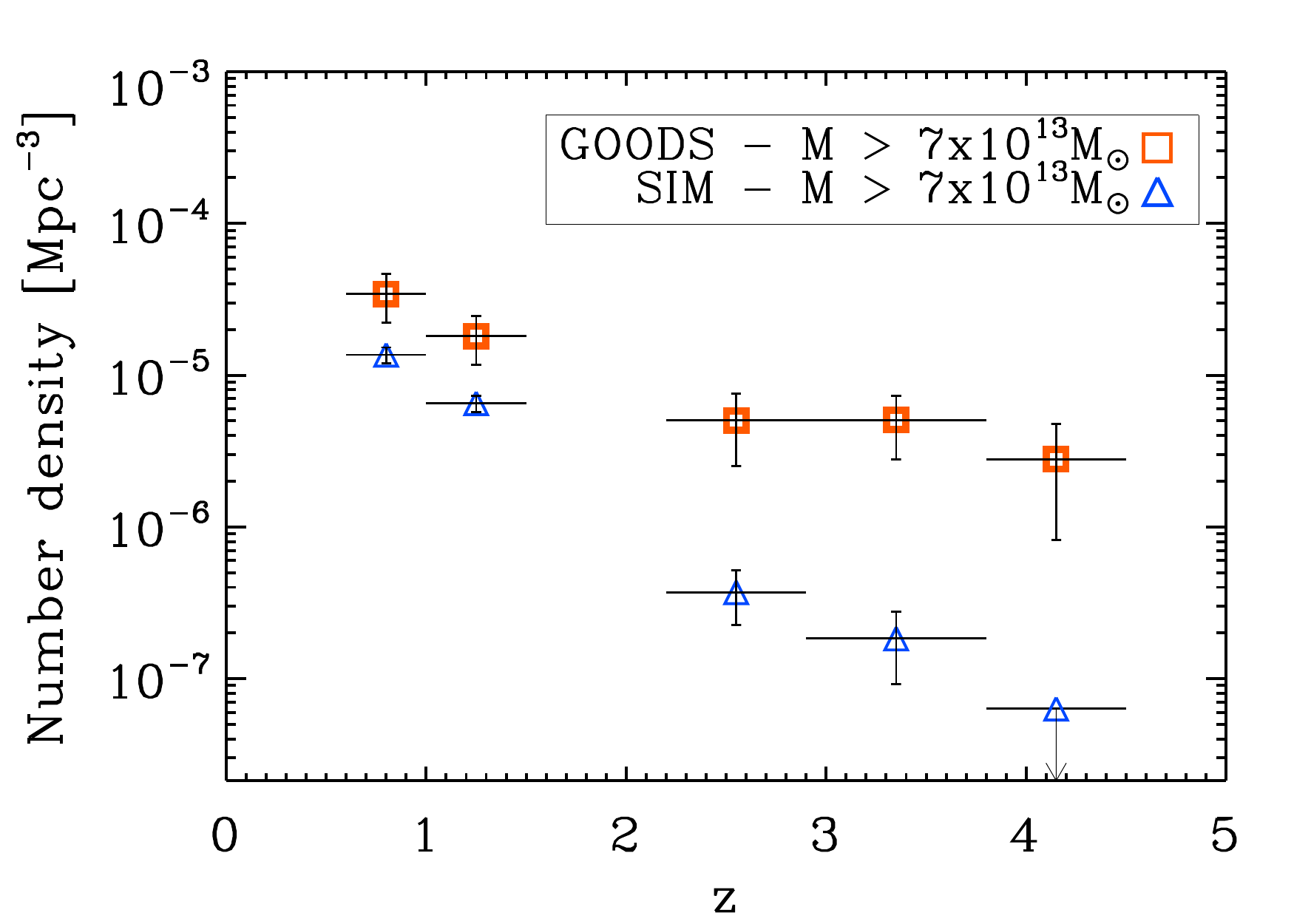} \\
\includegraphics[width=78mm]{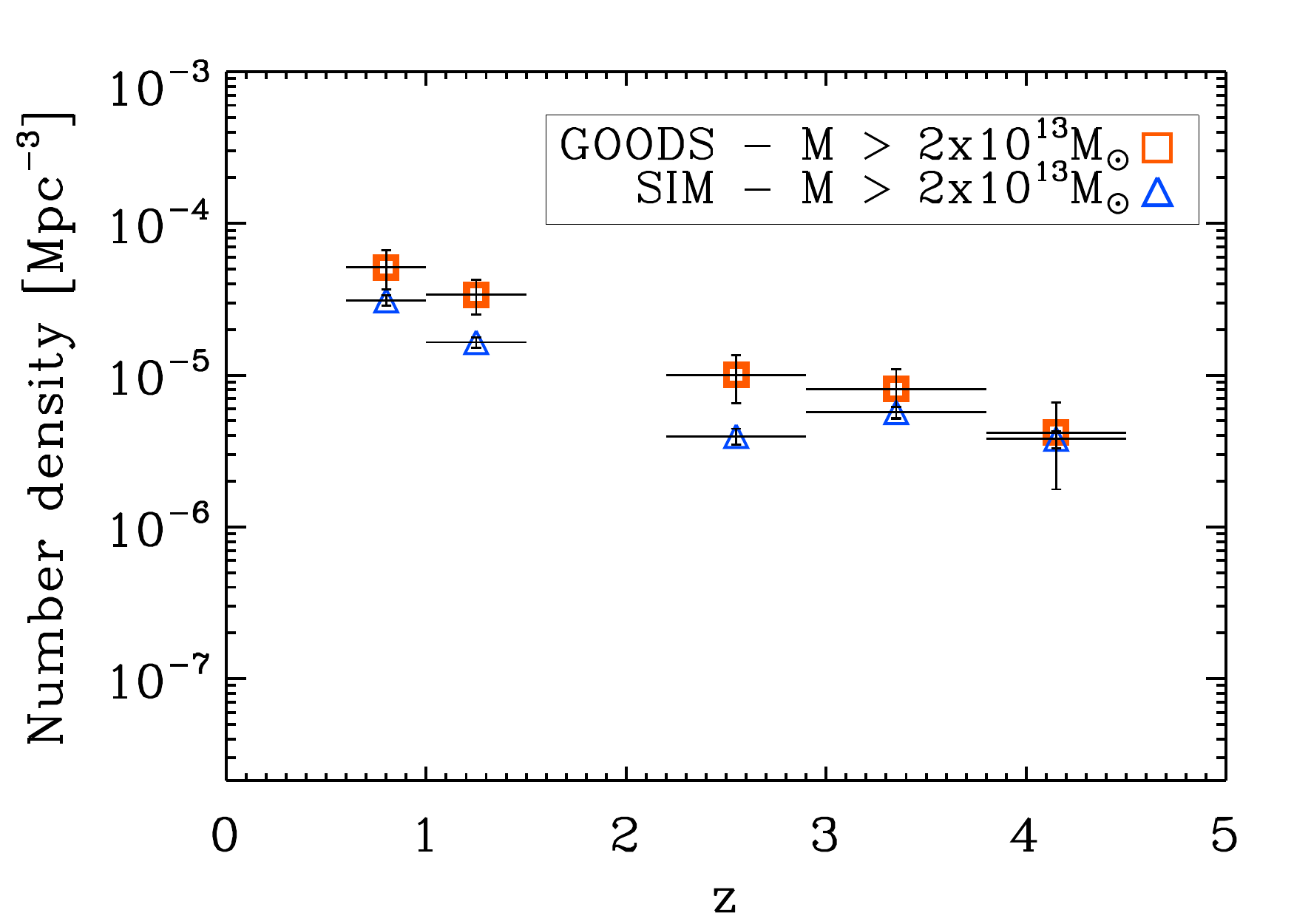} \includegraphics[width=78mm]{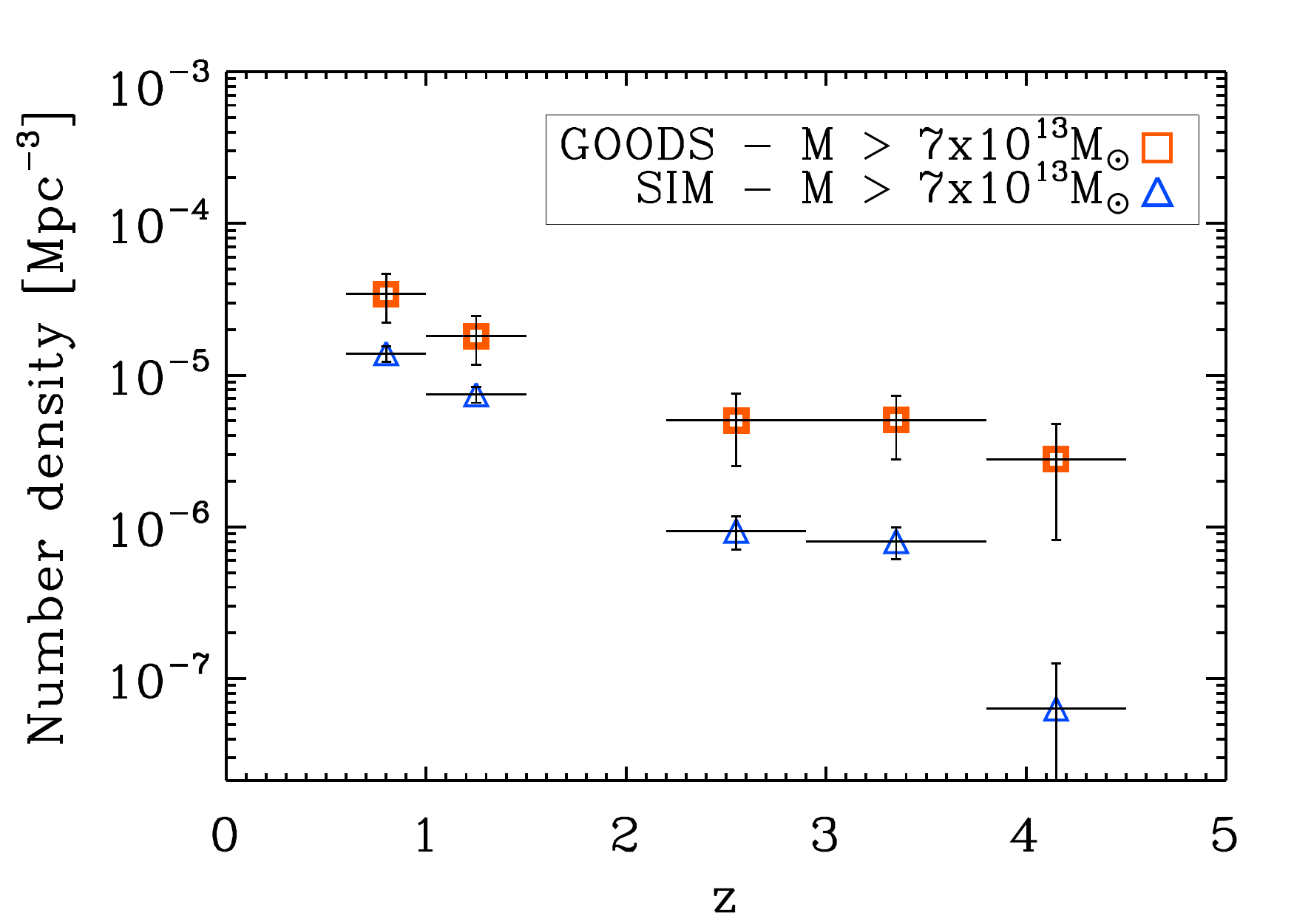} \\
\includegraphics[width=78mm]{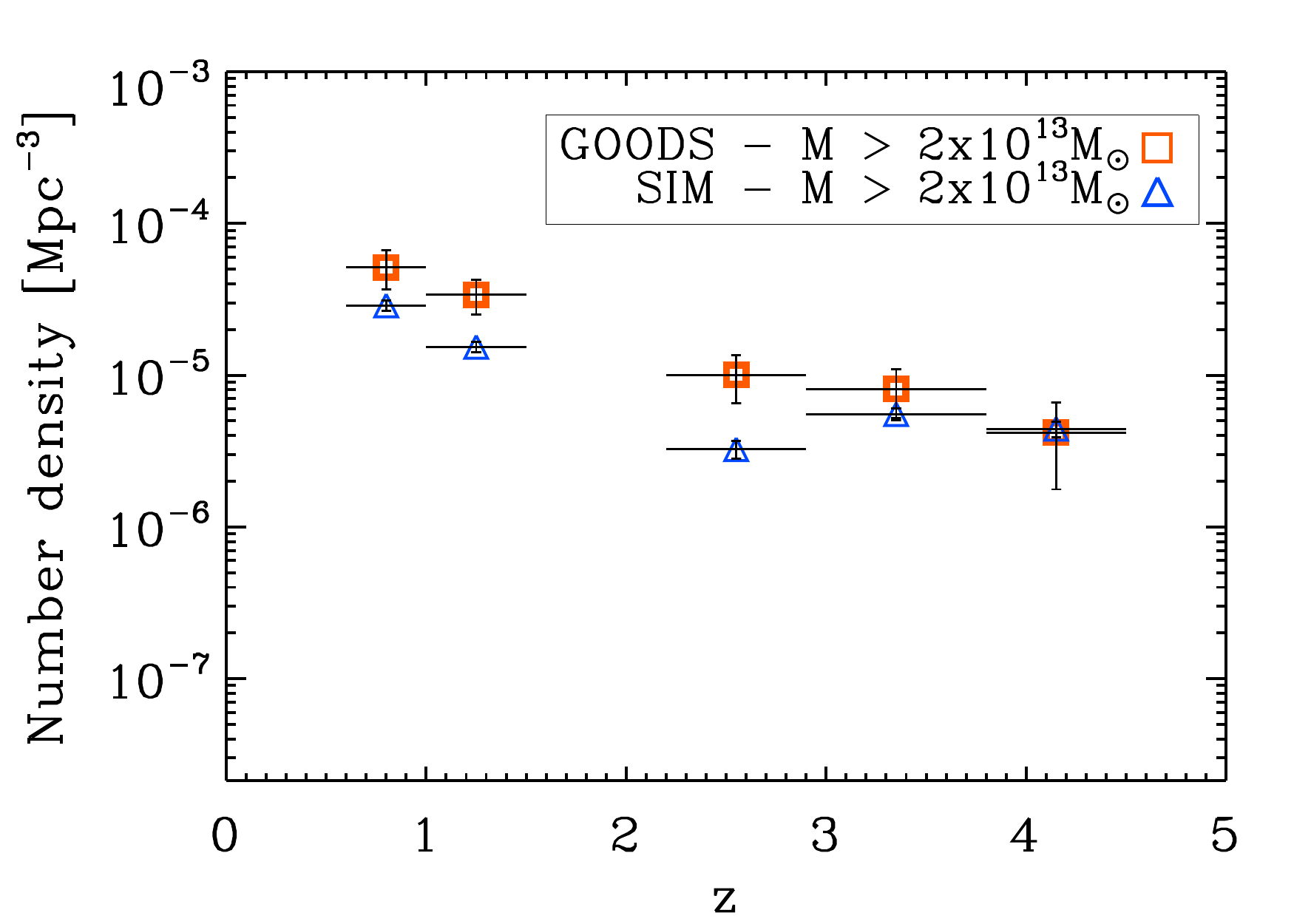} \includegraphics[width=78mm]{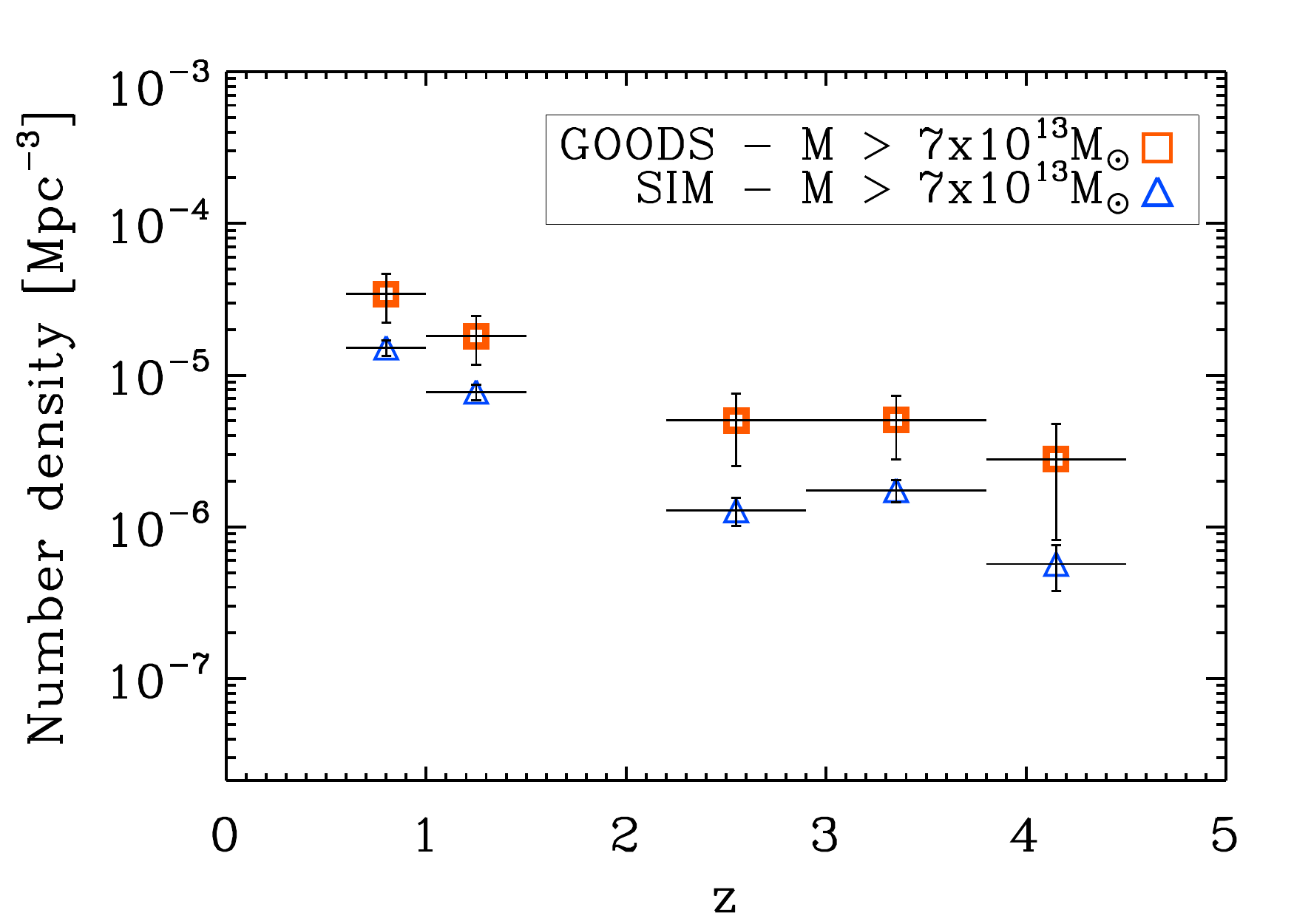} \\
\includegraphics[width=78mm]{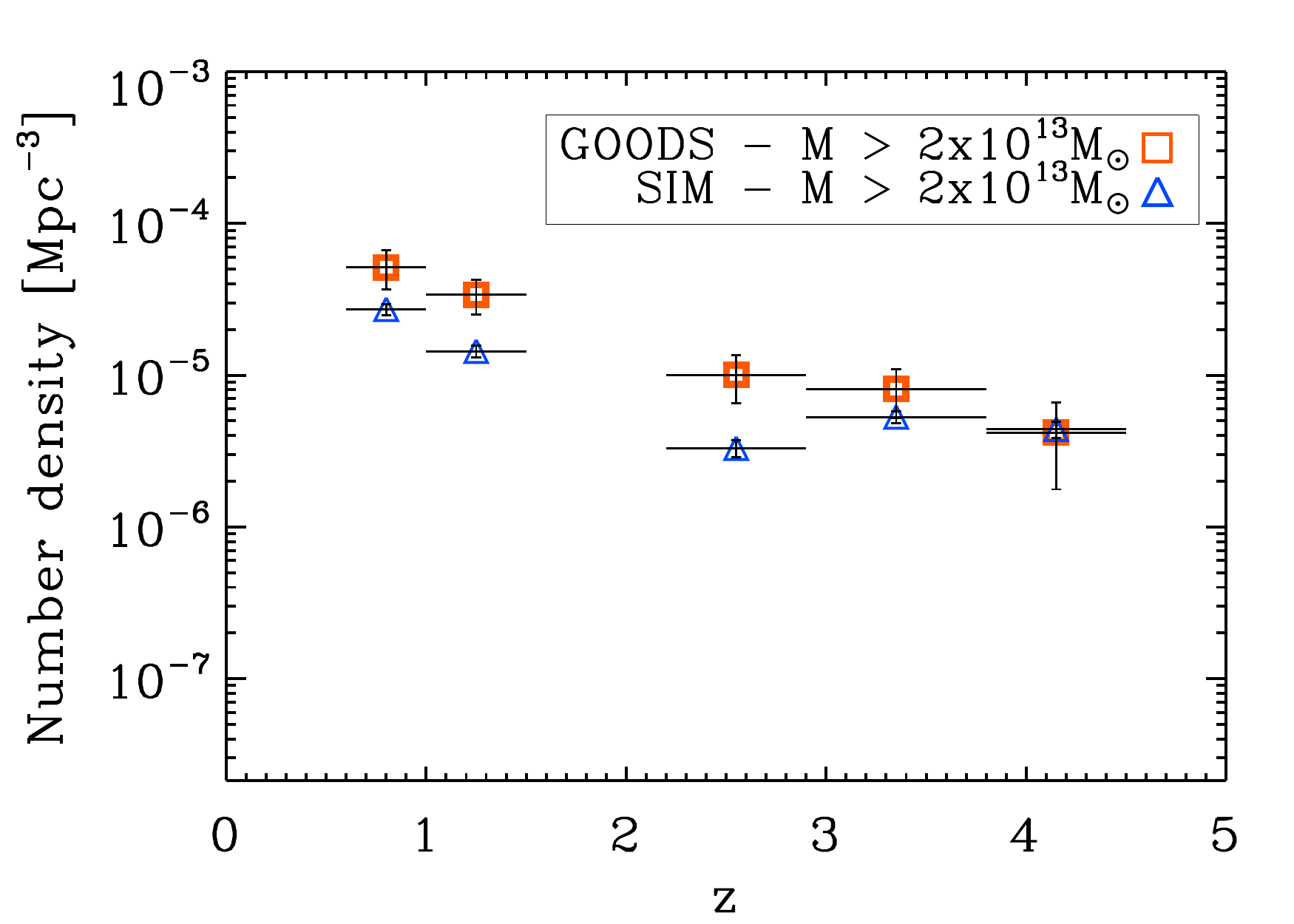} \includegraphics[width=78mm]{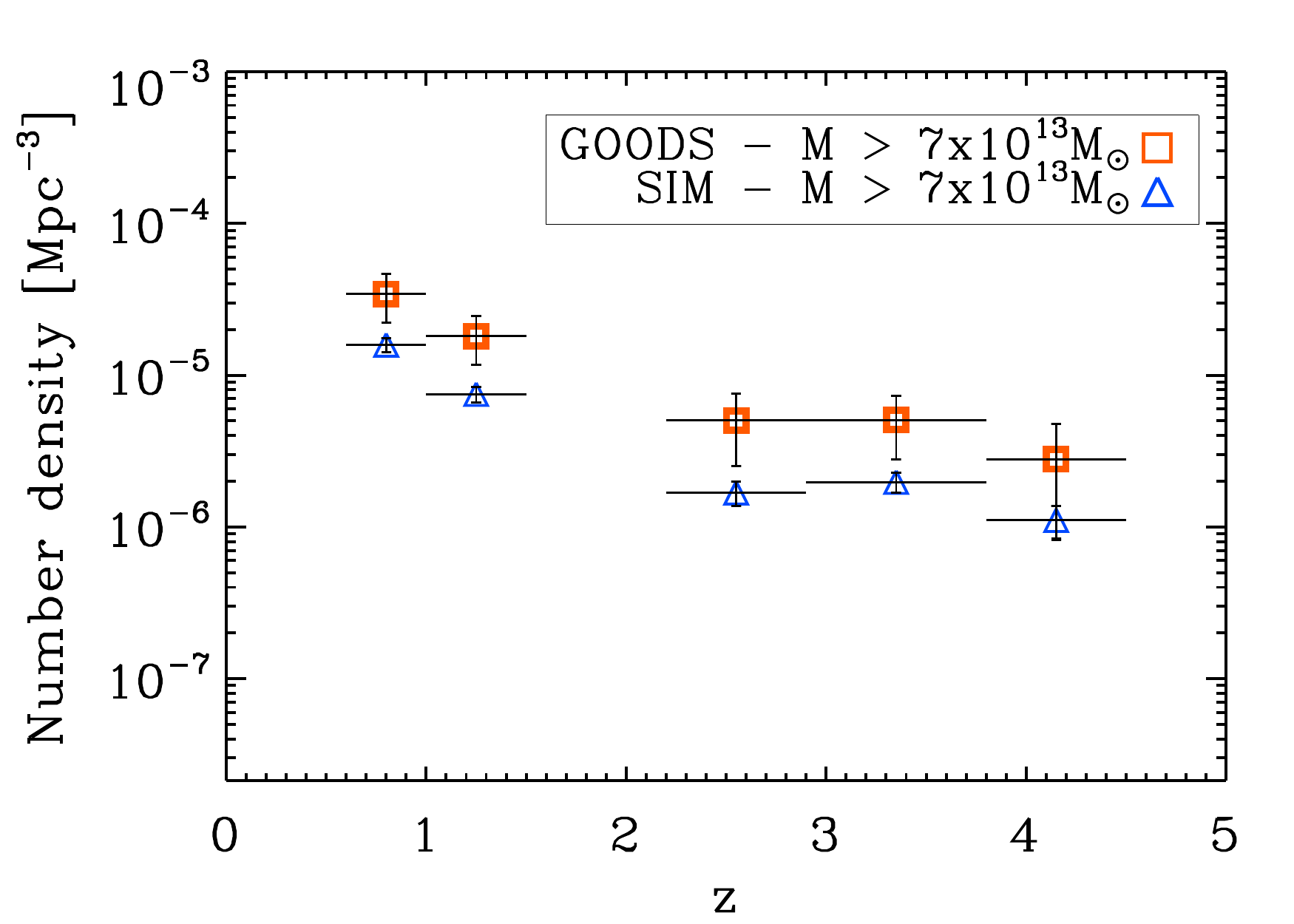}
\caption{ The number densities of MSGs with assumed scatters of 0, 0.21 dex, 0.35 dex, and 0.50 dex, from top to bottom panel, respectively, for the simulation results. The red squares and blue triangles indicate MSGs having a mass exceeding $ 2\times10^{13}M_\odot$ ($left$) and $7\times10^{13}M_\odot$ ($right$) in the GOODS fields and in the simulation data, respectively.
}
\label{f12a}
\end{figure*}

 Since the exact value of the error in the MSG mass estimates is uncertain, we examine in Figure \ref{f12a} how the result changes if we vary the error added in the MSG mass (in the simulation).
 In Figure \ref{f12a}, the random errors in MSG mass are changed to the values of
  0, 0.21 dex, 0.35 dex, and 0.5 dex.
 As the random error increases, the discrepancy between the observations and the simulation narrows, but the difference never decreases enough to remove the discrepancy.

\begin{figure*}[t!]
\centering
\includegraphics[width=83mm]{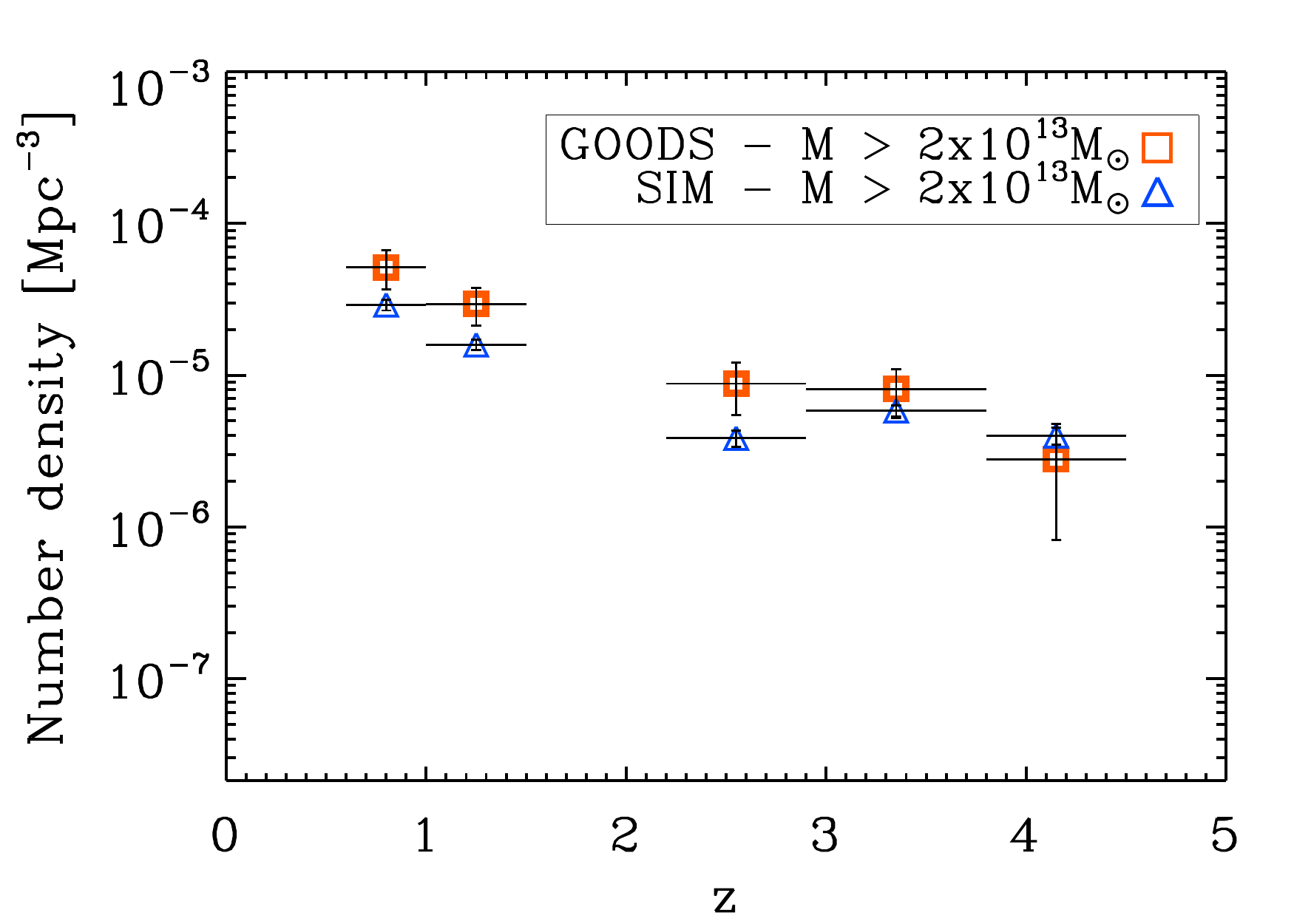} ~ \includegraphics[width=83mm]{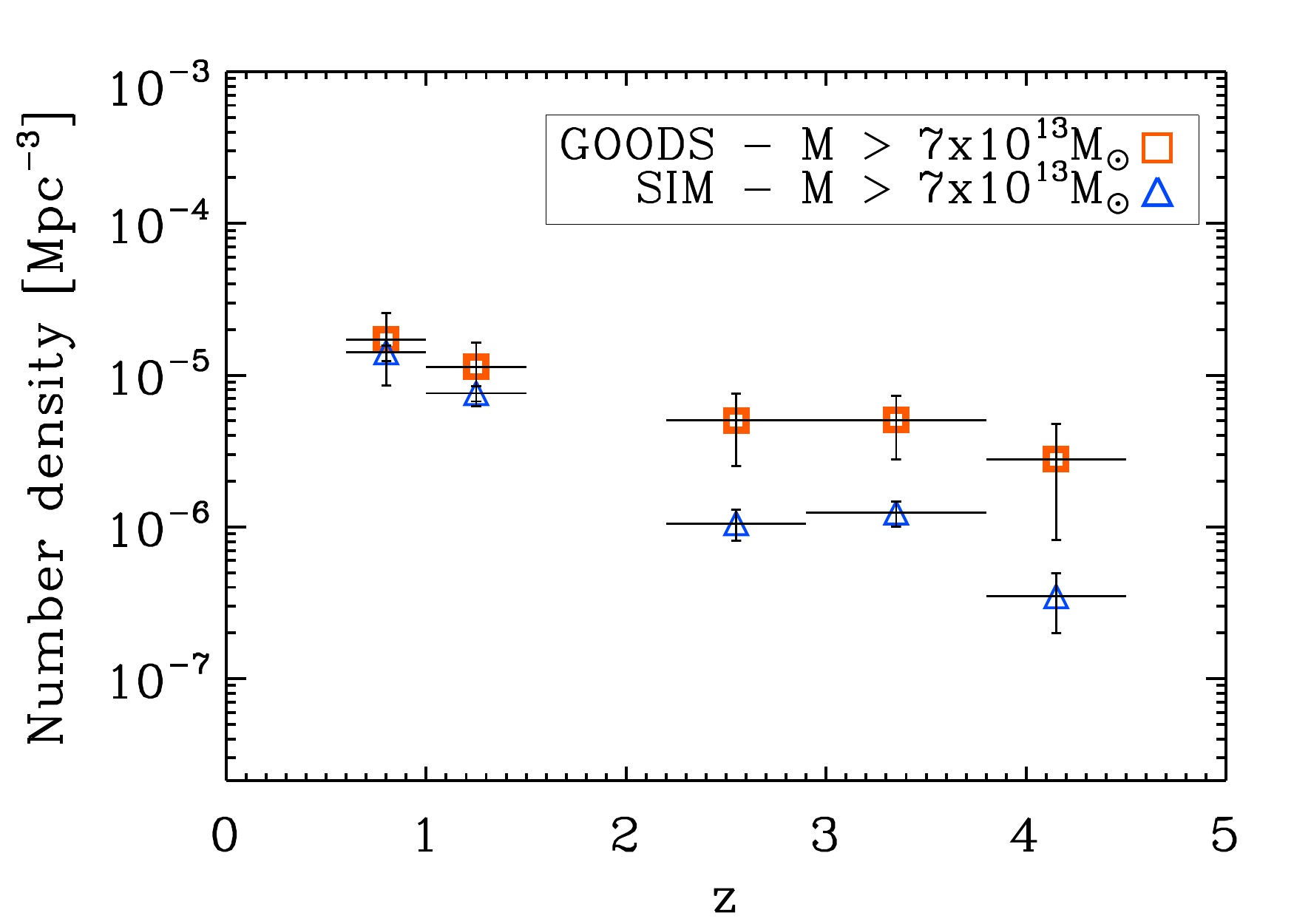} \\
\includegraphics[width=83mm]{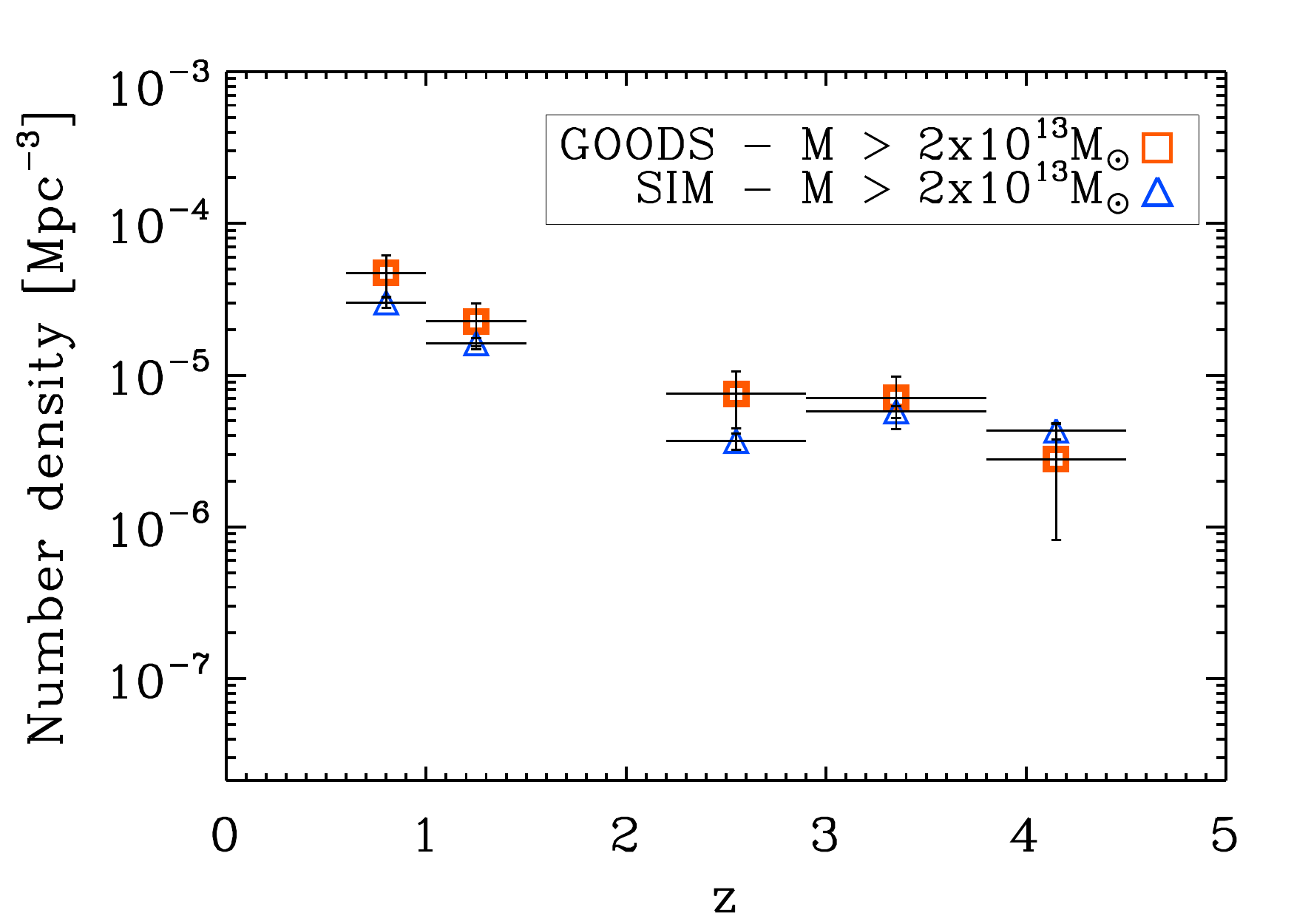} ~ \includegraphics[width=83mm]{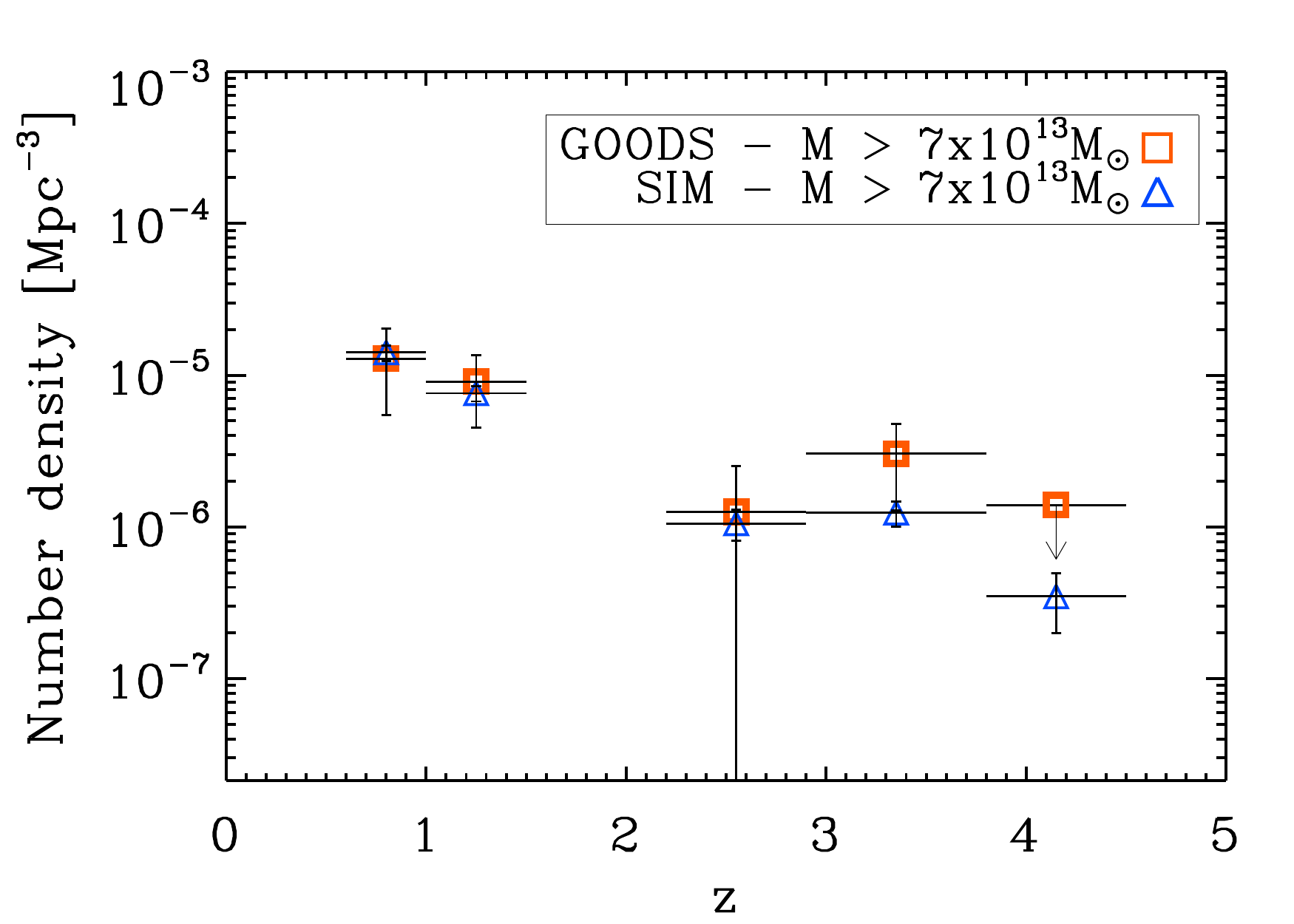} \\
\caption{ The number densities of MSGs with the systematic offsets with the reduction
 of the MSG masses from the observed data
 by 0.1 dex (upper panel) and 0.2 dex (bottom panel). A scatter of 0.28 dex is assumed for the simulation. The red squares and blue triangles indicate MSGs having a mass exceeding $ 2\times10^{13}M_\odot$ ($left$) and $7\times10^{13}M_\odot$ ($right$) in the GOODS fields
 and the simulation, respectively.
}
\label{f12b}
\end{figure*}

 We also investigate how systematic errors in the MSG mass estimates affect the result. Figure \ref{f12b} shows the results, where the systematic offset in the MSG mass measurements are varied from 0.1 dex to 0.2 dex. These systematic offsets are applied to the masses of MSGs from the observed data (i.e., MSG masses are reduced by this amount). As in the case for Figure \ref{f12}, the random errors of 0.28 dex are assumed for the simulation. When the observed MSG masses are systematically reduced by 0.2 dex, we find that the discrepancy in the MSG number densities  between the observation and the simulation disappears.
 However, we note that the systematic error we quoted earlier, when deriving the stellar masses through SED-fitting, does not affect the derived MSG mass. The conversion of the stellar mass to the halo mass also includes the same kind of systematics in the stellar mass, thus the systematic error cancels out when deriving the MSG mass.

Overall, we conclude that there is an over-abundance of massive MSGs with $M > 7 \times 10^{13}\, M_{\odot}$ at $z > 2$ by a factor of a few, compared with the Millennium Simulation.
One might be able to reconcile the observation and the simulation if there is a systematic offset in the derived MSG mass by 0.2 dex, but it is not clear how such a systematic bias could arise.

\section{Discussion}

\subsection{Cosmic Variance}

 We consider here the possibility that our finding of too many MSGs is due to cosmic variance effect.
  If the GOODS fields have an anomalously large number of MSGs due to cosmic variance, then, the difference between the simulation
 and the observation can be alleviated.
 The moderate field of view of 320 arcmin$^2$ covers a large volume ($\sim10^5$ Mpc$^3$)
  when the redshift bins are of order $\Delta z \sim 0.1$, which is sufficient to negate the cosmic variation.
 Nevertheless, we can estimate from the simulation the chance coincidence that a survey area as small as the GOODS fields could have anomalously large number of MSGs.
To do so, we divide the simulation area into 2500 sets of two 160 arcmin$^{2}$, and investigate how many such sets can have the MSG number densities as large as our result. No subdivided regions reveal as many MSGs as what we found from the GOODS field, giving a probability of $< 1/2500$ for the chance coincidence that the GOODS fields happened be highly
 overdense areas.

\subsection{Implications for Initial Density Fluctuations}

\begin{figure}[t!]
\centering
\includegraphics[trim=14mm 4mm 7mm 4mm, clip, width=83mm]{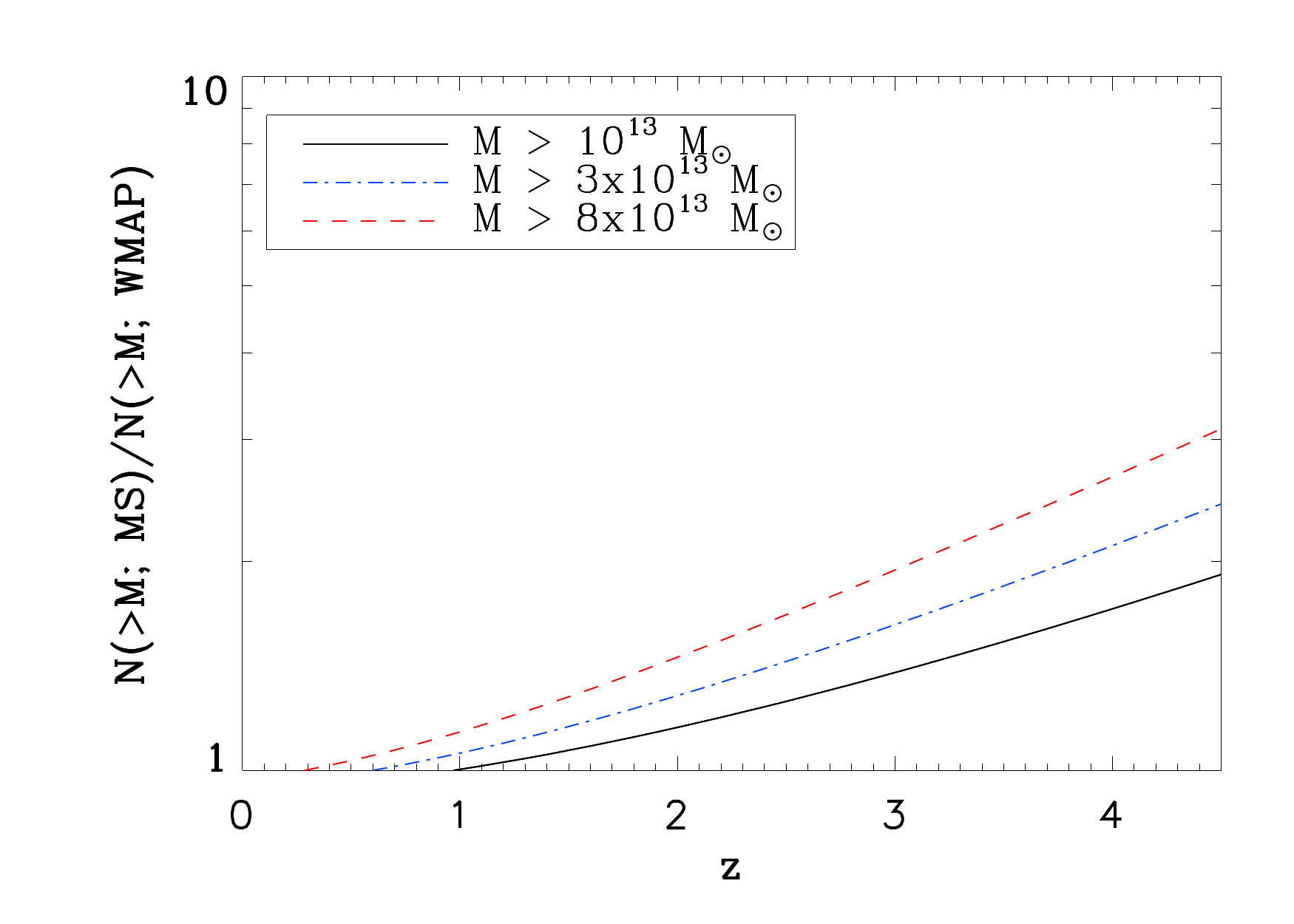}
\caption{The ratio of the abundance of halos as a function of redshift for the cosmological parameters of the $WMAP$ 7-year and of the Millennium Simulation with three different mass cuts. To create this figure, the Sheth-Tormen halo mass function is convolved with a lognormal distribution function in mass with a standard deviation of 0.28 dex.
}
\label{f13}
\end{figure}

Here, we discuss the implication of our results on the initial density fluctuations. We focus on $\sigma_{8}$ and $n_s$, and the non-Gaussianity.
  The initial density fluctuation assumed in the Millennium Simulation is not
 up-to-date with the more recent observational constraints. For $\sigma_8$ at a scale of $r = 8\,h^{-1}$ Mpc and the primordial power-law index $n_s$, they adopted the values of $\sigma_{8}=0.9$ and $n_s=1$ while the latest measurements from the $WMAP$ 7-year data suggest the values of $\sigma_{8}=0.816 \pm 0.024$ and $n_s=0.968 \pm 0.012$, respectively
 (\citealt{34, 35}).
 The adoption of $\sigma_{8}=0.9$ and $n_s=1$ introduces a factor of $\sim2-3$ increase of the halo number density in the Millennium Simulation compared to the halo number density from the Sheth-Tormen halo mass function with the $WMAP$ values of $\sigma_{8}$ and $n_{s}$ (Figure \ref{f13}).
 This shows that the expected number density of MSGs in the simulation
would become even less than what is found in the Millennium Simulation data, if we were to adopt the $WMAP$ $\sigma_{8}$ and $n_{s}$ values.

 Next, we consider non-Gaussianity in the initial density fluctuation as one possibility, since the abundance of massive halos at high redshifts is also sensitive to non-Gaussianity of initial density fluctuation spectrum. Recently, there is interest in the non-Gaussianity
 from a theoretical standpoint (\citealt{da1,lo1,sm1}).
  Deviations from Gaussian initial conditions are parameterized in terms of the dimensionless $f_{NL}$ parameter (\citealt{36, 37, 38, 39}).
 If the initial density fluctuation has a skewed distribution in which high density peaks are more abundant than what is expected in a Gaussian density fluctuation, one would expect that more massive halos can be more easily found early in the Universe. Such a case corresponds to a large positive value of $f_{NL}$ in the non-Gaussian framework.
Adopting a large non-linearity parameter of $f_{NL} \sim +500$ or $\sim -500$, one can boost or decrease the number density of massive halos  by a factor of few at $z > 2$ (\citealt{40}). Therefore, our result may imply that the $f_{NL}$ value could be as large as +500. On the other hand, several recent works provide constraints on $f_{NL}$ to be about $\sim30$
(\citealt{41, 42, 43}). If so, the number density of massive halos does not change more than a factor of 1.1, a negligible amount. Clearly, while further investigation is needed, but it seems that the non-Gaussianity cannot explain our result, assuming that the recent limits of
 $f_{NL} \sim 30$ or less are valid.

\subsection{Constraints on Hot Dark Matter}

  We discuss the implication of our result for hot dark matter.
 The streaming motion of hot dark matter such as massive neutrinos smooths out the density fluctuation, and suppresses the formation of massive structures leading to a reduction of the number of massive structures at high redshift. Therefore, massive hot dark matter particles are likely to reduce the number of MSGs in the simulation, rather than helping explain our result.
 Marulli et al. (2011) predict that the number density of halos with $M = 10^{14}\, M_{\odot}$, in comparison to the standard Sheth-Tormen halo mass function, is reduced by a factor of $\sim 10$ and $\sim 3$ for $M_{\nu} =0.6$ eV, and $M_{\nu} = 0.3$ eV respectively, where $M_{\nu}$ represents the total neutrino mass of different species.
 If the neutrinos have masses as heavy as $M_{\nu} \sim 0.3$ eV, the discrepancy between the observed MSG number density and the number density in the simulation will become even more severe.
 The current experimental lower limit is $M_{\nu} \sim 0.05$ eV (\citealt{lp1}), and our result suggests that $M_{\nu}$ is not likely to be much higher than the lower limit.

\subsection{Why Are There Too Many MSGs at {\itshape z}\,$>$\,2?}

 The discussions above narrow down the possible causes for the over-abundance of MSGs in the GOODS fields to three remaining possibilities: (1) some baryonic physics that are not implemented well in the simulation; (2) systematic overestimate of MSG mass; and (3)
 the assumption of the $\Lambda$CDM cosmology in the simulation.

 First, we consider a case that unknown baryonic physics in galaxy formation leads to play a role in the mismatch of the MSGs number density between the observation results
 and the simulation. Our search and mass estimation of MSGs rely on galaxies, therefore it is important that the simulation reproduces the observed galaxy properties well.
 The Millennium Simulation has been
 tested extensively against observational data, and in many cases, excellent agreement
 has been reported between the observation results and the simulation (e.g., Scoville et al.
 2013). However, it is noteworthy that the Millennium Simulation has difficulty explaining
 some of the galaxy properties such as the abundance of massive galaxies at
 high redshifts (Collins et al. 2009), and a certain aspect of the SFR-density relation (Elbaz et al. 2007).
  A more comprehensive comparison of the observed data and simulated data,
  including spectroscopic data and newer simulations, should help us understand what kind of baryonic physics has been missed in the simulation.

 Second, as we have shown earlier, systematic overestimation of MSG masses by a factor of 0.2 dex can explain why the MSG number density appears to be too high. This is perhaps the most likely explanation of our results, although quantification of this effect is difficult. Independent methods to estimate MSG masses would provide a useful consistency check of the MSG mass.
 We hope to produce such a study in the future, including more spectroscopic data for our MSG member candidates.

 Finally, if all of the above possibilities fail to explain the result, the over-abundance of MSGs at high redshifts could present a problem for the $\Lambda$CDM cosmology. A different simulation may reveal that this is not the case, but at this moment, we suggest that the $\Lambda$CDM cosmology may be in trouble if the above solutions do not succeed to explain our results.

\begin{table*}[t!p]
\centering
\setcounter{table}{3}
\caption{Properties of MSGs in the GOODS fields}
\begin{tabular}{ccccccccccc}
\toprule
 MSG ID & $z$ & $z_{\rm spec}$  & $\sigma$ &  N$_{zs}$ & N$_{zp}$ & $M_{*,\rm MSG}$  & Mass &  $\frac{M_{*,\rm MSG}}{M_{\rm halo}}$ & $M$(Lit.)  & $r_p$ \\
    &   &   &  &   &    & log ($M/M_\odot$) & log ($M/M_\odot$) &   & log ($M/M_\odot$) & Mpc \\ \addlinespace
(1) & (2) & (3) & (4) & (5) & (6) & (7) & (8) & (9) & (10) & (11) \\
\midrule
\textsc{GOODS North} &&&&&&&&&& \\ \addlinespace
  1236.1+6210 & $0.65$A & 0.64 &  4.55  & 13 & 16 & 12.43 & 13.95 & 0.030 &  --- & 0.36  \\
  1236.5+6215 & $0.65$B & 0.64 &  4.33 & 10 & 31  & 12.42 & 13.92 & 0.032 & --- & 0.48 \\
  1236.6+6215 & $0.70$ & 0.68 &  3.78  & 7 & 33   & 12.20 & 13.50 & 0.050 & --- & 0.43 \\
  1236.8+6215 & $0.75$ & 0.74  &  4.04 & 3 & 43  & 11.77 & 13.07 & 0.050 & --- & 0.51 \\
  1237.0+6220 & $0.80$ & 0.80  &  4.18 & 9 & 29  & 12.53 & 14.17 & 0.023 & --- & 0.44 \\
  1237.5+6215 & $0.84$ & 0.84 &  7.97  & 16 & 28 & 12.35 & 13.77 & 0.038 & --- & 0.47 \\
  1236.5+6216 & $0.85$ & 0.85  &  5.07 & 37 & 53   & 12.95 & 15.10 & 0.007 & --- & 0.56 \\
  1236.4+6208 & $0.96$ & 0.96 &  3.85  & 7 & 44 & 12.38 & 13.84 & 0.035 & --- & 0.49 \\
  1236.9+6211 & $1.02$A & 1.02  & 4.17 & 12 & 32  & 12.94 & 15.08 & 0.007 & --- & 0.44 \\
  1237.0+6212 & $1.02$B & 1.02  &  3.54 & 16 & 36 & 12.81 & 14.79 & 0.010 & --- & 0.48 \\
  1237.3+6220 & $1.05$ & 1.03  &  4.45 & 7 & 30  & 12.20 & 13.50 & 0.007 & --- & 0.52 \\
  1237.3+6213 & $1.15$ & 1.14  &  3.60 & 8 & 15 & 12.43 & 13.95 & 0.030 & --- & 0.55 \\
  1237.0+6212 & $1.20$ & 1.20  &  6.91 & 2 & 30  & 11.93 & 13.23 & 0.050 & --- & 0.45 \\
  1237.0+6212 & $1.25$ & 1.24  &  5.98 & 9 & 34  & 12.52 & 14.14 & 0.024 & --- & 0.45 \\
  1236.9+6211 & $1.27$A  & 1.27 &  4.28  & 2 & 33 & 12.22 & 13.52 & 0.050 & --- & 0.50 \\
  1236.5+6208 & $1.27$B  & 1.27 &  3.72 & 4 & 39 & 12.47 & 14.04 & 0.027 & --- & 0.57 \\
  1236.3+6208 & $1.36$A & 1.36  &  4.11 & 5 & 41  & 12.63 & 14.38 & 0.018 & --- & 0.51 \\
  1237.2+6221 & $1.36$B  & 1.36 &  4.20 & 1 & 46 & 12.43 & 13.94 & 0.031 & --- & 0.49 \\
  1236.2+6212 & $1.60$ & 1.60  &  4.19 & 2 & 111  & 12.81 & 14.78 & 0.011 & --- & 0.51 \\
  1237.2+6212 & $2.45$ & 2.43  &  4.70 & 1 & 31  & 12.57 & 14.25 & 0.021 & --- & 0.54 \\
  1237.5+6214 & $2.55$ & 2.55  &  5.37 & 2 & 27   & 12.54 & 14.19 & 0.022 & --- & 0.49 \\
  1236.9+6211 & $2.80$ & 2.80  &  4.93 & 1 & 20  & 12.70 & 14.53 & 0.014 & --- & 0.52 \\
  1236.9+6210 & $2.95$ & 2.95  &  5.86 & 4 & 17  & 12.62 & 14.37 & 0.018 & --- & 0.56 \\
  1236.8+6211 & $3.00$A & 3.00  &  5.11 & 1 & 12   & 12.25 & 13.54 & 0.050 & --- & 0.40 \\
  1236.7+6213 & $3.00$B & 3.00 &  4.19 & 3 & 17 & 12.16 & 13.46 & 0.050 & --- & 0.57 \\
  1236.7+6214 & $3.10$ & 3.10 &  4.88  & 2 & 11 & 12.67 & 14.47 & 0.016 & --- & 0.50 \\
  1236.6+6217 & $3.25$ & 3.23 &  4.75  & 2 & 11 & 12.36 & 13.79 & 0.037 & --- & 0.56 \\
  1236.6+6217 & $3.40$ & 3.40 &  4.82  & 1 & 13 & 13.18 & 15.61 & 0.004 & --- & 0.49 \\
  1236.3+6208 & $3.55$ & --- &  4.67  & 0 & 10 & 12.50 & 14.10 & 0.025 & --- & 0.47 \\
  1237.5+6212 & $3.70$ & --- &  4.94  & 0 & 13 & 11.94 & 13.24 & 0.050 & --- & 0.64 \\
  1237.2+6221 & $4.05$ & 4.05 &  5.05  & 0 & 14 & 12.49 & 14.08 & 0.026 & $\sim$14$^{\rm a}$ & 0.40 \\
  1237.7+6215 & $4.20$ & --- &  5.48  & 0 & 17 & 12.05 & 13.36 & 0.050 & --- & 0.38 \\
  1236.4+6209 & $4.45$A1 & --- &  3.91 & 0 & 12 & 11.69 & 12.99 & 0.050 & --- & 0.36 \\
  1236.3+6214 & $4.45$A2 & --- &  3.91 & 0 & 9 & 11.66 & 12.96 & 0.050 & --- & 0.29 \\
\midrule
 \textsc{GOODS South} &&&&&&&&&& \\ \addlinespace
  0332.3--2749 & $0.65$ & 0.67 & 3.54 & 9 & 51  & 12.39 & 13.86 & 0.034 & $13.30-13.60$ $^{\rm b}$ & 0.48  \\
  0332.3--2749 & $0.70$ & 0.74  & 3.63 & 19 & 60  & 12.59 & 14.31 & 0.019 & $13.95-14.48$ $^{\rm b}$ & 0.51 \\
  0332.6--2745 & $0.90$ & 0.90 &  4.02 & 3 & 86 & 12.34 & 13.75 & 0.039 & --- &  0.39 \\
  0332.5--2745 & $0.95$ & 0.95  &  4.03 & 7 & 79 & 12.45 & 14.00 & 0.029 & --- & 0.58 \\
  0332.5--2741 & $1.05 $ & 1.05 &  3.87 & 12 & 26 & 12.62 & 13.69 & 0.042 & --- &  0.52 \\
  0332.3--2752 & $1.10$A &  1.09 &  4.20 & 12 & 25  & 12.62 & 14.36 & 0.018 & $13.70-13.85$ $^{\rm c}$ & 0.50  \\
  {} & ${}$ & ${}$ & ${}$ & ${}$ & ${}$ & ${}$ & ${}$ & ${}$ &  $13.9$ $^{\rm b}$ & {} \\
  0332.6--2745 & $1.09$B & 1.09  &  3.68 & 7 & 35  & 12.08 & 13.38 & 0.050 & --- & 0.56 \\
  0332.8--2752 & $1.12 $ & 1.13 &  3.65 & 3 & 17  & 11.47 & 12.78 & 0.050 & --- &  0.48 \\
  0332.2--2749 & $1.20$ & 1.20 &  3.73 & 1 & 26  & 11.94 & 13.24 & 0.050 & --- & 0.43 \\
  0332.4--2746 & $1.23 $ & 1.23 &  3.65 & 9 & 28  & 12.20 & 13.50 & 0.050 & --- &  0.49 \\
  0332.8--2751 & $1.25$A & 1.25 &  4.06 & 4 & 28  & 12.20 & 13.50 & 0.050 & --- & 0.51 \\
  0332.6--2747 & $1.25$B & 1.25 &  3.70 & 1 & 23  & 11.65 & 12.96 & 0.050 & --- & 0.44 \\
  0332.6--2747 & $1.30$ & 1.30 &  3.95 & 4 & 31 & 12.02 & 13.32 & 0.050 & --- & 0.48 \\
  0332.6--2749 & $1.60$A & 1.61  & 4.37 & 4 & 105 & 12.53 & 14.17 & 0.023 & --- & 0.50 \\
  0332.6--2747 & $1.60$B & 1.61  &  4.04 & 1 & 113 & 12.47 & 14.04 & 0.027 & --- & 0.51 \\
  0332.5--2742 & $1.60$C & 1.61  &  3.72 & 5 & 116 & 12.68 & 14.49 & 0.015 & $13.78-13.95$ $^{\rm d}$ & 0.50 \\
  {} & ${}$ & ${}$ & ${}$ & ${}$ & ${}$ & ${}$ & ${}$ & ${}$ & $14.30-14.69$ $^{\rm b}$ & {} \\
  {} & ${}$ & ${}$ & ${}$ & ${}$ & ${}$ & ${}$ & ${}$ & ${}$ & $14.11-14.76$ $^{\rm e}$ & {} \\
\bottomrule
\label{t3}
\end{tabular}
\end{table*}


\begin{table*}[t!]
\centering
\setcounter{table}{3}
\caption{(Continued)}
\begin{tabular}{ccccccccccc}
\toprule
 MSG ID & $z$ & $z_{\rm spec}$  & $\sigma$ &  N$_{zs}$ & N$_{zp}$ & $M_{*,\rm MSG}$  & Mass &  $\frac{M_{*,\rm MSG}}{M_{\rm halo}}$ & $M$(Lit.)  & $r_p$ \\
    &   &   &  &   &    & log ($M/M_\odot$) & log ($M/M_\odot$) &   & log ($M/M_\odot$) & Mpc \\ \addlinespace
(1) & (2) & (3) & (4) & (5) & (6) & (7) & (8) & (9) & (10) & (11) \\
\midrule
    0332.3--2742 & $2.30$ & 2.30 &  4.01 & 2 & 17  & 12.38 & 13.84 & 0.035 & $13.90-14.15$ $^{\rm b}$ & 0.43 \\
    0332.6--2740 & $2.40$ & 2.40 &  4.65 & 1 & 7 & 12.08 & 13.38 & 0.050 & $13.78-14.11$ $^{\rm b}$ & 0.53 \\
    0332.4--2741 & $2.45$ & 2.43 &  4.69 & 1 & 20  & 12.52 & 14.14 & 0.024 & --- & 0.53 \\
    0332.6--2740 & $2.50$ & 2.50 &  3.62 & 1 & 10 & 12.19 & 13.49 & 0.050 & --- & 0.51 \\
    0332.5--2753 & $2.55$ & 2.55 &  3.71 & 1 & 17  & 12.34 & 13.74 & 0.040 & --- & 0.56 \\
    0332.5--2751 & $3.70$ & 3.70 &  4.42 & 2 & 14  & 12.50 & 14.10 & 0.040 & $14.32-14.66$ $^{\rm f}$ & 0.53 \\
    0332.3--2742 & $3.65 $ & 3.60 &  3.51 & 2 & 15  & 11.91 & 13.21 & 0.050 & --- &  0.44 \\
    0332.3--2747 & $4.30$ & --- &  4.10 & 1 & 23 & 12.52 & 14.15 & 0.023 & --- & 0.58 \\
    0332.4--2741 & $4.35$ & --- &  4.62 & 0 & 14 & 11.88 & 13.18 & 0.050 & --- & 0.49 \\
\bottomrule
\end{tabular}
\tabnote{
\textsc{Notes:} (1) MSG name based on arcmin-precision coordinates; (2) redshift of the MSG, based on
spectroscopic data; (3) redshift of MSG, based on the spectroscopic members, typical spectroscopic redshift accuracy is better than 0.01; (4) overdenisty factor; (5) number of spectroscopically confirmed member galaxies within $r_p\times1.5$ Mpc radius circle with $\Delta v < 2000 \, $km s$^{-1}$ from the surface number density peak of MSGs;
 (6) number of galaxies with photometric redshift within $r_p\times1.5$ Mpc radius circle from the surface number density peak of MSGs; (7) sum of stellar masses $\times 1/f_{\rm inter} \times f_{\rm MF}$ of each MSG; (8) mass of MSG; (9) ratio of stellar mass and total mass of MSG; (10) mass of MSGs from literature; (11) projected radius of MSG. \\
\textsc{References:}
a: Daddi et al. (2009);
b: Salimbeni et al. (2009);
c: D{\'{\i}}az-S{\'a}nchez et al. (2007);
d: Kurk et al. (2009);
e: Castellano et al. (2007);
f: Kang \& Im (2009)}
\label{to2}
\end{table*}

\section{Conclusions}

  Using the multi-wavelength data,
   we identified 59 MSGs from the combined areas of the GOODS-South and the GOODS-North fields
 with significances of $3.5-8\sigma$ from $z=0.6$ to $z \sim 4.5$.
  Among them, $\sim20$\% of MSGs show plausible associations with AGN/radio sources.

 In comparison with a simulation data set, we find a discrepancy
 between the observed number densities of MSGs and those from the simulation at $z > 1$
 ($M > 7 \times 10^{13}\,M_{\odot}$).
 The discrepancy becomes more significant at higher redshifts ($z > 2$) by a factor of $\sim 5$ or more.
   Even after considering possible systematic effects,
  our result implies that there are too many
 massive structures at $z > 2$ compared to the $\Lambda$CDM prediction.
  By tweaking the conditions for the initial density fluctuation or baryonic physics in galaxy formation, one may be
 able to explain the result within the $\Lambda$CDM cosmology framework, but as of now
 the overabundance of MSGs at $z > 2$ stands as a challenge to the models based on
 the $\Lambda$CDM cosmology.

\acknowledgments

 This work was supported by the Creative Initiative program, No. 2008-0060544, of the National Research Foundation of Korea (NRFK) funded by the Korean government (MSIP). We thank Ranga-Ram Chary for providing us with the IRAC catalog, and Lihwai Lin for the WIRCAM images of
 the GOODS-N field that were useful during the initial stage of the analysis.
  We thank Giulia Rodighiero, Ho Seong Hwang, Jae-Woo Kim, Juhan Kim, Marios Karouzos,
 and Rafael Gobat for useful communication.

\end{document}